\documentclass[useams,onecolumn,usenatbib,twoside]{Thesis}
\usepackage[T1]{fontenc}
\usepackage{fancyhdr}
\usepackage{titlesec}
\usepackage{graphicx}
\usepackage{longtable}
\usepackage{lettrine}
\usepackage{natbib}
\usepackage{amsmath}
\usepackage{setspace}
\usepackage[utf8,latin1]{inputenc}
\usepackage{setspace}
\usepackage{makeidx}
\usepackage{pdflscape,array,booktabs}
\usepackage{rotating}
\graphicspath{{./images/}}
\usepackage[table,xcdraw,svgnames]{xcolor}
\usepackage{amssymb}
\usepackage{dcolumn}
\usepackage{bm}
\usepackage{natbib}
\usepackage{graphicx}
\usepackage{subfigure}
\usepackage{diagbox}
\usepackage{hyperref}
\usepackage[hyphenbreaks]{breakurl}
\usepackage{rotating}
\usepackage{tikz}
\usepackage{setspace}
\usepackage{array} 
\usepackage{adjustbox}
\usepackage{siunitx}
\usepackage{comment}
\usepackage[final]{pdfpages}
\hypersetup{final}
\hypersetup{citecolor=CornflowerBlue}
\hypersetup{linkcolor=CornflowerBlue}
\hypersetup{urlcolor=CornflowerBlue}

\newcommand\aap{A\&A}                
\newcommand\aaps{A\&AS}              
\newcommand\aj{AJ}                   
\newcommand\ao{Appl. Opt.}           
\newcommand\apj{ApJ}                 
\newcommand\apjl{ApJ}                
\newcommand\apjs{ApJS}               
\newcommand\apss{Ap\&SS}             
\newcommand\araa{ARA\&A}             
\newcommand\mnras{MNRAS}             
\newcommand\nat{Nature}              
\newcommand\pasa{Publ. Astron. Soc. Australia}  
\newcommand\pasp{PASP}               
\newcommand\procspie{Proc.~SPIE}     

\newcommand{\HRule}{\rule{\linewidth}{0.5mm}}

\makeatletter
\def\thickhrulefill{\leavevmode \leaders \hrule height 1ex \hfill \kern \z@}
\def\@makechapterhead#1{%
	\vspace*{10\p@}%
	{\parindent \z@ \centering \reset@font
		\thickhrulefill\quad
		\scshape \@chapapp{} \thechapter
		\quad \thickhrulefill
		\par\nobreak
		\vspace*{10\p@}%
		\interlinepenalty\@M
		\hrule
		\vspace*{10\p@}%
		\Huge \bfseries #1\par\nobreak
		\par
		\vspace*{10\p@}%
		\hrule
		\vskip 100\p@
}}
\def\@makeschapterhead#1{%
	\vspace*{10\p@}%
	{\parindent \z@ \centering \reset@font
		\thickhrulefill
		\par\nobreak
		\vspace*{10\p@}%
		\interlinepenalty\@M
		\hrule
		\vspace*{10\p@}%
		\Huge \bfseries #1\par\nobreak
		\par
		\vspace*{10\p@}%
		\hrule
		\vskip 100\p@
}}
\makeindex{}
\begin{document}
\pagenumbering{roman}
\DeclareGraphicsExtensions{.pdf,.png,.gif,.jpg}
\begin{titlepage}
 	\begin{center}

  	\textsc {\LARGE The University of Western Australia}\\[0.5cm]
  	\textsc {\LARGE  \& }\\[0.5cm]
	\textsc {\LARGE International Centre for Radio}\\[0.2cm]
	\textsc {\LARGE Astronomy Research}\\[1.5cm]
 	\textsc{School of Physics, Mathematics and Computing}\\[0.5cm]
	\LARGE PhD Thesis\\[0.5cm]

	\HRule \\[0.4cm]
	{ \huge \bfseries
    The Emergence of Bulges and Disks in the Universe}\\[0.4cm]
	\HRule \\[0.5cm]

\emph{Author}\\
\Large
\textsc{Abdolhosein Hashemizadeh\vspace{0.4cm}}

{\small This thesis is presented for the degree of Doctor of Philosophy of The University of Western Australia}

\vspace{0.2cm}

 \begin{minipage}{\textwidth}
  \centering
  \includegraphics[width=0.8\textwidth]{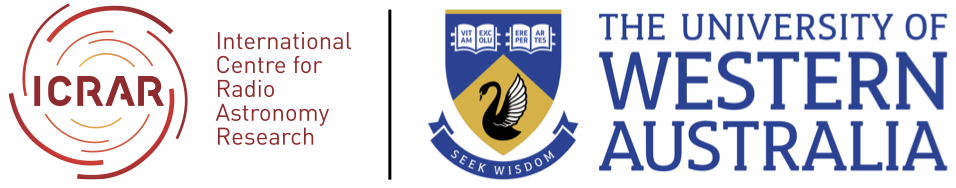}
 \end{minipage} \\[0.5cm]

	{\large October 2020}

	\end{center}
\end{titlepage}


\clearpage
%
%
%
%


\cleardoublepage
\begin{minipage}{13cm}
\addcontentsline{toc}{chapter}{Abstract}
\begin{center}
\vspace{1in}
{\Huge \bfseries Abstract}
\end{center}
{\small
This thesis makes use of the imaging data from the Advanced Camera for Surveys (ACS) of the Hubble Space Telescope (HST) in the Cosmic Evolution Survey (COSMOS) and the Deep Extragalactic VIsible Legacy Survey (DEVILS) field. We provide visual morphological classifications of 44,000 galaxies out to redshift $z = 1$ and above a stellar mass of $10^{9.5} M_\odot$ (D10/ACS sample). We perform a robust Bayesian bulge-disk decomposition analysis of the D10/ACS sample. This study forms one of the largest morphological classification and structural analyses catalogues in this field to date. 
Using these catalogues, we explore the evolution of the stellar mass function (SMF) and the stellar mass density (SMD) together with the stellar mass-size relations ($M_*-R_e$) of galaxies as a function of morphological type as well as for disks and bulges, separately. We quantify that one-third of the current stellar mass of the Universe was formed during the last 8 Gyr. We find that the moderate growth of the high-mass end of the SMF is dominated by the growth of elliptical systems and that the vast majority of the stellar mass of the Universe is locked up in disk+bulge systems at all epochs and that they increase their contribution to the total SMD with time. The contribution of the pure-disk morphology gradually decreases with time ($\sim40\%$), while ellipticals increase their contribution by a factor of $1.7$ since $z = 1$. By decomposing galaxies into disks and bulges we quantify that on average $\sim50\%$ of the total stellar mass of the Universe at all epochs is in disk structures with this contribution relatively unchanged since $z \sim 0.6$. With this comes more rapid growth of pseudo-bulges and spheroids (bulges and ellipticals) in mass. Furthermore, while the cosmic star-formation history is declining the Universe is transitioning from a disk dominated era to an epoch when pseudo- and classical-bulges are emerging. 
Finally, the $M_*-R_e$ relations of galaxies and their components show that different components follow distinct relations which may reflect their distinct evolutionary pathways through which these systems have been built. The evolution of the $M_*-R_e$ relations of galaxies and their components shows minimal variation. The unchanged scaling relations since $z = 1$ is consistent with the evolution along the $M_*-R_e$ trends.



{\em Words in text: 38,980

Words outside text: 5,194 

Number of floats/tables/figures: 85
}

}
\end{minipage}
\clearpage


\thispagestyle{empty}
\begin{minipage}{13cm}
\vspace{3in}
\begin{center}
{\Large\hfill``\textit{Galaxy formation is much less ``finished'' than we like to think! Our job is far from finished, too. For all of us students of galaxies, this is good news.}''\hfill }\\
{\large\hfill - J. Kormendy}\\
\end{center}
\end{minipage}
\clearpage
\thispagestyle{empty}


\clearpage
\begin{minipage}{13cm}
\addcontentsline{toc}{chapter}{Acknowledgements}
\begin{center}
\vspace{1in}
{\Huge \bfseries Acknowledgements}
\end{center}
{\small
PhD was not only a degree for me. It was also a chance to challenge myself to learn to think big and out of the box. PhD in Astronomy taught me that although with extraordinary abilities, we human beings are extremely small with a life period not even comparable to the age of the Universe. I learned that we are a dust grain in this Universe. A creature as small as grain cannot be the king, cannot be better or worse because of their colour, language or nationality.

In this pathway, there were many people who not only assisted me in achieving my goals but also accompanied me throughout the way. First and foremost, I would like to thank my supervisors: Simon, Aaron and Luke for all their expertise and patience. Their guidance helped me all the time throughout my research and led me to finish this thesis.
I also gratefully thank ICRAR and UWA. This research was supported by an Australian Government Research Training Program (RTP) Scholarship.      

My colleagues and friends at ICRAR brought joy to our academic life. I would like to sincerely appreciate my fellow PhD and Master's students and staffs. My appreciations also extend to the ICRAR executive team who sustained a positive environment to do science. I especially thank Lister and Renu for all their advice and great supports.  

I am indebted to my family, whose injected positive energy into my life grows as I do. My champion, and my love, Ellie, never let me stand unless she holds my hands. Her love keeps me continually going forward. I could not imagine how I would go through this alone. Thanks for your countless sacrifices to help me get to this point. I would like to acknowledge with gratitude, the spiritual support from my family: dad and mum, my sisters and my brother as well as my wife's family. Thanks for allowing me to explore the world and accepting to see and hear me only from a little screen so I can finish my PhD. Not all siblings, though, have blood relation with you; I would like to sincerely thank Maryam and Sambit (haj khanoom and haj agha!), my recent lovely sister and brother, who gave me enormous energy and for being around whenever I needed. 

My appreciation also goes to all my friends and volunteers in DASTAN Foundation, a charity organization that Ellie and I proudly founded two years ago. Our activities in the foundation revealed a deeper meaning of life to me and helped me to see far beyond my daily life.
}
\end{minipage}


\clearpage
\begin{minipage}{13cm}
\addcontentsline{toc}{chapter}{Author's Declaration}
\begin{center}
\vspace{1in}
{\Huge \bfseries Author's Declaration}
\end{center}

I, Abdolhosein Hashemizadeh, certify that:

This thesis has been substantially accomplished during enrolment in this degree.
This thesis does not contain material which has been submitted for the award of any other degree or diploma in my name, in any university or other tertiary institution.
In the future, no part of this thesis will be used in a submission in my name, for any other degree or diploma in any university or other tertiary institution without the prior approval of The University of Western Australia and where applicable, any partner institution responsible for the joint-award of this degree.
This thesis does not contain any material previously published or written by another person, except where due reference has been made in the text and, where relevant, in the Authorship Declaration that follows. 
This thesis does not violate or infringe any copyright, trademark, patent, or other rights whatsoever of any person.

This thesis contains works prepared for publication, some of which has been co-authored. 

\vspace{3cm}
Signature: 

\vspace{0.5cm}
Date: 14/04/2021

\end{minipage}


\clearpage
\addcontentsline{toc}{chapter}{AUTHORSHIP DECLARATION} 
\begin{center}
\vspace{1in}
{\Huge \bfseries AUTHORSHIP DECLARATION}  
\end{center}

This thesis contains works that have been submitted for publication in peer-reviewed journals and prepared for publication. 

\textbf{Deep Extragalactic VIsible Legacy Survey (DEVILS): Stellar Mass Growth by Morphological Type since $z = 1$.} \\
Abdolhosein Hashemizadeh, Simon P. Driver, Luke J.~M.~Davies, Aaron S. G. Robotham, Sabine Bellstedt, Rogier A. Windhorst, Malcolm Bremer, Steven Phillipps, Matt Jarvis, Benne W. Holwerda, Claudia del P. Lagos, Soheil Koushan, Malgorzata Siudek, Natasha Maddox, Jessica E. Thorne, Pascal Elahi

Published in the Monthly Notices of the Royal Astronomical Society (MNRAS)

\textbf{Section(s)}: Chapter 3 \\
\textbf{Contribution}: 95\%

\textbf{Deep Extragalactic VIsible Legacy Survey (DEVILS): The emergence of bulges and decline of disk growth since $z = 1$.} \\
Abdolhosein Hashemizadeh, Simon P. Driver, Luke J.~M.~Davies, Aaron S. G. Robotham, Sabine Bellstedt, Rogier A. Windhorst, Matt Jarvis, Benne W. Holwerda, Malgorzata Siudek, Caroline Foster, Steven Phillipps, Jessica E. Thorne, Christian Wolf

Submitted for publication in the Monthly Notices of the Royal Astronomical Society (MNRAS)

\textbf{Section(s)}: Chapter 4 \\
\textbf{Contribution}: 95\%

\textbf{Deep Extragalactic VIsible Legacy Survey (DEVILS): The evolution of the mass-size relation of bulges and disks since $z = 1$.} \\
Abdolhosein Hashemizadeh, Simon P. Driver, Luke J.~M.~Davies, Aaron S. G. Robotham, and DEVILS team

Prepared for publication in the Monthly Notices of the Royal Astronomical Society (MNRAS)

\textbf{Section(s)}: Chapter 5 \\
\textbf{Contribution}: 95\%

Student signature: \\

Date: 14/04/2021

I, Simon P. Driver certify that the student's statements regarding their contribution to each of the works listed above are correct. 

As all co-authors' signatures could not be obtained, I hereby authorise inclusion of the co-authored work in the thesis.

Coordinating supervisor signature: \\

Date: 14/04/2021



\clearpage
\tableofcontents


\clearpage
\listoffigures


\clearpage
\listoftables


\clearpage
\begin{minipage}{13cm}
\begin{center}
\addcontentsline{toc}{chapter}{Preface}
{\Huge \bfseries Preface}
\end{center}

Except where otherwise acknowledged, the work presented in this thesis is my own,
and no part of it has been submitted for a degree at this or any other university.

This thesis is constructed as a series of papers in compliance with the rules for
PhD thesis submission from the Graduate Research School at the University of
Western Australia. The publications arising from or related to it are as follows:

\begin{itemize}
\item Hashemizadeh et al., 2021, MNRAS (published)
\item Hashemizadeh et al., 2022a, MNRAS (submitted)
\item Hashemizadeh et al., 2022b, MNRAS (advance stages of prep.)
\end{itemize}

I have completed this work as part of DEVILS and GAMA surveys.
DEVILS is an Australian project based around a spectroscopic campaign using the Anglo-Australian Telescope. The DEVILS input catalogue is generated from data taken as part of the ESO VISTA-VIDEO and UltraVISTA surveys. DEVILS is part funded via Discovery Programs by the Australian Research Council and the participating institutions. The DEVILS website is \href{https://devilsurvey.org}{devilsurvey.org}. The DEVILS data is hosted and provided by AAO Data Central (\href{datacentral.aao.gov.au}{datacentral.aao.gov.au}). This work was supported by resources provided by The Pawsey Supercomputing Centre with funding from the Australian Government and the Government of Western Australia. This work is also part of the contribution of the entire COSMOS collaboration consisting of more than 200 scientists. The HST COSMOS Treasury program was supported through NASA grant HST-GO-09822. GAMA is a joint European-Australasian project
based around a spectroscopic campaign using the Anglo- Australian Telescope. The GAMA input catalogue is based on data taken from the Sloan Digital Sky Survey and the UKIRT Infrared Deep Sky Survey. Complementary imaging of the GAMA regions is being obtained by a number of independent survey programmes including GALEX MIS, VST KiDS, VISTA VIKING, WISE, Herschel-ATLAS, GMRT and ASKAP providing UV to radio coverage. GAMA is funded by the STFC (UK), the ARC (Australia), the AAO, and the participating institutions. The GAMA website is \href{http://www.gama-survey.org/}{gama-survey.org}.

\end{minipage}

\cleardoublepage
\pagenumbering{arabic}


\graphicspath{{images/ChapterOne/}}

\chapter{Introduction}

\section{Historical Background}
\label{sec:history}
Humanity started looking at the little shiny dots glued to the Earth's ceiling since the very origins of its history and recorded in the earlier cave paintings. The first known efforts to catalogue stars and constellations with cuneiform texts and artifacts, dates back roughly 6000 years. Our modern picture of the Universe, defined mostly by Greek and Roman mythologies was revolutionized with the invention of the telescope by Galileo in 1609 and his early observations of the Venus phases and Jupiter's moons \citep{Galilei10}. Only two hundred years later in the mid 18 century, French astronomer, Charles Messier published a catalogue of 110 visually diffuse celestial objects, known as the Messier objects \citep{Messier81}. Although the Messier catalogue was the largest catalogue of diffuse objects to date, he was not the first who recorded nebulae with Persian astronomer Abd al-Rahman al-Sufi, also known as Azophi, observing and recording the Andromeda and Large Magellanic clouds in 964 AD. He published his discoveries together with details of 48 constellations in his book, Kitab al-Kawatib al-Thabit al-Musawwar (also commonly known as the Book of Fixed Stars) \citep{Hafez11}.    

Since 1610 when Galileo opened the door to the sky by turning his little refractor telescope upward, telescopes have significantly grown in size expanding our knowledge of the Universe and the evolution of the galaxy population. Over time, thousands of questions arose about the nature of these smudges by observing the heavens using ever more powerful telescopes. Perhaps the most fundamental question at the time was, are these nebulae located within the Milky Way or are they ``island universes'' \citep{Curtis17}? The argument was settled in 1920s by the primary works of American astronomer Edwin Hubble who measured the distance to the Andromeda nebula (M31) using the flux periodicity of Cepheid stars (variable stars whose brightness varies periodically). Hubble continued measuring the distance to galaxies using Cepheids and in his seminal paper \cite{Hubble29} presented a plot showing the distance of galaxies against their line of sight recession velocity. Indicating that most galaxies are receding from us, this historic plot changed the face of our Universe forever and formed our modern cosmology. The plot showed that the velocity of galaxies correlates with their distance such that distant galaxies move away faster than nearby galaxies. We know this as Hubble's Law.

Galaxies then became accepted as the fundamental building blocks of our Universe in which most astrophysics occurs, stars are born, super massive black holes form and the elements of the periodic table are forged. If there were no galaxies there would be no life.    

In the last century, with the advent of large telescopes and advanced ground- and space-based facilities our samples of galaxies has increasingly grown, facilitating the statistical studies of galaxies. One of the primary ways in which the galaxy population is represented is via the distribution of luminosity and stellar mass, so called the luminosity, and stellar mass function (LF and SMF) representing the number density of galaxies within luminosity or stellar mass bins (e.g., \citealt{Schmidt68}; \citealt{Lynden-Bell71}; \citealt{Schechter76}; \citealt{Sandage79}; \citealt{Efstathiou88}; \citealt{Zwaan03}; \citealt{Cole11}; \citealt{Loveday15}; \citealt{Weigel16}; \citealt{Moffett16a}; \citealt{Obreschkow18}). Throughout this thesis we will explore the evolution of the SMF of a large sample of galaxies since high redshift down to the local Universe.

\section{Galaxies}
\label{sec:galaxies}
Galaxies are gravitationally bound systems containing thousands to a few trillion stars. Our host Milky Way contains $\sim250-500$ billion stars within a $\sim15$~kpc radius. Studying galaxy formation and evolution is not easy as the time-scales at which they form and evolve are much longer than the whole of human history. Therefore, it is next to impossible to directly track the evolution of individual galaxies. Thanks to the finite speed of light, however, we can observe and study the young Universe by looking at distant galaxies essentially building up a core sample through the history of the Universe.
 
\begin{figure*} 
	\centering
	\includegraphics[width=\textwidth]{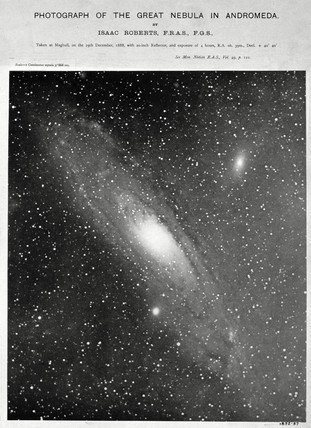}
	\caption{The first image of the Andromeda galaxy (M31) taken by Isaac Robert on 29 December 1888 with 4 hours exposure using his 20 inch reflector telescope. This image is considered a pioneer as, for the first time, showed the spiral features in this nearby galaxy. Credit: Science Museum/Science \& Society Picture Library; Image Ref. 10415880}
	\label{fig:M31_isaac}
\end{figure*}

Since the mid-19th century when the first images of galaxies were taken (Figure \ref{fig:M31_isaac}) astronomers recognized a great visual diversity in these structures. This diversity was then echoed in almost all physical properties that they measured.


The first notable property of galaxies is their visual appearance, i.e. their morphology. Galaxies are primarily seen as having two dominated forms: disks and ellipticals (e.g., \citealt{Hubble36}). Disk galaxies have flattened structures where stars are predominantly rotating and are also known as spiral galaxies, since the gas and stars in disks form patterns such as spiral structures (e.g., \citealt{Kormendy04}). Elliptical galaxies, on the other hand, represent systems with smooth light distribution dominated by random motions, with no bulk rotation (e.g., \citealt{Illingworth77}; \citealt{Binney78}). More recently though, we have found rotating elliptical systems thanks to the kinematic investigations using integral field spectroscopic (IFS) surveys such as ATLAS$^{3D}$ (see e.g., \citealt{Emsellem07}; \citealt{Bois11}; \citealt{Cappellari11}; \citealt{Foster13}; \citealt{Weijmans14}; \citealt{Li18} ).  

Galaxy morphologies are of course more complicated than the above distinct and simplistic dichotomy. However, most galaxies adhere to being a combination of a disk-like and elliptical-like structure, combined to varying degrees (e.g., \citealt{Kormendy93}; \citealt{Andredakis94}; \citealt{Andredakis95}). Lying at the centre of some disk galaxies, is an ellipsoidal component called the bulge and the bulge itself is reported as having two types of structures, known as pseudo- and classical-bulges (see e.g., \citealt{Kormendy04}; \citealt{Fisher06}; \citealt{Fisher08}; \citealt{Mendez-Abreu10}; \citealt{Krajnovic13}; \citealt{Zhu18}; \citealt{Schulze18}; \citealt{Gao20}). This distinction will be described in much greater detail later. 

Obvious morphological differences of galaxies soon raised an important question as to how these systems became very distinct in shape. The remarkable amount of effort that astronomers put to answer this question has formed a significant portion of the modern galaxy formation and evolution field on both observational and theoretical sides. The origin of this distinction is now better understood with studying high redshift galaxies suggesting that galaxies of different morphologies with varying prominency of structural components have likely gone through different evolutionary pathways. 
The first step in studying galaxies of different morphologies is, however, to propose a recipe for classifying these systems.      

\subsection{The Hubble Sequence} \label{subsec:HubbleFork}

\begin{figure*}
\centering
\begin{subfigure}
  \centering
  \includegraphics[width=0.98\textwidth, height = 10cm]{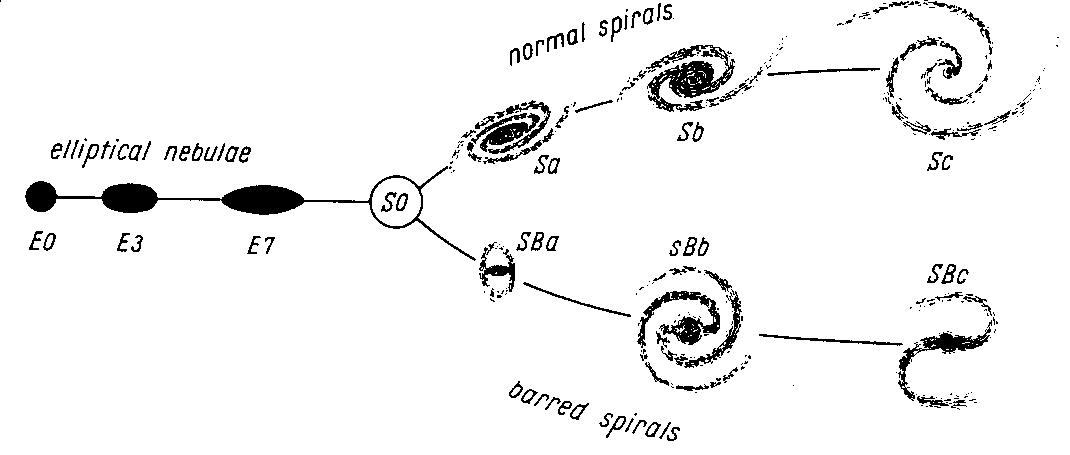}
\end{subfigure}

\begin{subfigure}
  \centering
  \includegraphics[width=0.98\textwidth]{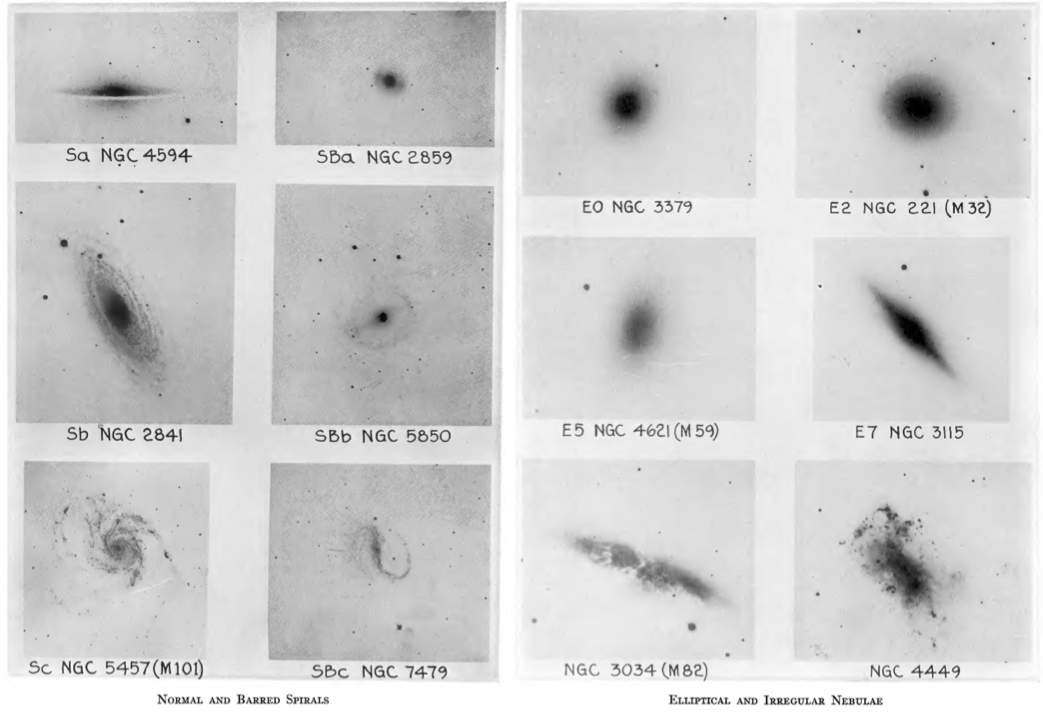}
\end{subfigure}
\caption{The original Hubble sequence as he noted in his book: Realm of the Nebulae (top: \citealt{Hubble36}), and his classification of galaxies (nebulae at the time) with his photography using 100-inch Mt. Wilson reflector telescope (bottom: \citealt{Hubble26})
}
\label{fig:Hubble_seq}
\end{figure*}

Edwin Hubble \citep{Hubble26} constructed the first classification scheme of galaxies according to their basic and most prominent visual features, i.e. disk, bulge and bar. Also known as Hubble tuning fork, this scheme ranges from early-type ellipticals to late-type spirals (Figure \ref{fig:Hubble_seq}). Along the handle of the tuning fork, ellipticals are classified from E0-E7 with the number indicating their ellipticity. According to the Hubble tuning fork, spirals are divided into two main classes based on the presence or absence of a bar (S vs SB). From left to right on both strands (Sa-c and SBa-c) galaxies are then separated into different classes based on the strength of the bulge, and how tightly/loosely wound the spiral arms are. However, an important caveat has been the morphological evolution of galaxies since high redshift and the significant increase of peculiar/irregular galaxies (Irr) at high-$z$ fail to fit into the Hubble sequence. See an earlier review by \cite{Abraham99}.   

The Hubble sequence was later extended by G\'erard de Vaucouleurs (\citealt{deVaucouleurs59}; \citealt{deVaucouleurs63}; \citealt{deVaucouleurs74}) and Allan Sandage (\citealt{Sandage61}; \citealt{Sandage75}). De Vaucouleurs' observations with the 30-inch Reynolds telescope - Mount Stromlo Observatory today- led to the largest Atlas of southern galaxies of the date (\citealt{deVaucouleurs56}). De Vaucouleurs argued that the Hubble sequence with only bar and spiral identifiers was not adequate to picture all observed galaxies. In fact, he expanded the Hubble's basic sequence by adding rings and lenses as two additional structures.

De Vaucouleurs, in addition, assigned numerical values to identify morphologies. This system is known as morphological T or T-type. T value extends from -6 to +10 with negative values identifying early-type galaxies (ellipticals and lenticulars) and positive values late type galaxies (spirals and irregulars).

In addition to above, other morphological classification systems exist which make use of additional parameters and provide other information (for example: Yerkes: \citealt{Morgan58, Morgan59}; DDO: \citealt{vandenBergh60}; RDDO: \citealt{vandenBergh76}).

The Hubble Sequence was originally motivated by observations of local galaxies; however, current high-resolution and high-sensitivity facilities such as HST enable us to observe substructures of high redshift galaxies with a far better spatial resolution (e.g., \citealt{Elmegreen05}; \citealt{Gargiulo11}; \citealt{Guo15}; \citealt{Shibuya16}; \citealt{Guo18}). In general, galaxies at $z > 1$ do not adhere to the Hubble schema, due to the increasing fraction of giant kpc-scale clumps in star-forming galaxies, and which are less frequent in massive low-$z$ galaxies (e.g, \citealt{Conselice04}; \citealt{Elmegreen08} \citealt{Bournaud08}; \citealt{Guo18}). For example, \cite{Guo15} showed that the fraction of low-mass star-forming galaxies with at least one off-centre clump is $60\%$ at $z \sim 3-0.5$ while this fraction varies $55-40\%$ and $55-15\%$ for intermediate- and high-mass galaxies throughout this redshift range.   
Therefore, deep and high-resolution rest-frame UV (e.g, \citealt{Conselice04}; \citealt{Guo12, Guo15, Guo18}) and rest-frame optical (e.g., \citealt{Elmegreen09}; \citealt{Schreiber11}) studies of these high-$z$ galaxies suggest that the Hubble Sequence likely breaks down at $z > 1$ (also see \citealt{Hashemizadeh21}), or at least can be considered incomplete. 

\section{Morphological Classification of Large Samples} \label{sec:morph_class}

In addition to above systems (T-type etc.), morphological classifications have been done in several ways, including visual inspection through which volunteer amateurs or expert astronomers classify objects according to their apparent features. The largest crowd-sourced program for morphological classifications is Galaxy Zoo (\citealt{Lintott08}) that originally provided the classification for one million galaxies using images of the Sloan Digital Sky Survey (SDSS) and later was expanded into HST images (Galaxy Zoo Hubble program, GZH, \citealt{Willett17}) and radio data (Radio Galaxy Zoo, \citealt{Banfield15}; \citealt{Willett16}).  

Semi-automatic techniques have been developed to do this intensive task. For example, \cite{Abraham94} proposed the central concentration of light as a parameter for automatic morphological classification of faint and high redshift galaxies.
Later, \cite{Abraham03} introduced another classification scheme based on the Gini coefficient to quantify galaxy morphological types and find that this parameter well correlates with central concentration, and mean surface brightness.

More recently, some studies used a combination of three parameters of the Concentration index, Asymmetry, and Clumpiness (CAS parameters) to identify the morphological classes (see e.g., \citealt{Scarlata07a}). However, this techniques has not achieved a high accuracy (e.g., \citealt{Huertas-Company14}; \citealt{Mager17}).
More recently, in efforts to find an automatic approach scalable to large numbers of objects ($> 10^4$) and applicable to future surveys such as the Large Survey of Space and Time (LSST) some works have used machine learning and convolutional neural networks to provide automated classification (e.g., \citealt{Dieleman15}; \citealt{Aniyan17}; \citealt{Ghosh19}; \citealt{Tadaki20}; \citeauthor{Cheng20}; \citealt{Walmsley20}). However, these networks are not applicable to all data sets as they are each trained on different data sets with different qualities with varying degrees of success. 

In this project, in addition to using GZH and artificial neural networks, we will perform our own visual inspection of 44,000 galaxies using HST imaging data. See Section \ref{sec:morph_class} for more details.

\begin{figure*} 
	\centering
	\includegraphics[width=\textwidth]{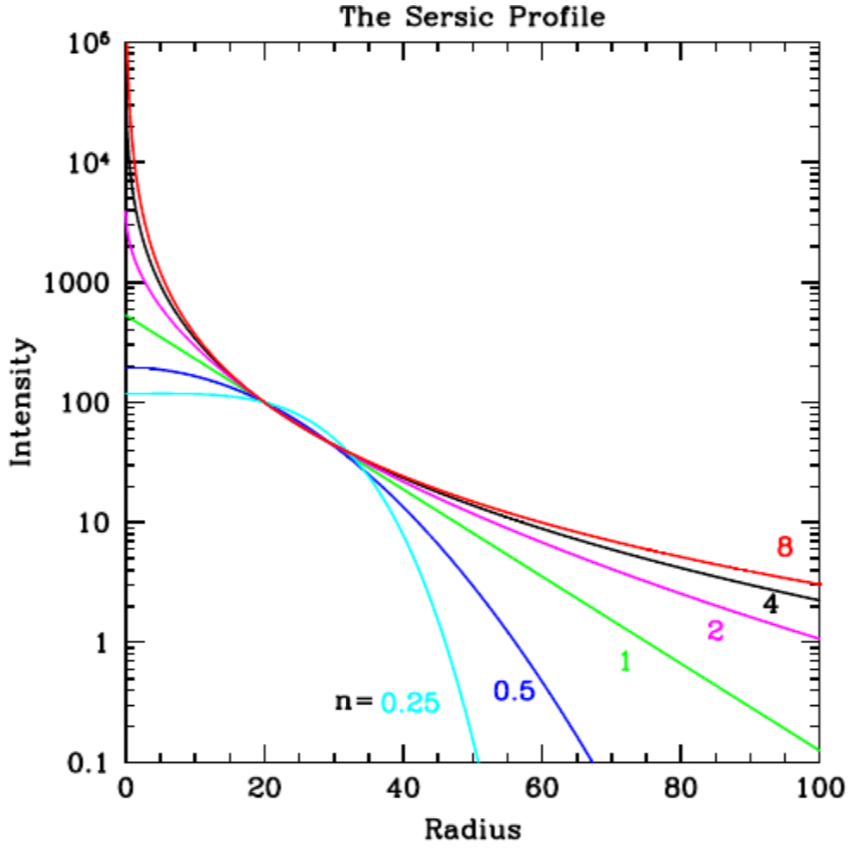}
	\caption{The S\'ersic profiles for different S\'ersic indices ($n = 0.25-8$, \citealt{PengC10}).}
	\label{fig:SersicPro}
\end{figure*}

\section{Galaxy Light Profiles} \label{sec:profiles}

All the above classification systems revealed that the variation of morphological types correlates with the light distribution in galaxies. In other words, the variation of light intensity as a function of radius behaves differently for galaxies of different morphological types making galaxy light profiling a crucial area of exploration along with morphological investigations.  

Following de Vaucouleurs' observations, Jos\'e-Luis S\'ersic who worked at 1.54-m telescope at the Astrophysical Station in Argentina, published ``Galaxies Australes'', his southern hemisphere galaxy Atlas (\citealt{Sersic68}). Both de Vaucouleurs and S\'ersic were looking beyond simple visual morphological classifications and highlighted the wealth of information that one can extract from the light distribution of galaxies. With a quantitative analysis S\'ersic fitted all galaxies in his Atlas with a $R^{1/n}$ model generalizing the de Vaucouleurs' $R^{1/4}$ model for elliptical galaxies \citep{Sersic63}. 

S\'ersic light profile as described in \cite{Sersic63} (see also \citealt{Graham05}) finally provided a quantitative analytic formula to describe the light intensity variation as a function of radius:

\begin{equation}
I(r)=I_e \exp\bigg[-b_n \bigg(\left(\frac{r}{r_e}\right)^{1/n} - 1\bigg)\bigg]
\end{equation}

\noindent where $r_e$ is the effective radius (the radius containing half of the total flux), and $I_e$ is the intensity at that radius. $n$ is the S\'ersic index that specifies the shape of the profile. As special cases, $n=0.5$, $n=1$ and $n=4$ represent Gaussian, pure exponential and de Vaucouleurs profiles, respectively. In general, it has been shown that stable disks are likely to follow an exponential profile (e.g., \citealt{Patterson40}; \citealt{deVaucouleurs59}; \citealt{Freeman70}; \citealt{Kormendy77}), as opposed to classical-bulges and spheroids which tend to follow a de Vaucouleurs profile - at least for the more luminous bulges. Figure \ref{fig:SersicPro} shows the S\'ersic profiles with various S\'ersic indices from $n = 0.25$ to $n = 8$.

Nowadays, the S\'ersic profile is routinely used in either single or double form. In case of a single component galaxy (pure disk or elliptical), or for low-resolution data where resolving bulge component is impossible, galaxies are fitted with a single S\'ersic profile (e.g., \citealt{Blanton03}; \citealt{Driver06}; \citealt{vanderWel08}; \citealt{Hoyos11}; \citealt{vanderWel12}; \citealt{Carollo13}; \citealt{vanderWel14}; \citealt{Straatman15}; \citealt{Tarsitano18}). Well-resolved disk and lenticular galaxies are often fitted with a combination of two S\'ersic profiles, or as an exponential disk plus a free-$n$ S\'ersic bulge (e.g., \citealt{Andredakis95}; \citealt{Khosroshahi00}; \citealt{Graham01}; \citealt{Prieto01}; \citealt{Simard11}; \citealt{Kelvin12}; \citealt{Mendel14}; \citealt{Meert15}; \citealt{Lange16}; \citealt{Kennedy16}; \citealt{Dimauro18}; \citealt{Cook19}; \citealt{dosReis20}).    

\section{Galaxy Profile Fitting (Bulge-Disk Decompositions)} 
\label{sec:galaxy_profile}

Galaxy fitting techniques have been broadly used to investigate the galaxy population and to infer the formation and evolutionary pathways for distinct components (e.g., \citealt{Kormendy04} and \citealt{Driver13}).
In general, one can model a galaxy via two main techniques: kinematic dynamical modeling and light profile fitting. Dynamical modeling uses internal kinematics and the full six dimensional phase space which can be reconstructed and utilized to decompose different components. \cite{Zhu18} first used stellar kinematics of galaxies in the CALIFA survey (\citealt{Sanchez12}) to reconstruct stellar orbits. They then further extracted kinematically cold, warm, hot, and counter-rotating components (\citealt{Zhu20}). However, kinematic decomposition have so far only been applied to relatively small samples of galaxies (e.g. \citealt{Johnston17}; \citealt{Zhu18}; \citealt{Tabor19}; \citealt{Zhu20} and \citealt{Oh20}) although this is rapidly changing changing with SAMI (Sydney-AAO Multi-object Integral-field spectroscopy Galaxy Survey; \citealt{Croom12}), MANGA (Mapping Nearby Galaxies at Apache Point Observatory; \citealt{Bundy15}), and other upcoming Hector surveys (\citealt{Bland-Hawthorn15}).     

On the other hand, light profile fitting relies on the projected light distribution tracing the underlying structure of galaxies. Light profiling is typically done using either 1D azimuthally-averaged profiles or 2D images. After early endeavors of de Vaucouleurs and S\'ersic in one-dimensional single profile fitting \cite{Kormendy77} and \cite{Kent85} brought the idea of one-dimensional bulge+disk decomposition into the literature (i.e., two component fitting).   

The first comprehensive studies of 2D galaxy profiling were started by \cite{Andredakis95}, \cite{Byun95}, \cite{deJong96} and \cite{Wadadekar99}. Their approach, which is very similar to modern techniques, was to fit the image of a galaxy pixel-by-pixel instead of fitting averaged ellipses.   

There are advantages and disadvantages to both 1D and 2D profiling approaches. In 1D profile fitting, the 2D galaxy image must be collapsed to a 1D profile, typically using {\sc ELLIPSE} task of {\sc IRAF}. This makes the approach challenging when the galaxy is not a pattern of smoothly varying isophotes. In addition, taking the effects of the point spread function (PSF) into the consideration is further unclear in 1D profiling, as for example convolving the profile with a circular PSF artificially changes central ellipses to circular profiles (\citealt{Robotham17}). 

1D profiling is more applicable to highly resolved galaxies where complicated isophotes are detectable. 2D profiling codes, however, are more automatic since they do not need a 2D to 1D collapse of data. Therefore, these codes are more popular for large sample of data (e.g., \citealt{Simard02}; \citealt{Allen06}; \citealt{Kelvin12}; \citealt{Lange16}).   

Building upon earlier methods (e.g., \citealt{Wadadekar99}), a number of 2D galaxy profiling codes have been developed over the last two decades and are publicly available, such as: {\sc GIM2D} \citep{Simard02}, {\sc BUDDA} \citep{deSouza04}, {\sc GALMORPH} \citep{Hyde09}, {\sc GALFIT} \citep{Peng10}, {\sc IMFIT} \citep{Erwin15}, and most recently {\sc PROFIT} \citep{Robotham17}. 

There are numerous complexities and uncertainties in the light profiling of galaxies, such as the impact of incorrect background sky subtraction (e.g., \citealt{deJong96}), errors in PSF modelling, and contamination from neighbouring objects (e.g., \citealt{Haussler07}). As shown in Figure \ref{fig:SersicPro}, profiles with high S\'ersic index systems ($n \gtrsim 2$) will contain more flux at larger radii making robust sky background subtraction crucial for fitting galaxies with high-$n$ profiles such as spheroids.
In addition, it has been shown that in the presence of a nearby companion, the aperture used is crucial. For example, one could use a large aperture to capture the maximum flux of the object. However, this method would increase the sky noise and the likelihood of contamination from neighbouring objects (\citealt{Haussler13}). Using profile optimised photometric apertures such as those used for Petrosian magnitude would not completely solve this problem as this method can miss 20-70\% of the flux for de Vaucouleurs profiles (\citealt{Graham05}). Masking and simultaneously modelling the companion galaxy are other popular techniques. \cite{Haussler07} showed that using masking in codes like {\sc GIM2D} \citep{Simard02} can also result in significant systematic errors, particularly for spheroidal systems with companions. Simultaneous fitting of companions implemented in codes such as {\sc GALFIT} \citep{Peng10} is more robust but more CPU costly and with more free parameters to fit.

In this study, we will use {\sc PROFIT} developed at UWA, as it:
\begin{description}

\item[$\bullet$] is open source and publicly available;

\item[$\bullet$] provides an extensive set of built-in models;

\item[$\bullet$] is fast in model generation;

\item[$\bullet$] very well documented;

\item[$\bullet$] has the image generation module separated from the optimization code;

\item[$\bullet$] supports a range of optimizers, including Markov Chain Monte Carlo (MCMC) and robust with full posterior error mapping

\item[$\bullet$] accepts any prior distributions for parameters (important in Bayesian evaluation);

\item[$\bullet$] can fit parameters in linear or log space

\item[$\bullet$] allows user to define different likelihoods; including: Normal, Student-T, Chi-Squared and Poisson;

\item[$\bullet$] can be conveniently parallelized.

\end{description}

We refer to \cite{Robotham17} for more info about {\sc PROFIT}\footnote{{\sc PROFIT} is publicly available at: \href{https://github.com/ICRAR/ProFit}{https://github.com/ICRAR/ProFit}}.

\section{The Revolutionary Hubble Space Telescope} \label{sec:Hubble_launch}

Before the era of space telescopes, our instruments were trapped within the Earth's atmosphere that blocks a part of the electromagnetic spectrum (e.g., the UV) and blurs the imaging data due to atmospheric seeing. The Hubble Space Telescope (HST) was launched and deployed into low Earth orbit in 1990. With its 2.4-m mirror accompanied by four main instruments, HST detects light in UV, visible and NIR parts of the spectrum. Located above the Earth's atmosphere HST is least suffered from the atmospheric seeing and hence able to observe to the Universe with incredible resolution facilitating deep investigations of galaxy structures and their evolution.   
    


\begin{figure*} 
	\centering
	\includegraphics[width=\textwidth]{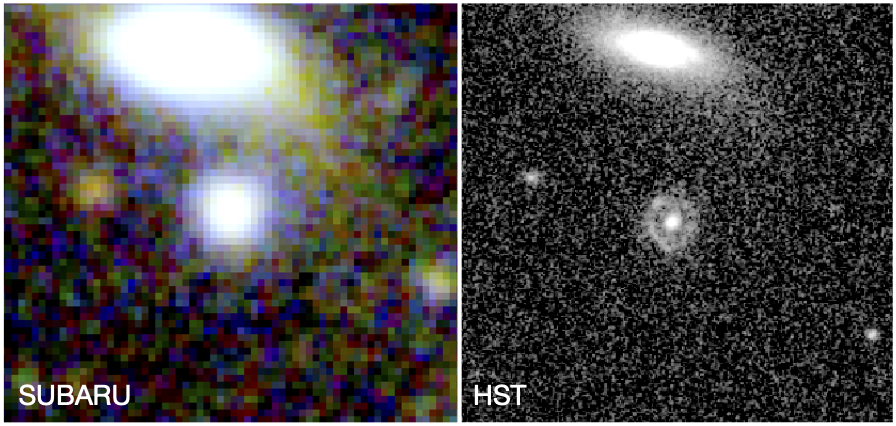}
	\caption{Comparing the resolution of Hubble Space Telescope with ground based telescopes. A distant ring galaxy at redshift 1.14 imaged with the Subaru 8-m telescope on Mauna Kea (left); and with the HST 2.4-m (right). }
	\label{fig:Hubble_vision}
\end{figure*}

The HST's outstanding sub-arcsecond resolution, after more than 1.3 million observations, has revolutionized our insight into galaxy morphology and structure, especially at high-$z$. As an exampke, to highlight the dramatic resolution of HST, we compare the image of a high redshift ($z = 1.4$) ring galaxy taken by HST with the one by Subaru 8-m ground-based telescope in Figure \ref{fig:Hubble_vision}. It can be seen how extraordinarily HST resolves the structures of distant galaxies.

Here, we briefly point to a few of the important discoveries using HST. For example, astronomers discovered that the number density of peculiar galaxies grows with increasing redshift (\citealt{Driver95}; \citealt{Glazebrook95}; \citealt{Brinchmann98}). HST observations also confirmed the Butcher-Oemler effect (the increase of the fraction of blue galaxies at intermediate-z) and further elaborated that it is associated with a growth of the fraction of spiral galaxies with redshift (e.g., \citealt{Couch94}). It also made the first observations of the host galaxies of low-$z$ quasi-stellar object (QSO). High resolution HST observations also found that $\sim 80\%$ of early-type galaxies with $M_B < -15$ are nucleated (\citealt{Grant05}; \citealt{Cote06}) and revealed the central properties of disky and boxy ellipticals (E) (cuspy centres of disky Es versus gently rising luminosity of boxy Es; \citealt{Ferrarese94}; \citealt{Rest01}).

\begin{figure*} 
	\centering
	\includegraphics[width=\textwidth]{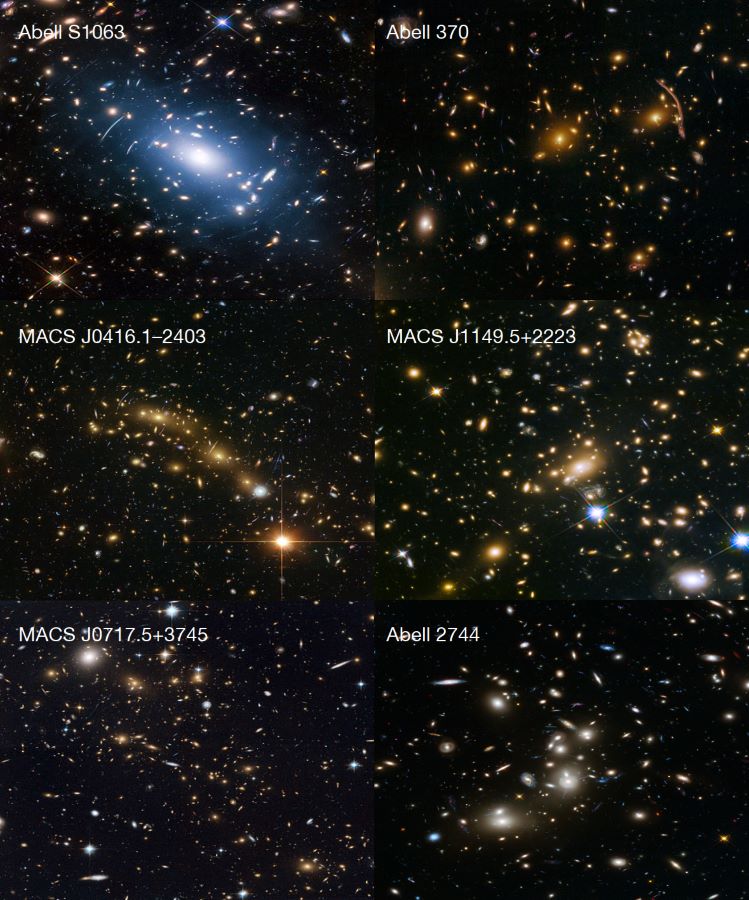}
	\caption{Hubble Frontier Fields. Source: \href{https://frontierfields.org}{https://frontierfields.org}  }
	\label{fig:HFF}
\end{figure*}

\begin{figure*} 
	\centering
	\includegraphics[width=\textwidth]{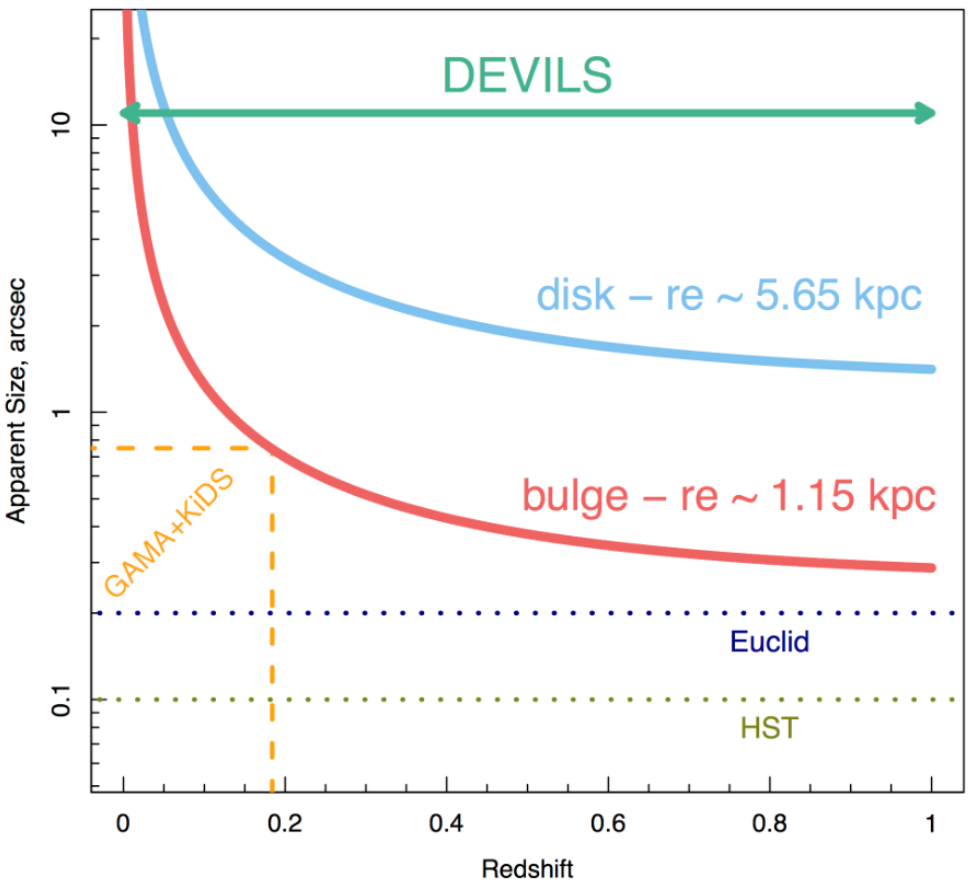}
	\caption{Comparing the angular resolution of different telescopes, including HST and Euclid. HST resolution allows for exploring galaxy structures until $z=1$. Credit: DEVILS website \href{https://devilsurvey.org/}{https://devilsurvey.org/}. }
	\label{fig:Resolution}
\end{figure*}

HST is used to observe the properties of galaxies in deep space, their morphologies in particular. Through the late 90's and early 00's the Hubble Deep Field (HDF), Hubble Ultra-Deep Field (HUDF), and Hubble eXtreme Deep Field (XDF) images opened three unique windows covering only 2.6, 2.4, 2.3 arcminutes, respectively, through which Hubble sampled the very distant galaxy population. More recently, the limits of the Hubble Space Telescope was pushed further with the Hubble Frontier Field Program capturing ultra deep colour images of six massive clusters of galaxies aiming to utilize gravitational lensing to study the earliest epoch of galaxy formation at $z = 5-10$ (Figure \ref{fig:HFF}). HST is also the primary tool that astronomers use to explore the earliest and most distant galaxies in the Universe. More recently, using HST \cite{Oesch16} discovered GN-z11 at $z=11.09$, the most distant galaxy.

\subsection{Large Astronomical Surveys Using HST}
\label{subsec:HSTsurveys}

To sample large numbers of galaxies with HST a number of surveys have been conducted with various instruments aboard HST. 
Amongst several surveys performed with HST, the Cosmic Assembly Near-infrared Deep Extragalactic Legacy Survey (CANDELS; \citealt{Grogin11}) and the Cosmic Evolution Survey (COSMOS; \citealt{Scoville07}) are the two largest. CANDELS explores galaxy evolution from $z = 8$ to $1.5$ and using HST's WFC3/IR (Wide Field Camera 3) and ACS (Advanced Camera for Surveys) reaches to a limiting surface brightness of 25.25 mag arcsec$^{-2}$ (Wide) and 26.25 mag arcsec$^{-2}$ (Deep). Commencing in 2006, COSMOS covers a two square degree equatorial field, the largest contiguous area of sky ever observed by HST. 

In addition, the COSMOS field comes with complementary data from spectroscopic campaigns conducted by multiple groups on some of the world's largest telescopes (e.g., \citealt{Davies18}) and X-ray to radio wavelength imaging data (\citealt{Scoville07}, see Section \ref{subsec:ACS/HST} for more details).

In this project, we will make use of the HST/ACS/F814W imaging data in the COSMOS field with a depth of 27.2 mag ($10 \sigma$ point source), also D10 region of Deep Extragalactic VIsible Legacy Survey (DEVILS, \citealt{Davies18}), an ongoing spectroscopic campaign at the Anglo-Australian Telescope probing intermediate redshifts ($0.3 < z < 1.0$), to investigate the formation and evolution of galaxies with different morphologies since $z \sim 1$ in an effort to bridge the current $z = 1$ to $z = 0.1$ divide. Figure \ref{fig:Resolution} compares the angular resolution of the HST with some other surveys/facilities indicating that with the HST's unprecedented spatial resolution we can resolve galaxy structures, even bulges with regular size of $1.15$ kpc. Therefore, HST enables us to explore the formation and evolution of the internal structures of galaxies since $z \sim 1.0$.   

\section{Galaxy Stellar Mass Function} \label{sec:SMF_evol_intro}

The galaxy stellar mass function and the luminosity function are crucial and informative statistical tools that represent the number density of galaxies within bins of stellar mass/luminosity (see e.g., \citealt{Schmidt68}; \citealt{Lynden-Bell71}; \citealt{Schechter76}; \citealt{Sandage79}; \citealt{Efstathiou88}; \citealt{Kennicutt89}; \citealt{White91}; \citealt{Zwaan03}; \citealt{Loveday15}). It has been shown that, within a certain volume, in going from faint/low-mass to luminous/high-mass end, the number of galaxies within mass/luminosity intervals decreases as a power law. It then decreases exponentially beyond a bright/massive cutoff (e.g., \citealt{Efstathiou88}; \citealt{Loveday92}; \citealt{Blanton01}; \citealt{Norberg02}).

\begin{figure*} 
	\centering
	\includegraphics[width=\textwidth]{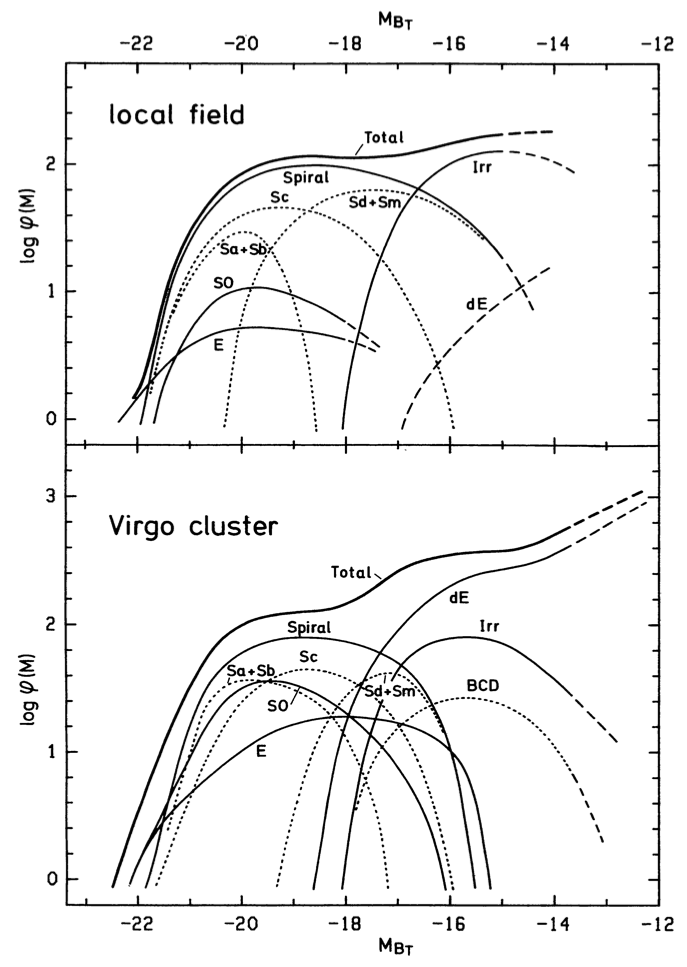}
	\caption{The luminosity function of local field galaxies (top panel) and Virgo cluster (bottom panel) as measured by \cite{Binggeli88}}
	\label{fig:Binggeli_LF}
\end{figure*}

\begin{figure*} 
	\centering
	\includegraphics[width=\textwidth]{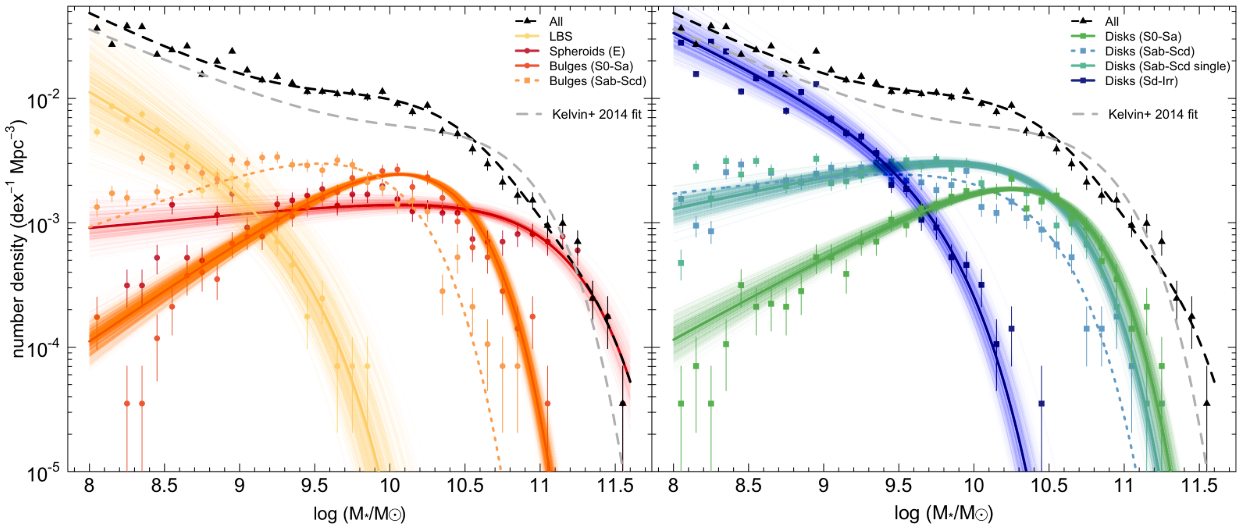}
	\caption{The SMF of spheroids (left) and disk (right) of different morphological types \citep{Moffett16b}.}
	\label{fig:Moffett_SMF}
\end{figure*}

Many studies have explored the luminosity function of galaxies (e.g., \citealt{Soifer87}; \citealt{Saunders90}; \citealt{Andredakis95}; \citealt{Rosati98}; \citealt{Willmer06}; \citealt{Reddy09}; \citealt{Grogin11}; \citealt{Cole11}; \citealt{Loveday15}; \citealt{Finkelstein15}; \citealt{Lehmer19}; \citealt{Ito20}) indicating how important this tool is to understand galaxy formation and evolution. An early important study though is the review by \cite{Binggeli88} where they investigate the LF of galaxies for different morphologies. Figure \ref{fig:Binggeli_LF} shows the LF of galaxies in the local field environment and compare with Virgo cluster to highlight that galaxies with different morphologies have different shapes of LF and therefore contribute to different parts of the total luminosity function.

With larger spectroscopically observed samples and more accurate multi-wavelength photometric data, calculating the stellar mass of large sample of galaxies became increasingly feasible, opening the doors to further investigation of the stellar mass function of galaxies. The stellar mass function (SMF) is one of the most important and informative statistical tools to constrain the stellar mass of the Universe and gain insight into galaxy formation (e.g., \citealt{Zwaan03}; \citealt{Baldry04}; \citealt{Pannella06}; \citealt{Baldry08}; \citealt{Baldry12}; \citealt{Weigel16}; \citealt{Davidzon17}; \citealt{Kawinwanichakij20}). The SMF of galaxies in the local Universe has been investigated by many studies (see e.g., \citealt{Cole01}; \citealt{Bell03}; \citealt{Baldry08}; \citealt{Wright17} and references therein). It is now well established that the global stellar mass distribution of local galaxies can be described with a double \cite{Schechter76} function with a characteristic mass of M$^* = 10^{10.6}$ to $10^{11} M_\odot$ (see e.g., \citealt{Panter07}; \citealt{Baldry08}; \citealt{Peng10}; \citealt{Baldry12}; \citealt{Wright17}; \citealt{Weigel16}). In the last two decades, large statistical samples of galaxies and more established methods for calculating the stellar mass (e.g., \citealt{Kauffmann03}; \citealt{Bell07}; \citealt{Driver07}; \citealt{Taylor11}) have led to the investigation of the SMF of two major distinct galaxy populations, ``blue'' and ``red'' or ``star-forming'' and ``passive'' or ``Late-type'' and ``Early-type'' leading to the emergence of quenching mechanisms and environmental and mass dependent galaxy formation scenarios (see e.g., \citealt{Robotham06}; \citealt{Peng10}; \citealt{Robotham10}; \citealt{Pozzetti10}; \citealt{Baldry12}; \citealt{Ilbert13}; \citealt{Taylor15}; \citealt{Davidzon17}).  

Analogous to \cite{Binggeli88} some works have studied the contribution of galaxies with different morphological types to the total stellar mass function (see e.g., \citealt{Pannella06}; \citealt{Fukugita07}; \citealt{Bernardi10}; \citealt{Bundy10}; \citealt{Vulcani11}). Two state-of-the-art studies by \cite{Kelvin14} and \cite{Moffett16a} presented the contribution of galaxies with different Hubble types to the local Universe SMF. By visually classifying $\sim 7,500$ of the Galaxy And Mass Assembly (GAMA) galaxies \cite{Moffett16a} quantified that the contribution of spheroid and disk dominated galaxies to the total stellar mass density of the Universe are 70\% and 30\%, respectively. 

By photometric bulge-disk decomposition further investigations have explored the contribution of disks and bulges in the total stellar mass of the Universe highlighting that the stellar mass of the Universe is roughly equally distributed between bulge and disk structures (e.g., \citealt{Driver07}; \citealt{Benson07}; \citealt{Gadotti09}; \citealt{Thanjavur16}).
\cite{Moffett16b} further performed a bulge disk decomposition of GAMA galaxies with a 2D profile fitting using {\sc GALFIT} \citep{Peng10} and further confirmed that 50\% of the local stellar mass density is bounded in spheroids (ellipticals and bulges) and 48\% in disk components (Figure \ref{fig:Moffett_SMF}). The rest of stellar mass is in the form of very low mass little blue spheroids (LBS). In efforts to study the galaxy population dichotomy, some works further investigated the evolution of the total SMF together with that of star-forming and passive galaxies (or similarly for late- and early-type galaxies) from high-$z$ to low-$z$ (e.g., \cite{Bundy05}; \citealt{Pannella06}; \citealt{Vergani08}; \citealt{Pozzetti10}; \citealt{Muzzin13}; \citealt{Whitaker14}; \citealt{Leja15}; \citealt{Mortlock15}; \citealt{Wright18}; \citealt{Kawinwanichakij20}). 

Spite of all the above investigations of the SMF of galaxies with different star-formation rates and colours (from high- to low-$z$) or morphologies (at low-$z$), further thorough exploration of the evolution of the SMF and stellar mass density of galaxies by morphological types and structural analysis is required to better understand galaxy formation and evolution.  
In addition to SMF, having large datasets with robust measurements of their stellar mass enables us to explore an important relation between the stellar mass and size of galaxies.  

\begin{landscape}
\begin{figure*} 
	\centering
	\includegraphics[width=1.5\textwidth]{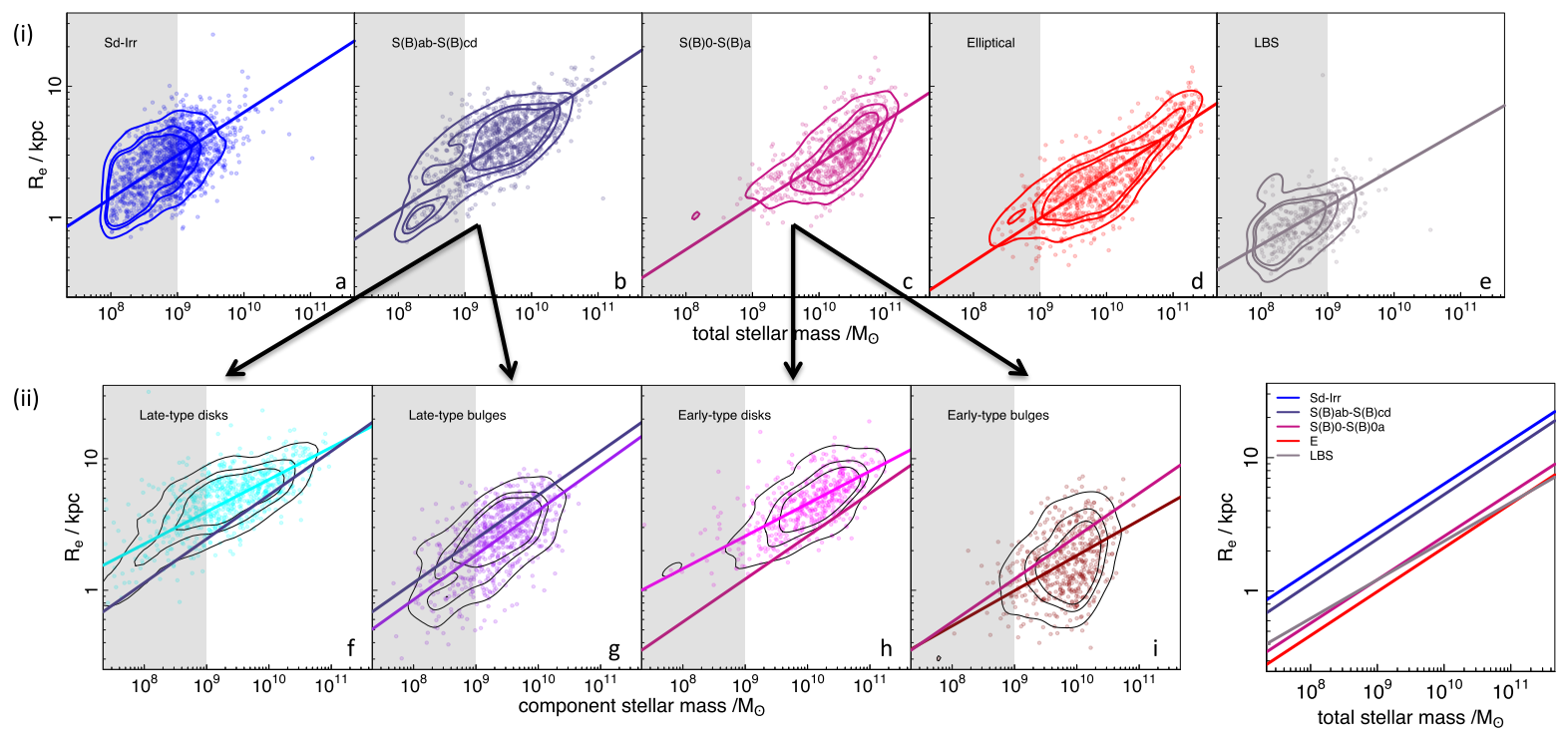}
	\caption{(i) Global $M_*-R_e$ relation subdivided into morphological classes. (ii) Structural $M_*-R_e$ relation for disks and bulges \citep{Lange16}.  }
	\label{fig:Lange_MRe}
\end{figure*}
\end{landscape}

\section{Mass-Size Relation} \label{sec:SMF_evol_intro}

Besides the stellar mass function, the relation between stellar mass and half-light radius (hereafter $M_*-R_e$) is argued to be a key fundamental plane to explain galaxy formation and evolution (e.g, \citealt{Bouwens04}; \citealt{Taylor20}; \citealt{Bernardi20}). The correlation of size and mass with the specific angular momentum of galaxies makes the $M_*-R_e$ relation further important as the angular momentum in turn gives insight into the dark matter halos (e.g., \citealt{Dalcanton97}; \citealt{Mo98}; \citealt{Romanowsky12}; \citealt{Obreschkow18}). 
Furthermore, the relation is of interest of simulations because recent hydrodynamical simulations generate galaxies with realistic sizes making direct comparison with observations feasible (see e.g., \citealt{Bahe16}; \citealt{Rodriguez-Gomez19}). The conservation of angular momentum during the primordial dark matter halo collapse could potentially connect the $M_*-R_e$ with dark matter halo properties (\citealt{Fall80}; \citealt{Dalcanton97}; \citealt{Mo98}). Hence the $M_*-R_e$ plane potentially represents a meeting ground between observations, theory and numerical simulations where simulations can be tuned according to observed $M_*-R_e$ relations (see e.g., \citealt{Guo16}).
In spite of significant recent progresses in hydrodynamical simulations, this comparisons have revealed that simulations have had difficulties in reproducing the observational $M_*-R_e$ relation (e.g., \citealt{Bahe16}; \citealt{vandeSande19}). For example, \cite{Rodriguez-Gomez19} showed the disk galaxies in the IllustrisTNG (\citealt{Marinacci18}) do not follow the morphology-size relation in observations where at a given stellar mass disks are observed to be larger than spheroids. 


\begin{figure*} 
	\centering
	\includegraphics[width=\textwidth]{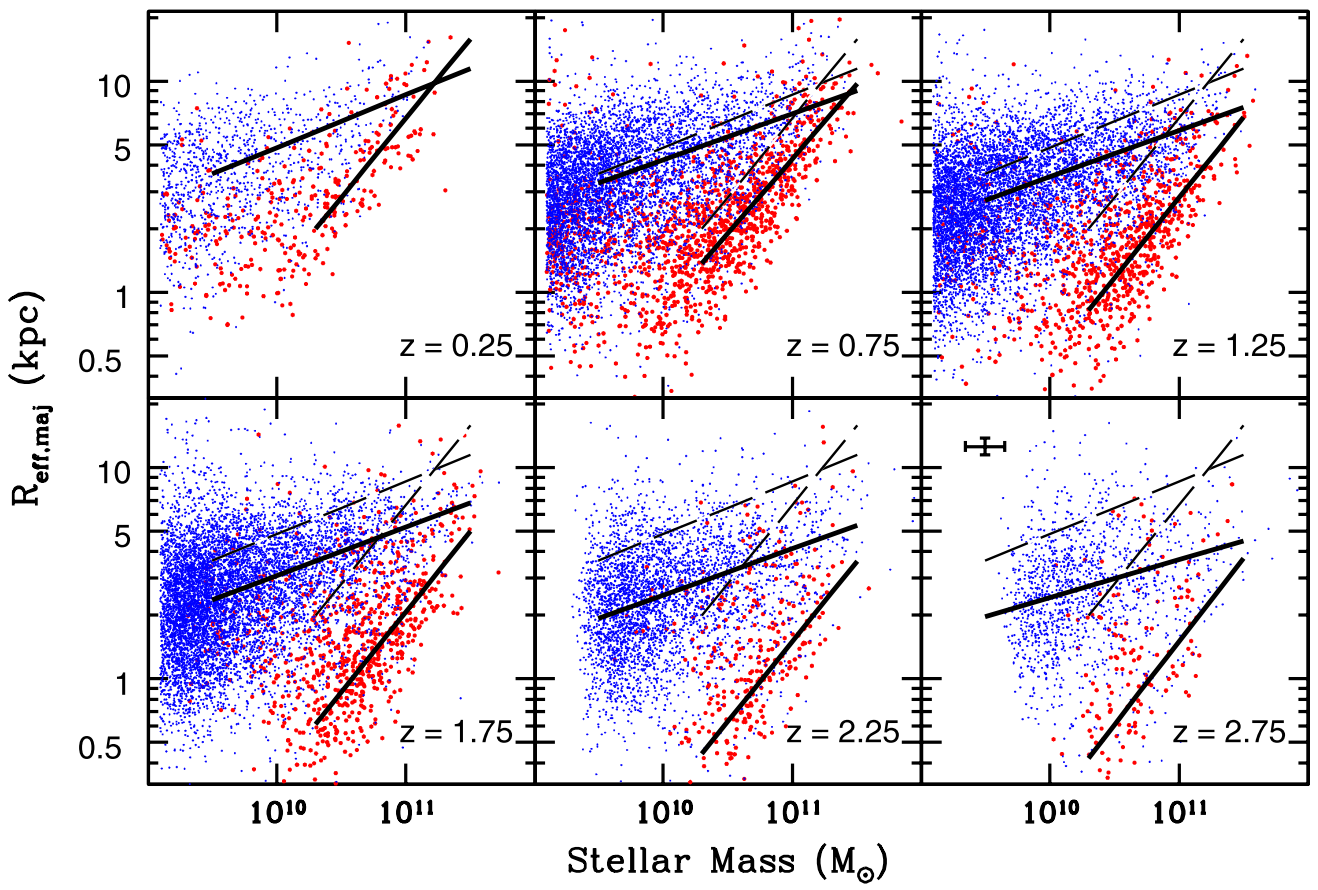}
	\caption{The evolution of $M_*-R_e$ relation for early- and late-type galaxies (red and blue symbols, respectively). Solid lines show the power law fit to the data while dashed lines represent the relation at lowest redshift bin \citep{vanderWel14}. }
	\label{fig:vanderWel_MRe}
\end{figure*}

In addition, it has been shown that $M_*-R_e$ is dependent on the morphological types. As shown in Figure \ref{fig:Lange_MRe}, early-type galaxies are smaller than late-type galaxies with steeper $M_*-R_e$ relation, although at high stellar masses ($\mathrm{log}(M_*/M_\odot) > 11$) early-types start to become larger (e.g., \citealt{Shen03}; \citealt{Lange16}). $M_*-R_e$ relation further gives clues about the surface mass density of different morphological types. The slope of the relation also correlates with the galaxy mass-halo mass relation. This relation has been also traced back in cosmic time which can give insight into the size and mass evolution of galaxies as well as the evolution of the dark matter halos. A state-of-the-art study of the evolution of $M_*-R_e$ relation of early- and late-type galaxies was done by \cite{vanderWel14} using 3D-HST (\citealt{Brammer12}) and CANDELS (\citealt{Grogin11}) data confirming that early-type galaxies are smaller than late-type galaxies at all epochs and experience a faster evolution than late-types. They also find that the slope of the relation varies little, implying that potentially the slope of the galaxy mass-halo mass is somewhat constant. We show their mass-size distribution in Figure \ref{fig:vanderWel_MRe}.

\subsection{Cosmic Dimming and morphological k-correction} \label{subsec:k-corr}

Although the exploration of galaxy morphological types at high redshift is important, there are some challenges involved here.
Studying the morphology of high redshift galaxies is increasingly impacted by cosmological expansion in two ways; cosmological surface brightness dimming and morphological k-corrections. Since this thesis is fundamentally a morphological investigation of high-$z$ galaxies it is important to highlight these effects. 

As first derived by \cite{Tolman30} and later stated by \cite{Phillipps90} and \cite{Lanzetta02} the apparent surface brightness of extended objects decreases with redshift as:   

\begin{equation}
I_0 \equiv I_{e}(1+z)^{-4}
\end{equation}

\noindent where $I_0$ and $I_e$ represent the observed and intrinsic surface brightnesses, respectively.

The second cosmological factor is that apparent galaxy structures vary with wavelengths, and this is known as morphological k-correction. As most deep imaging is performed in optical bandpasses, at high-$z$ it is no longer sampling the rest-frame optical wavelength due to redshifting of the distant galaxies light. Beyond $z \sim 1.5$ the rest-frame UV is redshifted into the optical window and dominated by young stars with ages $< 100$ Myrs. Even though starbursting systems are likely to stay unaffected, early-type spiral galaxies (such as Sa and Sb) with two distinct old and young stellar components will have different structures in optical and UV (\citealt{Windhorst02}; \citealt{Papovich03}; \citealt{Conselice04}). See Section \ref{subsec:k-corr} for more details on how morphological k-correction might affect our visual classifications. 

The effects of K-correction on our morphological classifications will be later discussed in more details in Section \ref{subsec:K-corr}.

\begin{figure*} 
	\centering
	\includegraphics[width=\textwidth]{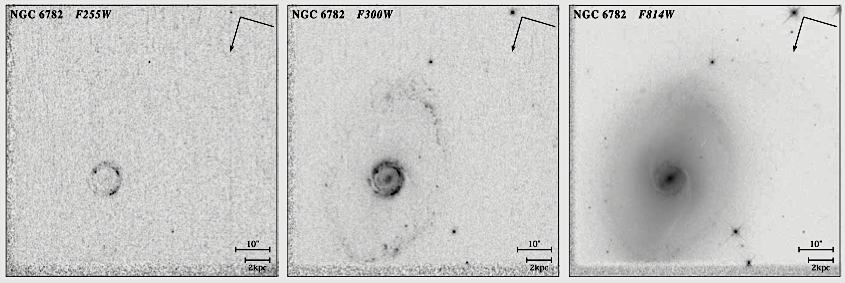}
	\caption{Morphological k-correction: an image of NGC 6782 in HST F255W (left), F300W (middle) and F814W (right) filters, respectively \citep{Windhorst02}. }
	\label{fig:Driver13TwoPhase}
\end{figure*}

\section{Galaxy Formation Overview} \label{sec:gal_formation}

\begin{figure*} 
	\centering
	\includegraphics[width=\textwidth]{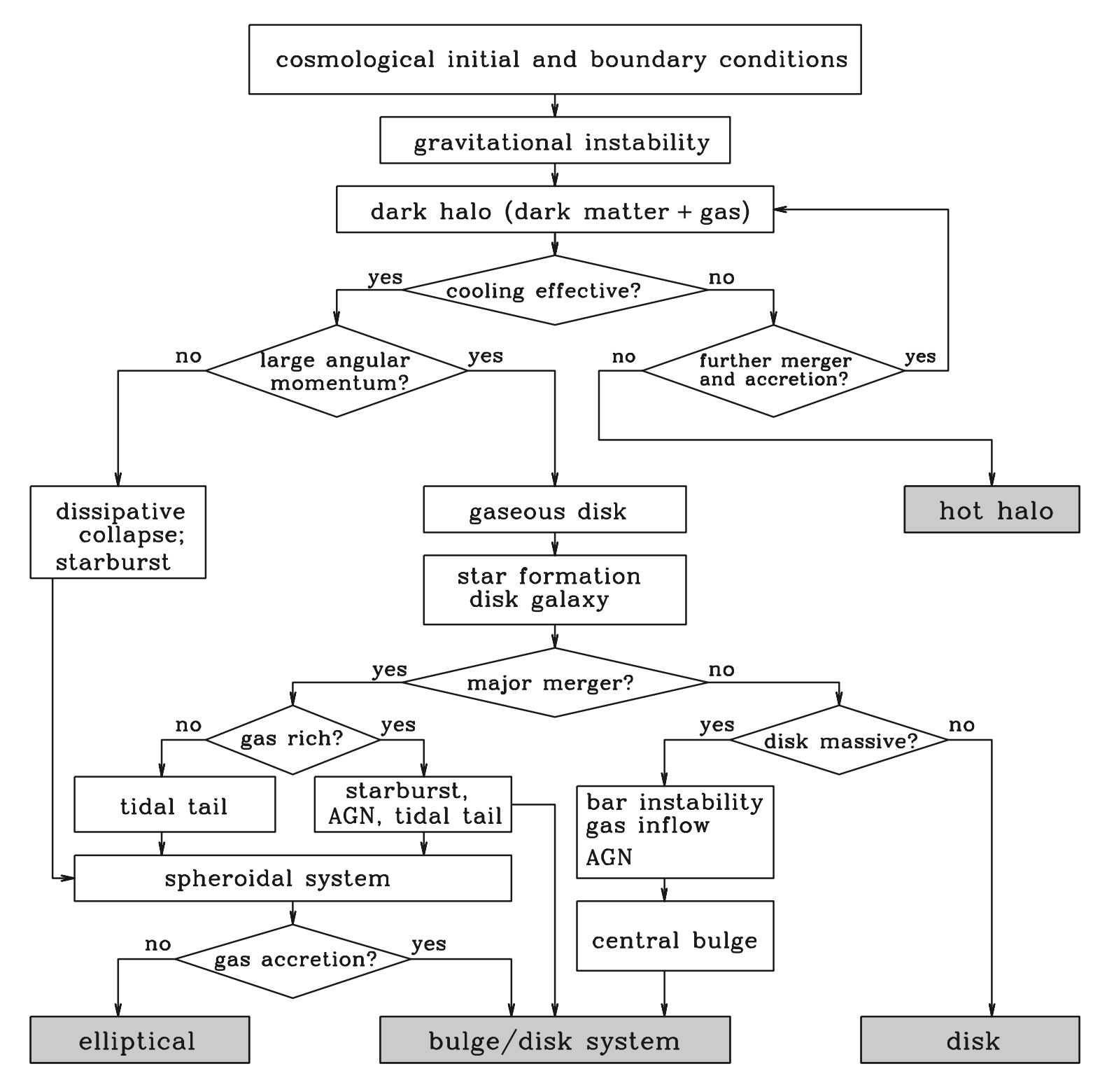}
	\caption{Flow diagram showing a simple view of our current picture of physical processes involved in galaxy formation and evolution \protect\citep{Mo10} }
	\label{fig:GF_flow}
\end{figure*}

During their history, galaxies go through various processes playing important role in forming the current morphology of galaxies. In this section, we briefly discuss the physical processes involved in galaxy formation and morphological evolution.
Before explaining these processes, shown in Figure \ref{fig:GF_flow}, we briefly explain the most recent and accepted cosmological paradigm, $\Lambda$CDM. 

\textbf{$\Lambda$CDM paradigm:} According to modern cosmology the Universe is homogeneous and isotropic and evolves based on the theory of general relativity (\citealt{Einstein17}), the mass distribution forms the structure of space-time in the Universe. ``Mass bends space-time and space-time tells mass how to move.'' Cosmologies are now capable of predicting the geometry and dynamics of the Universe based on the mass and energy content of the Universe. In essence, there are three main components that contribute to the mass-energy budget: baryonic and dark matter, radiation, and dark energy. Different cosmologies argue the relative contribution of these components as well as their nature. The most popular cosmology, $\Lambda$CDM, predicts a flat Universe containing $\sim 70$ percent dark energy, $\sim 25$ percent cold dark matter and $\sim 5$ percent baryonic matter, i.e. the protons, neutrons and electrons out of which our visible Universe is made up (for example see \citealt{Planck20} for recent Planck cosmological parameters).

Figure \ref{fig:GF_flow} adopted from \cite{Mo10}, shows a flow diagram of the relationships between various mechanisms that might cooperate in forming a galaxy of a specific type. In fact, our current observations and theories imply that depending on the pathway that a galaxy takes in this diagram it will eventually have different structures and physical properties. Here, we only briefly describe three important part of the diagram.   

\textbf{Gravitational instability and structure formation:} If the distribution of mass in the Universe began as a perfectly uniform distribution, there would be no structure today. However, in the very early Universe, quantum effects were in play. Inflation theory predicts that quantum fluctuations, could lead to the density perturbations required to build structure in the Universe. Exponentially growing perturbations, driven by inflation, eventually result in fluctuations in the density field at recombination, generating regions of over- and under-density. Gravitational instability then amplifies these over- and under-densities and is believed to be responsible for the eventual formation of galaxies and large-scale structures.

\textbf{Cooling and disk formation:} For stars and hence galaxies to be formed the gas needs to be cooled. There are different cooling mechanisms depending on the virial temperature, for example bremsstrahlung emission from free electrons is responsible for cooling gas in halos with $T_{vir} > 10^7$ K. The cooling gas flows inward to the centre of dark matter halos making the gas rotate due to the conservation of angular momentum and eventually forming a disk galaxy. While the cooling gas is rotating and falling towards the centre its self-gravity is likely to overcome the overall gravity of the dark halo. Such catastrophic collapse, under certain temperature and density circumstances, is likely to initiate a gas fragmentation and eventually forming stars.

\textbf{Mergers:} Galaxies and halos are rarely isolated and typically clustered; this clustering means they frequently gravitationally interact with each other through collisions, mergers and/or accreting or losing (baryonic and dark) material. All galaxies, independent of their halo mass, are predicted to continually grow through minor mergers, i.e. merging of progenitors with mass ratios of $< 1/3$ (e.g., \citealt{Li07}). $\Lambda$CDM cosmology predicts a hierarchical growth of dark matter halos in which larger halos form via the merging of smaller halos at earlier times. Nowadays, this picture is typically illustrated by a merger tree which shows the merger history of galaxies by tracing the coalescence of their progenitors (e.g., \citealt{De-Lucia07}; \citealt{Benson13}; \citealt{Rodriguez-Gomez15}). Models suggest that galaxies form and continually grow through successive mergers and gas and dark matter accretion (e.g., \citealt{Cooper15}; \citealt{Rodriguez-Gomez16}).
 
Cosmological and individual galaxy numerical simulations have shown that merging two or more galaxies/halos can lead to the formation of an entirely new system with different morphology and physical properties arguing that elliptical systems are likely merger remnants (for early works see for example \citealt{Toomre72}; \citealt{White78}; \citealt{Gerhard81}; \citealt{Negroponte83}; \citealt{Barnes88}). 

The outcome of mergers correlates with a few factors, including the number, gas richness and mass ratio ($q \equiv M_1/M_2$) of involved progenitor halos with $q \lesssim 3$ and $q \gtrsim 3$ representing major and minor mergers, respectively. 
Major mergers likely result in remnants that are quite different in morphological and dynamical properties, while in minor mergers, the process is less destructive.
As shown in Figure \ref{fig:GF_flow}, major mergers play a vital role in the morphological evolution of galaxies, typically resulting in an elliptical or bulge+disk system from disk progenitors. 
In the case of large $q$ ($q \gtrsim 10$), disk progenitors can survive a minor merger, although they might experience a disk thickening. Another essential property of mergers that impacts the outcome is the progenitors' gas mass fraction. 
As shown in Figure \ref{fig:GF_flow}, according to simulations, gas-rich or gas-poor mergers, also known as ``wet'' and ``dry'' mergers, can produce very different remnants  (e.g., \citealt{Kang07}; \citealt{Khochfar09}). 

Dry minor mergers are thought to play an essential role in the growth of high-$z$ (i.e., $z = 2-3$) early-type galaxies making them grow by a factor of four in size and two in mass (e.g., \citealt{Naab06}; \citealt{Khochfar09}; \citealt{Remus17}). As mentioned above, depending on the progenitors' mass ratio, these mergers might or might not change the host galaxy's inner structure. This scenario is also known as the ``two-phase formation scenario'' (e.g., \citealt{Oser10}) that we will explain in more details below. Mini-mergers with $q > 10$ have even less destructive footprints on the centre of ETGs (e.g., \citealt{Arnaboldi20}), leaving stars preferentially at the outskirts of the host galaxy, and resulting in significant size growth (e.g., \citealt{Karademir19}).

\begin{figure*} 
	\centering
	\includegraphics[width=\textwidth]{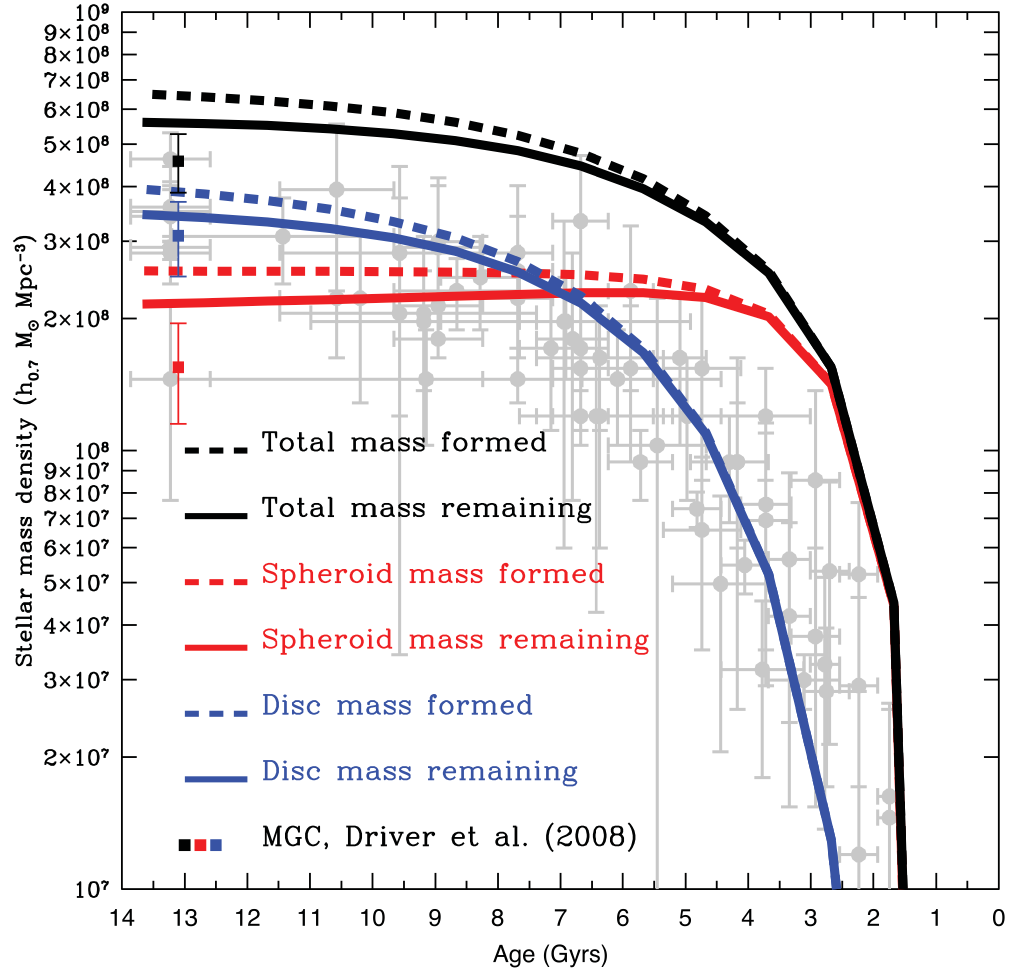}
	\caption{The two-phase formation scenario proposed by \protect\citep{Driver13}. }
	\label{fig:Driver13TwoPhase}
\end{figure*}

\section{Two-Phase Formation Scenarios} \label{sec:two-phase}

As mentioned earlier, one of the most important and not yet well answered questions in astronomy is whether the dichotomy in galaxy structures can be explained by distinct formation formalisms?
\cite{Oser10} performed high-resolution simulation of 39 individual galaxies to investigate whether the stars at the centre of modern galaxies formed freshly from gas within the galaxy or at the centre of another galaxy at earlier times and then accreted into the final galaxy. They find that the vast majority of stars at the central regions of massive galaxies formed at higher redshifts either near the centre of the system or accreted from outside, so called ``in-situ'' and ``ex-situ'', respectively. \cite{Oser10}, therefore, proposed a ``two-phase'' formation scenario in which phase one includes in-situ star formation at the centre of galaxies ($< 3~\mathrm{kpc}~h^{-1}$) at $z > 2$ and phase two is dominated by accretion and minor/mini mergers. According to this scenario phase one is responsible for the formation of central compact core of modern galaxies (e.g., see \citealt{vanDokkum08}) and phase two for more extended components ($> 8~\mathrm{kpc}~h^{-1}$) (\citealt{Bluck12}; \citealt{McLure13}; \citealt{Robotham14}; \citealt{Ferreras17}). Consequently, several works started to compare high-$z$ galaxies with low-$z$ ones seeking similarities between the central structure of galaxies. For example, \cite{Hopkins09} investigated the radial surface density profiles of massive high-$z$ spheroidal systems and highlighted that they are similar to the central profile of local spheroids. Another study by \cite{Bezanson09} also confirmed this conclusion indicating an inside-out growth scenario \cite{delaRosa16} further studied galaxies of Sloan Digital Sky Survey (SDSS) and reported the same result confirming that the core of $z=0$ massive galaxies are analogous to compact high-$z$ red nuggets (\citealt{Damjanov09}).  

In addition, by analysing the cosmic star-formation histories of disk galaxies and spheroids, \cite{Driver13} suggested two main phases of galaxy formation: ``hot mode'' and ``cold mode'' with a switching point at $z \approx 1.7$ (see Figure \ref{fig:Driver13TwoPhase}). In the hot mode evolution phase the Universe rapidly develops most spheroids through mergers, fragmentation or collapse, while during cold mode disks slowly start to form via minor mergers and cold gas in-fall (\citealt{Larson76}; \citealt{Tinsley78}).  

The two-phase formation scenario is somewhat plausible according to the fact that galaxies are a combination of disks and spheroids. However, this scenario requires dark halo mergers to happen at earlier times than current simulations predict. This low merger rate is corroborated as only $\sim 40\%$ of the local stellar mass of the Universe is found in spheroidal systems (\citealt{Gadotti09}; \citealt{Tasca11}; \citealt{Moffett16a}). Although some simulations have shown that gas-rich mergers can produce disk structures (e.g., \citealt{Robertson06}; \citealt{Governato07,Governato10}), major mergers are generally believed to more frequently destroy disks. Therefore, having the majority of mass residing in disks, indicates that the majority of the stellar mass of the Universe is not built by major mergers. Hence the dominant formation mechanism in the Universe is unlikely to be major merger-driven.
Therefore, more investigation is required to further assess whether a two-phase formation scenario is the pathway that galaxies have gone through.

\section{Bulge Formation} 
\label{sec:bulge_form}

Necessarily coupled with galaxy formation models is bulge formation scenarios explaining the growth of the central structure of disk galaxies. As a significant component, bulges play a key role in the general morphology and physical properties of galaxies such as stellar mass, angular momentum and star formation rate. This thesis looks into the formation of bulges in detail, therefore, here, we define bulges and shortly highlight different types of bulges and some popular formation scenarios.  

\textbf{Bulge definition:} 
The central region of disk galaxies are often seen not following the inward extrapolation of a disk's near-exponential light profile but rather raising in surface brightness more steeply. This extra light/mass is typically called a ``bulge''. 
\cite{Renzini99} concluded from the similarities of bulges with elliptical galaxies and stated: ``It appears legitimate to look at bulges as ellipticals that happen to have a prominent disk around them [and] ellipticals as bulges that for some reason have missed the opportunity to acquire or maintain a prominent disk.'' Bulges might also host and co-evolve with a central black hole as their masses are shown to correlate (see a review by \citealt{Graham16}) 

\begin{figure*} 
	\centering
	\includegraphics[width=\textwidth]{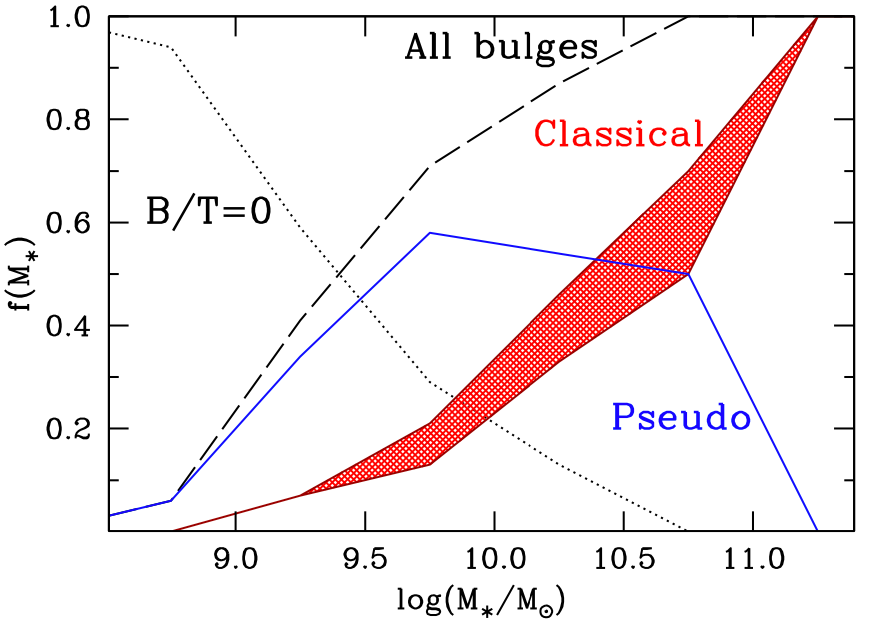}
	\caption{ The fraction of galaxies with different types of bulges as a function of their global stellar mass. Blue line and red filled region indicate pseudo- and classical-bulges and dotted line represent pure-disk galaxies containing no bulge. Low mass systems are dominated by pure-disk galaxies while pBs and cBs can be found in intermediate and high mass ends, respectively (adopted from \protect\citealt{Fisher16}). }
	\label{fig:Bulge_M_Fisher}
\end{figure*}

\textbf{Bulge dichotomy:} With the explosion of deep imaging and kinematic data, a dichotomy in the dense central structure of disk galaxies is being reported in many studies, suggesting these different structures may have formed through different formation pathways. Two different types of bulges, are: ``classical'' and ``pseudo'' bulge (hereafter: cB and pB). A number of works have considered these bulges distinct classes and studied their properties (e.g., \citealt{Kormendy04}; \citealt{Athanassoula05}; \citealt{Fisher06}; \citealt{Fisher08}; \citealt{Fisher10}; \citealt{Fisher11}; \citealt{Luo20}). Classical-bulges are observed as featureless, pressure supported, dynamically hot and similar to elliptical galaxies, while pseudo-bulges are disk-shaped, star-forming and rotational supported systems (\citealt{Kormendy82}; \citealt{Kormendy93}; \citealt{Gao20}). Some studies have investigated the colour distribution of pBs and cBs and concluded that pBs resemble their parent disks while cBs are more similar to ellipticals (\citealt{Fisher06}; \citealt{Du20}; \citealt{Gao20}). Comprehensive reviews on this interesting dichotomy are presented by \cite{Kormendy16} and \cite{Fisher16}.
As shown in Figure \ref{fig:Bulge_M_Fisher} (taken from \citealt{Fisher16}), bulge type correlates with the global stellar mass of galaxies and hence with galaxy evolution in general. Therefore, clearly studying different types of bulges will help us to better understand galaxy formation and evolution.           

\textbf{Bulge formation scenarios:} Currently, three main scenarios are in place for bulge formation. (i) merger driven (\citealt{Aguerri01}), (ii) secular formation (\citealt{Kormendy04}; \citealt{Athanassoula05}), (iii) clump migration (\citealt{Elmegreen08}; \citealt{Bournaud08}).

A broad range of surface brightness profiles are reported for bulges, reflected in their S\'ersic indices ($0.5 < n < 10$) (e.g. \citealt{Andredakis94}; \citealt{Andredakis95}; \citealt{Fisher08}; \citealt{Elmegreen08}) and indicating that, analogous to \textit{Elliptical} galaxies, classical bulges with $n > 2$ are likely to be formed via major mergers (e.g. \citealt{Hopkins09}; \citealt{Rodriguez-Gomez17}). However, a notable problem with this scenario is that it has been shown both theoretically and observationally that major mergers are rare in the Universe (see review by \citealt{Naab13}). In spite of that, some still argue that classical bulges and elliptical galaxies are rare too (\citealt{Kormendy16}). These authors argue that major mergers can still be the main process through which all spheroids in the Universe are generated. 

On the other hand, minor mergers are also shown to likely play a role in the formation and mass/size growth of classical bulges (\citealt{Eliche-Moral06}; \citealt{Hopkins10}; \citealt{Tacchella19}). Theoretical and observational studies have shown that low-mass ratio ($\le 1:10$) minor mergers are ubiquitous in the Universe (e.g. \citealt{Ostriker75}; \citealt{Woods07}; \citealt{Barton07}; \citealt{Stewart08}). The vast majority of galaxies experience such minor mergers in their life. Depending on the gas and mass fraction of merger companions, satellite galaxies could either survive and migrate to the centre of the disk and directly contribute in forming/growing classical bulges or destabilize the disk and trigger an in-situ bulge formation (\citealt{Aguerri01}; \citealt{Eliche-Moral06}; \citealt{Gao20}). The role of mergers in in-situ bulge formation is also recently shown in IllustrisTNG simulations by \cite{Tacchella19}.     

In addition, secular evolution in galaxies is presented as a candidate for bulge formation.
\cite{Kormendy79} first reported disky bulges and highlighted the importance of secular evolution (\citealt{Kormendy82}). A notion was born that the central dense component of disk galaxies might be formed slowly from disk material, i.e. through the formation of a disk-like central structure or pseudo-bulge. This theory further discusses that classical-bulges likely emerge as an outcome of this slow rearrangement of disk material, and eventually look like merger-driven bulges.  
Bars, spiral arms, oval disks and other non-axisymmetric structures can rearrange the disk's angular momentum, drive a gas inflow and eventually form a dense central component, pB (e.g. \citealt{vandenBosch98}; \citealt{Kormendy04}; \citealt{Avila-Reese05}; \citealt{Chown19}). These components keep some memories of their disk origin, such as star burst, rotation, disk-shaped near exponential light profile through which one can distinguish them from cBs build out of mergers (\citealt{Kormendy04}).

Another scenario for bulge formation is argued to be migration and coalescence of giant gas clumps in primitive disks at high redshifts (\citealt{Noguchi99}; \citealt{Carollo07}; \citealt{Elmegreen08}; \citealt{Bournaud08}). It has been shown that primordial gas clumps with mass of $10^8 \mathrm{M}_{\odot}$ could form through the fragmentation of the disks at high redshift. Clumps interact gravitationally and due to the dynamical friction migrate to the centre of the system and eventually coalesce to form a bulge if they survive of supernova explosions (e.g. \citealt{Noguchi99}; \citealt{Elmegreen08}). This whole process is expected to occur in a short time scale of a few $\times 10^8$ yr (e.g. \citealt{Noguchi99}). By decreasing redshift the disks become more stable and so less clumpy (\citealt{Elmegreen08}). 

\section{This Thesis}

In this thesis, we aim to explore the origin of the above dichotomy in galaxy morphology in general and also in galaxy structures by studying the evolution of the stellar mass and size of galaxies since $z \sim 1$. For this study, it is crucial to use resolved imaging data so one can detect the structure and morphology of galaxies at high-$z$. Accordingly, throughout this thesis, we make use of high-resolution imaging data from the Hubble Space Telescope in COSMOS region (ACS/COSMOS) and morphologically classify $\sim 44,000$ galaxies up to redshift $z = 1$ into broad Hubble sub-classes and study the evolution of their distinct stellar mass functions and resulting stellar mass densities (SMD). 

To study the evolution of bulges and disks, we further perform a robust bulge disk decomposition informed by our visual classifications and using the recently developed code {\sc ProFit} \citep{Robotham17} to quantify the evolution of the SMF and SMD of bulges and disks separately. 

We then explore the evolution of the $M_*-R_e$ relation of different morphological types as well as that of bulges and disks, separately using the results of our structural analysis.

Unlike other studies that broadly classify galaxies into early- and late-type or star-forming and quiescent, in this thesis we take advantage of both our morphological classification and structural analysis and explore the evolution of morphological types as well as bulges (pseudo and classical) and disks, separately. 

Ultimately, we aim to use our results to investigate current galaxy formation scenarios such as the two-phase formation scenario and take a step forward in understanding the formation of bulges and disks in the Universe.

\section{Tools and Data Availability} \label{sec:data_access}

The analysis in Chapter \ref{ch:GAMA} of this thesis is based on the visual morphological classification of GAMA DR4 (\citealt{Driver22}). In addition, our structural analysis of low-$z$ GAMA galaxies are build using \texttt{BDModelsAllRv03} DMU (\citealt{Casura-inprep}) available on GAMA website \href{http://www.gama-survey.org}{www.gama-survey.org}. This DMU is currently only available to GAMA team members.

The high redshift analysis of this thesis presented in Chapters \ref{ch:3},\ref{ch:4},\ref{ch:5} are based upon the DEVILS visual morphology catalogue, \texttt{DEVILS\_D10VisualMoprhologyCat\_v0.1} presented in \cite{Hashemizadeh21} and DEVILS structural analysis catalogue, \texttt{DEVILS\_D10BDDecomp\_v0.1} described in \cite{Hashemizadeh22}. These catalogues are held on the DEVILS database managed by AAO Data Central (\href{https://datacentral.org.au}{https://datacentral.org.au}) and are currently only for internal DEVILS team, but will be made publicly available in a future DEVILS data release. 

All the HST imaging data are in the public domain and were downloaded from the the NASA/IPAC Infrared Science Archive (IRSA) web-page: \href{https://irsa.ipac.caltech.edu/data/COSMOS/images/acs\_2.0/I/}{irsa.ipac.caltech.edu}. We were also provided an access to the the raw HST imaging frames of the COSMOS survey in a private collaboration with Anton Koekemoer. 

The main tool used in this study is our structural decomposition pipeline, {\sc GRAFit}, which is available at: \href{https://github.com/HoseinHashemi/GRAFit\_a}{https://github.com/HoseinHashemi/}. In {\sc GRAFit}, we use {\sc ProFit} \citep{Robotham17} version 1.3.3 for our structural decomposition (available at \href{https://github.com/ICRAR/ProFit}{https://github.com/ICRAR/ProFit}) and ProFound \citep{Robotham18} version 1.3.4 for photometry (available at \href{https://github.com/asgr/ProFound}{https://github.com/asgr/ProFound}). We use {\sc Tiny Tim} version 6.3 to generate the HST/ACS Point Spread Function (PSF). We further use {\sc LaplacesDemon} version 1.3.4 implemented in {\sc R} available at \href{https://github.com/LaplacesDemonR/LaplacesDemon}{https://github.com/LaplacesDemonR}. 

\graphicspath{{images/ChapterTwo/}}

\chapter[GAMA: The Redshift Zero Benchmark]{GAMA and setting the redshift zero benchmark}
\label{ch:GAMA}

\section{The Redshift Zero Benchmark}

The main focus of this thesis is the evolution of disks and bulges as well as galaxies of different morphological types, in general, since redshift $z = 1$ where we will explore the evolution of the stellar mass functions, stellar mass densities and mass-size relations. However, since we have a poor low-$z$ sample of galaxies in COSMOS field, we need to establish our redshift zero benchmark so we can compare our higher-$z$ evolution to local Universe. Therefore, in this chapter, we make use of GAMA galaxies and set up our $z \sim 0$ basement of stellar mass functions and mass-size relations ($M_*-R_e$).

The state-of-the-art study of the contribution of different morphological types in the total SMF in GAMA region is done by \cite{Moffett16a}. They find that spheroid dominated galaxies contribute 70\% to the total stellar mass density of the local Universe, while disk dominated systems contribute 30\%. Furthermore, \cite{Moffett16b} used the structural analysis of GAMA galaxies and studied the stellar mass budget of disks and spheroids in the local Universe. They confirmed that, in agreement with other studies (e.g., \citealt{Driver07}; \citealt{Benson07}; \citealt{Gadotti09}; \citealt{Thanjavur16}), disks and bulges contribute 50\% and 48\%, respectively, in the total stellar mass of the modern day Universe (see also \citealt{Allen06} for the same analysis using the Millennium Galaxy Catalogue, MGC).  

In addition, the state-of-the-art investigation of $M_*-R_e$ relation in the local Universe has been presented by \cite{Lange16}. They make use of the structural analysis of a sample of $7,500$ GAMA galaxies in redshift range of $0.002 < z < 0.06$ to study the $M_*-R_e$ relation of different morphological types as well as bulges and disks.   

In this work, we make use of the recently released GAMA DR4 data (\citealt{Driver11} and \citealt{Driver22}) to study the stellar mass budget of the local Universe by morphological types. Using this new data, we further investigate the contribution of galaxy structures (bulges and disks) to the total stellar mass budget of the local Universe. Taking advantage of higher quality VST KiDS\footnote{Killo Degree Survey} DR4 (\citealt{deJong13}; \citealt{Kuijken19}) imaging data instead of SDSS (previous GAMA works), this work provides an update to the previously published $z=0$ stellar mass functions by morphological types (\citealt{Moffett16a} and \citealt{Kelvin14}). We further use the new bulge-disk decomposition of GAMA galaxies using a newer software {\sc ProFit} (\citealt{Robotham17}) to extend and update the \cite{Moffett16b} work on the structural SMF and the \cite{Lange16} work on the $M_*-R_e$ relation. 

This chapter is organized as follows. In Section \ref{sec:GAMA_DR4_data} we discuss the data that we use in this work. In Sections \ref{sec:GAMA_SMF_morph} and \ref{sec:GAMA_SMF_str} we show our morphological and structural stellar mass functions. Section \ref{sec:GAMA_MRe} presents our stellar mass-size relation and finally we summarize our results in Section \ref{Sec:GAMA_Summ}.

Throughout this work, we use a flat standard $\Lambda$CDM cosmology of $\Omega_{\mathrm{M}} = 0.3$, $\Omega_\Lambda = 0.7$ with $H_0 = 70 \mathrm{km}\mathrm{s}^{-1}\mathrm{Mpc}^{-1}$ (\citealt{Planck20}). Magnitudes are given in the AB system \citep{Oke83}.
\begin{figure} 
	\centering
	\includegraphics[width = \columnwidth]{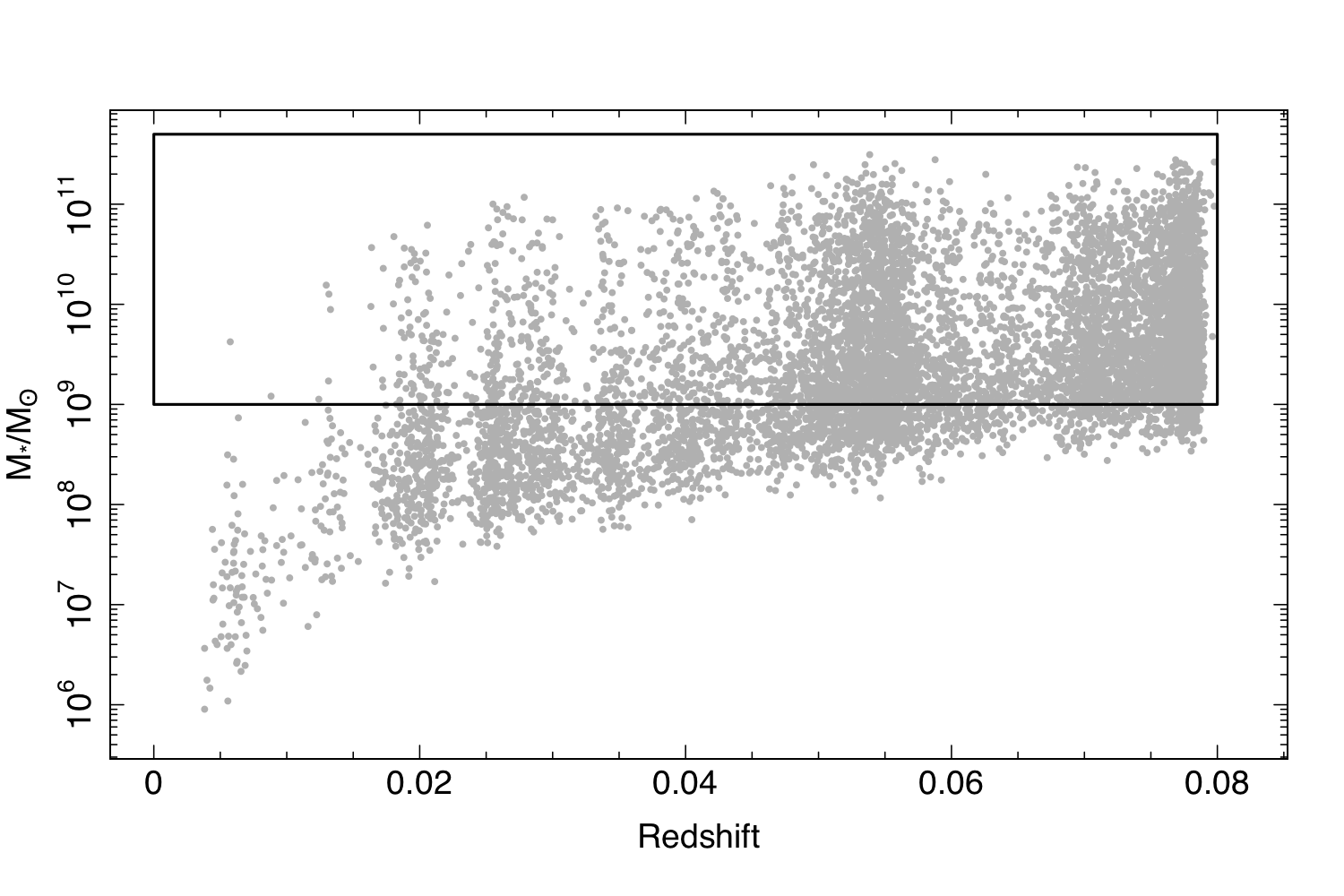}
	\caption{The GAMA sample selection. We select a sample of galaxies up to $z \leq 0.08$ and $\mathrm{log(M}_*/\mathrm{M}_\odot) \geq 9$ for which we have visual morphological classification as well as bulge-disk decomposition.}
	\label{fig:GAMA_SmplSel}
\end{figure}

\begin{figure} 
	\centering
	\includegraphics[width = \columnwidth]{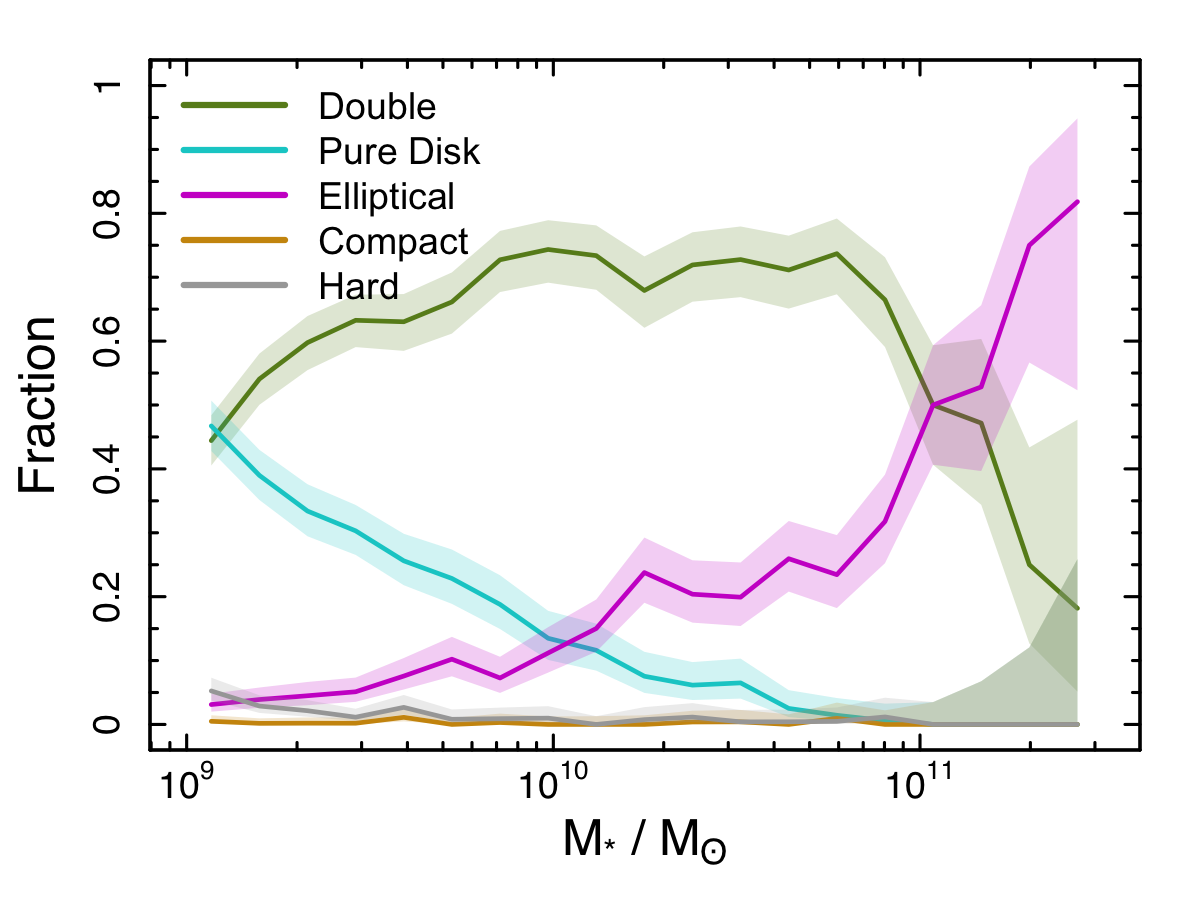}
	\caption{Fraction of galaxies in each morphological class as a function of their stellar mass. Shaded regions display 95\% confidence intervals calculated using beta distribution method estimated using {\fontfamily{qcr}\selectfont prop.test} in R package {\fontfamily{qcr}\selectfont stats}.}
	\label{fig:GAMA_Frac}
\end{figure}

\section{The GAMA DR4 Data}
\label{sec:GAMA_DR4_data}

In this study, we use GAMA DR4 \citep{Driver22} and select a sample of galaxies in $9^h$ (G09), $12^h$ (G12), and $15^h$ (G15) regions covering 169.3 square degrees (\citealt{Bellstedt20a}) up to redshift $z \leq 0.08$ and stellar mass $\mathrm{log(M}_*/\mathrm{M}_\odot) \geq 9$. We then cross-match our sample with the structural decomposition catalogue in this field \citep{Casura-inprep}. As a result, our final sample contains $\sim 5,581$ galaxies for which we have both visual morphological classification and structural analysis. Figure \ref{fig:GAMA_SmplSel} shows our sample selection. We then make use of the visual morphological classifications that we performed in \cite{Driver22} to investigate the contribution of each morphological type as well as divided into bulges and disks to the total stellar mass density of the local Universe. Briefly, in \cite{Driver22} we first use Galaxy Zoo decision trees to pre-classify the sample, then use KiDS DR4 (\citealt{Kuijken19}) imaging data in the GAMA fields to further visually classify galaxies. We generate 40kpc$\times$40kpc cutouts of grZ colour stamp mapped within the surface brightness range of 15 to 25 mag/arcsec$^{2}$ and subdivide galaxies into five classes of double-component (bulge+disk), pure-disk (bulgeless), elliptical, compact (low angular-size objects), and hard (irregulars, mergers). Note that we independently classify the sample to achieve higher accuracy and also estimate our classification error. See \ref{sec:GAMA_err_budget} for more details about our classification error estimation.  

Figure \ref{fig:GAMA_Frac} shows the fraction of each morphological class as a function of stellar mass. As expected, pure-disk and elliptical systems dominate the extreme low and high stellar mass regimes, respectively, while double-component galaxies occupy intermediate masses. In total, we have 3,690 galaxies in the double class, 1,215 in the pure-disk, 491 ellipticals, 16 in compact, and 101 in the hard subclass (irregulars and merger systems). 

\subsection{Distinguishing pseudo-bulges from classical-bulges}
\label{subsec:GAMA_pB_cB_disting}
In \cite{Driver22} we attempted to visually separate disk galaxies containing a pseudo-bulge (pB) from those containing a classical bulge (cB). This classification is based on the central structure of the disk galaxies' visual morphology as to whether it is an extended lower surface brightness structure or a compact high surface brightness structure. The separation of pBs from cBs is a challenging and controversial subject in astronomy (See \cite{Kormendy04} for a review). Attempts to distinguish these components has led some studies to use the Kormendy relation (see e.g., \citealt{Gadotti09}), and some others direct visual inspection (see e.g., \citealt{Fisher08}) and more recently, kinematic decomposition using Integral Field Spectroscopy (IFS, e.g. \citealt{Zhu18}). For this work, we use the \cite{Driver22} visual classification. Later in this thesis, we will discuss using the bulges' stellar surface density as an indicator for separating pBs and cBs (see Section \ref{sec:pB_cB_dist} for more details). 

\section{Stellar Mass Function by Morphological Types} 
\label{sec:GAMA_SMF_morph}
 
Having our morphologically classified sample we now investigate the contribution of each morphology in the total stellar mass density.
To explore the morphologically decomposed SMF in the local Universe we use the galaxy stellar mass estimates of \cite{Bellstedt20b} for GAMA galaxies, which uses the {\sc ProSpect} \citep{Robotham20} SED fitting code with \cite{BC03} stellar synthesis models, assuming a \cite{Chabrier03} Initial Mass Function (IMF).

\begin{figure*} 
	\centering
	\includegraphics[width = \textwidth, height = 1.1\textwidth]{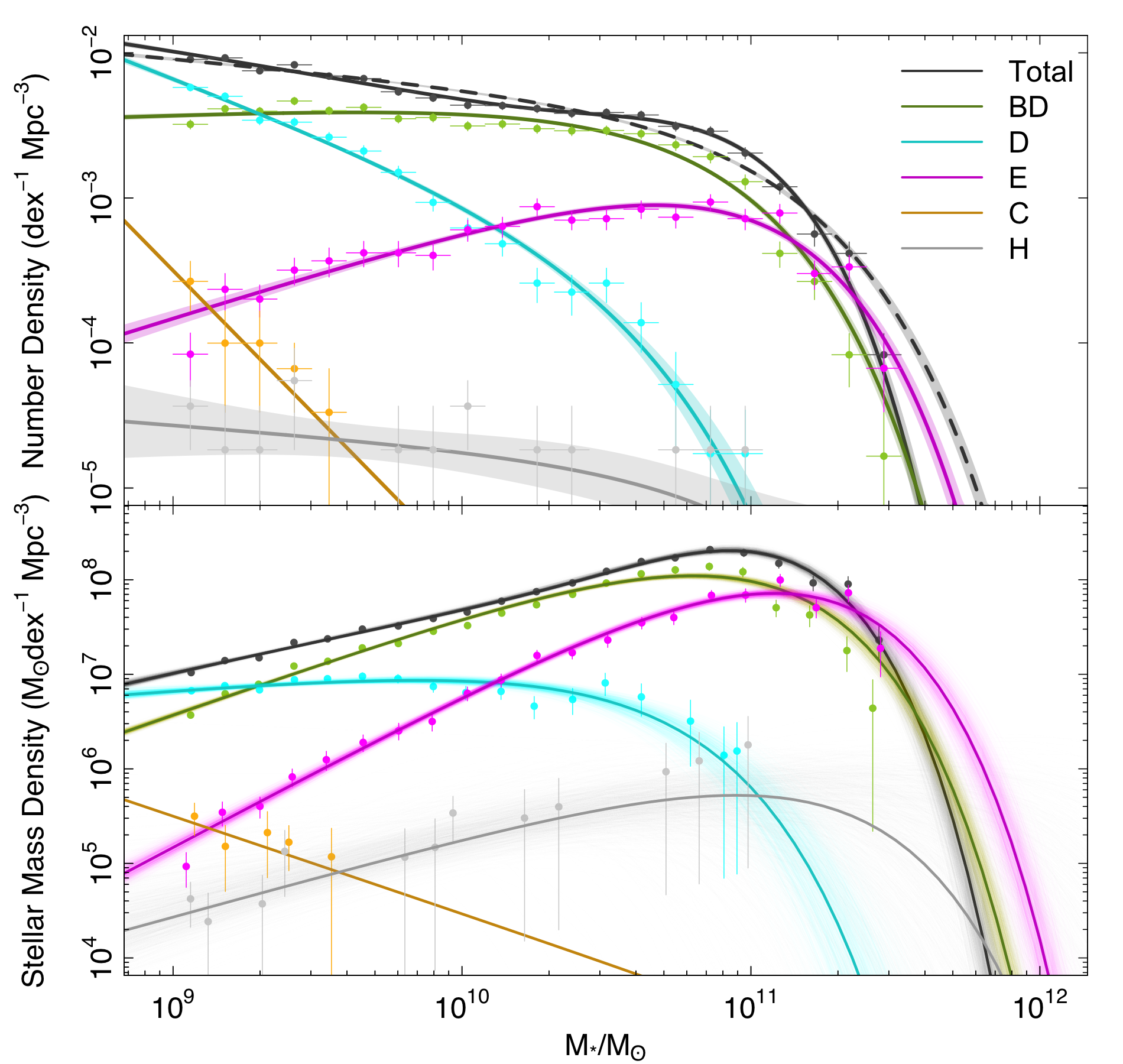}
	\caption{Top: the stellar mass function of GAMA galaxies subdivided into different morphological types. Data points are galaxy counts in each of equal-size stellar mass bins. Width of stellar mass bins are shown as horizontal bars on data points. Vertical bars represent poisson errors. Shaded regions show 68 per cent confidence regions. The dashed black line represents our single Schechter function fit to the total SMF, while the solid black line shows our double Schechter fit.} Bottom: the distribution of the stellar mass density. The transparent shade regions represent the error range calculated by 1000 times sampling of the full posterior probability distribution of the fit parameters.
	\label{fig:GAMA_SMF_morph}
\end{figure*}

We then utilize the {\sc dftools} package \citep{Obreschkow18} to fit single and double Schechter functions to our stellar mass distribution, which uses a modified maximum likelihood (MML) method. The top panel of Figure \ref{fig:GAMA_SMF_morph} shows our total SMF together with the subdivision into the distinct morphological classes. We fit the total stellar mass function with both single and double Schechter functions and find that in agreement with previous studies (see e.g., \citealt{Baldry08}; \citealt{Peng10}; \citealt{Baldry12}; \citealt{Wright17}) the double Schechter form better fits the data. We then subdivide the total SMF into different morphological types. Double-component systems dominate the SMF through most of the stellar mass regime except for the very high-mass end which is dominated by elliptical systems. The bottom panel of Figure \ref{fig:GAMA_SMF_morph} shows the total and morphological distribution of the stellar mass density (SMD). Our total, and almost all morphological mass densities, are bounded within our stellar mass range except for the compact class indicating that integrating under the curves will capture most of the stellar mass in each subclass. We note that although our best fit elliptical SMF slightly exceeds our Total double Schechter fit (face value unphysical), it is only a fitting issue due to the nature of the double Schechter function.
Our best fit line of ellipticals, however, does not exceed our Total single Schechter function. Note that elliptical data points never exceed the Total data points.

We calculate the total SMD by integrating under the SMF best fits and derive the total SMD for the entire galaxy populations of $\rho_* = 10^{8.32} \,\mathrm{M}_\odot\mathrm{Mpc}^{-3}$. Our morphological SMDs are summed in a similar way and result in the values reported in Table \ref{tab:GAMA_MF_par}. We find that $11.6\%$ of the total stellar mass is in pure-disk galaxies while $58.2\%$ and $\sim 30\%$ are in double-component and elliptical systems, respectively. The rest of the stellar mass density is in compact and hard subcategories. Therefore, bulge+disk systems (double-component) dominate the stellar mass of the local Universe. An interesting question directly correlated with the galaxy formation and evolution history is that whether the stellar mass density of the Universe is dominated by double-component systems at earlier epochs or they have evolved significantly and recently token over the stellar mass. On the hand tracing the evolution of the stellar mass locked in pure-disk and elliptical systems will give insight into the mass and morphological transformations in the Universe. We will investigate these questions in Chapters \ref{ch:3} and \ref{ch:4}.  

\begin{figure*} 
	\centering
	\includegraphics[width = \textwidth, height = 1.1\textwidth]{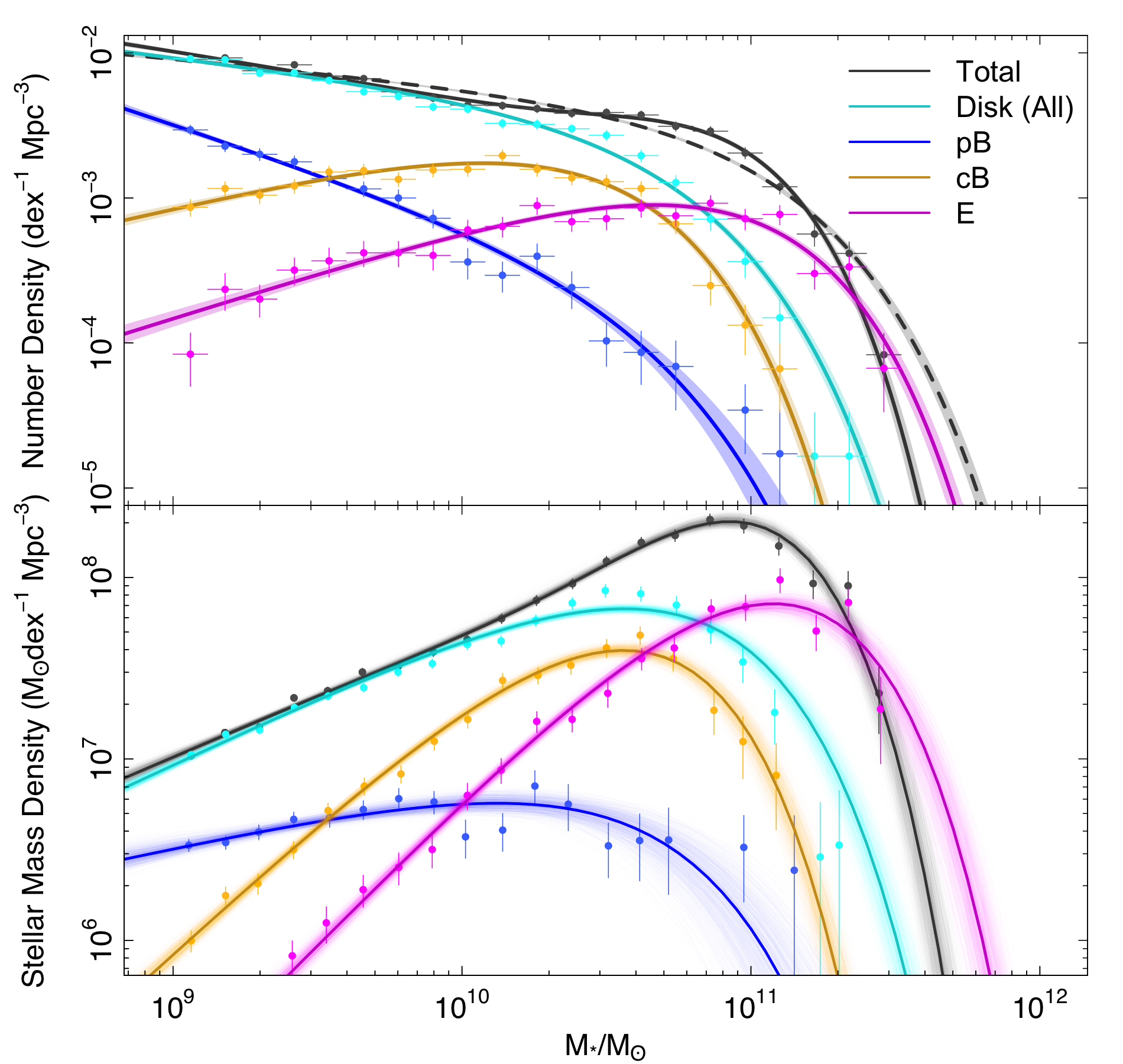}
	\caption{Top: the structural stellar mass function of GAMA galaxies. Disk (all) represent all disks, including pure-disk systems (D) and disk component of the BD systems, while we separate bulge components into pseudo- and classical-bulges (blue and golden colours). The dashed black line represents our single Schechter function fit to the total SMF, while the solid black line shows our double Schechter fit. Bottom: the distribution of the stellar mass density.}
	\label{fig:GAMA_SMF_str}
\end{figure*}

\begin{landscape}
\begin{table}
\centering
\caption{Best Schechter fit parameters as well as associated errors of total and structural SMF of GAMA galaxies.}
\begin{adjustbox}{scale = 0.9, angle = 0}
\begin{tabular}{lcccccccc}
\firsthline \firsthline \\
           &  $\mathrm{log}_{10}(\Phi^*)$ &  $\mathrm{log}_{10}(M^*)$ & $\alpha$ & & &  $\mathrm{log}_{10}(\rho_*) \pm$ FitErr $\pm$ ClassErr $\pm$ CV  \\ \\ \hline \\
  
Total (single Schechter)           & $-2.78\pm0.03$   & $11.03\pm0.03$ & $-1.18\pm0.01$ & & & $8.31\pm0.03\pm0.0\pm6.05$ \\ \\
Pure-Disk        & $-3.63\pm0.13$   & $10.46\pm0.09$ & $-1.76\pm0.05$ & & &  $7.40\pm0.15\pm6.35\pm6.90$ \\ \\
Double           & $-2.66\pm0.02$   & $10.77\pm0.02$ & $-0.93\pm0.02$ & & &  $8.10\pm0.03\pm6.38\pm6.35$ \\ \\  
Elliptical       & $-3.02\pm0.03$   & $10.87\pm0.04$ & $-0.37\pm0.05$ & & &  $7.81\pm0.06\pm5.92\pm6.76$ \\ \\  
Compact          & $-9.74\pm -^*$   & $11.90\pm -^*$ & $-3.03\pm -^*$ & & &  $3.70\pm -^* \pm5.84\pm6.80$ \\ \\  
Hard             & $-5.23\pm0.47$   & $11.02\pm0.47$ & $-1.15\pm0.28$ & & & $5.83\pm1.33\pm5.64\pm5.72$  \\ \\  
Disks (All)      & $-2.81\pm0.034$  & $10.68\pm0.030$ & $-1.25\pm0.021$ & & & $7.96\pm0.03\pm5.6\pm6.14$  \\ \\
pseudo-Bulge     & $-3.72\pm0.013$  & $10.50\pm0.097$ & $-1.56\pm0.065$ & & &  $7.07\pm0.07\pm5.22\pm6.46$ \\ \\
classical-Bulge  & $-2.78\pm0.029$  & $10.43\pm0.031$ & $-0.56\pm0.045$ & & &  $7.59\pm0.05\pm5.90\pm6.21$ \\ \\  
\hline \\
           &  $\mathrm{log}_{10}(\Phi^*_1)$   &  $\mathrm{log}_{10}(M^*)$ &  $\alpha_1$ &  $\mathrm{log}_{10}(\Phi^*_2)$ & $\alpha_2$ &  $\mathrm{log}_{10}(\rho_*)$  \\ \\ \hline  \\

Total (double Schechter)           & $-2.57\pm0.04$  & $10.60\pm0.04$  & $0.39\pm0.2$ & $-2.82\pm0.06$ & $-1.30\pm0.04$ & $8.32\pm0.03\pm0.0\pm5.52$  \\ \\           
           
\lasthline
\end{tabular}
\end{adjustbox}
$^*$As can be seen in Figure \ref{fig:GAMA_SMF_morph} due to the lack of data in the compact subclass within our mass range we have an unconstrained fit so the fitting process fails to return error on parameters.
\label{tab:GAMA_MF_par}
\label{tab:rho}
\end{table}
\end{landscape}

\section{Stellar Mass Function of Disks and Bulges} 
\label{sec:GAMA_SMF_str}

So far, we have found that the majority of the stellar mass of the local Universe is in bulge+disk systems (BD). One can now explore how this mass is distributed between disks and bulges.  
In this section, we make use of the new structural analysis of GAMA galaxies of \cite{Casura-inprep} to investigate this. We use the \texttt{BDDecomp} Data Management Unit (DMU) version 3 (\texttt{BDModelsAllRv03}) available on the GAMA website\footnote{\href{http://www.gama-survey.org}{www.gama-survey.org}}. This DMU consists of single-S\'ersic fits and bulge-disk decompositions of $\sim 13,000$ GAMA galaxies up to $z < 0.08$ obtained from running {\sc ProFit} \citep{Robotham17} on KiDS r-band.
In order to estimate the stellar mass of bulges and disks we make use of the bulge-to-total flux ratio (B/T), i.e., $M_*^\mathrm{Bulge} = \mathrm{B/T} \times M_*^\mathrm{Total}$ and $M_*^\mathrm{disk} = (1-\mathrm{B/T}) \times M_*^\mathrm{Total}$. Note that GAMA structural decomposition catalogue \citep{Casura-inprep} provides different versions of structural parameters (e.g., flux, B/T and $R_e$). Besides the original values directly obtained from their {\sc ProFit} analysis, they also measure the quantities contained within $10 \times R_e$ and within their segmentation isophotes. In this work, as they recommend, we use the latter measurements of B/T, i.e. within segmentation isophotes.

Figure \ref{fig:GAMA_SMF_str} shows the SMF (top) and the distribution of the SMD (bottom) of structures, including all disks (pure disk systems+disk components; cyan), elliptical systems (E) and bulge components separated into pBs (blue) and cBs (golden).
As shown in the top panel of Figure \ref{fig:GAMA_SMF_str}, we find that, as seen in Figure \ref{fig:GAMA_SMF_morph} the high mass end of the SMF is dominated by elliptical galaxies while intermediate and lower masses ($\mathrm{log}(M_*/M_\odot) < 10.85$) are dominated by stellar mass locked within disk structures. Classical bulges (visually distinguished) occupy relatively higher stellar masses than pseudo-bulges. Interestingly, the shape of the SMF of pseudo- and classical-bulges roughly follow that of disks and ellipticals, respectively, implying that they probably had the same evolutionary pathways. 
We note that, as can be seen in Figure \ref{fig:GAMA_SMF_str}, we seem to have one pseudo-bulge+disk galaxy (pBD) with bulge stellar mass of $\mathrm{log(M_{pB}}/\mathrm{M}_\odot) > 11$, i.e., UIDs = 181990423805762 with B/T of 0.7. While highlighting that such high B/T could be unusual for pB systems, we do not rule out possible misclassification of these galaxies. This GAMA paper is still in preparation, so such objects might be reinvestigated (Driver et al. in prep.).
The bottom panel of Figure \ref{fig:GAMA_SMF_str} shows the distribution of the stellar mass density and the contribution of bulges and disks to the total SMD. All of the distributions of SMDs are bounded within our stellar mass range indicating that integrating under these curves will capture the most of stellar mass. We report our best fit Schechter parameters as well as the integrated SMDs of disks and bulges in Table \ref{tab:GAMA_MF_par}. 

We find that $\sim 44\%\pm0.024$ of the total stellar mass of the local Universe is locked in disks while $5.7\%\pm0.058$ in pBs and $\sim 18.8\%\pm0.190$ in cBs and $31\%\pm0.04$ in elliptical systems, and the rest is in hard and compacts subcategories as mentioned in Section \ref{sec:GAMA_SMF_morph}.   

\begin{figure*} 
	\centering
	\includegraphics[width = \textwidth]{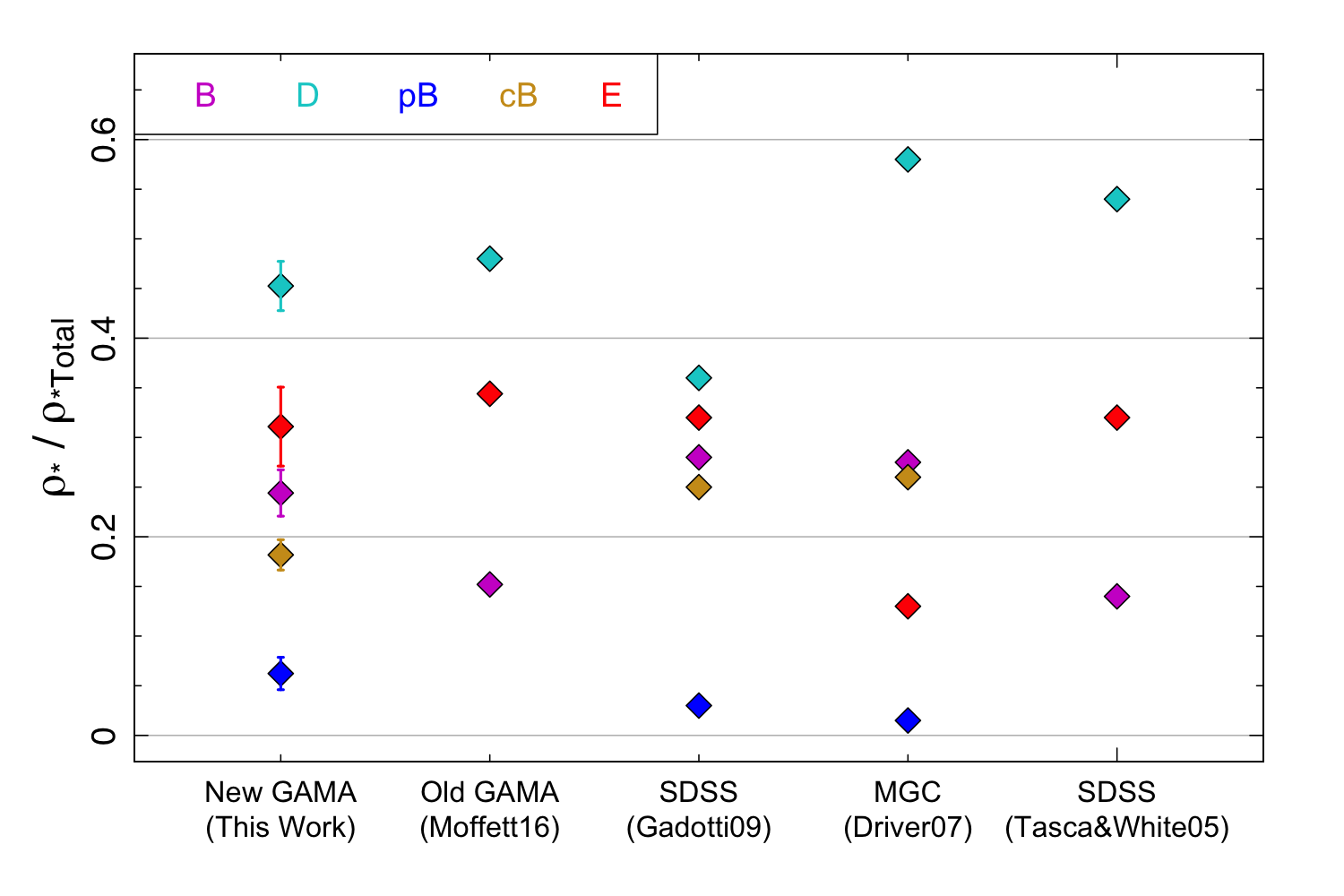}
	\caption{A comparison of the relative contribution of different components in total stellar mass density as calculated in some literature. B, D, pB, cB and E stand for bulge (all), disk (all), pseudo-bulge, classical-bulge and elliptical. The error bars on the New GAMA data represent the combination of the fit error, cosmic variance and classification error. The statistical errors are negligible. See the text for more details.  }
	\label{fig:GAMA_SMD_fraction}
\end{figure*}

In Figure \ref{fig:GAMA_SMD_fraction}, we compare our measurements of the contribution of different galaxy structures of our GAMA sample (noted in the figure as New GAMA) with the SMDs of previous studies including: GAMA (\citealt{Moffett16a}); SDSS (\citealt{Gadotti09} and \citealt{Tasca05}); and the Millennium Galaxy Catalogue (MGC, \citealt{Driver07}). The difference in measurements indicates the uncertainties in these studies reflecting how challenging and subjective these kind of studies can be and discussed in the next sections. 

\section{Error Budget}
\label{sec:GAMA_err_budget}

The analysis of the stellar mass density is affected by different sources of uncertainties. 
These errors include four main sources of error: statistical random error, fitting error, classification error and cosmic variance (CV) out of which the statistical error is negligible and the cosmic variance is expected to be the dominant error. We report our estimation of all errors in Table \ref{tab:GAMA_MF_par}. 

Below, we briefly explain how we estimate each error: 

\textbf{Fit error:} 
We calculate the fit error by 1000 times random sampling of the full posterior probability distribution of our Schechter fit parameters. This leads to the shaded region in the lower panels of Figures \ref{fig:GAMA_SMF_morph} and \ref{fig:GAMA_SMF_str} from which we calculate the fit error on our SMD measurements. As expected these are extremely small.

\textbf{Classification error:} 
Visual morphological classification inevitably incorporates a level of uncertainty into our measurements. We account for this source by repeating our analysis for different iterations of our morphological classifications \cite{Driver22}. In brief, we have classified our sample independently by three experts and finally select the most voted class as the final morphology. We now recalculate the SMFs and SMDs for each of our classifiers and use the spread to obtain a classification error. Figures \ref{fig:GAMA_morph_ClassErr}-\ref{fig:GAMA_str_ClassErr} shows our measurements of SMFs and SMDs with three classifications.        

\textbf{Cosmic variance:}
Cosmic variance (CV) is expected to be the dominant source of error in our analysis. 
One might estimate the CV error using equation 3 of \cite{Driver10} using to the survey area of independent fields, and the redshift range. In case of this work probing 3 regions of G09, G12, and G15 with a total effective area of 169.3 square degrees (G09:54.93; G12: 57.44; G15: 56.93, \citealt{Bellstedt20a}) in the redshift range of $0 < z < 0.08$. This method estimates a $22.77\%$ cosmic variance. We can also empirically calculate the cosmic variance by estimating the SMF for each GAMA region independently and exploring the spread and find a CV error of $\sim 16\%$. Figures \ref{subfig:GAMA_morph_3reg}-\ref{subfig:GAMA_str_3reg} show the morphological and structural SMFs and distribution of SMDs in GAMA regions. For this analysis, we adopt the variation of the integrated SMD between regions as our estimation of the cosmic variance. 

We highlight that structural decomposition of low-mass galaxies is increasingly challenging, making the bulge component parameters for low-mass systems uncertain, including B/T and hence this uncertainty is propagated into our bulge stellar mass measurements. To overcome this problem, one could raise the sample's total stellar mass limit from $10^9 \mathrm{M}_\odot$ to, for example, $10^{10}\mathrm{M}_\odot$. We, however, decided to keep our total stellar mass limit at $10^9$ and instead, as mentioned above, exclude very low mass bulges of $< 10^9 \mathrm{M}_\odot$ by imposing this mass limit on all structure. Hence ensure our structural SMF's are complete. We note that without this cut, we would have bulges (pBs, mainly) with the stellar masses down to $10^5 \mathrm{M}_\odot$.    

\section{Mass-size relation} 
\label{sec:GAMA_MRe}

\begin{figure*} 
	\centering
	\includegraphics[width = \textwidth, height = 1.2\textwidth]{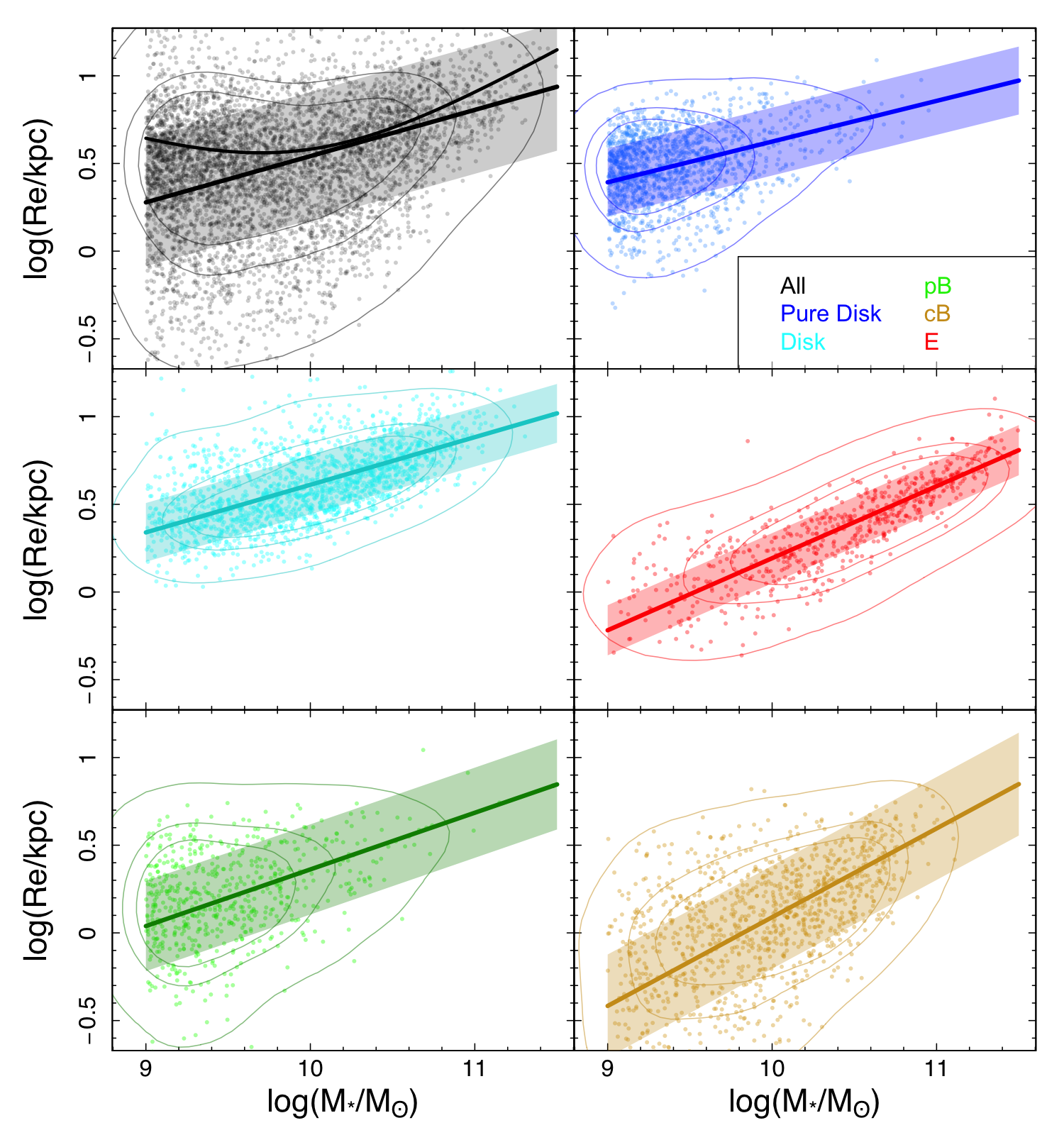}
	\caption{The distribution of the $M_*-R_e$ relation of GAMA galaxies subdivided into different structures. Note that $M_*$ represents the component stellar mass not the total stellar mass. Lines represent power law fits using an MCMC minimization to the data points. The faint regions around regression line represent 1 $\sigma$ of intrinsic scatter.
	The curved line in the top left panel shows the double power law fit to the $M_*-R_e$ distribution of all galaxies. See text for details. Contours represent 50\%, 68\% and 98\% quantiles.}
	\label{fig:GAMA_MRe}
\end{figure*}

In this section, we explore the mass-size relation ($M_*-R_e$) of GAMA galaxies divided into different structures. 
Figure \ref{fig:GAMA_MRe} shows the distribution and fits to our structural $M_*-R_e$ relations.
As mentioned in Section \ref{sec:GAMA_SMF_str} GAMA structural decomposition catalogue \citep{Casura-inprep} provides different size measurements ($R_e$). In this section, we use their sizes measurements contained within the segments. 
Following \cite{Lange15}, \cite{Lange16} and \cite{Shen03} we adopt a power law to fit the $M_*-R_e$ relation:

\begin{equation}
    \mathrm{log}(R_e/kpc) = a\,\mathrm{log}(M_*/M_\odot)-b,
    \label{eq:MRe}
\end{equation}

\noindent where $R_e$ is the half-light radius in kpc and $M_*$ is the stellar mass. To perform the fitting process we make use of the {\sc HyperFit} package \citep{Robotham15} developed in {\sc R} and a Markov Chain Monte Carlo (MCMC) optimization with CHARM algorithm and 10,000 iterations. We report our best fit regression parameters in Table \ref{tab:GAMA_MRe}.  

As shown in Figure \ref{fig:GAMA_MRe}, we fit the total $M_*-R_e$ relation with the above power law function. However, in the low-mass end ($M/M_\odot < 10^{10}$) we find a flattening making the total relation inconsistent with a single power law. Therefore, we provide also a fit with a double power law combining two power law functions as below:

\begin{equation}
R_{e}=\gamma\left(\frac{M_{*}}{M_{\odot}}\right)^{\alpha}\left(1+\frac{M_{*}}{M_{0}}\right)^{\beta-\alpha}
\label{eq:MRe_double_plaw}
\end{equation}

Figure \ref{fig:GAMA_MRe} indicates that above function seems to perform better in lower mass regime of the total $M_*-R_e$ plane than a single power law (Equation \ref{eq:MRe}). However, single power law works reasonably well for our structural $M_*-R_e$ relations. 

\begin{figure*} 
	\centering
	\includegraphics[width = \textwidth, height = 1.2\textwidth]{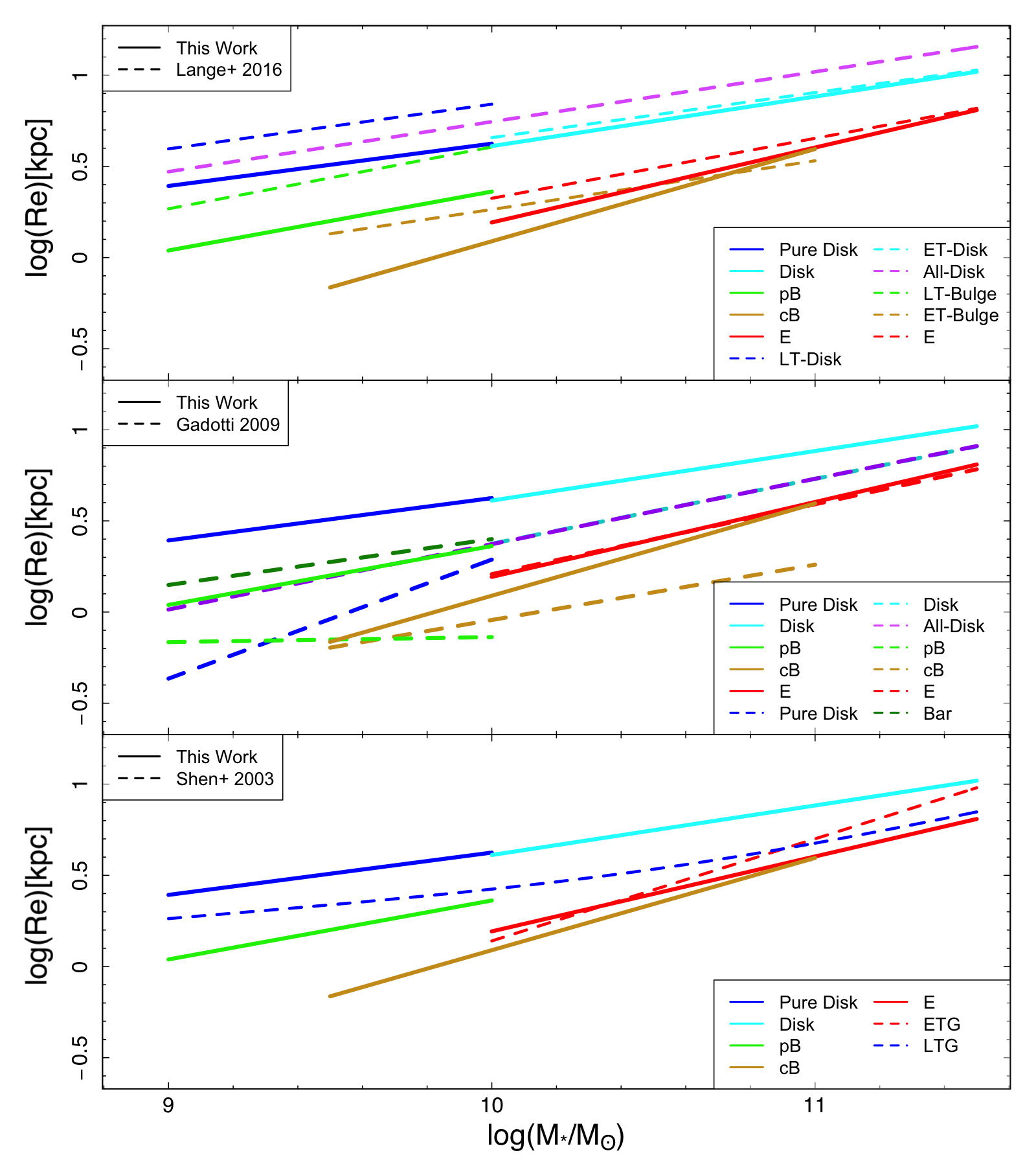}
	\caption{A comparison of our $M_*-R_e$ relations (solid lines) with \protect\cite{Lange16} (dashed lines) and \protect\cite{Shen03} (dotted lines).}
	\label{fig:GAMA_MRe_compare}
\end{figure*}

In Figure \ref{fig:GAMA_MRe_compare} we compare our $M_*-R_e$ relations with the previous GAMA study by \cite{Lange16} (top panel). This was based on a {\sc GALFIT} \citep{Peng02} analysis of lower resolution and shallower SDSS (\citealt{York00}) data. The middle panel shows our results compared with the \cite{Gadotti09} analysis of 1000 nearby well-resolved SDSS galaxies represents the current local gold standard. For completeness, we also show the somewhat older SDSS relations \citep{Shen03} for early- and late-type galaxies (lower panel). We note that not all of the \cite{Lange16} structural classifications are directly comparable to this study. They provide the $M_*-R_e$ relations for all disks subdivided into late- and early-disks (LTD and ETD referring to the disk component of late- and early type galaxies, respectively), as well as bulges subdivided into late- and early-type bulges (LTB and ETD referring to the bulge component of late- and early-type galaxies, respectively). Our total disk relation is consistent with \cite{Lange16}. As can be seen, our disk component is also in good agreement with their early-type disks while their late-type disk are estimated larger than our pure disk systems even though their slopes are in good agreement ($a_{LTD} = 0.245\pm0.008$ versus $a_{pure-disk} = 0.25\pm0.028$). We also show their early- and late-type bulges and find that the slope of our pB is consistent with their LTBs. However, as can be seen in Figure \ref{fig:GAMA_MRe_compare} their LTB/ETB will not be directly comparable to our pB/cB classification. Inconsistent with our result, \cite{Lange16} find on average smaller ellipticals leading to a flatter $M_*-R_e$ relation. This could be due to the fact that our new GAMA imaging data, based on KiDS is deeper than previous GAMA data, SDSS. We also do not rule out the effects of different morphological classifications and possible human error.

Comparing with \cite{Gadotti09}, we find that their $M_*-R_e$ relation for bars (dark green) is consistent with our pB relation. They however find flatter relations for their disk component as well as bulges (both pseudo- and classical-bulges). We note that their classification of pB/cB is based on the Kormendy relation while ours is, as mentioned in Section \ref{subsec:GAMA_pB_cB_disting}, based on a visual inspection. Our results on average lead to larger structures than \cite{Gadotti09}. 

We find that disks are, in general, the largest structures in the Universe, while at the extreme high mass end elliptical systems start to become larger than other structures. Elliptical systems on average occupy the high-mass large-size end of the parameter space. Figure \ref{fig:GAMA_MRe} also highlights that for a given effective radius ($R_e$), disks located within a bulge+disk system are more massive than bulge-less disks (pure-disks; comparing cyan and blue symbols). In addition, this figure reveals that classical-bulges are on average more massive than pseudo-bulges. Further, at a given stellar mass, pBs are larger than cBs indicating higher concentration of mass in cB systems. 

Other studies using higher resolution data (e.g., \citealt{Fisher08, Fisher16} using HST) show that the typical size of pBs is $\sim 0.5$ kpc. However, the $M_*-R_e$ plane of pBs show that we indeed have some bulges with the size of 0.3-1 kpc. However, we believe that the seeing is the main reason why our pB sizes might be unreliable. In Chapter \ref{ch:5}, where we investigate the $M_*-R_e$ planes using our HST data (way better seeing), we find the typical size of our pBs to be ~0.5 kpc (see Figure \ref{fig:MRe}).

As mentioned in Table \ref{tab:GAMA_MRe}, we find the slope of the total $M_*-R_e$ relation to be $0.35\pm0.010$. Pure-disk systems and disk components follow roughly the same slopes, i.e. $0.25\pm0.028$ and $0.27\pm0.010$. Pure-disk systems represent the flattest relation. This is plausible as disks regardless of hosting or not hosting a bulge have essentially a similar nature. cBs have the steepest $M_*-R_e$ relation, slope of $0.60\pm0.025$ even steeper than ellipticals ($a = 0.53\pm0.020$).

\begin{table}
\centering
\caption{The best regression fit parameters of Equation \ref{eq:MRe} to the $M_*-R_e$ data for different structures (see Figure \ref{fig:GAMA_MRe}). Last column represents the intrinsic scatter defined as the distance along the y-axis.}
\begin{adjustbox}{scale = 1}
\begin{tabular}{lcccccccc}
\firsthline \firsthline \\
           &  $a$  & $b$  & scatter \\ \\ \hline \\
  
Total            & $0.264\pm0.032$  & $2.095\pm0.313$   & $0.365\pm0.004$ \\ \\
Pure-Disk        & $0.232\pm0.014$  & $1.695\pm0.135$   & $0.194\pm0.003$ \\ \\
Disk             & $0.272\pm0.015$  & $2.105\pm0.149$   & $0.168\pm0.004$  \\ \\
pseudo-Bulge     & $0.323\pm0.036$  & $2.870\pm0.343$   & $0.257\pm0.007$ \\ \\
classical-Bulge  & $0.506\pm0.054$  & $4.971\pm0.533$   & $0.294\pm0.008$ \\ \\  
Elliptical       & $0.411\pm0.014$  & $3.916\pm0.149$   & $0.143\pm0.004$ \\ \\ \hline \\
                 & $\gamma$ &  $\alpha$ & $\mathrm{log}(M_0/M_\odot)$ & $\beta$ \\ \\ \hline \\
Total            & $1511.712\pm3350$   & $-0.286\pm0.1008$ & $9.961\pm0.176$ & $0.523\pm0.062$ \\ \\           

\lasthline
\end{tabular}
\end{adjustbox}
\label{tab:GAMA_MRe}
\end{table}

  


\section{Summary and Conclusion}
\label{Sec:GAMA_Summ}

In this chapter, we make use of the latest GAMA morphological classification (\citealt{Driver22}) and structural analysis (\citealt{Casura-inprep}) of GAMA galaxies to provide updated stellar mass functions by morphology and structure. We also present an updated stellar mass-size relation of local galaxies. 

We summarize our $z = 0$ benchmark based on GAMA survey as follows:

\begin{itemize}
    \item the total stellar mass function can be well described by a double Schechter function with the characteristic mass of $M^*/M_\odot = 10^{10.6}$.
    
    \item elliptical galaxies dominate the SMF at the high mass end, i.e. $M_*/M_\odot \gtrsim 10^{11}$. 
    
    \item double-component systems dominate the SMF at the stellar mass range of $2\times 10^{9} \lesssim M_*/M_\odot \lesssim 10^{11}$.
    
    \item pure-disk systems occupy the lower mass end of SMF and dominate over double-component galaxies in $M_*/M_\odot \lesssim 2\times 10^{9}$.
    
    \item the analysis of our structural SMF implies that in $M_*/M_\odot \lesssim 7\times 10^{10}$ the vast majority of the stellar mass of the local Universe is locked in disk structures while beyond this point the stellar mass is dominantly accumulated in elliptical systems.
    
    \item we utilize the visual classification of bulge component into pseudo- and classical-bulges (\citealt{Driver22}) and find that as expected cBs occupy higher stellar mass range while pBs dominate lower mass end of the SMF. In fact, we find that bulges within the stellar mass regime of $M_*/M_\odot \gtrsim 4\times 10^{9}$ are dominantly cBs while below this point pBs are dominant. 
    
    \item we analyse the stellar mass-size relation of GAMA galaxies and their bulges and disks separately and find that pure-disks and disk components exactly follow the same relation with the slope of 0.27.
    
    \item the total $M_*-R_e$ relation can be better described by a double power law function. 
    
    \item at a given stellar mass, disks are in general larger than other structures except for the very high stellar mass end where ellipticals start to become larger than other structures. 
    
    \item at a given half-light radius ($R_e$), disks located within a bulge+disk system are more massive than bulgeless disk systems (pure-disks).
    
    \item at a given $R_e$, cBs are more massive than pBs indicating that cBs are more dense structures. 
    
    \item cBs follow steeper relation compared with pBs; slope of 0.60 versus 0.42.
    
\end{itemize}

We quantify that the stellar mass density of the Universe is largely dominated by bulge+disk systems ($\sim 60\%$) while elliptical and pure disk systems occupy $\sim 31\%$ and $\sim 12\%$ of the total SMD. In addition, our results indicate that $\sim 44\%$ of the total stellar mass density of the Universe is in disk structures while only $24.2\%$ is in the bulge component with $18.8\%$ and $5.7\%$ in cBs and pBs, respectively. Hence D+pB = $50\%$ and E+cB = $50\%$, i.e. the dichotomy between disk and spheroid structures is perfectly balanced and the origin of this dichotomy as yet unknown.

We conclude that although studying the stellar mass budget of the local Universe is crucial and informative, further investigations of higher redshift galaxies is required to better explain the origin of this dichotomy and evolution of galaxies of different morphological types. Furthermore, exploring the evolution of galaxies and their components since high-$z$ will give more insight into the emergence and growth of bulges and disks in the Universe.


\graphicspath{{images/ChapterThree/}}

\chapter[DEVILS: Mass Growth of Morphological Types]{Deep Extragalactic VIsible Legacy Survey (DEVILS): Stellar Mass Growth by Morphological Type since $z = 1$}
\label{ch:3}

This Chapter has been accepted as a paper in the Monthly Notices of the Royal Astronomical Society with myself as the first author.

\section{Abstract}
Using high-resolution Hubble Space Telescope imaging data, we perform a visual morphological classification of $\sim 36,000$ galaxies at $z < 1$ in the DEVILS/COSMOS region. As the main goal of this study, we derive the stellar mass function (SMF) and stellar mass density (SMD) sub-divided by morphological types. We find that visual morphological classification using optical imaging is increasingly difficult at $z > 1$ as the fraction of irregular galaxies and merger systems (when observed at rest-frame UV/blue wavelengths) dramatically increases.
We determine that roughly two-thirds of the total stellar mass of the Universe today was in place by $z \sim 1$. Double-component galaxies dominate the SMD at all epochs and increase in their contribution to the stellar mass budget to the present day. 
Elliptical galaxies are the second most dominant morphological type and increase their SMD by $\sim 2.5$ times, while by contrast, the pure-disk population significantly decreases by $\sim 85\%$. According to the evolution of both high- and low-mass ends of the SMF, we find that mergers and in-situ evolution in disks are both present at $z < 1$, and conclude that double-component galaxies are predominantly being built by the in-situ evolution in disks (apparent as the growth of the low-mass end with time), while mergers are likely responsible for the growth of ellipticals (apparent as the increase of intermediate/high-mass end).

\setlength{\extrarowheight}{0pt}

\section{Introduction}

The galaxy population in the local Universe is observed to be bimodal. This bimodality manifests in multiple properties such as colour, morphology, metallicity, light profile shape and environment (e.g. \citealt{Kauffmann03}; \citealt{Baldry04}; \citealt{Brinchmann04}). This bimodality is also found to extend to earlier epochs (see e.g., \citealt{Strateva01}; \citealt{Hogg02}; \citealt{Bell04}; \citealt{Driver06}; \citealt{Taylor09}; \citealt{Brammer09}; \citealt{Williams09}; \citealt{Brammer11}) across many measurable parameters, for example in colour (e.g. \citealt{Cirasuolo07}; \citealt{Cassata08}; \citealt{Taylor15}), morphological type (e.g. \citealt{Kelvin14}; \citealt{Whitaker15}; \citealt{Moffett16a}; \citealt{Krywult17}), size (e.g. \citealt{Lange15}), and specific star formation rate (e.g. \citealt{Whitaker14}; \citealt{Renzini15} and the references therein). The inference from these observations is that there are likely two evolutionary pathways regulated by mass and environment giving rise to this bimodality (\citealt{Driver06}; \citealt{Scarlata07b}; \citealt{DeLucia07}; \citealt{Peng10}; \citealt{Trayford16}). However, it is unclear as to whether studying the global properties of galaxies or their individual morphological components can better explain the origin of this bimodality \citep{Driver13} or whether more complex astrophysics is required.  

In this study, we explore the origin of this bimodality by using the global properties of galaxies to study the evolution of their stellar mass function (SMF). The SMF, a statistical tool for measuring and constraining the evolution of the galaxy population, is defined as the number density of galaxies per logarithmic mass interval (\citealt{Schechter76}). The SMF of galaxies in the local Universe is now very well studied and found to be described by a two-component \cite{Schechter76} function with a characteristic cut-off mass of between $10^{10.6}$-$10^{11}\mathrm{M}_\odot$ and a steepening to lower masses (for example see: \citealt{Baldry08}; \citealt{Peng10}; \citealt{Baldry12}; \citealt{Kelvin14}; \citealt{Weigel16}; \citealt{Moffett16a}; \citealt{Wright17} ). Several studies have also investigated the evolution of the SMF at higher redshifts (\citealt{Pozzetti10}; \citealt{Muzzin13}; \citealt{Whitaker14}; \citealt{Leja15}; \citealt{Mortlock15}; \citealt{Wright18}; \citealt{Kawinwanichakij20}). A fingerprint of this bimodality is also observed in the double-component Schechter function required to fit the local SMF (e.g. \citealt{Baldry12}; \citealt{Wright18}). At least two distinct galaxy populations corresponding to \textit{star-forming} and \textit{passive} systems are thought to be the origin of this bimodal shape, with \textit{star-forming} systems dominating the low-mass tail and \textit{passive} galaxies dominating the high-mass ``hump'' of the SMF (\citealt{Baldry12}; \citealt{Muzzin13}; \citealt{Wright18}). 

Many previous studies have separated galaxies into two main populations of star forming and passive (or a proxy thereof), and measured their individual stellar mass assemblies (e.g. \citealt{Pozzetti10}; \citealt{Tomczak14}; \citealt{Leja15}; \citealt{Davidzon17}). For example, by separating their sample into early- and late-type galaxies based on colour, \cite{Vergani08} studied the SMF of the samples and confirmed that $\sim 50\%$ of the red sequence galaxies were already formed by $z \sim 1$. 
\cite{Pannella06} used a sample of $\sim 1600$ galaxies split into early- intermediate- and late-type galaxies classified according to their S\'ersic index. Similar work was carried out by \cite{Bundy05} and \cite{Ilbert10}. In the local Universe, some studies have classified small samples of galaxies and investigated the SMF of different morphological types (\citealt{Fukugita07} and \citealt{Bernardi10} using The Sloan Digital Sky Survey, SDSS). For example \cite{Bernardi10} reported that elliptical galaxies contain 25\% of the total stellar mass density and 20\% of the luminosity density of the local Universe.

Splitting galaxies into only two broad populations of star-forming and passive (or early- and late-type) galaxies might be inadequate to comprehensively study all aspects of galaxy evolution (e.g., \citealt{Siudek18}). Most previous studies at higher redshifts have separated galaxies into early- and late-type according to their colour distribution mainly due to the lack of spatial resolution as at high redshifts galaxies become unresolved in ground-based imaging. This has restricted true morphological comparisons in most studies to the local Universe (see e.g., \citealt{Fontana04}; \citealt{Baldry04}). A disadvantage of a simple colour-based classification is at higher redshifts, where for example regular disks occupied by old stellar populations are likely to be classified as early-type systems (see e.g., \citealt{Pannella06}). This is however unlikely to have an impact on the visual morphological classification, particularly when using high resolution imaging such as the Hubble Space Telescope (HST) data (\citealt{Huertas-Company15}). As such, we need better ways of quantifying the structure of galaxies throughout the history of the Universe. 
Consequently, some studies have started to probe visually classified morphological types (e.g., \citealt{Sandage05}; \citealt{Fukugita07}; \citealt{Nair10}; \citealt{Lintott11} and the references therein). In effect, there are two morphologies one might track: the end-point morphology at $z \sim 0$ or the instantaneous morphology at the observed epoch. Observing galaxies at higher redshifts we find a different distribution of morphological types. For example, in earlier works using HST, \cite{vandenBergh96} and \cite{Abraham96} presented a morphological catalogue of galaxies in the Hubble Deep Field and found significantly more interacting, merger and asymmetric galaxies than in the nearby Universe.

By studying the SMF of various morphological types in the local Universe ($z < 0.06$), visually classified in the Galaxy and Mass Assembly survey (GAMA, \citealt{Driver11}), \cite{Kelvin14} and \cite{Moffett16a} found that the local stellar mass density is dominated by spheroidal galaxies, defined as E/S0/Sa. They reported the contribution of spheroid- (E/S0ab) and disk-dominated (Sbcd/Irr) galaxies of approximately 70\% and 30\%, respectively, towards the total stellar mass budget of the local Universe. These studies, however, are limited to the very local Universe. For a better understanding of the galaxy formation and evolution processes at play in the evolution of different morphological types, and their rates of action, we need to extend similar analyses to higher redshifts. This will allow us to explore the contribution of different morphological types to the stellar mass budget as a function of time and elucidate the galaxy evolution process.

In the present work, we perform a visual morphological classification of galaxies in the DEVILS-D10/COSMOS field (\citealt{Scoville07}; \citealt{Davies18}). We make use of the high resolution HST imaging and use a sample of galaxies within $0 \le z \le 1$ and $\mathrm{M}_* > 10^{9.5}$, selected from the DEVILS sample and analysis, for which classification is reliable. This intermediate redshift range is a key phase in the evolution of the Universe where a large fraction ($\sim 50\%$) of the present-day stellar mass is formed and large structures such as groups, clusters and filaments undergo a significant evolution (\citealt{Davies18}). During this period the sizes and masses of galaxies also appear to undergo significant evolution (e.g. \citealt{vanderWel12}). 

By investigating the evolution of the SMF for different morphological types across cosmic time we can probe the various systems within which stars are located, and how they evolve. In this paper, we investigate the evolution of the stellar mass function of different morphological types and explore the contribution of each morphology to the global stellar mass build-up of the Universe. Shedding light on the redistribution of stellar mass in the Universe and also the transformation and redistribution of the stellar mass between different morphologies likely explains the origin of the bimodality in galaxy populations observed in the local Universe. 

The ultimate goal of this study and its companion papers is to not only probe the evolution of the morphological types but also study the formation and evolution of the galaxy structures, including bulges and disks. This will be presented in a companion paper (Hashemizadeh et al. in prep.), while in this paper we explore the visual morphological evolution of galaxies and conduct an assessment into the possibility of the bulge formation scenarios by constructing the global stellar mass distributions and densities of various morphological types. 

In Sections \ref{sec:DEVILS}-\ref{subsec:SampleSel}, we define our sample and sample selection. Section \ref{sec:MorphClass} presents different methods that we explore for the morphological classification. In Section \ref{sec:FitSMF}, we describe the parameterisation of the SMF. We then show the \textit{total} SMF at low-$z$ in Section \ref{subsec:MfunZ0}. Section \ref{subsec:LSS_cor} describes the effects of the cosmic large scale structure due to the limiting size of the DEVILS/COSMOS field, and the technique we use to correct for this. The evolution of the SMF and the SMD and their subdivision by morphological type are discussed in Sections \ref{subsec:MfunEvol}-\ref{sec:rho}. We finally discuss and summarize our results in Sections \ref{sec:ch3_discussion}-\ref{sec:ch3_summary}.

Throughout this paper, we use a flat standard $\Lambda$CDM cosmology of $\Omega_{\mathrm{M}} = 0.3$, $\Omega_\Lambda = 0.7$ with $H_0 = 70 \mathrm{km}\mathrm{s}^{-1}\mathrm{Mpc}^{-1}$. Magnitudes are given in the AB system.

\subsection{DEVILS: Deep Extragalactic VIsible Legacy Survey}
\label{sec:DEVILS}

The Deep Extragalactic VIsible Legacy Survey (DEVILS) \citep{Davies18}, is an ongoing magnitude-limited spectroscopic and multi-wavelength survey. Spectroscopic observations are currently being undertaken at the Anglo-Australian Telescope (AAT), providing spectroscopic redshift completeness of $> 95\%$ to Y-mag $< 21.2$ mag. 
This spectroscopic sample is supplemented with robust photometric redshifts, newly derived photometric catalogues (Davies et al. in prep.) and derived physical properties (\citealt{Thorne20}, and the work here) all undertaken as part of the DEVILS project. These catalogues extend to Y$\sim25$ mag in the D10 region.

The objective of the DEVILS campaign is to obtain a sample with high spectroscopic completeness extending over intermediate redshifts ($0.3 < z < 1.0$). At present, there is a lack of high completeness spectroscopic data in this redshift range (e.g., \citealt{Davies18}; \citealt{Scodeggio18}). DEVILS will fill this gap and allow for the construction of group, pair and filamentary catalogues to fold in environmental metrics. The DEVILS campaign covers three well-known fields: the XMM-Newton Large-Scale Structure field (D02: XMM-LSS), the Extended Chandra Deep Field-South (D03: ECDFS), and the Cosmological Evolution Survey field (D10: COSMOS). In this work, we only explore the D10 region as this overlaps with the HST COSMOS imaging. The spectroscopic redshifts used in this paper are from the DEVILS combined spectroscopic catalogue which includes all available redshifts in the COSMOS region, including zCOSMOS (\citealt{Lilly07}, \citealt{Davies15a}), hCOSMOS \citep{Damjanov18} and DEVILS \citep{Davies18}. As the DEVILS survey is ongoing the spectroscopic observations for our full sample are still incomplete (the completeness of the DEVILS combined data is currently $\sim 90$ per cent to Y-mag $= 20$). Those objects without spectroscopic redshifts are assigned photometric redshifts in the DEVILS master redshift catalogue (\texttt{DEVILS\_D10MasterRedshiftCat\_v0.2} catalogue), described in detail in \cite{Thorne20}. In this work, we also use the stellar mass measurements for the D10 region (\texttt{DEVILS\_D10ProSpectCat\_v0.3} catalogue) reported by \cite{Thorne20}.   
Briefly, to estimate stellar masses they used the {\sc ProSpect} SED fitting code \citep{Robotham20} adopting \cite{BC03} stellar libraries, the \cite{Chabrier03} IMF together with \cite{Charlot00} to model dust attenuation and \cite{Dale14} to model dust emission. This study makes use of the new multiwavelength photometry catalogue in the D10 field (\texttt{DEVILS\_PhotomCat\_v0.4}; Davies et al. in prep.) and finds stellar masses $\sim 0.2$ dex higher than in COSMOS2015 \citep{Laigle16} due to differences in modelling an evolving gas phase metallicity. See \cite{Thorne20} for more details.

\subsection{COSMOS ACS/WFC imaging data}
\label{subsec:ACS/HST}

The Cosmic Evolution Survey (COSMOS) is one of the most comprehensive deep-field surveys to date, covering almost 2 contiguous square degrees of sky, designed to explore large scale structures and the evolution of galaxies, AGN and dark matter \citep{Scoville07}. The high resolution Hubble Space Telescope (HST) F814W imaging in COSMOS allows for the study of galaxy morphology and structure out to the detection limits. In total COSMOS detects $\sim 2$ million galaxies at a resolution of $< 100$ pc \citep{Scoville07}.

The COSMOS region is centred at RA = $150.121$ ($10:00:28.600$) and DEC = $+2.21$ ($+02:12:21.00$) (J2000) and is supplemented by 1.7 square degrees of imaging with the Advanced Camera for Surveys (ACS\footnote{ACS Hand Book: \href{http://www.stsci.edu/hst/acs/documents/handbooks/current/c05\_imaging7.html\#357803}{www.stsci.edu/hst/acs/documents/}}) on HST. This 1.7 square degree region was observed during 590 orbits in the F814W (I-band) filter and also, 0.03 square degrees with F475W (g-band). In this study we exclusively use the F814W filter, not only providing coverage but also suitable rest-frame wavelength for the study of optical morphology of galaxies out to $z\sim1$ \citep{Koekemoer07}. The original ACS pixel scale is 0.05 arcsec and consists of a series of overlapping pointings. These have been drizzled and re-sampled to 0.03 arcsec resolution using the MultiDrizzle code (\citealt{Koekemoer03}), which is the imaging data we use in this work. The frames were downloaded from the public NASA/IPAC Infrared Science Archive (IRSA) webpage\footnote{\href{https://irsa.ipac.caltech.edu/data/COSMOS/images/acs\_2.0/I/}{irsa.ipac.caltech.edu/data/COSMOS/images/acs\_2.0/I/} } as fits images.

\begin{figure}
	\centering
	\includegraphics[width = \textwidth]{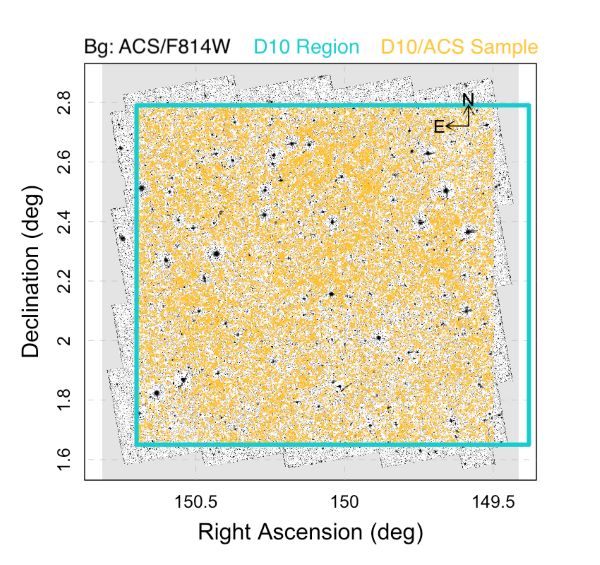}
	\caption{ Background shows the ACS/F814W mosaic image of the COSMOS field. The cyan rectangle represents the D10 region in the DEVILS survey. Gold points are the D10/ACS sources used in this work consisting of $\sim 36$k galaxies. See section \ref{subsec:SampleSel} for more detail. 
    }
	\label{fig:F814W_DEVILS_smp_z1}
\end{figure}

\begin{figure} 
	\centering
	\includegraphics[width = 0.6\textwidth]{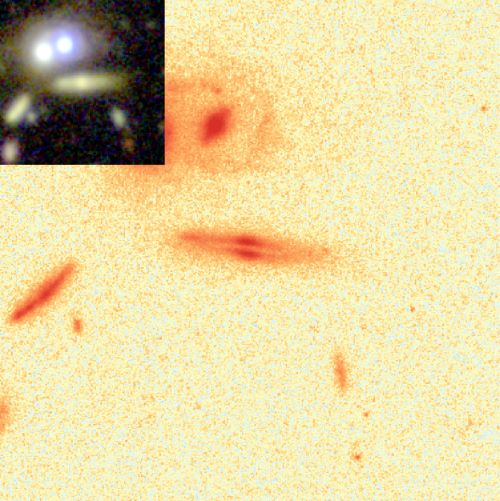}
	\caption{ A random galaxy in our sample at redshift $z \sim 0.47$ showing an example of the postage stamps we generate to perform our visual inspection. Main image is the HST ACS/F814W and the inset is SUBARU $gri$ colour image. The image shows a cutout of $5\times \mathrm{R90}$ on each side, where $\mathrm{R90} = 3.23$ arcsec measured from UltraVISTA Y-band \citep{Davies18}. }
	\label{fig:smplPostageStamp}
\end{figure}

\subsection{Sample selection: D10/ACS Sample} 
\label{subsec:SampleSel}

As part of the DEVILS survey we have selected an HST imaging sample with which to perform various morphological and structural science projects. 
Figure \ref{fig:F814W_DEVILS_smp_z1}, shows the ACS mosaic imaging of the COSMOS field with the full D10 region overlaid as a cyan rectangle (defined from the Ultra VISTA imaging). Our final sample, D10/ACS, is the common subset of sources from ACS and D10. The position of the D10/ACS sample on the plane of RA and DEC is overplotted on the same figure as yellow dots. Note that as shown in Figure \ref{fig:F814W_DEVILS_smp_z1}, we exclude objects in the jagged area of the ACS imaging leading to a rectangular effective area of $1.3467$ square degrees. 

Although our HST imaging data is high resolution (see Figure \ref{fig:smplPostageStamp} for a random galaxy at redshift $z \sim 0.47$), we are still unable to explore galaxy morphologies beyond a certain redshift and stellar mass in our analysis as galaxies become too small in angular scale or too dim in surface brightness to identify morphological substructures.

\begin{landscape}
\begin{figure} 
	\centering
	\includegraphics[width=1.3\textheight]{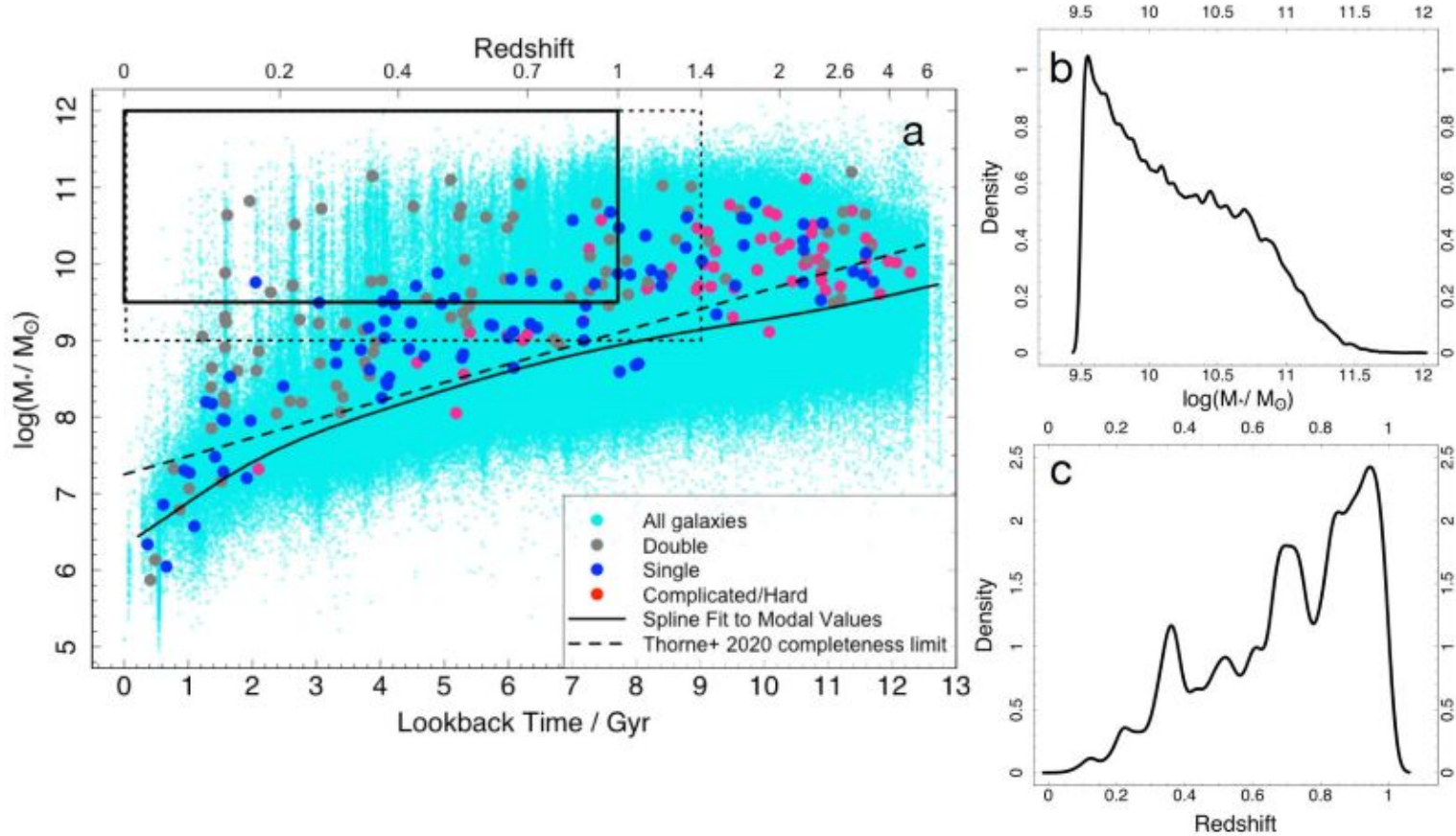}
	\caption{ (a) The relation between stellar mass and lookback time (redshift) of our sample. All galaxies are shown in cyan in background. Circles represent 284 galaxies that we randomly sample for initial visual inspection. Grey and blue circles show double and single component galaxies, respectively. Red symbols are complicated galaxies which consist of merging systems, perturbed galaxies, low S/N or high redshift clumpy galaxies. The dotted rectangle corresponds to our initial sample region ($z < 1.4$ and log$(\mathrm{M}_*/\mathrm{M}_{\odot}) > 9$). The solid rectangle shows our final sample region which covers galaxies up to $z < 1.0$ and log$(\mathrm{M}_*/\mathrm{M}_{\odot}) > 9.5$. The solid black line represents the spline fit to the modal value of the stellar mass in bins of lookback time indicating the stellar mass completeness. Black dashed line shows \protect\cite{Thorne20} completeness limit. See text for more details. Panels (b) and (c) display the distribution of stellar mass and redshift of our final sample (i.e. within the solid rectangle). Note that the PDFs are smoothed by a kernel with standard deviation of 0.02. }
	\label{fig:mzsmplsel}
\end{figure}
\end{landscape}

We first try to select a complete galaxy sample from the DEVILS-D10/COSMOS data to define the redshift and stellar mass range for which we can perform robust morphological classification and structural decomposition (Hashemizadeh et al. in prep.). Note that we make use of a combination of photometric and available spectroscopic redshifts as well as stellar masses as described in Section \ref{sec:DEVILS}. In total, at the time of writing this paper, $23,264$ spectroscopic redshifts are available in the D10 region (excluding the jagged edges) out of which $2,903$ redshifts are observed by DEVILS. See the DEVILS website\footnote{\href{https://devilsurvey.org}{https://devilsurvey.org}} for a full description of these data.
We select 284 random galaxies drawn from across the entire redshift and stellar mass distribution, and visually inspect them. These 284 galaxies are shown as circles in Figure \ref{fig:mzsmplsel}. From our visual inspection, we identify the boundaries within which we believe the majority of galaxies are sufficiently resolved that morphological classifications should be possible (i.e., not too small or faint). Our visual assessments are indicated by colour in Figure \ref{fig:mzsmplsel} showing two-component (grey), single component (blue), and problematic cases (red; merger, disrupted, and low S/N). Unsurprisingly, in agreement with other studies (e.g. \citealt{Conselice05}), we find the fraction of problematic galaxies increases drastically at high redshifts ($z > 1.4$). Beyond this redshift an increasing number of galaxies ($\sim 50\%$) become complex and no longer adhere to a simplistic picture of a central bulge plus disk system. A large fraction of galaxies appear interacting, clumpy, very faint and/or extremely \textit{compact}, hence the notion of galaxies as predominantly bulge plus disk systems becomes untenable. Note that the vast majority of galaxies in this redshift range are disturbed interacting systems, however our observation of a fraction of the clumpy galaxies could be due to the fact that the bluer rest-frame emission is more sensitive to star forming regions dominating the flux. The clumps and potential bulges at this epoch, are mostly comparable or smaller than the HST PSF, hence conventional 2D bulge+disk fits are also unlikely to be credible even at HST resolution. 

\begin{figure*}
	\centering
	\includegraphics[width = \textwidth]{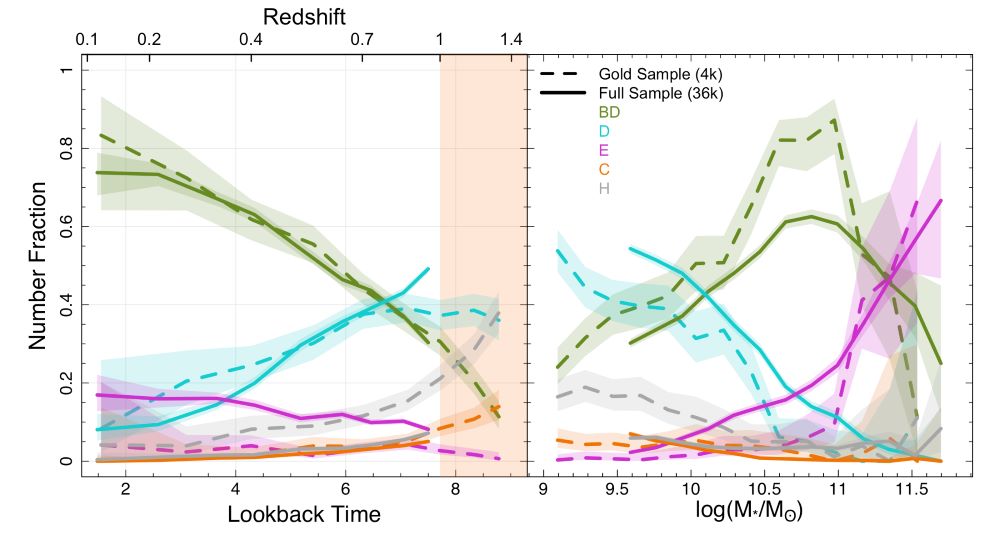}
	\caption{ The number fraction of each morphological type (BD: \textit{bulge+disk}, D: \textit{pure-disk}, E: \textit{elliptical}, C: \textit{compact}, H: \textit{hard}) in bins of lookback time (redshift) and total stellar mass, left and right panels, respectively. Each colour represents a morphology as indicated in the inset legend. Dashed lines are the $\sim4000$ random sample that we visually classified ($z < 1.4$) while solid lines are our full visual inspection of the final sample. Shaded stripes around lines display 95\% confidence intervals from beta distribution method as calculated using {\fontfamily{qcr}\selectfont prop.test} in R package {\fontfamily{qcr}\selectfont stats}.}
	\label{fig:frac_z1}
\end{figure*}

Note that as can be seen in Figure \ref{fig:mzsmplsel}, we also start to suffer significant mass incompleteness at low redshifts due to the COSMOS F814W sensitivity limit. The sample completeness limit is shown here as a smooth spline fitted to the modal values of stellar mass in bins of lookback time (shown as a solid line), i.e., peaks of the stellar mass histograms in narrow redshift slices. We also show the completeness limit reported by \cite{Thorne20} and shown as a dashed line in Figure \ref{fig:mzsmplsel}, which is based on a $g-i$ colour analysis and is formulated as $\mathrm{log(M_*)} = 0.25t + 7.25$, where $t$ is lookback time in Gyr. From this initial inspection we define our provisional window of 105,185 galaxies at $z < 1.4$ and log$(\mathrm{M}_*/\mathrm{M}_{\odot}) > 9$, shown as a dotted rectangle in Figure \ref{fig:mzsmplsel}.

To further tune this selection, we then generated postage stamps of $4,000$ random galaxies (now initial selection) using the HST F814W imaging and colour $gri$ insets from the Subaru Suprime-Cam data \citep{Taniguchi07}. Figure \ref{fig:smplPostageStamp} shows an example of the cutouts we generated for our visual inspection. The postage stamps are generated with $5 \times \mathrm{R90}$ on each side, where $\mathrm{R90}$ is the radius enclosing 90\% of the total flux in the UltraVISTA Y-band (soon to be presented in DEVILS photometry catalogue, Davies et al. in prep.).  
These stamps were independently reviewed by five authors (Simon Driver, Luke Davies, Aaron Robotham, Sabine Belstedt and Abdolhosein Hashemiozadeh) and classified into single component, double component and complicated systems (hereafter: \textit{hard}). The single component systems were later subdivided into disk or elliptical systems. Note that the \textit{hard} class consists of asymmetric, merging, clumpy, extremely compact and low-S/N systems, for which 2D structural decomposition would be unlikely to yield meaningful output. Objects with three or more votes in one category were adopted and more disparate outcomes discussed and debated until an agreement was obtained. In this way, we established a ``gold calibration sample'' of 4k galaxies to justify our final redshift and stellar mass range, and for later use as a training sample in our automated-classification process, see Section \ref{sec:MorphClass} for more details. 

Figure \ref{fig:frac_z1} shows the fraction of each of the above classifications versus redshift and total stellar mass (dashed lines). As the left panel shows, the fraction of \textit{hard} galaxies (gray dashed line) drastically increases at $z > 1$. At the highest redshift of our sample, $z \sim 1.4$, 40 percent of the galaxies are deemed unfittable, or at least inconsistent with the notion of a classical/pseudo-bulge plus disk systems (this is consistent with \citealt{Abraham96b} and \citealt{Conselice05}). Also see the review by \cite{Abraham01}. We therefore further restrict our redshift range to $z \leq 1$ in our full-sample analysis. Additionally, we increase our stellar mass limit to $\mathrm{log(M}_*/\mathrm{M}_{\odot}) = 9.5$ to reduce the effects of incompleteness (see Figure \ref{fig:mzsmplsel}) and to restrict our sample to a manageable number for morphological classifications. 
In Figure \ref{fig:mzsmplsel}, our final sample selection region is now shown as a solid rectangle. The distribution of the stellar mass and redshift of our final sample is also displayed in the right panels (b and c) of the same figure.  

Overall, our analysis of galaxies in different regions of the $M_*$-$z$ parameter space leads us to a final sample of galaxies for which we can confidently study their morphology and structure. We conclude that we can study the structure of galaxies up to $z \sim 1$ and down to log$(\mathrm{M}_{*}/\mathrm{M}_{\odot}) \geq 9.5$. Within this selection, our sample consists of $35,803$ galaxies with $14,036$ available spectroscopic redshifts.

\begin{figure*}
	\centering
	\includegraphics[width=\textwidth, height=11.5cm]{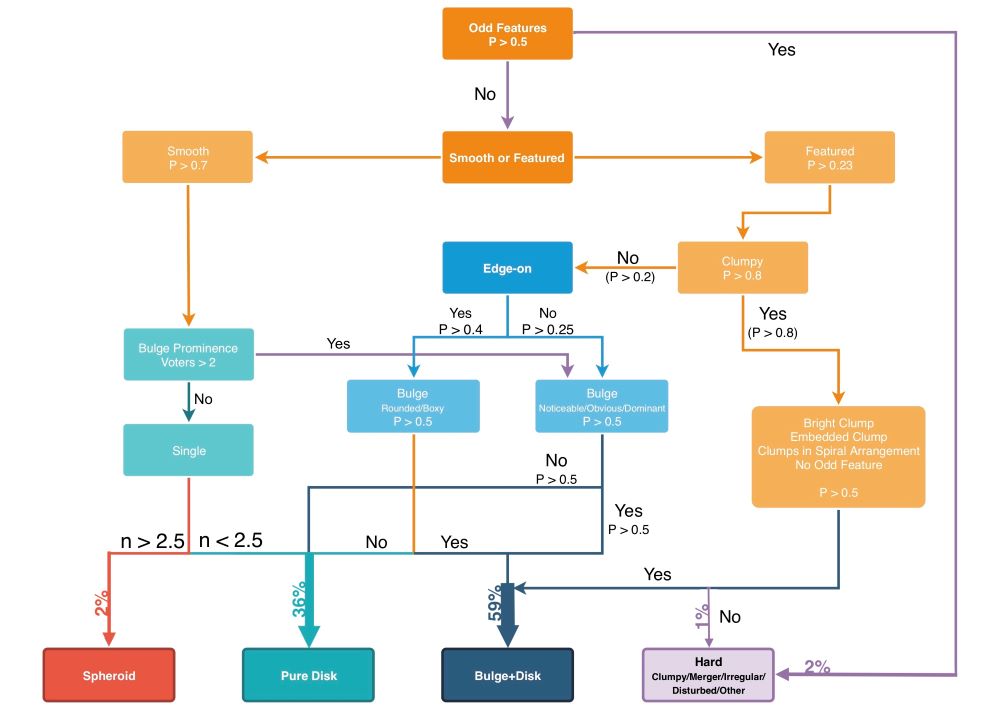}
	\caption{ This flowchart shows the decision tree that we adopt to translate Galaxy Zoo: Hubble (GZH) tasks and outputs into the desired morphologies to be used in this study. The weight of final arrows are proportional to the number of galaxies following those paths. In addition, the fraction of galaxies falling into each morphological category is annotated.}
	\label{fig:GZH}
\end{figure*}

\section{Final Morphological Classifications} 
\label{sec:MorphClass}

In this section, we initially aim to develop a semi-automatic method for morphological classification. Our ultimate goal would be to reach a fully automatic algorithm for classifying galaxies into various morphological classes. While this ultimately proves unsuitable we discuss it here to explain why we finally visually inspect all systems.
These classes are as mentioned above \textit{bulge+disk} (BD; double-component), \textit{pure-disk} (D), \textit{elliptical} (E) and \textit{hard} (H) systems. In order to overcome this problem we test various methods including: cross-matching with Galaxy Zoo, Hubble catalogue (GZH) and Zurich Estimator of Structural Type (ZEST) catalogues. In the end, none of the methods proved to be robust and we resort to full visual inspection.      

\subsection{Using Galaxy Zoo: Hubble} 
\label{subsec:GZH}

A large fraction of COSMOS galaxies with I$_{F814W} < 23.5$ have been classified in the Galaxy Zoo: Hubble (GZH) project \citep{Willett17}. Like other Galaxy Zoo projects, this study classifies galaxies using crowdsourced visual inspections. They make use of images taken by the ACS in various wavelengths including 84,954 COSMOS galaxies in the F814W filter. Our sample has more than 80\% overlap with GZH, so we can cross-match and use their classifications to partially improve our final sample. We translate the GZH classifications into our desired morphological classes by using the decision tree shown in figure 4 of \cite{Willett17}, as well as the suggested thresholds for morphological selection presented in Table 11 of the same paper. Our final decision tree is displayed in the flowchart shown in Figure \ref{fig:GZH}. We refer to \cite{Willett17} for a detailed description of each task. The thresholds are shown as P values in the flowchart. In addition to using different combinations and permutations of tasks as shown in Figure \ref{fig:GZH}, the only part that we add to the GZH tasks is where an object is voted to have a smooth light profile. In this case, the object is likely to be an E or S0 or a smooth \textit{pure-disk} galaxy. To distinguish between them, we make use of the single S\'ersic index (n). As shown in the flowchart, $n > 2.5$ and $n \leq 2.5$ are classified as spheroid and \textit{pure-disk}, respectively. The S\'ersic indices are taken from our structural analysis which will be described in Hashemizadeh et al. (in prep.). In order to capture the S0 galaxies or \textit{double-component} systems with smooth disk profiles we add an extra condition as to whether there are at least 2 Galaxy Zoo votes for a prominent bulge. If so then it is likely that the galaxy is a \textit{double-component} system.
An advantage of using the GZH decision tree is that it can identify a vast majority of galaxies with odd features such as merger-induced asymmetry etc. 
Table \ref{tab:GZH_cofmat} shows a confusion matrix comparing the GZH predictions with the visual inspection of our 4k gold sample as the ground truth. Double component galaxies are predicted by the GZH with maximum accuracy 81\%. Single component (i.e. \textit{pure-disk} and \textit{elliptical} galaxies) and \textit{hard} galaxies are predicted at a significantly lower accuracy with high misclassification rates. We further visually inspected misclassified galaxies and do not find good agreement between our classification and those of GZH. As such, we do not trust the GZH classification for our sample.

\begin{table}
\centering
\caption{The confusion matrix comparing the morphological predictions of the GZH with our visual inspection of 4k gold calibration sample as the ground truth. For example, 0.81 means 81\% of double component galaxies (in our visual classification) are also correctly classified by the GZH.  }
\begin{tabular}{c|ccc}
\hline \hline
\backslashbox[45mm]{GZH Pred}{Ground Truth}
            & Double     & Hard          & Single \\ \hline \\
Double  & \textbf{0.81}  & 0.28          & 0.32         \\
Hard        & 0.0038         & \textbf{0.18} & 0.011      \\
Single      & 0.18           & 0.53 & \textbf{0.67}         \\ \\

\lasthline
\end{tabular}
\label{tab:GZH_cofmat}
\end{table}

\subsection{Using Zurich Estimator of Structural Type (ZEST)} 
\label{subsec:ZEST}

Another available morphological catalogue for COSMOS galaxies is the Zurich Estimator of Structural Type (ZEST) \citep{Scarlata07a}. In ZEST, \cite{Scarlata07a} use their single-S\'ersic index (n) as well as five other diagnostics: asymmetry, concentration, Gini coefficient, second-order moment of the brightest 20\% of galaxy flux, and ellipticity. ZEST includes a sample of $\sim 56,000$ galaxies with I$_{F814W}\leq 24$. More than 90\% of our sample is cross-matched with ZEST which we use as a complementary morphological classification. We use the ZEST TYPE flag with four values of 1,2,3,9 representing early type galaxy, disk galaxy, irregular galaxy and no classification. For Type 2 (i.e. disk galaxy) we make use of the additional flag BULG, which indicates the level of bulge prominence. BULG is flagged by five integers as follows: 0 = bulge dominated, 1,2 = intermediate-bulge, 3 = \textit{pure-disk} and 9 = no classification. 
We present the accuracy of ZEST predictions in a confusion matrix in Table \ref{tab:ZEST_cofmat} which shows we do not find an accurate morphological prediction from ZEST in comparison to our gold calibration sample. While \textit{double-component} systems are classified with an accuracy of 74\%, other classes are classified poorly with high error ratios. We confirm this by visual inspection of the misclassified objects where we still favour our visual classification.
\\[2\baselineskip]

Overall, by analysing both of the above catalogues we do not find them to be sufficiently accurate for predicting the proper morphologies of galaxies when we compare their estimates with our 4k gold calibration visual classification.   

\begin{table}
\centering
\caption{The confusion matrix comparing the morphological predictions of the ZEST catalogue with our visual inspection of 4k gold calibration sample as the ground truth. }
\begin{tabular}{c|ccc}
\hline \hline
\backslashbox[45mm]{ZEST Pred}{Ground Truth}
            & Double  & Hard & Single \\ \hline \\
Double  & \textbf{0.74}  & 0.34 & 0.42         \\
Hard        & 0.047  & \textbf{0.37} & 0.036       \\
Single      & 0.21  & 0.28 & \textbf{0.55}         \\ \\

\lasthline
\end{tabular}
\label{tab:ZEST_cofmat}
\end{table}

\subsection{Visual inspection of full D10/ACS sample} 
\label{subsec:VisInsp_full}

As no automatic classification robustly matches our gold calibration sample we opt to visually inspect all galaxies in our full sample. 
However, we can use the predictions from GZH and ZEST as a pre-classification decision and put galaxies in master directories according to their prediction. The \textit{hard} class is adopted from the GZH prediction as it is shown to perform well in detecting galaxies with odd features. In addition, from analysing the distribution of the half-light radius (R50) of our 4k gold calibration sample, extracted from our DEVILS photometric analysis, we know that resolving the structure of galaxies with a spatial size of R50 $\leq 0.15$ arcsec is nearly impossible. R50 is measured by using the {\sc ProFound} package \citep{Robotham18}, a tool written in the {\sc R} language for source finding and photometric analysis. So, we put these galaxies into a separate \textit{compact} class (C).

Having pre-classified galaxies, we now assign each class to one of our team members (AH, SPD, ASGR, LJMD, SB) so each of the authors is independently a specialist in, and responsible for, only one morphological class. Initially, we inspect galaxies and relocate incorrectly classified galaxies from our master directories to transition folders for further inspections by the associated responsible person. In the second iteration, the incorrectly classified systems will be reclassified and moved back into master directories. The left-over sources in the transition directories are therefore ambiguous. For these galaxies, all classifiers voted and we selected the most voted class as the final morphology. As a final assessment and quality check three of our authors (SPD, LJMD, SB) independently reviewed the entire classifications ($\sim 15k$ each) and identified $\sim 7k$ that were still felt to be questionable. These three classifiers then independently classified these 7k objects. The final classification was either the majority decision or in the case of a 3-way divergence the classification of SPD.

\begin{figure*}
	\centering
	\includegraphics[width=\textwidth]{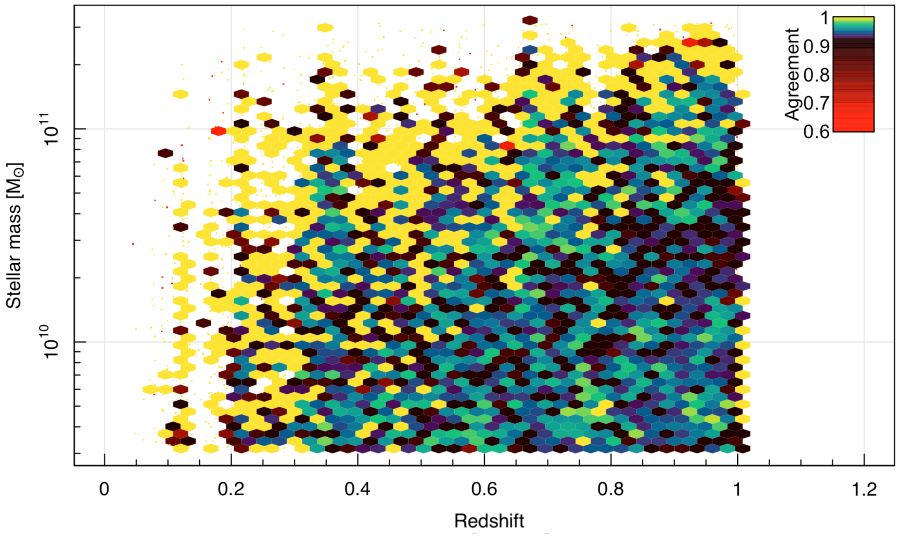}
	\caption{Stellar mass versus redshift, colour coded by the level of agreement in our final classifications (by three co-authors), i.e. the percentage of objects in each cell consistently classified by at least two of our classifiers. Note that we increase the resolution of the colour map of agreement $\gtrsim 0.92$ to highlight the variation of agreement in this level. See the text for more details.}
	\label{fig:agreent}
\end{figure*}

Figure \ref{fig:agreent} shows the stellar mass versus redshift plane colour coded according to the level of agreement between our 3 classifiers. Colour indicates the percentage of objects in each cell consistently classified by at least two classifiers, obtained as follows:

\begin{equation}
\mathrm{\scriptstyle Agreement} = \frac{\mathrm{\scriptstyle Number~of~objects~with~two~or~more~agreement~in~the~cell}}{\mathrm{\scriptstyle Total~number~of~objects~in~the~cell}},    
\label{eq:agree}
\end{equation}

This figure implies that we have the highest agreement ($\sim 100\%$) for high stellar mass objects at low redshifts. The agreement in our classification decreases to $\sim 90\%-95\%$ towards lower stellar mass regime at high redshift. This figure indicates that, on average, our visual classifications performed independently by different team members agree by $\sim95\%$ over the complete sample.

We report the number of objects in each morphological class in Table \ref{tab:FinalClassStats}. $44.4\%$ of our sample (15,931) are classified as \textit{double-component} (BD) systems. We find 13,212 \textit{pure-disk} (D) galaxies ($37\%$) while only 3,890 ($11\%$) \textit{elliptical} (E) galaxies. The \textit{Compact} (C) and \textit{hard} (H) systems, in total, occupy $7.6\%$ of our D10/ACS sample.

The visual morphological classification is released in the team-internal DEVILS data release one (DR1) in \texttt{D10VisualMoprhologyCat} data management unit (DMU). 

\subsection{Possible effects from the ``Morphological K-correction'' on our galaxy
classifications}
\label{subsec:K-corr}

The morphology of a subset of nearby galaxy classes shows some significant dependence on rest-frame wavelengths, especially below the Balmer break and towards the UV (e.g. \citealt{Hibbard97}; \citealt{Windhorst02}; \citealt{Papovich03}; \citealt{Taylor05}; \citealt{Taylor-Mager07}; \citealt{Huertas-Company09}; \citealt{Rawat09}; \citealt{Holwerda12}; \citealt{Mager18}). This ``Morphological K-correction'' can be quite significant, even between the B- and near-IR bands (e.g., \citealt{Knapen96}), and must be quantified in order to distinguish genuine evolutionary effects from simple bandpass shifting. Hence, the results of faint galaxy classifications may, to some extent, depend on the rest-frame wavelength sampled. 

Our D10/ACS classifications are done in the F814W filter (I-band), and the largest redshift in our sample, $z \sim$ 1, samples $\sim$412 nm (B-band), so for our particular case, the main question is, to what extent galaxy rest-frame morphology changes significantly from 412--823 nm across the BVRI filters. Here we briefly summarize how any such effects may affect our classifications. 

\cite{Windhorst02} imaged 37 nearby galaxies of all types with the Hubble Space Telescope (HST), gathering available data mostly at 150, 255, 300, 336, 450, 555, 680, and/or 814 nm, including some ground based images to complement filters missing with HST. These nearby galaxies are all large enough that a ground-based V-band image yields the same classification as an HST F555W
or F606W image. These authors conclude that the change in rest-frame morphology going from the red to the mid--far UV is more pronounced for early type galaxies (as defined at the traditional optical wavelengths or V-band), but not necessarily negligible for all mid-type spirals or star-forming galaxies. Late-type galaxies generally look more similar in morphology from the mid-UV to the red due to their
more uniform and pervasive SF that shows similar structure for young--old stars in all filters from the mid-UV through the red. \cite{Windhorst02} conclude {\it qualitatively} that in the rest-frame mid-UV, early- to mid-type galaxies are more likely to be misclassified as later types than late-type galaxies are likely to be misclassified as earlier types. 

To {\it quantify} these earlier qualitative findings regarding the morphological K-correction, much larger sample of 199 nearby galaxies across all Hubble types (as defined in V-band) was observed by \cite{Taylor05} and \cite{Taylor-Mager07}, and 2071 nearby galaxies were similarly analyzed by \cite{Mager18}. They determined their SB-profiles, radial light-profiles, color gradients, and CAS parameters (Concentration index, Asymmetry, and Clumpiness; e.g. \citealt{Conselice04}) as a function of rest-frame wavelength from 150-823 nm.

\cite{Taylor-Mager07} and \cite{Mager18} conclude that early-type galaxies (E--S0) have CAS parameters that appear, within their errors, to be similar at all wavelengths longwards of the Balmer break, but that in the far-UV, E--S0 galaxies have concentrations and asymmetries that more closely resemble those of spiral and peculiar/merging galaxies in the optical. This may be explained by extended disks containing recent star formation. The CAS parameters for galaxy types later than S0 show a mild but significant wavelength dependence, even over the wavelength range 436-823 nm, and a little more significantly so for the earlier spiral galaxy types (Sa--Sc). These galaxies generally become less concentrated and more asymmetric and somewhat more clumpy towards shorter wavelengths. The same is true for mergers when they progress from pre-merger via major-merger to merger-remnant stages.

While these trends are mostly small and within the statistical error bars for most galaxy types from 436--823 nm, this is not the case for the Concentration index and Asymmetry of Sa--Sc galaxies. For these galaxies, the Concentration
index decreases and the Asymmetry increases going from 823 nm to 436 nm (Fig. 17 of \citealt{Taylor-Mager07} and Fig. 5 of \citealt{Mager18}). Hence, to the extent that our visual classifications of apparent Sa--Sc galaxies depended on their Concentration index and Asymmetry, it is possible that some of these objects may have been misclassified. E-S0 galaxies show no such trend in Concentration with wavelength for 436--823 nm, and have Concentration indices much higher than Sa--Sc galaxies and Asymmetry and Clumpiness parameters generally much lower than Sa--Sc galaxies. Hence, it is not likely that a significant fraction of E--S0 galaxies are misclassified as Sa--Sc galaxies at $z \lesssim 0.8$ in the 436--823 nm filters due to the Morphological K-correction.

Sd-Im galaxies show milder trends in their CAS parameters and in the same direction as the Sa--Sc galaxies. It is thus possible that a small fraction of Sa-Sc galaxies --- if visually relying heavily on Concentration and Asymmetry index --- may have been visually misclassified as Sd--Im galaxies, while a smaller fraction of Sd--Im galaxies may have been misclassified as Sa--Sc galaxies. The available data on the rest-frame wavelength dependence of the morphology of nearby galaxies does not currently permit us to make more precise statements. In any case, these CAS trends with restframe wavelengths as observed at 436--823 nm for redshifts $z\simeq$ 0.8--0 in the F814W filter are shallow enough that the fraction of misclassifications between Sa--Sc and Sd--Im galaxies, and vice versa, is likely small. 

\begin{figure}
  \centering
  \includegraphics[width=\textwidth]{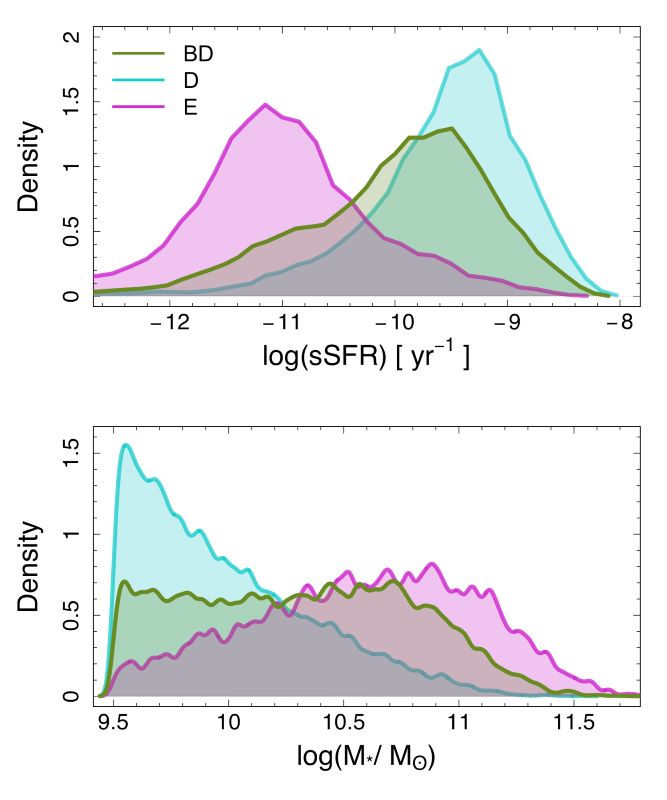}
  \label{subfig:stMass_PDF}
\caption{Top panel: the PDF of the specific star formation rate (sSFR = SFR/M$_*$) of the same morphologies. Bottom panel: the stellar mass probability density function (PDF) of three morphologies in our sample (all redshifts), i.e. \textit{pure-disk} (cyan), \textit{double-component} (green) and \textit{elliptical} (magenta) galaxies. Note that for clarity, the PDFs are slightly smoothed by a kernel with standard deviation of 0.02. }
\label{fig:stMass_sSFR_PDF}
\end{figure}

\subsection{Review} 
\label{subsec:VisInsp_verify}

In Figure \ref{fig:frac_z1}, we show the associated fractions of our final visual inspection as solid lines. The global trends are in good agreement with our initial 4k gold calibration sample. 
We find that, although our inspection procedure may have slightly changed from the 4k gold calibration sample to the full sample, the outcome classifications are consistent in the three primary classes making up more than $92\%$ of our sample.
 
\begin{table}
\centering
\caption{Final number of objects in each of our morphological classes. }
\begin{tabular}{ccc}
\firsthline \firsthline 
Morphology  & Object Number  & Percentage \\ \hline \\
Double  & 15,931  & 44.4\%  \\
Pure Disk & 13,212 & 37\%     \\
Elliptical & 3,890 & 11\%      \\
Compact & 1,124 & 3.1\%       \\
Hard & 1,600 & 4.5\%          \\ \\

\lasthline
\end{tabular}
\label{tab:FinalClassStats}
\end{table}

Figure \ref{fig:stMass_sSFR_PDF} shows the probability density function (PDF) of the sSFR (upper panel) and total stellar mass (lower panel) for galaxies classified as D, BD and E. We use the SFR from the {\sc ProSpect} SED fits described in \cite{Thorne20}. These figures indicate that, as one would expect, D galaxies dominate lower stellar mass and higher sSFR regime, opposite to Es which occupy the high stellar mass end and low sSFR. BD galaxies are systems located in-between these two classes in terms of both stellar mass and sSFR. These results are as expected and provide some confidence that our classifications are sensible.

\begin{figure*}
	\centering
	\includegraphics[width = \textwidth, height = 11.7cm]{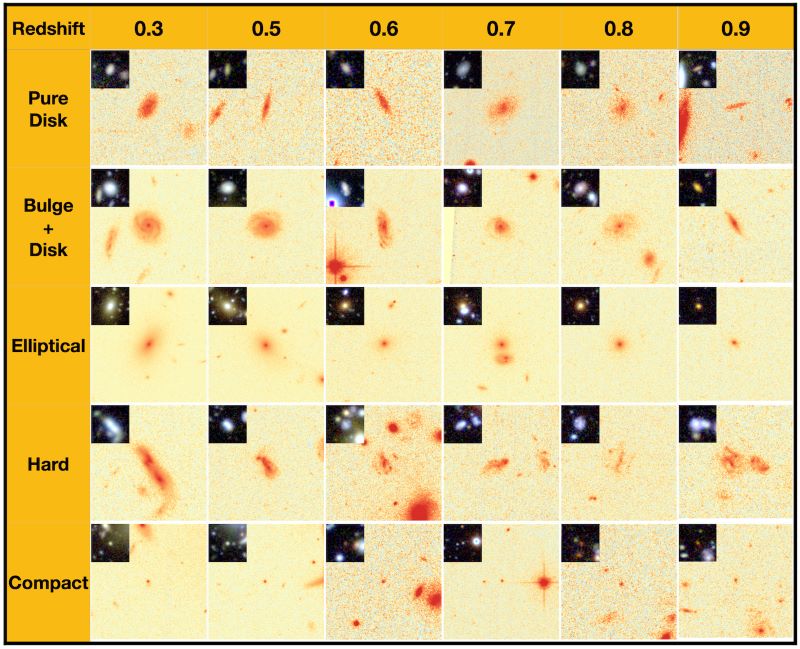}
	\caption{ Random postage stamps of various morphologies that we generate to perform our visual inspection. Background images are HST ACS/F814W while insets are combined SUBARU $gri$ colour image. Rows represent different morphologies. Galaxies are randomly selected within different redshift bins shown in columns. Redshifts annotated in the first row are the mean of the associated redshift bins. Cutouts indicate $5\times \mathrm{R90}$ on each side, where $\mathrm{R90}$ is measured from UltraVISTA Y-band in the DEVILS photometry catalogue (Davies et al. in prep.). }
	\label{fig:PostageStamp1}
\end{figure*}

Figure \ref{fig:PostageStamp1} displays a set of random galaxies in each of the morphological classes within different redshift intervals. In addition, we present 50 random galaxies of each morphology in Figures \ref{fig:contact_sheets_double}-\ref{fig:contact_sheets_comp}. 

Having finalised our morphological classification, we now investigate the stellar mass functions for different morphologies and their evolution from $z = 1$.

\section{Fitting galaxy stellar mass function} 
\label{sec:FitSMF}

For parameterizing the SMF, we assume a fitting function that can describe the galaxy number density, $\Phi\left(\mathrm{M}\right)$. The typical function adopted is that described by \cite{Schechter76} as:  

\begin{equation}
\Phi(M)\mathrm{d}M=\Phi^{*} e^{-M / M^{*}}\left(\frac{M}{M^{*}}\right)^{\alpha} \mathrm{d}M,
\label{eq:SingleSchechter}
\end{equation}

\noindent where the three key parameters of the function are, $\alpha$, the power low slope or the slope of the low-mass end, $\Phi^{*}$, the normalization, and, $M^{*}$, the characteristic mass (also known as the break mass or the knee of the Schechter function). 

At very low redshifts a number of studies have argued that the shape of the SMF is better described by a double Schechter function (\citealt{Baldry08}; \citealt{Pozzetti10}; \citealt{Baldry12}), i.e. a combination of two single Schechter functions, parameterized by a single break mass ($M^{*}$), and given by: 

\begin{equation}
\Phi(M) \mathrm{d} M=e^{-M / M^{*}}\left[\Phi_{1}^{*}\left(\frac{M}{M^{*}}\right)^{\alpha_{1}}+\Phi_{2}^{*}\left(\frac{M}{M^{*}}\right)^{\alpha_{2}}\right] \frac{\mathrm{d} M}{M^{*}},
\label{eq:DoubleSchechter}
\end{equation}

\noindent where $\alpha_2 < \alpha_1$, indicating that the second term predominantly drives the lower stellar mass range.

To fit our SMFs, we use a modified maximum likelihood (MML) method (i.e., not $1/V_{\mathrm{max}}$) as implemented in {\fontfamily{qcr}\selectfont dftools}\footnote{\href{https://github.com/obreschkow/dftools}{https://github.com/obreschkow/dftools}} \citep{Obreschkow18}. This technique has multiple advantages including: it is free of binning, accounts for small statistics and Eddington bias. {\fontfamily{qcr}\selectfont dftools} recovers the mass function (MF) while simultaneously handling any complex selection function, Eddington bias and the cosmic large-scale structure (LSS). Eddington bias tends to change the distribution of galaxies, particularly in the low-mass regime which is more sensitive to the survey depth and S/N, as well as high-mass regime due to the exponential cut-off which is sensitive to the scatters by noise (e.g. \citealt{Ilbert13}; \citealt{Caputi15}). Eddington bias occurs because of the observational/photometric errors and often can dominate over the shot noise \citep{Obreschkow18}. Photometric uncertainties, which are introduced in the redshift estimation, as well as the stellar mass measurements are fundamentally the source of this bias (\citealt{Davidzon17}). We account for Eddington bias in {\fontfamily{qcr}\selectfont dftools} by providing the errors on the stellar masses from the {\sc ProSpect} analysis by \cite{Thorne20}.

The R language implementation and MML method make {\fontfamily{qcr}\selectfont dftools} very fast. In the fitting procedures described in this work, we use the inbuilt {\fontfamily{qcr}\selectfont optim} function with the default optimization algorithm of \cite{Nelder65} for maximizing the likelihood function. In order to account for the volume corrections, we use the effective unmasked area of the D10/ACS region which we calculate to be $1.3467$ square degrees. This is exclusive of the masked areas from bright stars and the non-uniform edges of the ACS mosaic (see Figure \ref{fig:F814W_DEVILS_smp_z1}) and this is calculated using the process outlined in Davies et al. (in prep.). Our selection function, required for a proper volume corrected distribution function, is essentially a volume limited sample, i.e. a constant volume across adopted mass range. We refer the reader to \cite{Obreschkow18} for full details regarding {\fontfamily{qcr}\selectfont dftools} and its methodology. In this paper, we fit both single and double Schechter functions to the SMF. We examine both functions and will discuss further in Section \ref{subsec:MfunZ0}.

\begin{figure*}
	\centering
	\includegraphics[width=\textwidth]{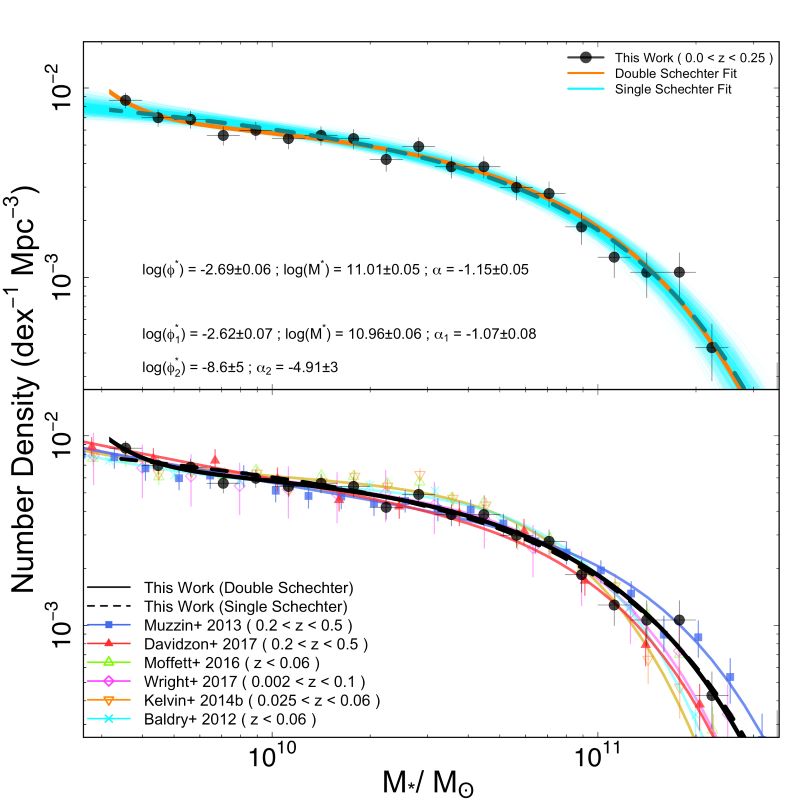}
	\caption{Top panel: Single and double Schechter functions fitted to a low-$z$ sample of D10/ACS ($0.0 < z < 0.25$). The fits are performed by using {\fontfamily{qcr}\selectfont dftools} down to the stellar mass limit of the sample, i.e. $10^{9.5} M_\odot$. 
    Transparent region shows the error range calculated by 1000 times sampling of the full posterior probability distribution of the single Schechter fit parameters. Black data points show the binned galaxy counts for which the stellar mass range from minimum to maximum ($9.5 \le$ log$(\mathrm{M}_{*}/\mathrm{M}_{\odot}) \le 12$) is binned into 25 equal-sized bins. Note that the Schechter functions are not fitted directly to the galaxy counts. Bottom panel: Comparison of our SMF at $0.0 < z < 0.25$ with \protect\cite{Muzzin13} and \protect\cite{Davidzon17}. For completeness, the SMF of the GAMA local galaxies from several studies are shown in different colours. Solid and dashed black lines represent the double and single Schechter function fits to our data, respectively (eq. \ref{eq:SingleSchechter} \& \ref{eq:DoubleSchechter}). Note that, the fits to other studies (colour solid lines) are double Schechter fits.}
	\label{fig:Mfunc_z0}
\end{figure*}

\subsection{Verification of our SMF fitting process at low redshift} 
\label{subsec:MfunZ0}  

We first validate our \textit{total} SMF fitting process at low-$z$ and compare with the known literature, prior to splitting by redshift and morphology.

To achieve this, we compare the SMF of D10/ACS galaxies in our lowest-$z$ bin, i.e. $0 < z < 0.25$, with literature studies. We primarily choose this redshift range to compare with the SMF \cite{Muzzin13} and \protect\cite{Davidzon17} at $0.2 < z < 0.5$ in the COSMOS/UltraVISTA field and local GAMA galaxies at $z < 0.06$ (\citealt{Baldry12}; \citealt{Kelvin14}; \citealt{Moffett16a}; \citealt{Wright17}). We fit the SMF within this redshift range using both single and double Schechter functions (Equations \ref{eq:SingleSchechter} and \ref{eq:DoubleSchechter}, respectively). The upper panel of Figure \ref{fig:Mfunc_z0}, shows our single and double Schechter fit to this low-$z$ sample. As annotated in the figure, we report the best fit single Schechter parameters of log$(\mathrm{M}^{*}/\mathrm{M}_{\odot}) = 11.01\pm0.05$, $\alpha = -1.15\pm0.05$, log$(\Phi^*) = -2.69\pm0.06$ and log$(\mathrm{M}^{*}/\mathrm{M}_{\odot}) = 10.96\pm0.06$ $\alpha_1 = -1.07\pm0.08$, $\alpha_2 = -4.91\pm3$, log$(\Phi^*_1) = -2.62\pm0.07$ and log$(\Phi^*_2) = -8.6\pm5$ in our double Schechter fit. Therefore, the double Schechter fit involves a broad range of uncertainty, in particular at the low-mass range. 
Larger uncertainty is likely due to the fact that the stellar mass range of the D10/ACS sample does not extend to log$(M_*/M_\odot) < 9.5$, where the pronounced upturn in the SMF occurs. 
The error ranges shown in the upper panel of Figure \ref{fig:Mfunc_z0} as transparent curves are obtained from 1000 samples drawn from the full posterior probability distribution of all the single Schechter parameters. 

Despite larger errors on the double Schechter parameters, we elect to use this function for our future analysis at all redshifts as it has been shown that a double Schechter can better describe the stellar mass distribution even at higher-$z$ (e.g., \citealt{Wright18}).

In the lower panel of Figure \ref{fig:Mfunc_z0}, we show that our SMF is in good agreement with \cite{Muzzin13} and \protect\cite{Davidzon17} within the quoted errors.
Overall, we see a good agreement with the mass function of galaxies at low-$z$. We observe no significant flattening of the SMF at the intermediate stellar masses ($10^{9.5}-10^{10.5}$) as reported in the local Universe by e.g., \cite{Moffett16a}. We are unsurprised that we do not see this flattening as the D10/ACS contains only 3 galaxies in the redshift range comparable to GAMA ($z < 0.06$). 

One might expect an evolution of the SMF within $0 < z < 0.25$ that would explain the slightly higher number density in the intermediate mass range. Our fitted Schechter functions however also deviate from the GAMA SMFs in the high stellar mass end. We find that this is systematically due to the Schechter function fitting process. Unlike D10/ACS, as shown in Figure \ref{fig:Mfunc_z0}, the local literature data extend to lower stellar mass regimes (log$(\mathrm{M}_{*}/\mathrm{M}_{\odot}) = 8$ and $7.5$ in the case of \citealt{Wright17}), influencing the bright end fit and the position of $\mathrm{M}^*$. In this regime the upturn of the stellar mass distribution is remarkably more pronounced. This strong upturn drives the optimization fitting and impacts the high-mass end. This highlights the difficulty in directly comparing fitted Schechter values if fitted over different mass ranges. 

\subsection{Cosmic Large Scale Structure Correction} 
\label{subsec:LSS_cor}

\begin{figure}
	\centering
	\includegraphics[width=\textwidth]{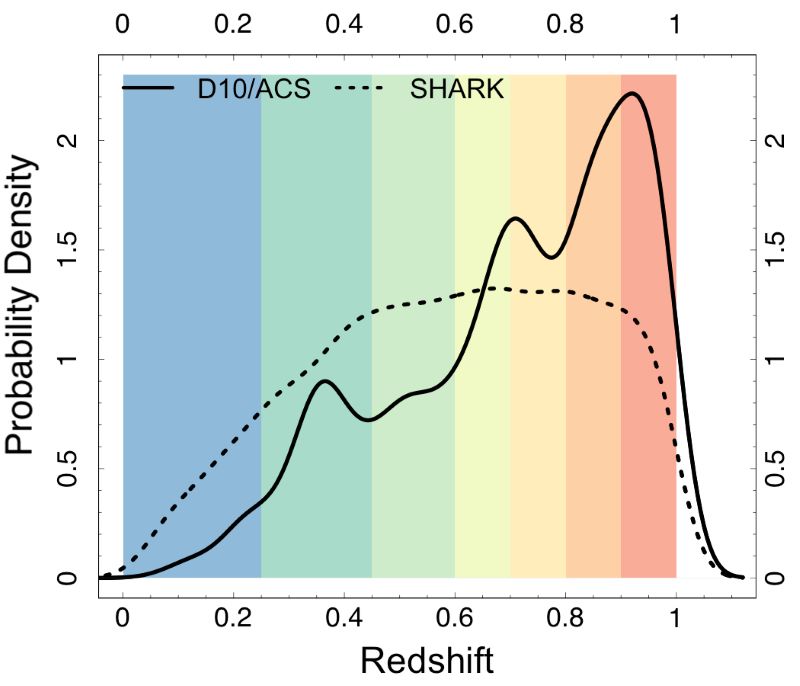}
	\caption{ The $\mathrm{N}(z)$ distribution of the D10/ACS sample (solid line) compared with the SHARK semi-analytic model prediction (dashed line). SHARK data represent a light-cone covering 107.889 square degrees with Y-mag $< 23.5$.  Colour bands represent the redshift bins that we consider in this work. Note that the PDFs are smoothed by a kernel with standard deviation of 0.3 so are non-zero beyond the nominal limits.}
	\label{fig:zdens}
\end{figure}

All galaxy surveys are to some extent influenced by the cosmic large scale structure (LSS, \citealt{Obreschkow18}). Generally, the LSS produces local over- and under-densities of the galaxies at particular redshifts in comparison to the mean density of the Universe at that epoch. For example, GAMA regions are $\sim 10\%$ underdense compared with SDSS (\citealt{Driver11}).  

We observe this phenomenon in the nonuniform D10/ACS redshift distribution, $\mathrm{N}(z)$.
To highlight this, Figure \ref{fig:zdens} compares the $\mathrm{N}(z)$ distribution of the D10/ACS sample with the prediction of the SHARK semi-analytic model (\citealt{Lagos18b}; \citealt{Lagos19}). SHARK data in this figure represent a light-cone covering 107.889 square degrees with Y-mag $< 23.5$, and because of the much larger simulated volume, is less susceptible to LSS. In this figure the redshift bins we shall use later in our analysis are shown as background colour bands indicating redshift intervals of (0, 0.25, 0.45, 0.6, 0.7, 0.8, 0.9 and 1.0). SHARK predicts a nearly uniform galaxy distribution with no significant over- and under-density regions, while the empirical D10/ACS sample shows a nonuniform $\mathrm{N}(z)$ with overdensities and underdensities.
These density fluctuations can introduce systematic errors in the construction of the SMF by, for example, overestimating the number density of very low-mass galaxies which are only detectable at lower redshifts \citep{Obreschkow18}. In other words, the LSS can artificially change the shape/normalization of the SMF.

Using the distance distribution of galaxies, {\fontfamily{qcr}\selectfont dftools} (\citealt{Obreschkow18}) internally accounts for the LSS by modeling the relative density in the survey volume, $g(r)$, i.e. it measures the mean galaxy density of the survey at the comoving distance $r$, relative to the mean density of the Universe. Incorporating this modification into the effective volume, the MML formalism works well for a sensitivity-limited sample (see \citealt{Obreschkow18} for details). However, for our volume limited sample, this method is unable to thoroughly model the density fluctuations. Therefore, to take this non-uniformity into account, we perform a manual correction as follows. 

\begin{figure}
	\centering
	\includegraphics[width = \textwidth, angle = 0]{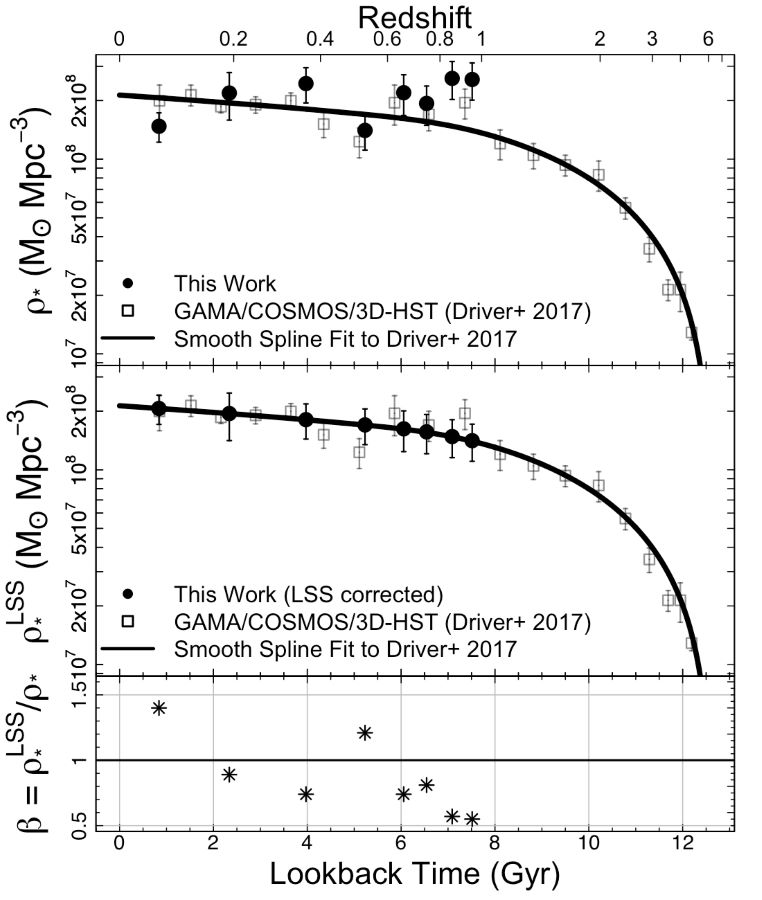}
	\caption{ Top panel: our measurement of the evolution of the \textit{total} SMD (per comoving Mpc$^{3}$) compared with a compilation of GAMA, COSMOS and 3D-HST by \protect\cite{Driver18}. The black curve represents a spline fit to the \protect\cite{Driver18} data. Middle panel: the large scale structure correction factor applied to our SMD in order to meet the predictions of the spline fit. Bottom panel: the residual of the SMDs before and after the LSS correction indicating the correction coefficient we apply to each redshift bin. }
\label{fig:LSS_cor}
\end{figure}

In a comprehensive study of the cosmic star formation history, \cite{Driver18} compiled GAMA, COSMOS and 3D-HST data to estimate the total stellar mass density from high redshifts to the local Universe ($0 < z < 5$). Figure \ref{fig:LSS_cor} (upper panel) shows these data. As we know the total stellar mass density must grow smoothly and hence we assume that perturbations around a smooth fit represent the underlying LSS. 
In Figure \ref{fig:LSS_cor} (middle panel), we fit a smooth spline with the degree of freedom $3.5$ to the \cite{Driver18} data to determine a uniform evolution of the total $\rho_*$. We then introduce a set of correction factors ($\beta$) to our empirical measurements of the \textit{total} SMDs to match the prediction of the spline fit (middle panel of Figure \ref{fig:LSS_cor}). We then apply these correction factors (bottom panel of Figure \ref{fig:LSS_cor}) to our estimations of the morphological SMD values.
This ignores any coupling between morphology and LSS which we consider a second order effect.
We report the $\beta$ correction factors in Table \ref{tab:rho} and show them in the bottom panel of Figure \ref{fig:LSS_cor}.

\subsection{Evolution of the SMF since \lowercase{$z$} $= 1$} 
\label{subsec:MfunEvol}

With the LSS correction in place, we now split the sample into 7 bins of redshift (0-0.25, 0.25-0.45, 0.45-0.6, 0.6-0.7, 0.7-0.8, 0.8-0.9 and 0.9-1.0) so that we have enough numbers of objects in each bin and are not too wide to incorporate a significant evolution (these redshift ranges are shown by the colour bands in Figure \ref{fig:zdens}). Each bin contain 1,108, 5,041, 4,216, 4,867, 5,277, 7,090, 8,207 galaxies, respectively.

\begin{landscape}
\begin{figure}
	\centering
	\includegraphics[width = 1.5\textwidth, height = 0.95\textwidth, angle = 0]{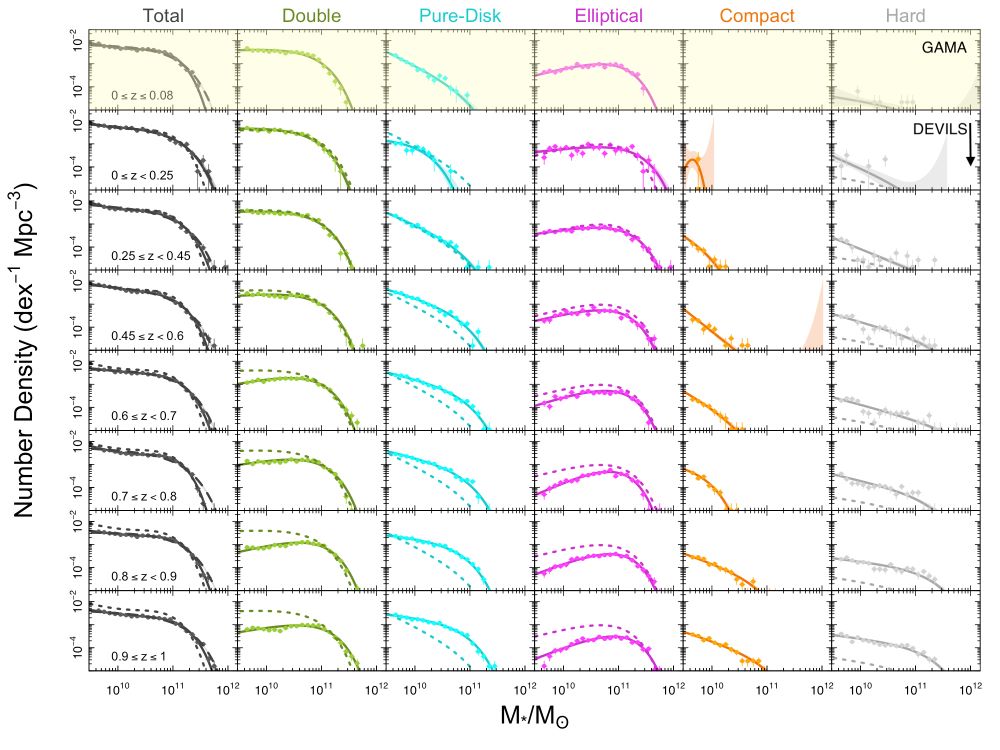}
	\caption{ The \textit{total} and morphological SMFs in eight redshift bins. Top row highlighted by yellow colour represents the GAMA SMFs ($0 \le z \le 0.08$). Data points are galaxy counts in each of equal-size stellar mass bins. Width of stellar mass bins are shown as horizontal bars on data points. Vertical bars are poisson errors. Shaded regions around the best fit curves are 68 per cent confidence regions. Black solid and dashed curves demonstrate double and single Schechter functions of all galaxies, respectively, while dotted curves over-plotted on higher-$z$ bins are the GAMA $z=0$ SMFs to highlight the evolution of the SMF.}
	\label{fig:Mfunc_6z}
\end{figure}
\end{landscape}

Figure \ref{fig:Mfunc_6z} shows the SMF in each redshift bin for the full sample, as well as for different morphological types of: \textit{double-component} (BD), \textit{pure-disk} (D), \textit{elliptical} (E), \textit{compact} (C) and \textit{hard} (H). These SMF measurements include the correction for LSS as discussed in Section \ref{subsec:LSS_cor}. Note that the first row of Figure \ref{fig:Mfunc_6z} highlighted by yellow shade shows GAMA $z = 0$ SMFs (Driver et al., in prep.).

As shown in Figure \ref{fig:Mfunc_6z}, we fit the \textit{total} SMF within all 8 redshift bins (including GAMA) by both single and double Schechter functions (displayed as dashed and solid black curves, respectively), while the morphological SMFs are well fit by single Schechter function at all epochs. The difference between double and single Schechter fits is insignificant, compared to the error on individual points. However, as noted in Section \ref{subsec:MfunZ0}, for calculating the stellar mass density we will use our double Schechter fits.
As we will see later, the effect of this choice on the measurement of our total stellar mass density is negligible. In Figure \ref{fig:Mfunc_6z}, we over-plot the GAMA $z = 0$ SMFs (Driver et al., in prep.) on higher-$z$ to highlight the evolution (dotted curves). 
In the \textit{total} SMF we see a growth at the low-mass end and a relative stability at high-mass end, particularly when comparing high-$z$ with the D10/ACS low-$z$ ($0 < z < 0.25$). 
At face value, this suggests that since $z = 1$ in-situ star formation and minor mergers (low-mass end) play an important role in forming or transforming mass within galaxies \citep{Robotham14}.

Looking at the SMF of the morphological sub-classes we find that the BD systems show no significant growth with time at their high-mass end, while at the low-mass end we see a noticeable increase in number density. 
In D galaxies, at both low- and high-mass ends we find a variation in the number density with the low-mass end increasing and extreme steepening at the lower-$z$ and high-mass end decreasing with time. Note that we do not rule out the possible effects of some degree of incompleteness in this mass regime on the evolution of the low-mass end. 

For E galaxies, however, we report a modest growth in their high-mass end (again when comparing with D10/ACS low-$z$) and a significant growth in intermediate- and low-mass regimes from $z = 1$ to $z = 0$. 
Finally, H and C systems become less prominent with declining redshift as fewer galaxies occupy these classes. The physical implications of these trends will be discussed in Section \ref{sec:discussion}.  

\begin{figure*}
	\centering
	\includegraphics[width=\textwidth]{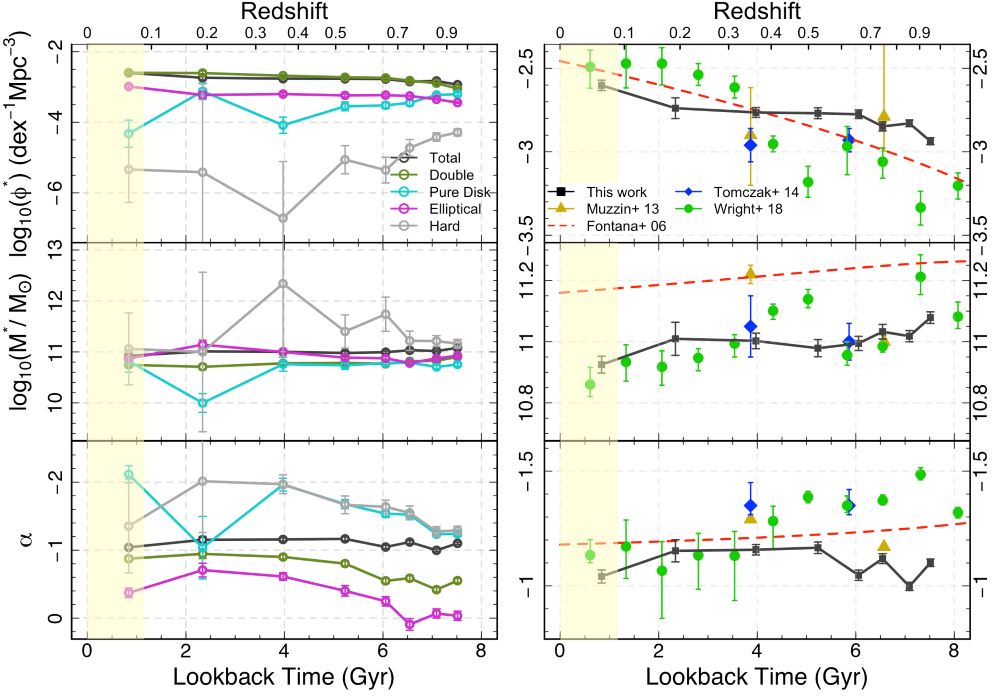}
	\caption{ Left: Evolution of the single Schechter best fit parameters, $\Phi^*$, $M^*$ and $\alpha$. Error bars are the standard deviations on each of the most likely parameters. Black line is the \textit{total} SMF while colour coded data represent the best fit parameters of the Schechter function of different morphologies. Note that for simplicity we remove \textit{compacts} as their trends are very noisy and washes out the trends of other morphological types. Right: Zoomed-in plot showing the evolution of our \textit{total} Schechter function parameters compared with a compilation of other studies. Highlighted region shows the epoch covered by the GAMA data.}
	\label{fig:Mfunc_6z_par_evol}
\end{figure*}

Figure \ref{fig:Mfunc_6z_par_evol} further summarizes the trends in Figure \ref{fig:Mfunc_6z}, showing the evolution of our best fit single Schechter parameters $\Phi^*$, $M^*$ and $\alpha$. We report the best Schechter fit parameters in Table \ref{tab:Morph_MF_par}. For comparison, we also show a compilation of literature values in the right panel. The literature data show only single Schechter parameters of the \textit{total} SMF. Comparing our \textit{total} SMF with other studies, including \cite{Muzzin13}, we find a good agreement within the quoted errors. 

Note that the three parameters are, of course, highly correlated and the mass limits to which they are fitted vary.    

The $\Phi^*$ of BD systems has the largest value and almost mimics the trend of the \textit{total} $\Phi^*$ which is largely consistent with no evolution. This is expected as BD galaxies dominate the sample at almost all redshifts. We report a slight decrease in the $\Phi^*$ value of D galaxies, while almost no evolution in E systems. Note that as mentioned earlier, our lowest redshift bin ($0.0 < z < 0.25$) contains only $1,108$ galaxies of which only $102$ are D systems incorporating an uncertainty in our fitting process. This can be seen in the deviation of the second data point of D galaxies (cyan lines) from other redshifts in Figure \ref{fig:Mfunc_6z_par_evol}.  
H systems also contribute more at higher redshifts (higher $\Phi^*$ values) owing to the fact that the abundance of mergers, clumpy and disrupted systems dramatically increases at higher redshifts (despite lower resolution). We note that due to our LSS corrections (Section \ref{subsec:LSS_cor}) the $\Phi^*$ reported here is not the directly measured parameter but modified according to our LSS correction coefficients (reported in Table \ref{tab:rho}). This normalization smooths out the evolutionary trends, otherwise fluctuating due to the large scale structures, but does not impact their global trends.

The characteristic mass, $M^*$, of the BD systems presents a stable trend since $z~=~1$. We interpret this behaviour to be a result of the lack of significant evolution of the massive/intermediate-mass regime of the SMF of this morphological class. D galaxies also show slight evidence of evolution with some fluctuations in $M^*$ value. E galaxies, likewise, evolve only a small amount. Overall, we observe a behaviour consistent with no evolution in $M^*$ for all morphological types. Note that large variation of the $M^*$ in H and C types, at low-$z$ in particular, is not physical. This is massively driven by the lack of H and C galaxies at low-$z$ resulting in an unconstrained turnover and as can be seen in Figure \ref{fig:Mfunc_6z_par_evol} large uncertainties.

The low-mass end slope, $\alpha$, of different morphological classes also shows some evolution. Similar to D systems, BD galaxies show a marked steepening increase in their slope at later times, indicating that the SMF steepens at lower redshifts. The steepest mass function at almost all times is for the D galaxies. E galaxies occupy the lowest steepening but constantly growing from $\alpha = 0.34$ to $-0.75$ at $z \sim 0.2$.

\begin{table*}
\centering
\caption{Best Schechter fit parameters of \textit{total} and different morphological types in different redshift bins.}
\begin{adjustbox}{scale = 0.6}
\begin{tabular}{lcccccccc}
\firsthline \firsthline \\
$z$-bins         &  $0.0 \leq z < 0.08$  &  $0.0 \leq z < 0.25$ &  $0.25 \leq z < 0.45$ &  $0.45 \leq z < 0.60$ &  $0.60 \leq z < 0.70$ &  $0.70 \leq z < 0.80$ &  $0.80 \leq z < 0.90$ &  $0.90 \leq z \leq 1.00$ \\ \\ \hline 
                   &   \multicolumn{8}{c}{\textbf{Total (Double Schechter)}}  \\
\cline{2-9} \\
$\mathrm{log}_{10}\Phi^*_1$      & $-2.43\pm0.04$  & $-2.68\pm0.07$   & $-2.55\pm0.03$ & $-2.66\pm0.07$ & $-2.71\pm0.08$ & $-2.61\pm0.03$ & $-2.73\pm0.04$ & $-2.81\pm0.04$    \\ \\
$\mathrm{log}_{10}M^*$           & $10.65\pm0.04$  & $10.96\pm0.06$   & $10.83\pm0.03$ & $10.76\pm0.06$ & $10.79\pm0.05$ & $10.66\pm0.04$ & $10.76\pm0.04$ & $10.84\pm0.04$    \\ \\
$\alpha_1$                       & $-0.04\pm0.22$  & $-1.08\pm0.08$   & $-0.79\pm0.09$ & $-0.32\pm0.30$ & $-0.22\pm0.33$ & $0.15\pm0.20$  & $-0.02\pm0.21$ & $-0.23\pm0.20$     \\ \\
$\mathrm{log}_{10}\Phi^*_2$      & $-3.11\pm0.21$  & $-8.90\pm4.85$   & $-4.76\pm0.55$ & $-3.11\pm0.23$ & $-3.14\pm0.26$ & $-3.26\pm0.14$ & $-3.22\pm0.15$ & $-3.32\pm0.16$    \\ \\
$\alpha_2$                       & $-1.56\pm0.16$  & $-5.06\pm3.27$   & $-2.62\pm0.36$ & $-1.48\pm0.15$ & $-1.36\pm0.17$ & $-1.61\pm0.11$ & $-1.36\pm0.11$ & $-1.43\pm0.10$     \\ \\ \hline
                    &   \multicolumn{8}{c}{\textbf{Total (Single Schechter)}}  \\
\cline{2-9} \\
$\mathrm{log}_{10}\Phi^*$         & $-2.60\pm0.03$   & $-2.74\pm0.06$   & $-2.77\pm0.03$ & $-2.77\pm0.03$ & $-2.78\pm0.03$ & $-2.85\pm0.03$ & $-2.83\pm0.02$ & $-2.94\pm0.2$ \\ \\
$\mathrm{log}_{10}M^*$            & $10.93\pm0.03$   & $11.01\pm0.05$   & $11.00\pm0.03$ & $10.98\pm0.03$ & $10.99\pm0.02$ & $11.03\pm0.02$ & $11.02\pm0.02$ & $11.08\pm0.02$ \\ \\
$\alpha$                          & $-1.04\pm0.03$   & $-1.15\pm0.05$   & $-1.16\pm0.02$ & $-1.17\pm0.03$ & $-1.05\pm0.02$ & $-1.12\pm0.02$ & $-1.00\pm0.02$ & $-1.10\pm0.02$ \\ \\ \hline 
                    &   \multicolumn{8}{c}{\textbf{Double}}  \\
 \cline{2-9} \\
$\mathrm{log}_{10}\Phi^*$         & $-2.60\pm0.03$  & $-2.61\pm0.06$   & $-2.68\pm0.03$ & $-2.73\pm0.03$ & $-2.75\pm0.02$ & $-2.82\pm0.02$ & $-2.90\pm0.02$ & $-3.05\pm0.02$ \\ \\
$\mathrm{log}_{10}M^*$            & $10.75\pm0.03$  & $10.71\pm0.05$   & $10.78\pm0.03$ & $10.78\pm0.03$ & $10.77\pm0.02$ & $10.82\pm0.02$ & $10.83\pm0.02$ & $10.90\pm0.02$ \\ \\
$\alpha$                          & $-0.87\pm0.04$  & $-0.95\pm0.07$   & $-0.90\pm0.03$ & $-0.80\pm0.04$ & $-0.55\pm0.04$ & $-0.59\pm0.04$ & $-0.42\pm0.03$ & $-0.55\pm0.03$ \\ \\ \hline
                    &   \multicolumn{8}{c}{\textbf{Pure-Disk}}  \\
 \cline{2-9} \\
$\mathrm{log}_{10}\Phi^*$         & $-4.32\pm0.38$  & $-3.12\pm0.23$   & $-4.09\pm0.23$ & $-3.55\pm0.12$ & $-3.52\pm0.09$ & $-3.45\pm0.08$ & $-3.23\pm0.04$ & $-3.21\pm0.04$ \\ \\
$\mathrm{log}_{10}M^*$            & $10.82\pm0.22$  & $10.00\pm0.18$   & $10.76\pm0.14$ & $10.74\pm0.08$ & $10.78\pm0.06$ & $10.79\pm0.06$ & $10.71\pm0.03$ & $10.75\pm0.03$ \\ \\
$\alpha$                          & $-2.12\pm0.13$  & $-1.04\pm0.46$   & $-1.96\pm0.09$ & $-1.68\pm0.07$ & $-1.54\pm0.05$ & $-1.52\pm0.05$ & $-1.24\pm0.04$ & $-1.24\pm0.03$ \\ \\ \hline
                    &   \multicolumn{8}{c}{\textbf{Elliptical}}  \\
 \cline{2-9} \\
$\mathrm{log}_{10}\Phi^*$         & $-2.99\pm0.03$  & $-3.23\pm0.08$   & $-3.20\pm0.04$ & $-3.24\pm0.04$ & $-3.23\pm0.03$ & $-3.26\pm0.02$ & $-3.35\pm0.02$ & $-3.45\pm0.02$ \\ \\
$\mathrm{log}_{10}M^*$            & $10.87\pm0.04$  & $11.14\pm0.09$   & $11.00\pm0.04$ & $10.89\pm0.05$ & $10.88\pm0.04$ & $10.78\pm0.04$ & $10.87\pm0.03$ & $10.93\pm0.03$ \\ \\
$\alpha$                          & $-0.37\pm0.07$  & $-0.71\pm0.10$   & $-0.61\pm0.05$ & $-0.40\pm0.08$ & $-0.25\pm0.07$ & $0.09\pm0.09$  & $-0.07\pm0.07$ & $-0.03\pm0.07$ \\ \\ \hline
                    &   \multicolumn{8}{c}{\textbf{Compact}}  \\
 \cline{2-9} \\
$\mathrm{log}_{10}\Phi^*$         & $-$  & $-19.91\pm45.11$   & $-4.01\pm1.14$ & $-9.70\pm62.46$ & $-4.76\pm0.89$ & $-3.31\pm0.20$ & $-4.72\pm0.32$ & $-4.69\pm0.24$ \\ \\
$\mathrm{log}_{10}M^*$            & $-$  & $8.37\pm0.74$      & $9.79\pm0.51$  & $12.71\pm32.57$ & $10.37\pm0.41$ & $9.67\pm0.17$  & $10.73\pm0.20$ & $10.86\pm0.16$ \\ \\
$\alpha$                          & $-$  & $17.85\pm35.50$    & $-2.09\pm1.37$ & $-2.90\pm0.28$  & $-2.29\pm0.42$ & $-1.16\pm0.62$ & $-1.79\pm0.15$ & $-1.75\pm0.11$ \\ \\ \hline
                    &   \multicolumn{8}{c}{\textbf{Hard}}  \\
 \cline{2-9} \\
$\mathrm{log}_{10}\Phi^*$         & $-5.34\pm0.94$  & $-5.42\pm2.50$   & $-6.72\pm1.60$ & $-5.06\pm0.40$ & $-5.36\pm0.37$ & $-4.73\pm0.24$ & $-4.42\pm0.12$ & $-4.29\pm0.10$ \\ \\
$\mathrm{log}_{10}M^*$            & $11.06\pm0.70$  & $11.00\pm1.56$   & $12.34\pm1.43$ & $11.40\pm0.33$ & $11.74\pm0.33$ & $11.22\pm0.19$ & $11.21\pm0.11$ & $11.16\pm0.09$ \\ \\
$\alpha$                          & $-1.35\pm0.69$  & $-2.01\pm0.75$.  & $-1.97\pm0.14$ & $-1.67\pm0.13$ & $-1.64\pm0.10$ & $-1.55\pm0.10$ & $-1.27\pm0.07$ & $-1.29\pm0.06$ \\ \\

\lasthline
\end{tabular}
\end{adjustbox}
\label{tab:Morph_MF_par}
\end{table*}

\section{The Evolution of the Stellar Mass Density Since \lowercase{$z$} $= 1$} 
\label{sec:rho}

\begin{landscape}
 \begin{figure}
	\centering
	\includegraphics[width = 1.5\textwidth, height = 0.95\textwidth, angle = 0]{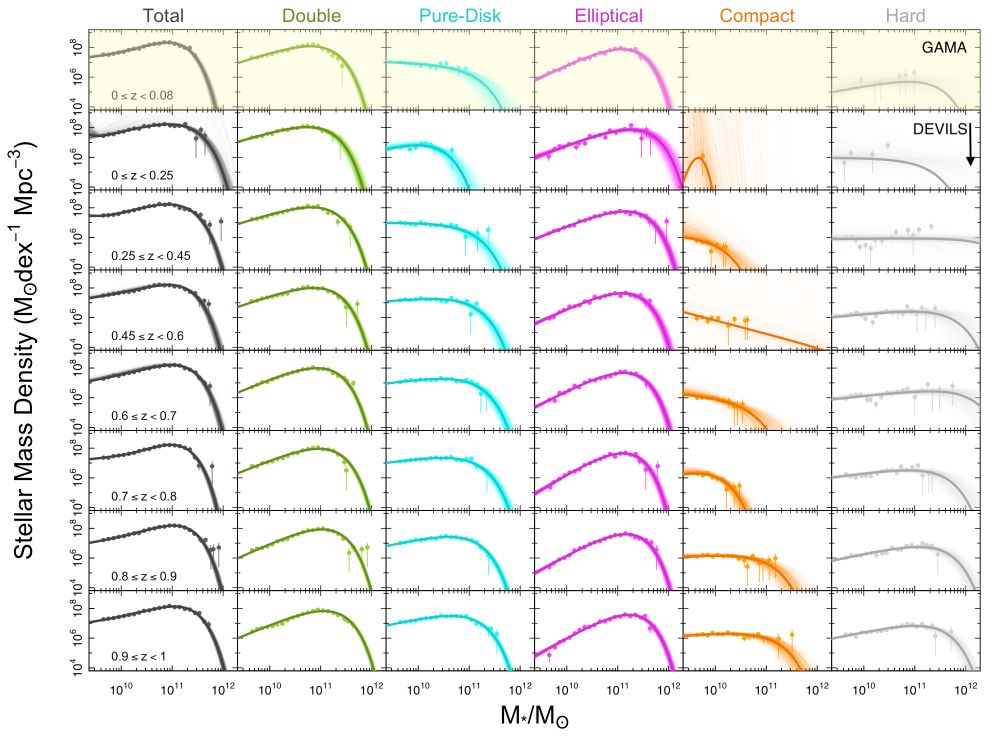}
	\caption{The distribution of \textit{total} and morphological stellar mass density in different redshifts. Redshift bins are the same as Figure \ref{fig:Mfunc_6z}. Points and lines indicate $M^\prime \Phi(M^\prime)$ in Equation \ref{eq:massDens}, where $\Phi(M^\prime)$ is our Schechter function fit. The shaded transparent regions represent the error range calculated by 1000 times sampling of the full posterior probability distribution of the fit parameters. The distribution of the stellar mass density of all individual morphological types in almost all redshift bins is bounded. Top row highlighted by yellow colour represents the GAMA data ($0 \le z \le 0.08$).}
	\label{fig:Mdens_6z}
 \end{figure}
\end{landscape}

In this section, we investigate the evolution of the Stellar Mass Density (SMD) as a function of morphological types. To determine the SMD, we integrate under the best fit Schechter functions over all stellar masses from $10^{9.5}$ to $\infty$. This integral can be expressed as a gamma function:

\begin{equation}
\rho_* = \int_{M=10^{9.5}}^{\infty} M^\prime \Phi(M^\prime) dM^\prime = \Phi^* M^* \Gamma(\alpha+2,10^{9.5}/M^*),
\label{eq:massDens}
\end{equation}

\noindent where $\Phi^*$, $M^*$ and $\alpha$ are the best regressed Schechter parameters. Figure \ref{fig:Mdens_6z} shows the evolution of the distribution of the SMDs, term $M^\prime \Phi(M^\prime)$, for different morphologies. We also illustrate the fitting errors by sampling 1000 times the full posterior probability distribution of the fit parameters. These are shown in Figure \ref{fig:Mdens_6z} as transparent shaded regions around the best fit curves. The standard deviations of the integrated SMD calculated from each of these functions are reported as the fit error on $\rho_*$ in Table \ref{tab:rho}. As can be seen in Figure \ref{fig:Mdens_6z}, the distribution of the stellar mass density of all individual morphological types in almost all redshift bins is bounded, implying that integrating under these curves will capture the majority of the stellar mass for each class. 

\begin{figure*}
	\centering
	\includegraphics[width = \textwidth, angle = 0]{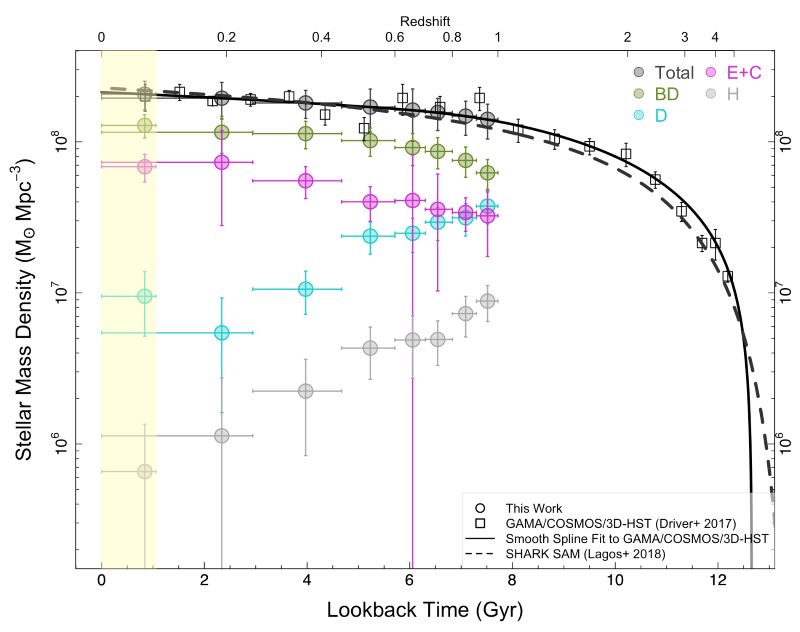}
	\caption{The evolution of the \textit{total} and morphological stellar mass density, $\rho_{*}$, in the last 8 Gyr of the cosmic age. $\rho_{*}$ expresses our measurements of the analytical integration under the best Schechter function fits in 8 redshift bins, i.e. Equation \ref{eq:massDens}. Highlighted region shows the epoch covered by the GAMA data ($0 \le z \le 0.08$). This figure includes the correction for the large scale structure, LSS, that we apply by fitting a smooth spline to the empirical data of GAMA/G10COSMOS/3D-HST by \protect\cite{Driver18}. See the text for more details on our method for this correction. The prediction of the SHARK semi-analytic model is also overlaid as dashed line. Colour codes are similar to Figure \ref{fig:Mfunc_6z}. Vertical bars on the points show the total error budget on each data point including: cosmic variance within the associated redshift bin taken from \protect\cite{Driver10}, classification error, fit error and Poisson error. See Section \ref{sebsec:morph_err_budget} for more details about our error analysis. Horizontal bars indicate the redshift ranges while the data points are plotted at the mean redshift.}
	\label{fig:MassBuildUp_LSS}
\end{figure*}

Figure \ref{fig:MassBuildUp_LSS} then shows the evolution of the integrated SMD, $\rho_*$, in the Universe between $z = 0-1$. 
This includes the LSS correction by forcing our \textit{total} SMD values to match the smooth spline fit to the \cite{Driver18} data, as described in Section \ref{subsec:LSS_cor}. The uncorrected evolutionary path of $\rho_*$ is shown in the upper panel of Figure \ref{fig:LSS_cor}.

We report the empirical LSS corrected $\rho_*$ values in Table \ref{tab:rho}. This table also provides the LSS correction factor, $\beta$, so one might obtain the original values by $\rho_*^{Orig} = \rho_*^{corr}/\beta$. 

For completeness, we show the evolution of the SMDs before we apply the LSS corrections in Figure \ref{fig:SMD_evol_noLSS} of Appendix \ref{sec:SMD_evol_noLSS}. In this Figure, the trends are not as smooth as one would expect without the LSS correction being applied, given the evident structure in the $N(z)$ distribution from Figure \ref{fig:zdens}. However, even without the LSS corrections, the main trends are still present, albeit not as strong. This highlights that the LSS correction is an important aspect and also the need for much wider deep coverage than that currently provided by HST. This will become possible in the coming Euclid and Roman era.

Before we analyse the evolution of the SMDs in Figure \ref{fig:MassBuildUp_LSS}, below we investigate the errors that are involved in this calculation.

\subsection{Analysing the Error Budget on the SMD}
\label{sebsec:morph_err_budget}

The error budget on our analysis of the SMD includes \textit{cosmic variance } (CV), \textit{fit error}, \textit{classification error} and \textit{Poisson error}. Cosmic variance and classification are the dominant sources of error. 

We make use of \cite{Driver10} equation 3 to calculate the \textit{cosmic variance} in the volume encompassed within each redshift bin. This calculation is implemented in the R package: {\fontfamily{qcr}\selectfont celestial}\footnote{ The online version of this cosmology calculator is available at: \href{http://cosmocalc.icrar.org/}{http://cosmocalc.icrar.org/} }. Note that for our low-$z$ GAMA data, instead of using \cite{Driver10} equation that estimates a $\sim 22.8\%$ cosmic variance, we empirically calculate the CV using the variation of the SMD between 3 different GAMA regions of G09, G12, and G15 with a total effective area of 169.3 square degrees (G09:54.93; G12: 57.44; G15: 56.93, \citealt{Bellstedt20a}) and find CV to be $\sim 16\%$. 

We measure the \textit{fit error} by 1000 times sampling of the full posterior probability distribution of the Schechter parameters (shown as shaded error regions in Figure \ref{fig:Mdens_6z}) and calculating the associated $\rho_*$ in each iteration. 
We calculate the \textit{classification error} by measuring the stellar mass density associated with each of our 3 independent morphological classifications. The range of variation of the SMD between classifiers gives the error of our classification. 
The \textit{Poisson error} is calculated by using the number of objects in each morphology per redshift bin.

The combination of all the above error sources will provide us with the total error that is reported in Table \ref{tab:rho}.
\\[2\baselineskip]


As can be seen in Figure \ref{fig:MassBuildUp_LSS}, the extrapolation of our D10/ACS $\rho_*$ to $z = 0$ agrees well with the our local GAMA estimations. Note slight difference in D, but consistent in errors.

\begin{figure}
	\centering
	\includegraphics[width=\textwidth]{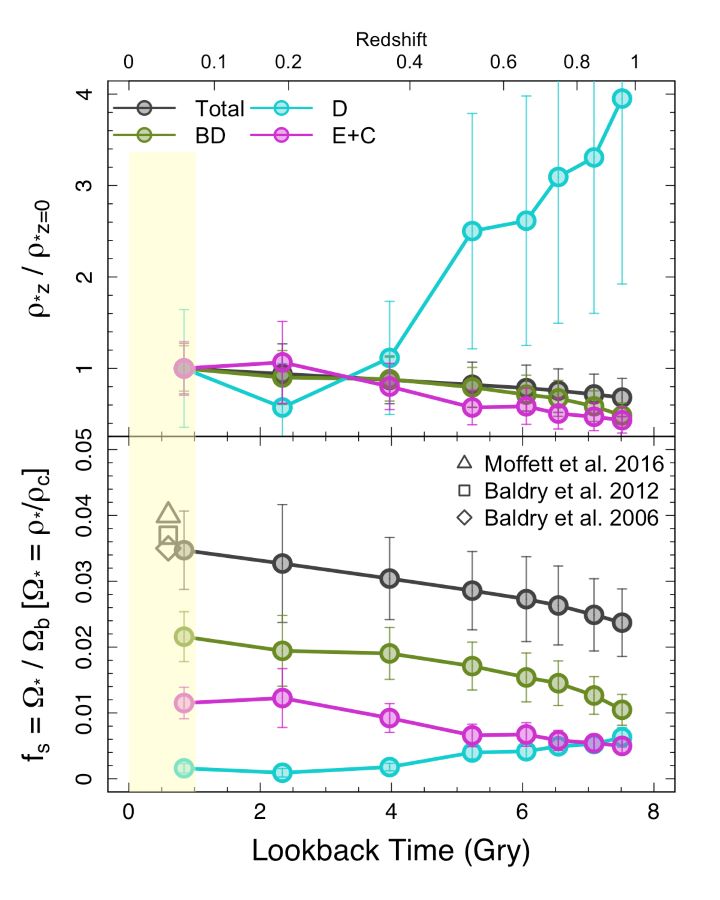}
	\caption{Top panel: the variation of the stellar mass density showing the fraction of final stellar mass density assembled or lost by each redshift, i.e. $\rho_{*z}/\rho_{*z=0}$. $\rho_{*z=0}$ is taken from GAMA estimations of the local Universe. Bottom panel: the evolution of the baryon fraction in form of stars ($f_s$). For simplicity, in this figure we only show the main morphological types and remove the \textit{hard} and \textit{compact} sub-classes. Highlighted region shows the epoch covered by the GAMA data.}
	\label{fig:MassDensVar_fs}
\end{figure}

The total change in the stellar mass is consistent with observed SFR evolution (e.g., \citealt{Madau14}; \citealt{Driver18}) as we will discuss more in Section \ref{sec:discussion}. Analysing the evolution of the SMD (Figure \ref{fig:MassBuildUp_LSS}), we find that in total (black symbols), 68\% of the current stellar mass in galaxies was in place $\sim 8$ Gyr ago ($z \sim 1.0$). The top panel of Figure \ref{fig:MassDensVar_fs} illustrates the variation of the $\rho_*$ ($\rho_{z}^{*} / \rho_{z=0}^{*}$), where $\rho_{z}^{*}$ is the SMD at redshift $z$ while $\rho_{z=0}^{*}$ represents the final SMD at $z~=~0$. 

According to our visual inspections C types are closer to Es than other subcategories. We, therefore, combine C types with E galaxies that shows a smooth growth with time of a factor of $\sim 2.5$ up to $z = 0.25$ and flattens out since then ($0.0 < z < 0.25$). Note that as reported in Table \ref{tab:rho} the amount of mass in the C class is very little. This demonstrates a significant mass build-up in E galaxies over this epoch ($\sim 150\%$). We also note that the large error bars on some of E+C data points are dominantly driven by the large errors in C SMDs.  

Having analytically integrated the SMD, we now measure the fraction of baryons in the form of stars ($f_s = \Omega_*/\Omega_b$) locked in each of our morphological types. We adopt $\Omega_b = 0.0493$ as estimated from Planck by \cite{Planck20} and the critical density of the Universe at the median redshift of GAMA, i.e. $z = 0.06$ to be $\rho_c = 1.21 \times 10^{11} \mathrm{M}_{\odot} \mathrm{Mpc}^{-3}$ in a 737 cosmology.

As shown in the bottom panel of Figure \ref{fig:MassDensVar_fs}, at our D10/ACS lowest redshift bin $\overline{z} \sim 0.18$ we find the fraction of baryons in stars $f_s \sim 0.033\pm0.009$. Including our GAMA measurements at $\overline{z} \sim 0.06$ we find this fraction to be $f_s \sim 0.035\pm0.006$. As shown in Figure \ref{fig:MassDensVar_fs}, this result is consistent with other studies within the quoted errors for example: \cite{Baldry06}, \cite{Baldry12} and \cite{Moffett16a}.
The evolution of the fraction of baryons locked in stars, $f_s$, shows that it has increased from $(2.4\pm0.5)$\% at $z \sim 1$ to $(3.5\pm0.6)$\% at $z \sim 0$ indicating an increase of a factor of $\sim 1.5$ during last 8 Gyr.   

Figure \ref{fig:MassDensVar_fs} also shows the breakdown of the total $f_s$ to each of our morphological subcategories highlighting that as expected BD systems contribute the most to the stellar baryon fraction increasing from $f_s = 0.011\pm0.006$ to $f_s = 0.022\pm0.005$ at $0 < z < 1$.
E+C systems take less percentage of the total $f_s$ but increase their contribution from $0.005\pm0.001$ to $0.012\pm0.002$ while D galaxies decrease from $f_s = 0.006\pm0.002$ to $f_s = 0.001\pm0.005$ between $0 < z < 1$.
We report our full $f_s$ values for all morphological types at all redshifts in Table. \ref{tab:fs}.

In summary, over the last 8 Gyr, double component galaxies clearly dominate the overall stellar mass density of the Universe at all epochs. The second dominant system is E galaxies. However the extrapolation of the trends to higher redshifts in Figures \ref{fig:MassBuildUp_LSS} and \ref{fig:MassDensVar_fs} indicates that D systems are likely to dominate over Es in the very high-$z$ regime ($z > 1$), which is reasonable according to the rise of the cosmic star formation history and the association of star-formation with disks. 

\begin{landscape}
\begin{table*}
\centering
\caption{The \textit{total} and morphological stellar mass density in different redshift ranges displayed in Figure \ref{fig:MassBuildUp_LSS} (left panel). $\rho_*$ values are calculated from the integration under the best Schechter function fits. Note that the $\rho_*$ are presented after applying the LSS correction. Columns represent: $z-$bin: redshift bins, $\overline{z}$: mean redshift, LBT: lookback time, T (S-Schechter): \textit{total with single Schechter}, T (D-Schechter): \textit{total with double Schechter}, BD: \textit{bulge+disk}, pD: \textit{pure-disk}, E: \textit{elliptical}, C: \textit{compact}, H: \textit{hard}, $\beta$: the large scale structure correction factor. Errors incorporate all error sources including, cosmic variance, fit error, Poisson error and classification error (see Section \ref{sebsec:morph_err_budget} for details). }
\begin{adjustbox}{angle = 0, scale = 0.8}
\begin{tabular}{ccccccccccccc}
\firsthline \firsthline \\
                      &     &   &  \multicolumn{6}{c}{log10$(\rho_*/M_\odot) $} \\ \\
\cline{4-9} \\
$z-$bin      & $\overline{z}$ & LBT (Gyr) & T (S-Schechter)  & T (D-Schechter)  & BD               & pD               & E                & C                & H                & $\beta$ \\ \\ \hline \\
$0.00 \le z < 0.08$   &  $0.06$ & $0.82$  &  $8.315\pm0.08$  &  $8.316\pm0.11$  &  $8.109\pm0.08$  &  $6.977\pm0.26$  &  $7.835\pm0.10$  &  $-$             &  $5.818\pm1.25$  &  $1.40$        \\
$0.00 \le z < 0.25$   &  $0.18$ & $2.27$  &  $8.289\pm0.14$  &  $8.289\pm0.14$  &  $8.063\pm0.14$  &  $6.735\pm0.53$  &  $7.863\pm0.20$  &  $5.332\pm0.83$  &  $6.054\pm0.38$  &  $0.89$        \\
$0.25 \le z < 0.45$   &  $0.36$ & $3.95$  &  $8.258\pm0.10$  &  $8.257\pm0.10$  &  $8.054\pm0.10$  &  $7.025\pm0.17$  &  $7.740\pm0.12$  &  $5.513\pm1.23$  &  $6.350\pm0.43$  &  $0.74$        \\
$0.45 \le z < 0.60$   &  $0.53$ & $5.23$  &  $8.231\pm0.10$  &  $8.230\pm0.16$  &  $8.008\pm0.10$  &  $7.376\pm0.12$  &  $7.594\pm0.13$  &  $5.936\pm1.01$  &  $6.635\pm0.21$  &  $1.21$       \\
$0.60 \le z < 0.70$   &  $0.65$ & $6.05$  &  $8.210\pm0.12$  &  $8.210\pm0.21$  &  $7.962\pm0.12$  &  $7.395\pm0.13$  &  $7.603\pm0.13$  &  $5.922\pm0.60$  &  $6.688\pm0.25$  &  $0.74$       \\
$0.70 \le z < 0.80$   &  $0.74$ & $6.55$  &  $8.195\pm0.11$  &  $8.194\pm0.12$  &  $7.936\pm0.12$  &  $7.468\pm0.12$  &  $7.540\pm0.13$  &  $6.032\pm0.95$  &  $6.692\pm0.17$  &  $0.81$       \\
$0.80 \le z < 0.90$   &  $0.85$ & $7.07$  &  $8.171\pm0.11$  &  $8.170\pm0.13$  &  $7.877\pm0.11$  &  $7.497\pm0.12$  &  $7.508\pm0.12$  &  $6.258\pm0.55$  &  $6.863\pm0.15$  &  $0.57$       \\
$0.90 \le z \le 1.00$ &  $0.95$ & $7.51$  &  $8.150\pm0.11$  &  $8.149\pm0.13$  &  $7.795\pm0.11$  &  $7.574\pm0.12$  &  $7.473\pm0.12$  &  $6.430\pm0.43$  &  $6.946\pm0.14$  &  $0.55$       \\
\\

\lasthline

\end{tabular}
\end{adjustbox}
\label{tab:rho}
\end{table*}
\end{landscape}

\section{Discussion} 
\label{sec:ch3_discussion}

Making use of our morphological classifications, we explore the evolution of the stellar mass function at $0 \le z \le 1$, to assess the physical processes that are likely affecting the individual morphological SMF. In particular, major mergers are thought to primarily occur between comparable mass companions (1:3) resulting in the fast growth of the SMF \citep{Robotham14}. Conversely, secular activities, minor mergers and tidal interactions will primarily alter the number density at the low-stellar mass end of the SMF \citep{Robotham14}. Investigating the \textit{total} SMF (Figure \ref{fig:Mfunc_6z}) shows that unlike high-mass end the low-mass end grows significantly. This suggests that since $z~\sim~1$, the galaxy population goes through mainly in-situ/secular processes at the low-mass end.

An important caveat, in what follows, is that although we have undertaken multiple tests of our visual classifications, possible uncertainties due to human classification error or inconsistencies will be present. Nevertheless some clear trends, which we believe are resilient to the classifications uncertainties are evident. Furthermore, we do not rule out the effects of dust in distinguishing bulges, particularly at high-$z$ leading to overestimating the number of \textit{pure-disk} systems at high redshifts. Although, high level of agreement in our visual classifications (see Figure \ref{fig:agreent}) gives us some confidence that our evolutionary trends are unlikely dominated by incorrect classifications, it is possible that all classifiers agree on the incorrect classification.

Firstly, double component galaxies display a modest decrease with redshift in the high-stellar mass end of their SMF (Figure \ref{fig:Mfunc_6z}), whilst their low-stellar mass end steepens significantly. This could be interpreted as most of the stellar mass in double component systems evolving via lower mass weighted star formation or secular activity, rather than major merging, i.e., BD systems are not merging together to form higher-mass BD systems. 

Secondly, \textit{pure-disk} systems show a strong increase at the low-mass end and a {\it decrease} at their high-mass end. The low-mass end evolution suggests in-situ star formation of the disk and/or the formation of new disks. However, the decrease at the high-mass end is to some extent unphysical, unless these systems are undergoing a transformation from the D class to another class. The most likely prospect is the secular formation of a central bulge component, resulting in a morphological transformation into the BD class. Hence, as time progresses and the second component forms, galaxies exit the D class leading to a mass-deficit at the high-mass end.

The new component forming through such an in-situ process is most likely a pseudo-bulge (pB), resulting in mass loss from the high-mass D SMF and a corresponding mass gain at comparable mass in the BD class. 

Finally, \textit{elliptical} galaxies, generally thought to be inert, show little growth in their high-mass end but a significant growth of low- and intermediate mass end presumably due to mergers.  

The evolution of the global and morphological SMDs indicates that BD galaxies dominate ($\sim 60\%$ on average) the stellar mass density of the Universe since at least at $z < 1$. This morphological class also shows a constant mass growth, compared to other morphologies. As mentioned, D systems slightly decrease their stellar mass density with time. Presumably, the mass transfer out of the D class out-weighs the mass gain due to in-situ star-formation for this class. 

We remark that the \textit{rate} of mass growth in the BD systems also decreases with time. This is reflected in the decrease of their SMD slope although it is still steeper than the \textit{total} SMD evolution and consistent with the general decline in the cosmic star-formation history. On the other hand, the E galaxies experience an initial growth until $z\sim0.2$ and a recent flattening in their stellar-mass growth from $z=1-0$ (see Figure \ref{fig:MassBuildUp_LSS} and the top panel of Figure \ref{fig:MassDensVar_fs}). 

\begin{landscape}
\begin{table*}
\centering
\caption{\textit{Total} and morphological stellar baryon fraction ($f_s$) in different times. See the text for details.}
\begin{adjustbox}{angle = 0, scale = 0.8}
\begin{tabular}{lcccccccc}
\firsthline \firsthline \\
                      &   \multicolumn{7}{c}{Redshift}  \\ \\
\cline{2-8} \\

Morphology Type         & $0.00 \le z < 0.08$    & $0.00 \le z < 0.25$ & $0.25 \le z < 0.45$ & $0.45 \le z < 0.60$ & $0.60 \le z < 0.70$ & $0.70 \le z < 0.80$ & $0.80 \le z < 0.90$ & $0.90 \le z < 1.0$ \\ \hline \\
Total                & $0.035\pm0.006$ & $0.033\pm0.009$ & $0.030\pm0.006$ & $0.029\pm0.006$ & $0.027\pm0.006$ & $0.026\pm0.006$ & $0.025\pm0.005$ & $0.024\pm0.005$ \\
Double               & $0.022\pm0.004$ & $0.019\pm0.005$ & $0.019\pm0.004$ & $0.017\pm0.004$ & $0.015\pm0.004$ & $0.015\pm0.003$ & $0.013\pm0.003$ & $0.010\pm0.002$ \\
Pure-Disk            & $0.002\pm0.001$ & $0.001\pm0.001$ & $0.002\pm0.001$ & $0.004\pm0.001$ & $0.004\pm0.001$ & $0.005\pm0.001$ & $0.005\pm0.001$ & $0.006\pm0.001$ \\
Elliptical           & $0.012\pm0.002$ & $0.012\pm0.004$ & $0.009\pm0.002$ & $0.007\pm0.002$ & $0.007\pm0.002$ & $0.006\pm0.002$ & $0.005\pm0.001$ & $0.005\pm0.001$ \\
Compact              & $-$             & $0.000\pm0.000$ & $0.000\pm0.000$ & $0.000\pm0.000$ & $0.000\pm0.000$ & $0.000\pm0.000$ & $0.000\pm0.000$ & $0.000\pm0.000$ \\
E+C                  & $0.012\pm0.002$ & $0.012\pm0.004$ & $0.009\pm0.002$ & $0.007\pm0.002$ & $0.007\pm0.002$ & $0.006\pm0.002$ & $0.006\pm0.001$ & $0.005\pm0.001$ \\
Hard                 & $0.000\pm0.000$ & $0.000\pm0.000$ & $0.000\pm0.000$ & $0.001\pm0.000$ & $0.001\pm0.000$ & $0.001\pm0.000$ & $0.001\pm0.000$ & $0.001\pm0.000$ \\

\lasthline
\end{tabular}
\end{adjustbox}
\label{tab:fs}
\end{table*}
\end{landscape}

\section{Summary and Conclusions} 
\label{sec:ch3_summary}

We have presented a visual morphological classification of a sample of galaxies in the DEVILS/COSMOS survey with HST imaging, from the D10/ACS sample. The quality of the imaging data (HST/ACS) provides arguably the best current insight into galaxy structures and therefore the best pathway with which to explore morphological evolution. 

We summarize our results as below: 

\begin{description}

  \item[$\bullet$] By visually inspecting galaxies out to $z \sim 1.5$, we find that morphological classification becomes far more challenging at $z > 1.0$ as many galaxies ($> 40\%$) no longer adhere to the classical notion of spheroids, bulge+disk or disk systems (\citealt{Abraham01}). We see a dramatic increase in strongly disrupted systems, presumably due to interactions and gas clumps. Nevertheless, at all redshifts we find that more mass is involved in star-formation than in merging and most likely it is the high star-formation rates that are driving the irregular morphologies.
  
  \item[$\bullet$] The SMF of the D10/ACS sample in our lowest redshift bin ($z<0.25$) is consistent with previous measurements from the local Universe. 

  \item[$\bullet$] The evolution of the global SMF shows enhancement in both low- and slightly in high-mass ends. We interpret this as suggesting that at least two evolutionary pathways are in play, and that both are significantly impacting the SMF.
  
  \item[$\bullet$] Despite a slight decrease in their high-mass end, BD systems demonstrate a non-negligible growth in the low-mass end of their SMF. In the D type, we witness a significant variation in both high- and low-mass ends of the SMF. We interpret the high-mass end decrease in D systems, which is at first sight unphysical, as an indication of significant secular mass transfer through the formation of pseudo-bulges and hence an apparent mass loss as galaxies transit to the BD category.
  
  \item[$\bullet$] E galaxies experience a modest growth in their high-mass end as well as an enhancement in their low/intermediate-mass end which we interpret as a consequence of major mergers resulting in the relentless stellar mass growth of this class. 
  
  \item[$\bullet$] Despite the above shuffling of mass we find that the best regressed Schechter function parameters in the \textit{total} SMF are observed to be relatively stable from $z = 1$. This is consistent with previous studies (\citealt{Muzzin13}; \citealt{Tomczak14}; \citealt{Wright18}). Conversely, the component SMFs show significant evolution. This implication is that while stellar-mass growth is slowing, mass-transformation processes via merging and in-situ evolution are shuffling mass between types behind the scenes.
  
  \item[$\bullet$] We measured the integrated total stellar mass density and its evolution since $z = 1$ and find that approximately $32\%$ of the current stellar mass in the galaxy population was formed during the last 8 Gyr.

  \item[$\bullet$] As shown in Figure \ref{fig:MassBuildUp_LSS}, we find that the BD population dominates the SMD of the Universe within $0 \le z \le 1$ and has constantly grown by a factor of $\sim 2$ over this timeframe. On the other hand, the SMD of Ds declines slowly and eventually loses $\sim 85\%$ of it's original value until $z \sim 0.2$. On the contrary, the E population experiences constant growth of a factor of $\sim 2.5$ since $z = 1$. We observe that the extrapolation of the trends of all of our morphological SMD estimations meets GAMA measurements at $z = 0$ (see Figure \ref{fig:MassBuildUp_LSS}) except for the \textit{pure-disk} systems which is likely due to unbound distribution of their SMD (see Figure \ref{fig:Mdens_6z}). 
  
\end{description}
 
One clear outcome of our analysis is that the late Universe ($z<1$) appears to be a time of profound spheroid and bulge growth/emergence. To move forward and explore this further we think it appropriate that to move forward it is key to decompose the double component morphological type, which significantly dominates the stellar mass density, into disks, classical bulges, and pseudo-bulges. To do this, requires robust structural decomposition which we will describe in Hashemizadeh et al. (in prep.) using our morphological classifications to guide the decomposition process.

\section{Acknowledgements}
DEVILS is an Australian project based around a spectroscopic campaign using the Anglo-Australian Telescope. The DEVILS input catalogue is generated from data taken as part of the ESO VISTA-VIDEO \citep{Jarvis13} and UltraVISTA \citep{McCracken12} surveys. DEVILS is part funded via Discovery Programs by the Australian Research Council and the participating institutions. The DEVILS website is \href{https://devilsurvey.org}{https://devilsurvey.org}. The DEVILS data is hosted and provided by AAO Data Central (\href{https://datacentral.org.au}{https://datacentral.org.au}). This work was supported by resources provided by The Pawsey Supercomputing Centre with funding from the Australian Government and the Government of Western Australia. We also gratefully acknowledge the contribution of the entire COSMOS collaboration consisting of more than 200 scientists. The HST COSMOS Treasury program was supported through NASA grant HST-GO-09822. SB and SPD acknowledge support by the Australian Research Council's funding scheme DP180103740. MS has been supported by the European Union's  Horizon 2020 research and innovation programme under the Maria Sklodowska-Curie (grant agreement No 754510), the National Science Centre of Poland (grant UMO-2016/23/N/ST9/02963) and by the Spanish Ministry of Science and Innovation through Juan de la Cierva-formacion program (reference FJC2018-038792-I). ASGR and LJMD acknowledge support from the Australian Research Council's Future Fellowship scheme (FT200100375 and FT200100055, respectively).

This work was made possible by the free and open R software (\citealt{R-Core-Team}).
A number of figures in this paper were generated using the R \texttt{magicaxis} package (\citealt{Robotham16b}). This work also makes use of the \texttt{celestial} package (\citealt{Robotham16a}) and \texttt{dftools} (\citealt{Obreschkow18}).

\section{Data Availability} 
\label{sec:AvailData}
In this work, we draw upon two datasets; the established HST imaging of the COSMOS region (\citealt{Scoville07}, \citealt{Koekemoer07}), and multiple data products produced as part of the DEVILS survey \citep{Davies18}, consisting of a spectroscopic campaign currently being conducted on the Anglo-Australian Telescope, photometry (Davies et al. in prep.), and deep redshift catalogues, and stellar mass measurements from \cite{Thorne20}.
These datasets are briefly described below.

\graphicspath{{images/ChapterFour/}}

\chapter[DEVILS: Mass Growth of Bulges and disks]{Deep Extragalactic VIsible Legacy Survey (DEVILS): The emergence of bulges and decline of disk growth since $z = 1$.}
\label{ch:4}

This Chapter is submitted to the Monthly Notices of the Royal Astronomical Society with myself as the first author.

\section{Abstract}
The majority of the stellar mass in the Universe today resides in galaxies with two primary structural components (bulge and disk). In this work, we use the largest contiguous HST imaging region (COSMOS), the Deep Extragalactic VIsible Legacy Survey (DEVILS), and the galaxy structural fitting code {\sc ProFit}, to deconstruct $\sim 35,000$ galaxies into their sub-structures. We use this sample to determine the stellar mass density (SMD) sub-divided by structural components and its evolution since $z = 1$. We find that the majority of stellar mass at all epochs lies in disk-like structures. The SMD in the disk population increases in $z = 1-0.35$ and stabilizes/decreases to $z = 0$ (contributing $\sim 60\%$ to the total SMD at $z = 1$ and declining to $\sim 30\%$ at $z = 0$). This decline is countered by a rapid rise of the SMD in pseudo-bulge population and a consistent growth of spheroidal structures (classical bulges and ellipticals) with significant stellar mass growth in $z = 1-0.35$ and a somewhat flattened trend to $z = 0$. While the physical mechanisms for this are not obvious, the results are consistent with a transition from a Universe dominated by disk growth, to a Universe in which pseudo-bulges are emerging and spheroids are growing. Further study is required before this can be apportioned to internal secular processes, gas accretion, minor and major mergers. However, it is clear that since $z = 0.35$ the processes mentioned above now dominate over quiescent disk growth as the cosmic star-formation history declines.

\section{Introduction}
\label{sec:intro}

Galaxies can experience significant morphological and structural transformation over cosmic time, from clumpy high redshift star-forming disks to smooth red spheroidal systems at the present day (e.g., \citealt{Trujillo07}; \citealt{vanDokkum10}; \citealt{Huertas-Company15}; \citealt{dosReis20}; \citealt{Hashemizadeh21}). However, there are still many open questions as to how galaxies build-up their stellar mass, how it is distributed to form the various structural components, and how these substructures evolve, resulting in the plethora of morphological types observed in the local Universe.  

Two-dimensional photometric decompositions of galaxies have been used in numerous studies to understand the formation pathways of different galaxy types. The earliest 2D decomposition endeavours came from \cite{Byun95}; \cite{Andredakis95} and \cite{deJong96}, which gave us our first understanding of the light distribution variation across different galaxy types. Historically, due to the computational complexity, single S\'ersic profiles (\citealt{Sersic63}) have been used for profile fitting of large sample of galaxies (e.g., \citealt{Simard02}; \citealt{Wuyts11} and \citealt{vanderWel12}; \citealt{Kelvin14}). Galaxies, however, are often more complex requiring extra components such as bulge, bar, etc., to be robustly fit. For example, a two-component model consisting of a spheroidal bulge (\citealt{deVaucouleurs48}) and an extended near exponential disk (\citealt{Freeman70}) have been used to great success in describing the light profile of galaxies (e.g., \citealt{Allen06}; \citealt{Simard11}; \citealt{Mendel14};\citealt{Salo15}; \citealt{Lange16}; \citealt{Dimauro18}; \citealt{Cook19}; \citealt{dosReis20}). Going further, several studies have developed kinematic structural decomposition methods using advanced IFU spectroscopy (e.g., \citealt{Emsellem07}; \citealt{Taranu17}). However, these have so far only been applied to relatively small samples of galaxies, $< 1000$, mostly at low redshifts due to the required high signal to noise ratio (e.g., \citealt{Johnston17}; \citealt{Zhu18}; \citealt{Tabor19}; \citealt{Zhu20}; and \citealt{Oh20}).  

The evolution of galaxies, and particularly multi-component systems, are inevitably tied to the disk and bulge formation scenarios. Currently two leading possible bulge-formation scenarios have been proposed. First, the ``early-bulge formation'' scenario predicts that mergers of small systems in the early Universe resulted in the formation of a spheroidal, pressure-supported system (e.g., \citealt{Aguerri01}; \citealt{Driver13}). Following this a disk grows around the bulge through various gas accretion events. A bulge that has formed in this manner is known as a \textit{classical bulge} (cB) and is dynamically hot, featureless, and similar to a dry major merger remnant, an \textit{elliptical} galaxy (\citealt{Fisher08}). Second, the ``late-bulge formation'' scenario proposes that disks form first and then bulges form through in-situ events within the disk, such as disk instabilities and epicyclic motions (\citealt{Elmegreen08}). In this scenario, disk instabilities can lead to the flow of gas towards the centre of the gravitational potential and epicyclic motions amplify over time once the disk is stable, causing centralised star-formation and the growth of a bulge inside the already established disk. This type of bulge is traditionally known as a \textit{pseudo-bulge} (pB) \citep{Kormendy04}. Unlike cBs, pBs are dynamically cold and rotationally supported (\citealt{Kormendy93}; \citealt{Gao20}). In terms of colour, stellar population and metallicity, pBs are more similar to the outer disk than cBs or ellipticals (\citealt{Fisher06}; \citealt{Du20}; \citealt{Gao20}). Morphologically, pB and cB are argued to be distinguishable through their S\'ersic indices, with pBs having S\'ersic indices close to unity ($n \sim 1$), i.e., a near-exponential surface brightness profile, and cBs having a higher S\'ersic index ($n > 2$)  more akin to that of spheroids (\citealt{Andredakis94}; \citealt{Andredakis95}; \citealt{Fisher06}; \citealt{Mendez-Abreu10}). However, recent kinematic decomposition studies find that S\'ersic index is not a good indicator of different types of bulge (\citealt{Krajnovic13}; \citealt{Zhu18}; \citealt{Schulze18}).  

A popular galaxy formation model called the two-phase scenario involves two periods of (i) a rapid high redshift in-situ star-formation at $2 < z < 6$ \citep{Oser10} and (ii) a successive phase dominated by minor mergers that are thought to form today's spheroidal structures (\citealt{Bluck12}; \citealt{McLure13}; \citealt{Robotham14}; \citealt{Ferreras17}; \citealt{Harmsen17}; \citealt{{DSouza18}}). Following this scenario, several studies compared the central surface brightness of massive high-$z$ spheroids with local galaxies and confirmed that they are structurally similar (\citealt{Hopkins09}; \citealt{Bezanson09}; \citealt{delaRosa16}). These studies reveal that high-$z$ ($z \simeq 1.5$) compact galaxies, also known as red nuggets \citep{Damjanov09}, are possibly at the centre of massive modern galaxies. While the \cite{Oser10} model mainly explains massive galaxies, more generally, by analysing the cosmic star-formation histories of disk galaxies and spheroids, \cite{Driver13} also proposed a two-phase galaxy evolution model. According to this model, compact bulges form first, and then from $z \approx 1.7$ disks grow around the bulges in low density environments and major mergers drive the formation of ellipticals in high-density environments. Note that in reality the above processes (mergers and disk instabilities) will both happen at all cosmic epochs but one process may dominate at high- or low-$z$.

By probing the dominant epochs of bulge and disk formation and the relative contribution of both pB and cB in the galaxy population as a function of time, we can begin to disentangle their likely structural formation and evolution scenarios. While this is of paramount importance to our understanding of galaxy formation mechanisms, previous studies exploring the evolution of galaxy components on large evolutionary baselines have been hampered on a number of fronts. First, stellar populations cause colour gradients, so that measured parameters would vary due to bandpass shifting when comparing high-$z$ with low-$z$ images in the same wavelength band (e.g., \citealt{Kelvin14}; \citealt{Vulcani14}; \citealt{Kennedy16}). Second, dust is argued to distort our structural measurements including S\'ersic index and effective radius. Therefore, due to dust attenuation it is often impossible to measure the true profiles (e.g., \citealt{Pastrav13}). Third, galaxies are often more complicated than only a bulge+disk, so that it is not always obvious how to determine the appropriate number of components to fit (e.g., \citealt{Salo15}; \citealt{Lange16}). 

Motivated by this and recent software development in both source identification and structural fitting routines, we now revisit the structural decomposition of galaxies from $z \simeq 1$ to the present day.  We perform a robust 2D photometric decomposition of galaxies in the Deep Extragalactic VIsible Legacy Survey (DEVILS; \citealt{Davies18}) $10^h$ region (D10) using the Hubble Space Telescope (HST) imaging dataset of the Cosmic Evolution Survey (COSMOS). For our modeling, we make use of the state-of-the-art galaxy fitting software {\sc ProFit} (\citealt{Robotham17}). In this study, we adopt the perspective of fitting a disk and bulge complex, where the complex might be a cB, pB, and in some cases a combination. Using these decompositions, we explore the evolution of the stellar mass density contribution of structural components, and use this to propose a solution to the competing bulge-formation scenarios.

This work is structured as follows. Section \ref{sec:SampleSel} discusses the D10/ACS sample, in Section \ref{sec:ProfileFitting} we outline our fitting pipeline ({\sc GRAFit}) and the tools used therein, as well as the HST PSF modelling. The verification of our structural analysis as well as our method for distinguishing between pB and cB are described in Section \ref{sec:profselection} and we then explain the evolution of the SMF and SMD in Sections \ref{Sec:SMF_evol} and \ref{sec:rho}, respectively. Finally, we discuss and summarize our results in Section \ref{sec:discussion} and \ref{sec:summary}. 

Throughout this paper, we use a flat standard $\Lambda$CDM cosmology of $\Omega_{\mathrm{M}} = 0.3$, $\Omega_\Lambda = 0.7$ with $H_0 = 70 \mathrm{km}\mathrm{s}^{-1}\mathrm{Mpc}^{-1}$ (\citealt{Planck20}).
Magnitudes are given in the AB system \citep{Oke83}.

\section{D10/ACS Sample and HST imaging data} 
\label{sec:SampleSel}

In this study, we use the D10/ACS sample constructed in \cite{Hashemizadeh21}, where we provide a morphological classification of the sample into single- and double-component categories. Briefly, D10/ACS is a sample of galaxies in the $10^h$ the Cosmic Evolution Survey (COSMOS; \citealt{Scoville07}) region of the Deep Extragalactic VIsible Legacy Survey (DEVILS, \citealt{Davies18}). It consists of $35,803$ galaxies with multi-wavelength photometry from FUV to far-IR wavelengths (i.e., 0.2 to 500 micron; Davies et al. in prep.) and the sample extends up to $z = 1$ for systems with log$(\mathrm{M}_*/\mathrm{M}_\odot) \geq 9.5$. The redshift and stellar mass limits were set from visually inspecting galaxies drawn from the $\mathrm{M}_*-z$ plane and identifying the region where visual classification and 2D structural analysis was deemed viable, see \cite{Hashemizadeh21} for more details.

In order to perform our structural decomposition, we make use of the main imaging data of COSMOS, taken with the Advanced Camera for Surveys (ACS\footnote{ACS Hand Book: \href{http://www.stsci.edu/hst/acs/documents/handbooks/current/c05\_imaging7.html\#357803}{www.stsci.edu/hst/acs/documents/}}) on the Hubble Space Telescope (HST). It covers 1.7 square degrees centred at RA $150.12$ ($10:00:28.600$), and DEC $+2.21$ ($+02:12:21.00$) (J2000). The ACS observations used the F814W filter (I-band), providing good depth and flux measurements mostly red-ward of the $4000$\AA \, break out to $z = 1$, i.e., one is sampling red-ward of the Balmer and $4000$\AA \, break out to $z = 1$ at 814nm (\citealt{Hashemizadeh21}).
We use the drizzled COSMOS HST images for our bulge-disk decomposition analysis, which utilises the MultiDrizzle code \citep{Koekemoer03}. These data have been re-sampled to a pixel scale of $0.03$ arcsec from the original ACS pixel size of $0.05$ arcsec. 
The redshift and stellar masses used in the present work are taken from the DEVILS/D10 master redshift catalogue (\texttt{DEVILS\_D10MasterRedshiftCat\_v0.2}) and the \texttt{DEVILS\_D10ProSpectCat\_v0.3} catalogue, described in detail in \cite{Thorne20}.
For their stellar mass measurements, they perform SED fitting with {\sc ProSpect} code \citep{Robotham20} using \cite{BC03} stellar libraries, the \cite{Chabrier03} IMF, \cite{Charlot00} to model dust attenuation and \cite{Dale14} to model dust emission. \cite{Thorne20} uses the new multiwavelength photometry measurements in the D10 field (\texttt{DEVILS\_PhotomCat\_v0.4}; Davies et al. in prep.). They report stellar masses $\sim 0.2$ dex higher than in COSMOS2015 catalogue \citep{Laigle16} due to different modelling of the evolving gas phase metallicity. See \cite{Thorne20} for more details.




\section{Profile Fitting} 
\label{sec:ProfileFitting}

In order to perform bulge-disk decompositions we need to consider a number of elements, which include: the pixels that are used for the fitting ({\sc ProFound}, \citealt{Robotham18}), the code for fitting the structural parameters ({\sc ProFit}, \citealt{Robotham17}), and our management of the end-to-end process including modelling of the {\it Hubble Space Telescope} point-spread function ({\sc GRAFit}). These are described below in full detail and the non-technical reader may wish to move forward to Section \ref{sec:profselection}, where we show and validate our resulting fits.

\subsection{ProFound}
Critical for a robust structural analysis is appropriate selection of the pixels used in the fitting. This process needs to ensure neighbouring objects are removed or flagged, but also aims to maximize the number of true pixels associated with the object. To achieve this we make use of {\sc ProFound} \citep{Robotham18}, an open source astronomical image analysis package. {\sc ProFound} analyses the image pixels, identifies all distinct sources, and provides a segmentation map for use in our fitting process. In addition, the code provides basic object size, and flux information that is used to define the initial parameters for the fitting code (this is non-essential but reduces the burn-in time of the MCMC fitting). The {\sc ProFound} segmentation map is a fundamental input for running {\sc ProFit} and specifies those pixels associated with the source, and from which the likelihood is computed. 
In addition to the segmentation, we also make use of {\sc ProFound}'s photometric measurements to provide initial estimates of the half-light radii, magnitudes, flux centers, axial ratios, and angle of rotation for the disk to be passed to {\sc ProFit}. 

The key distinction of {\sc ProFound} from previous source-detection codes is that it constructs segments that trace the outline of the galaxy as opposed to circles or ellipses. This is critical, as galaxies are not perfect ellipses, and elliptical apertures will not always accurately represent their flux distribution. Moreover, in complex regions, ellipses of neighbouring objects may overlap or intersect and disentangling the flux is complex. {\sc ProFound}'s solution is to define segments, based on the outer isophote, and to dilate these segments until they contain $95\%$ of the source's flux, essentially performing a curve of growth analysis. Notably, the dilation process does not allow segments to ever overlap and, therefore, each pixel is allocated entirely to a single object or left unallocated. This avoids the need to disentangle flux from objects, but can include some intervening light from neighbouring sources. On the whole, the dilation process is more aggressive for more luminous objects, and so pixels should end up assigned to the object that dominates the light. This aspect is somewhat of a trade-off between the errors associated with the dominant flux versus allowing for some cross-contamination. In our analysis we take the decision that the latter is less liable to gross error.

\subsection{ProFit}
\label{subsec:ProFit}

To determine bulge-disk decompositions, we use the Profile Fitting package, {\sc ProFit} (\citealt{Robotham17}). This was specifically designed for 2D structural analysis and can use a wide range of minimisation algorithms, essentially any of those available in {\sc R}, to obtain reliable solutions with robust error analysis, which are independent of the initial parameters. {\sc ProFit} and the low-level C++ library ({\sc libprofit}) are combined with a high-level R interface. Several profiles are in-built in {\sc ProFit} and any combination, as well as user defined profiles, can be used to model galaxy images. The in-built profiles are: S\'ersic, Core-S\'ersic, broken-exponential, Ferrer, Moffat, empirical King, point-source, and sky. We use a S\'ersic profile for both the disk and bulge components, i.e., a double S\'ersic fit.

This profile is described in \cite{Sersic63}; (also see \citealt{Graham05}) and provides an analytic formula for the light intensity profile as a function of radius:

\begin{equation}
I(r)=I_e \exp\bigg[-b_n \bigg(\left(\frac{r}{r_e}\right)^{1/n} - 1\bigg)\bigg],
\end{equation}

\noindent where $r_e$ is the effective radius, the radius containing half of the total flux, $I_e$ is the intensity at that radius and $n$ is known as the S\'ersic index that specifies the shape of the profile. For example, $n=0.5$, $n=1$ and $n=4$ represent Gaussian, exponential and de Vaucouleurs profiles (\citealt{deVaucouleurs48}), respectively. In general, it has been shown that disks are likely to follow an exponential profile, as opposed to classical-bulges and spheroids which tend to follow a near-de Vaucouleurs profile - at least for the more luminous bulges (e.g., \citealt{Patterson40}; \citealt{deVaucouleurs59}; \citealt{Freeman70}; \citealt{Kormendy77}).

Compared to {\sc GALFIT} (\citealt{Peng02,Peng10}), {\sc ProFit} is more robust to the effects of local minima due to its compatibility with several optimization algorithms such as Markov Chain Monte Carlo (MCMC); as was shown in \cite{Robotham17}. In this work, we use the Componentwise Hit-And-Run Metropolis (CHARM) algorithm in our MCMC sampling. We refer the reader to \cite{Robotham17} for further details regarding {\sc ProFit}.

\begin{figure*}
	\centering
	\includegraphics[width=\linewidth]{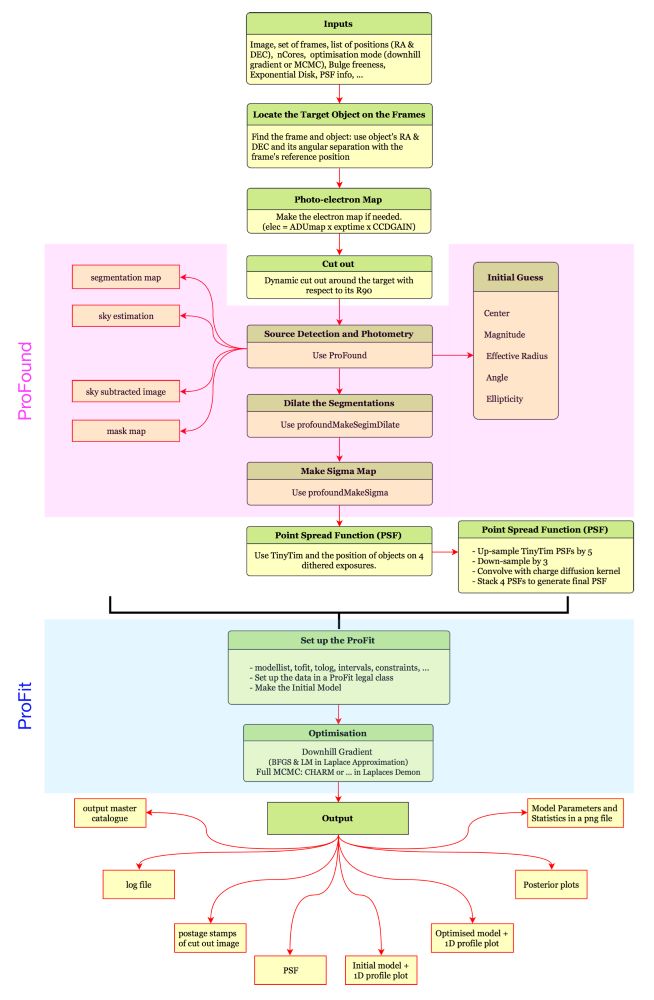}
	\caption{Flow diagram of {\sc GRAFit} containing five main parts; (i) inputs and cut-out generation, (ii) running ProFound, (iii) PSF generation (iv) running ProFit and (v) outputs. }
	\label{fig:grafitflowdiagram}
\end{figure*}

\subsection{Pipeline: {\sc GRAFit}}
\label{sec:GRAFit}
In order to manage the full end-to-end process, including HST point-spread function measurement at the location of each galaxy, we developed an automatic galaxy decomposition pipeline, {\sc GRAFit}. {\sc GRAFit} is a series of modules and functions developed in {\sc R} with calls to {\sc ProFit}, {\sc ProFound} and other astronomical tools. {\sc GRAFit} is principally designed to operate on HST ACS data, however it can be used with any imaging survey. The full process is reasonably complex and hence the flow diagram for {\sc GRAFit} is shown in Figure \ref{fig:grafitflowdiagram}.

The minimum requirement to run {\sc GRAFit} is either a galaxy image in standard format (e.g., a FITS file) and a list of RA \& DEC positions indicating the location of the objects to be profiled, or a directory of pre-cutout postage-stamp images. In the case of the latter, {\sc GRAFit} identifies the correct image with which to extract the target object(s).   
By default, {\sc GRAFit} allocates both a bulge and a disk to the galaxy by performing double S\'ersic modelling, which distributes the total flux into bulge and disk. However, by altering the \texttt{nComp} flag, the user can also model a single S\'ersic profile. Since {\sc GRAFit} is efficiently programmed as parallel code, one can spread the tasks over multiple cores/nodes using the flags \texttt{nCores} and \texttt{ThreadMode}. It is hence supercomputer friendly, and has now been actively used on a number of supercomputer architectures. There are some other additional parameters that can be added (see Figure \ref{fig:grafitflowdiagram}).

{\sc GRAFit} is a modular-based script with a central master script, \texttt{GRAFitMaster}, that calls other modules internally. At the very first step, {\sc GRAFit} locates the object(s) by searching all the frames, runs {\sc ProFound}, identifies the segment associated with the desired object (position matching) and then makes a dynamic cutout around the galaxy. See Section \ref{subsec:Sky est} for more details. Initial estimation of the structural parameters are made, and a sigma (noise) map generated that indicates the errors in pixels across the image using \texttt{profoundMakeSigma}. This noise map includes a pixel-by-pixel mapping of the combined (in quadrature) sky noise (skyRMS), read noise and the RMS of the dark current noise, where pixels associated with interloping objects are masked out.

\begin{figure*} 
	\centering
	\includegraphics[width=\textwidth]{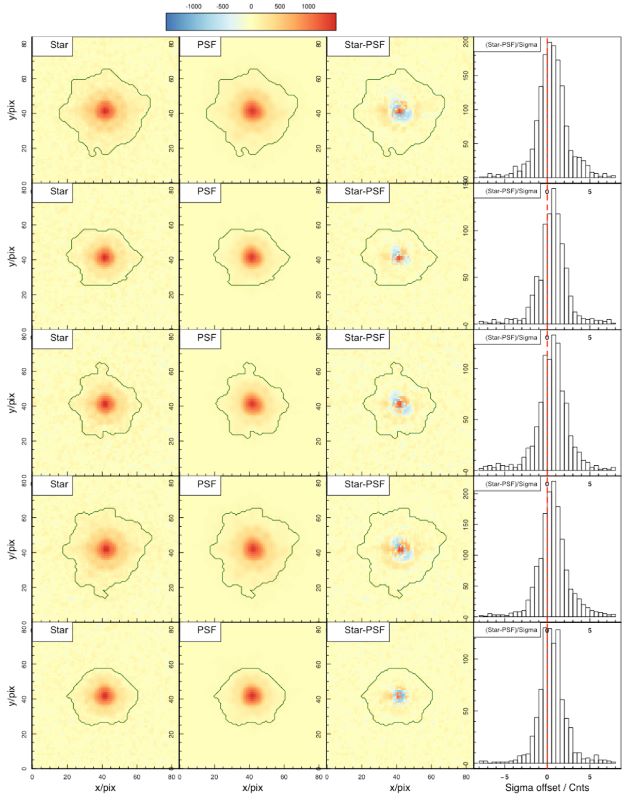}
	\caption{Five stars that were selected to subtract from associated PSFs (see the text for selection method). First column shows the stars in the drizzled images. Second and third columns display the PSF and the residual (Star-PSF), respectively. Fourth column represents the distribution of residual pixel values (Star-PSF/Sigma). }
	\label{fig:starsub_5star}
\end{figure*}

\subsubsection{Modelling the HST/ACS PSF} 
\label{sec:PSF}

With the raw pixel cutout, the associated segment map, the initial parameter guess, and the noise map all prepared, the final -- and perhaps most complex -- aspect of the decomposition process is the {\it Hubble Space Telescope} ACS point spread function (PSF) modelling. Having an accurate PSF is obviously critical for modelling the central structures of galaxies. However, due to the off-axis location of the HST's ACS optics, the HST PSF is geometrically distorted and asymmetric along both X and Y directions \citep{Anderson06}. For this reason, we use the publicly available software {\sc Tiny Tim} \citep{Krist11} to model the PSF for different pixel positions on the ACS detections as observed through the F814W filter and also implement the function \texttt{tiny3} to apply the final ACS PSF geometrical distortion. 

In the mosaiced COSMOS HST/ACS imaging data, each pointing has been constructed with 4 distinct exposures (each 507 seconds), dithered by a few tens of pixels in both X and Y directions, to allow cosmic ray and bad pixel rejection. The dithering also compensates for the gap between the two ACS chips. We therefore revert to the four raw exposure frames to locate the position of our target object on each exposure. This enables us to generate four PSFs that we combine (i.e., stack) to produce the representative PSF for each object at 0.05 arcsec resolution. The mosaiced COSMOS HST/ACS imaging data is ultimately provided re-sampled to a pixel size of 0.03 arcsec (\citealt{Koekemoer07}). We therefore also re-sample the final stacked PSF to a 0.03 arcsec pixel scale (1.6 factor). {\sc Tiny Tim} only allows integer sub-sampling, so we first up-sample the PSF by a factor of 5 in the final stage of the {\sc Tiny Tim} process (by selecting SUB=5 in the \texttt{tiny3} function). We then down-sample the output PSF by a factor of 3 in an external step, leading us to the desired pixel size (0.03 arcsec/pix). 

{\sc Tiny Tim} does not automatically convolve the sub-sampled PSF with the CCD charge diffusion kernel. This is required, as point sources experience a slight blurring due to the charge diffusion into adjacent pixels. This reduces the sharpness of the PSF and causes a $\sim 0.5$ magnitude loss in WFC imaging at short wavelengths. Such blurring, which is also known as the pixel response function (PRF), is also field dependent due to the non-constant CCD thickness (12 to 17 microns for the WFC). See the ACS handbook\footnote{ACS Hand Book: \href{http://www.stsci.edu/hst/acs/documents/handbooks/current/c05\_imaging7.html\#357803}{www.stsci.edu/hst/acs/documents/}} 
for more detail. To simulate this blurring effect, {\sc Tiny Tim} provides the charge diffusion kernel as a $3\times3$ matrix in the PSF's header. This kernel is specific to the PSF's location and we use this kernel matrix and convolve it with our final re-sampled PSF. 

\subsubsection{Testing the HST PSF modelling} 
\label{sec:PSF_test}

To evaluate the accuracy of our PSF modelling, we perform a star subtraction test using the HST/COSMOS images. For this, we randomly select 5 bright unsaturated stars with half light radii of $R50\sim0.07$ arcsec (the typical radius seen), and with axial ratio of $> 0.9$ to ensure that the stars are unlikely to be binary systems. Note that R50 is obtained from our {\sc ProFound} analysis. See Appendix \ref{sec:star_sel} for more details on our star selection.

In Figure \ref{fig:starsub_5star}, the first column shows the star as observed (with segment boundary), the second column shows our modeled PSF to the same scale, and the third shows the residual having subtracted the PSF from the star. The rightmost column is the distribution of the pixel residual. Note that when subtracting the PSFs from stars, their centers must be accurately matched to the sub-pixel level to guarantee that there is no offset between the centres of the star and the PSF. We use {\sc ProFit} to interpolate the flux and find the sub-pixel center. For this we model a point source with the magnitude of the real star and convolve it with the PSF. We then run an optimization with the BFGS \footnote{Broyden-Fletcher-Goldfarb-Shanno} algorithm \citep{Broyden70} to find the accurate sub-pixel center and magnitude, and perform star subtraction precisely. We then analyze the residuals and the goodness of fits (GOF) calculated as GOF $=$ (PSF$-$star)/star and find $\mathrm{GOF} \sim 80\%$ implying that our PSFs simulate on average $\sim 80\%$ of the real stars' pixels with the most significant residual evident for the central pixel. 

For an additional quality check we apply a similar process to a star in the raw exposure frames. This is necessary to demonstrate that our PSF generation process such as re-sampling and charge diffusion kernel convolution is not affecting the PSF's profile, particularly for the central pixels. We present the result of this test in Figure \ref{fig:starsub_raw}. Here, the PSF is not required to be re-sampled as it is already matched with the original pixel scale of $0.05$ arcsec identical to the raw ACS imaging data. Again, a residual can be seen at the centre and the spread is in agreement with our previous conclusion. We therefore note that while {\sc Tiny Tim} represents the best model of the HST PSF, it comes with limitations. An aspect of the Hubble Space Telescope PSF not accounted for is the periodic ``breathing'' of the telescope referring to the small changes of the telescope's focus due to micron-scale movements of the secondary mirror \citep{Hasan94}. Currently this is outside the bounds of {\sc Tiny Tim} to model, and would require shifting to an empirical database of PSFs, currently under development at Space Telescope Science Institute (STSci). 

\begin{table}
\centering
\caption{{\sc ProFound} setting arguments. The rest of massive {\sc ProFound}'s arguments were left as default.}
\begin{adjustbox}{scale = 0.8}
\begin{tabular}{ll}
\firsthline \firsthline \\

{\sc ProFound} argument & Value \\
\firsthline \\
\texttt{tolerance} & 7 \\
\texttt{sigma} & 7 \\
\texttt{pixcut} & 3 \\
\texttt{skycut} & 1.1 \\
\texttt{smooth} & TRUE \\
\texttt{size} (dilation) & 51 \\
\texttt{ext} & 2 \\
\texttt{box}1 (initial run) & $5\times \mathrm{R90}$ (Y-UltraVista) \\
\texttt{box}2 (final run) & $3\times \mathrm{R90}$ (I-ACS) \\
\texttt{type} & "bicubic" \\
SD of Gaussian Priors & (2,5,0.3,0.3,0.1,30,0.3,$\infty$) \\
(position,mag,Re,bulge-$n$,disk-$n$,angle,axial ratio, boxiness) & \\

\lasthline
\end{tabular}
\end{adjustbox}
\label{tab:ProFound_par}
\end{table}

\begin{figure}
	\centering
	\includegraphics[width=\textwidth]{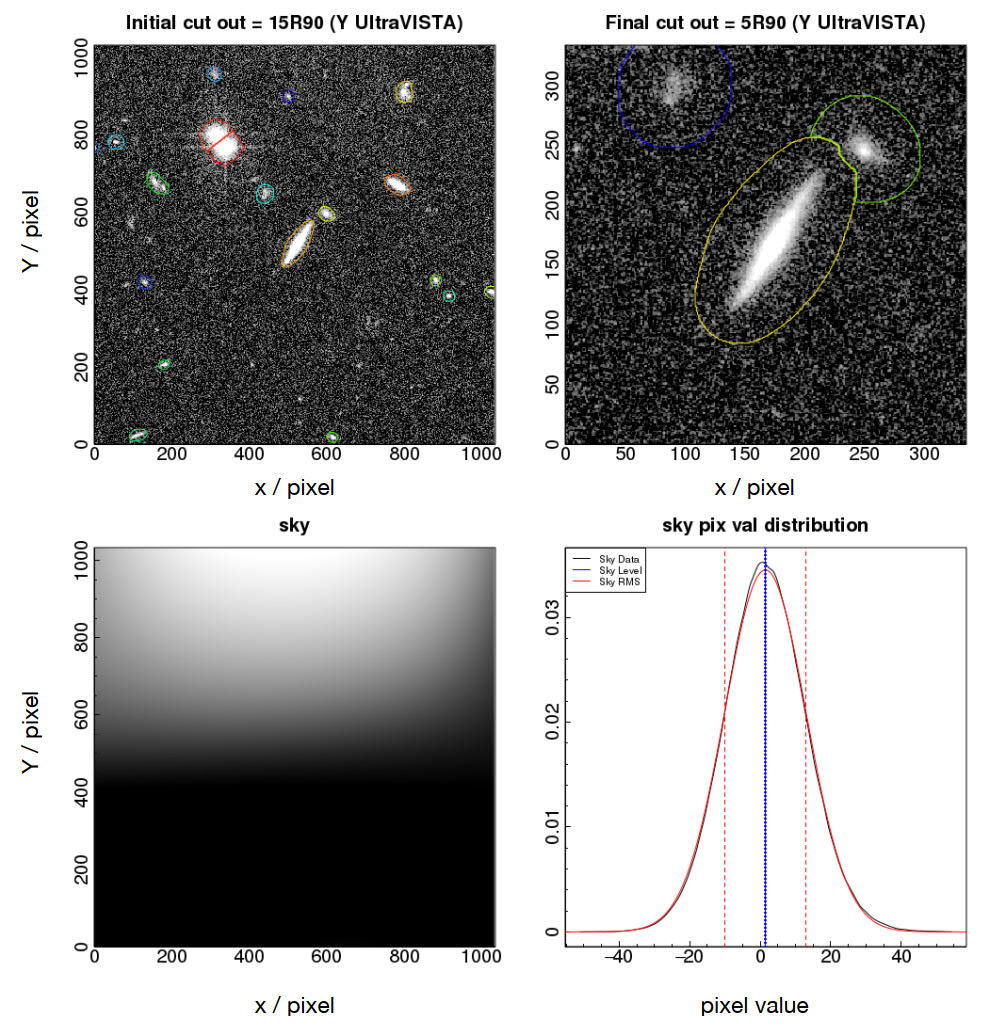}
	\caption{ A set of postage stamps showing the initial cut out of 15$\times$R90 around central galaxy in the top left panel. We perform the source detection and sky estimation on this large cut out. The extracted sources by {\sc ProFound} are over-plotted as isophotal contours. Top right panel displays the final cut out of 5$\times$R90. Bottom left and right panels show the sky image and sky pixel value distribution, respectively. In bottom right panel, the black and red curves represent the distribution of the real sky and sky RMS, respectively. The blue vertical line shows the measured sky level.  }
	\label{fig:postageStams}
    \end{figure}
    
\subsubsection{Dynamic cutout and sky estimation} 
\label{subsec:Sky est} 

A robust sky estimation is also crucial in accurately determining source properties in {\sc GRAFit}. For this, we run {\sc ProFound} on our desired image cutout. Table \ref{tab:ProFound_par} shows the {\sc ProFound} settings and argument values that we find work appropriately on our data. To measure the sky and sky RMS, we must use a sufficiently large cutout around the object without creating overly large images. Typically, we can only obtain a reliable sky estimate when $>50 \%$ of the pixels in the cutout region are sky, however, one has to be mindful of extraneous low surface brightness features. As shown in Figure \ref{fig:postageStams}, we therefore use a box car filter with a size scaled to that of the target object and set this to 15$\times$R90 and run {\sc ProFound} on this large cutout (top left panel of Figure \ref{fig:postageStams}). R90 is the radius encompassing 90 percent of the main object's flux. {\sc GRAFit} then sets the box car size for sky estimation as to be 5$\times$R90. The R90 of the objects are taken from the DEVILS input catalog UltraVISTA (\citealt{McCracken12}) Y-band using {\sc ProFound} (see \citealt{Davies18} for more details). Our choice therefore guarantees that more than 50\% of the pixels are real sky pixels. We then use the median and quantile range to estimate a constant sky level as these have been shown to be more stable than the sky mean and standard deviation. For more details on this see the {\sc ProFound} package description\footnote{\href{https://github.com/asgr/ProFound}{https://github.com/asgr/ProFound}}.

Following the sky measurement we perform a final dynamic cutout with 5$\times$R90 (top right panel of Figure \ref{fig:postageStams}) to reduce the number of pixels for a faster run time but {\sc GRAFit} allows the final cutout size to grow if the main source's segmentation touches the outer edges of the cut out (e.g., for extreme inclination objects). An example of this process is shown in Figure \ref{fig:postageStams}. 

The bottom left panel of Figure \ref{fig:postageStams} displays the distribution of the sky pixel values, and the bottom right panel shows the sky and sky RMS in black and red curves, respectively. Therefore, the closer the sky RMS distribution is to the sky distribution, the more accurate sky measurement we achieve. {\sc GRAFit} handles all the above processes using the module \texttt{GRAFitDynamo}.

\begin{figure*}
	\centering
	\includegraphics[width=\linewidth]{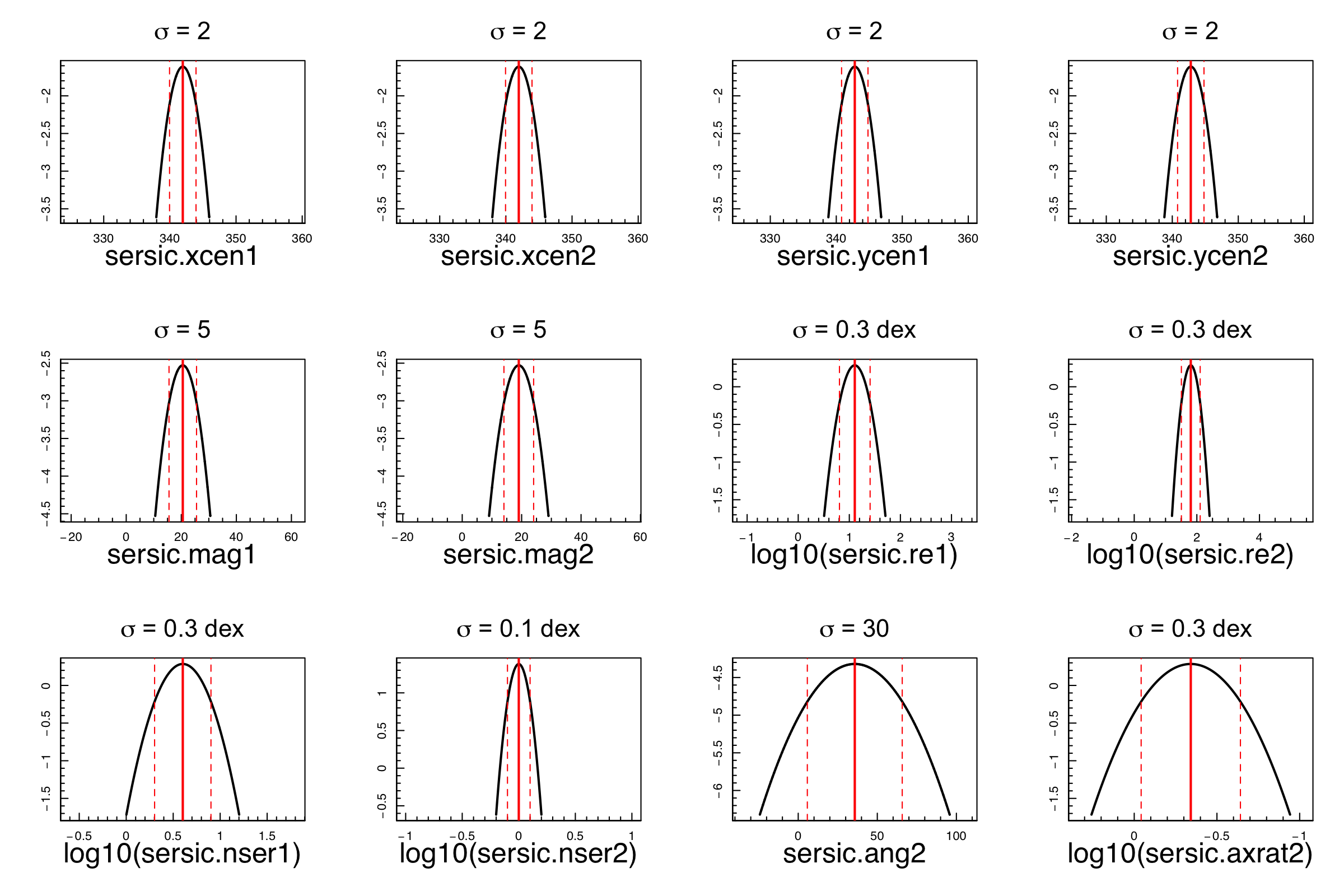}
	\caption{The prior distributions for all the parameters that we fit for a sample galaxy of D10/ACS. We consider Gaussian priors with appropriate mean and standard deviation. The mean values, shown as red solid vertical lines, are generally our initial guesses coming from photometric measurements by {\sc ProFound}. Dashed red vertical lines represent 1$\sigma$ region of the Gaussian distributions. Note that indices 1 and 2 refer to bulge and disk, respectively. Also, re is in the unit of pixel and angle in degree.}
	\label{fig:priorplot}
\end{figure*}

\subsubsection{Other input priors}
\label{subsubsec:priors}

 {\sc GRAFit} makes use of the ProFit function \texttt{profitSetupData} to set up all the data in a ProFit-understandable class, and then provides a PSF convolved model image given a set of structural parameters. These include the half-light radius ( $\rm{R}_e$), the S\'ersic index ($n$), the ellipticity ($e$), and the x and y coordinates of the central pixel. 
{\sc GRAFit} also imposes limits on the S\'ersic indices of: 0.5 $\leq$n$\leq$ 1.5 for disks and 0.5 $\leq$n$\leq$ 20 for bulges. 
{\sc ProFit} also accepts prior distributions for each of the parameters. Within {\sc GRAFit} we define Gaussian prior distributions with the mean value of the initial {\sc ProFound} measurements, and standard deviations of $2,5,0.3,0.3,0.1,30,0.3,\infty$ for the central position, magnitude, $\rm{R}_e$, bulge's S\'ersic index, disk's S\'ersic index, position angle, ellipticity and boxiness, respectively. Note $\infty$ corresponds to a flat distribution and S\'ersic indices are fitted in log space, and hence the standard deviation is in dex. Figure \ref{fig:priorplot} displays the prior distribution for a random galaxy in our sample. 
Unlike the parameters' limits (uninformed priors) which allow a wide exploration of the parameter space, the priors are tighter and based on our prior knowledge of galaxy parameters from our {\sc ProFound} photometry. The priors are also not too restrictive and solutions can be found outside the prior range if the data requires it.

{\sc GRAFit} also gives the user the flexibility to choose whether the centres of disk and bulge should be tied together, or allowed to roam to some pixel tolerance, which is sometimes necessary due to disk asymmetry and the presence of dust. In this work, by testing different offsets and performing visual inspections we select the maximum offset of the bulge from the disk to be 11 pixels ($0.33$ arc seconds). 






\begin{figure*}
	\centering
	\includegraphics[width=\textwidth]{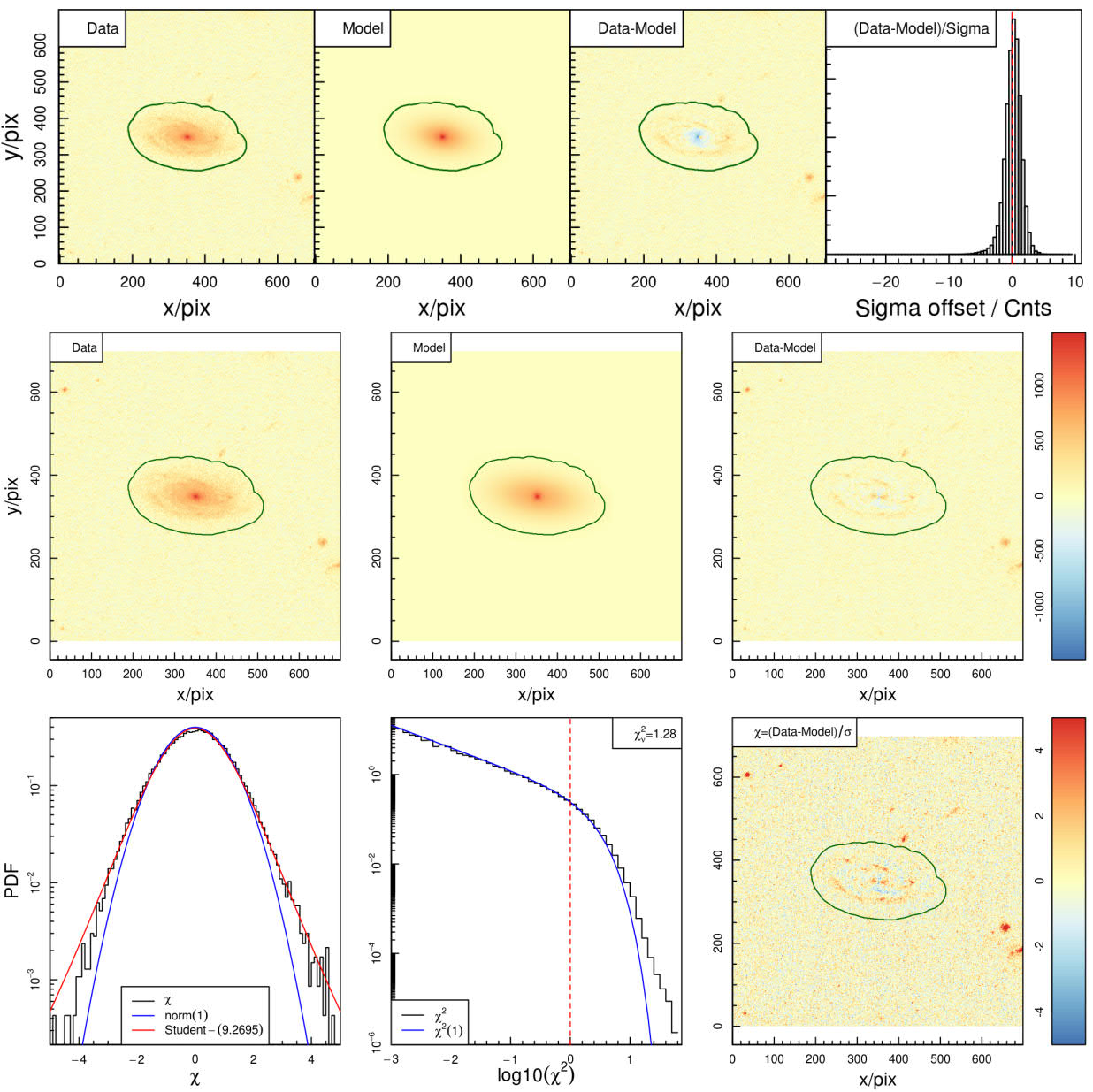}
	\caption{Our double S\'ersic decomposition of galaxy D103274337 highlighting the standard {\sc ProFit} outputs. Top row shows our initial model inferred from initial conditions from {\sc ProFound} photometry. First, second and third panels show the galaxy image (data), initial model and residual (Data-Model), respectively. Forth panel shows the distribution of the residual pixel values ($\chi = \mathrm{Data-Model}/\sigma$), indicating the sigma offset of the model's pixel value from actual image within the segmentation area (green isophotal contours).  
	Middle and lower rows show our final MCMC optimised model. Left, middle and right panels of the middle row show data, final model and residual, respectively. Lower row indicates the residual pixel value distribution (left), the $\chi^2$ distribution with an overlaid $\chi^2$ with one degree of freedom (middle) and the 2D residual pixel map scaled by $\sigma$ (right). The residuals indicate non-smooth structures in this galaxy, in this case spiral arms.  }
	\label{fig:output_init_final}
\end{figure*}

\begin{figure}
\centering
\includegraphics[width=0.9\textwidth]{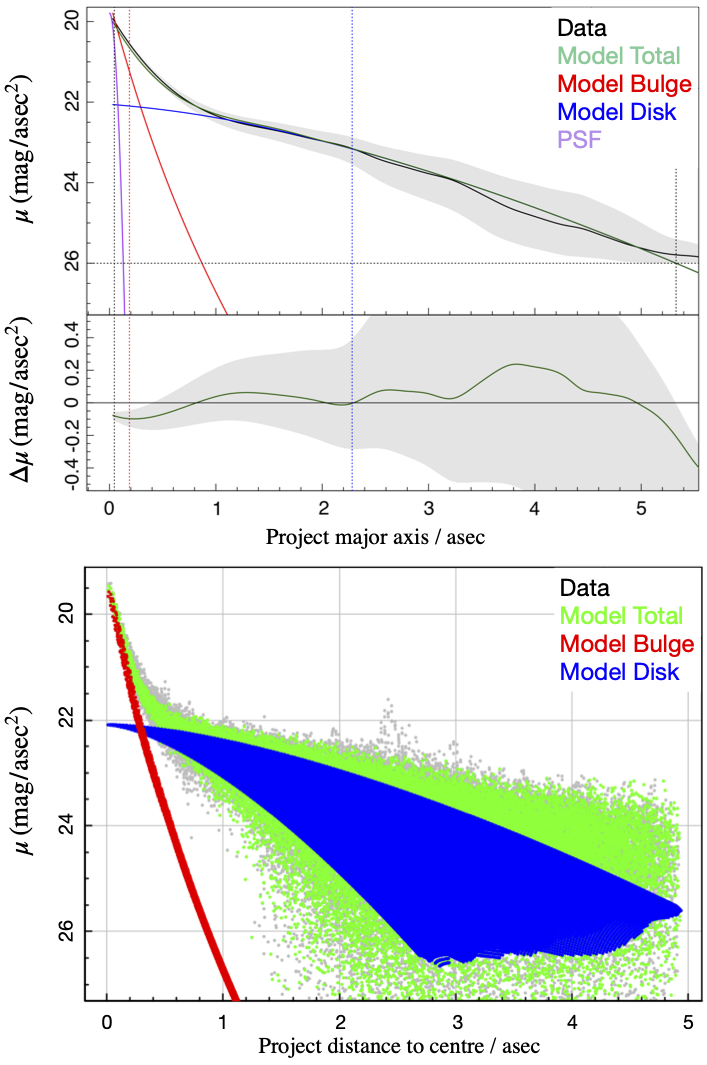}
\label{subfig:stMass_PDF}
\caption{ Top panel: Radial profile of D103274337 highlighting the variation of the surface brightness as a function of projected major axis. Black curve shows the data with its $1\sigma$ region shown as gray area while green, red and blue represent the total, bulge and disk optimized models, respectively. The purple profile shows the PSF. Middle panel: indicates the one-dimensional residual of the image and model profiles. Dashed vertical lines show the half-light-radius, R50, of each profile. Bottom panel: pixel by pixel surface brightness as a function of their distance to the centre of the galaxy.}
\label{fig:1D_pro}
\end{figure}

\subsection{{\sc GRAFit} Outputs} 
\label{sec:GRAFit_out}

The top panels of Figure \ref{fig:output_init_final} show an example of the initial model for galaxy UID = 101494996111806000 at $z = 0.53$. In this Figure, we show the galaxy image (data), our initial model using {\sc ProFound} parameters and the residual. From the residual pixel values and its histogram (right-most panel), we see that our initial parameters provide to first order a relatively good model of the galaxy, where the over- and under-subtracted regions in the residual are mapped with blue and red colours, respectively. 

We then run our initial model through an MCMC optimization process using the CHARM algorithm within the \texttt{LaplacesDemon} package implemented in {\sc R}. We show our final optimized model for the above galaxy in the middle and bottom rows of Figure \ref{fig:output_init_final}. Now, we see that the residual pixel maps (right panels) are significantly improved with most of the non-zero residual pixels highlighting small scale features, such as spiral arms, star-forming clumps and a bar.

Figure \ref{fig:1D_pro} shows the collapsed one-dimensional radial profile of our final model for UID = 101494996111806000. The bottom panel shows the surface brightness values of the actual pixels together with total, disk and bulge model pixels. Finally, in Figure \ref{fig:triPlot}, we display the corner plot of the stationary MCMC chain of our double S\'ersic model for UID = 101494996111806000.
Alongside the graphical outputs, {\sc GRAFit} returns a comprehensive catalogue including all the inputs and final model parameters. 

We now run GRAFit over our full sample of $35,803$ galaxies electing to fit three models in each case: 
\begin{enumerate}
\item a single S\'ersic model with free S\'ersic index $n$ ($0.5 < n < 20$).
\item a double S\'ersic model with near exponential disk ($0.5 < n_d < 1.5$) and free S\'ersic index for bulge ($0.5 < n_b < 20$).
\item a double S\'ersic model but with a fixed exponential disk ($n_d = 1$) and free S\'ersic index for bulge ($0.5 < n_b < 20$).
\end{enumerate}

\begin{figure*}
	\centering
	\includegraphics[width = \textwidth, angle = 0]{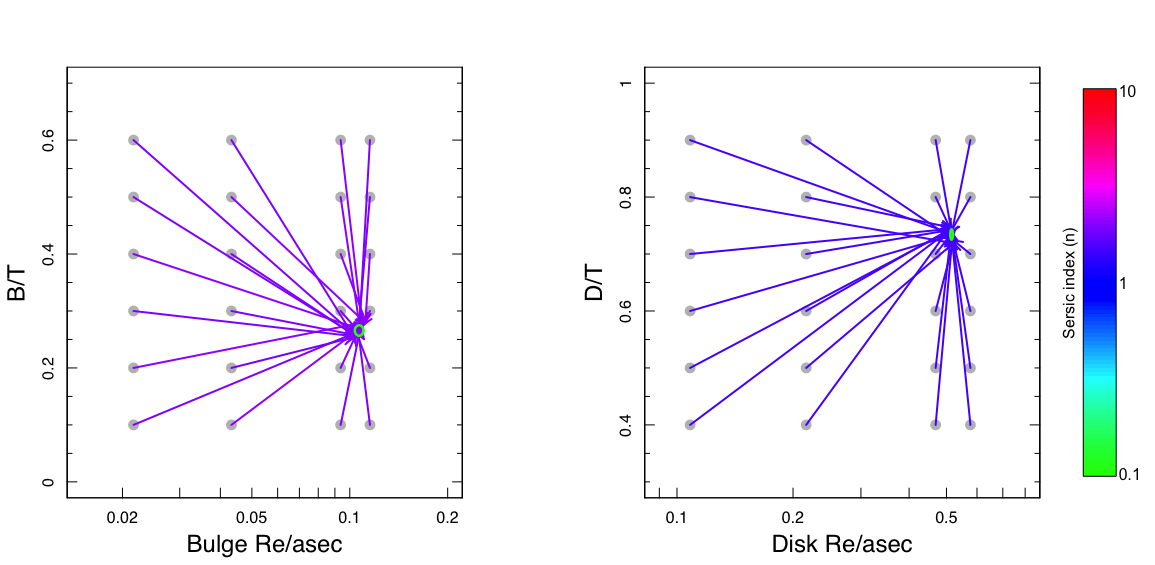}
	\caption{A convergence test of ProFit performance on galaxy D102465848. We start the model from different initial conditions and show that the final model is converged to an identical answer. Left panel shows the relation of B/T with Bulge $\rm{R}_e$ and bulge's S\'ersic index. Arrows indicate the initial and final values colour coded according to the S\'ersic index. Right panel displays the D/T as a function of disk $\rm{R}_e$ and disk's S\'ersic index.}
	\label{fig:Conv_test_HST}
\end{figure*}

\subsection{Convergence test of the MCMC minimisation} 
\label{sec:MCMC}

As a final check of the {\sc ProFit} MCMC approach, we consider the possibility of our algorithm becoming trapped in local minima of the parameter space. To explore this, we address the issue raised by \cite{Lange16} who used {\sc GALFIT3} \citep{Peng10} to fit nearby galaxies from the GAMA survey \citep{Driver11}, and found in many cases a strong dependence of the recovered parameters on the initial input parameters. This was demonstrated by running a grid of input parameters, and in many cases finding convergence was not consistent. We repeated this experiment for a random sample of galaxies, and found that consistently, for a grid of initial conditions, we recover the same solution. An example of this is shown as Figure \ref{fig:Conv_test_HST} for galaxy UID = 101501367451880000 which reflects the results seen for all the galaxies tested in this fashion. We therefore conclude that, in general, the MCMC is finding unique global solutions, and not getting trapped in local minima. To further highlight this point, we specifically explore an extreme failure highlighted by \citealp{Lange16}, i.e., galaxy G32362 (see figure 6 of \citealp{Lange16}), for which we find convergence using {\sc ProFit}, but in this instance applied to the identical Sloan Digital Sky Data as used by \cite{Lange16} (see Appendix \ref{sec:conv_GAMA}).

\section{One-component or two-component profile selection} 
\label{sec:profselection}

The GRAFit package produces viable outputs and three different models for all $\sim 35$k systems with only 33 cases ($< 0.07 \%$ of the sample) failing due to exceeding the computation (wall) time. We now need to determine which of our three fits is the most appropriate representation for each galaxy. 
As reported in \cite{Hashemizadeh21}, we explored the prospect of using different methods including cross matching with other available morphological catalogues to try to determine whether a galaxy contained a single dominant component or two distinguishable components.
Ultimately, we found no suitable solution that aligned well with our visual classifications. 
For this reason, we select either a S\'ersic (1C) or S\'ersic+S\'ersic (2C) profile based on our prior visual classifications. For elliptical (E) and pure-disk systems (D), we adopt the 1C profile, and for bulge+disk systems (BD) we adopt a 2C model. 
In a small fraction of cases 2C fits were poor due to an un-physical fit (e.g., $\mathrm{R}_{e, bulge} \gg \mathrm{R}_{e, disk}$). For these objects, we assess whether a S\'ersic+exponential disk profile solves the problem and find that for $\sim 3\%$ of the sample (1,072 objects) this profile describes the light distribution better than S\'ersic+S\'ersic. The rest of un-physical fits are flagged as poorly fitted in the final catalogue ($\sim 5\%$ of the full sample). This resulted in $3,812$ 1C elliptical systems ($\sim 11\%$), $15,608$ 2C two-component systems ($\sim 45\%$), $12,882$ 1C pure-disk systems ($\sim 37\%$), and $2,615$ unclassifiable systems ($\sim7\%$; representing objects visually identified as hard -interacting and visually disturbed systems etc.- or compact, or the aforementioned failed fits). Our fractions are to first order consistent with those \cite{Cook19} found for their xGASS sample.

\subsection{Distinguishing between pseudo- and classical-bulges} 
\label{sec:pB_cB_dist}

As a final step, we now attempt to separate our bulge components into ``classical''- and ``pseudo''-bulges, as they likely have different formation and evolutionary histories. However, we highlight that this distinction is problematic and increasingly challenging. This classification would be optimally done with kinematic data, but such large sample of kinematic data do not yet exist, especially not at these redshift ranges.  

Many studies have shown, by photometric and/or kinematic means, that the central regions of disk galaxies are often occupied by two types of structures (pseudo- and/or classical-bulges; pB and cB, respectively), see, e.g., the review of \cite{Kormendy04}. The definition of a pB is varied within the literature, and often depends on the information at hand, which can vary from a single-band image to full kinematic analysis. Here, for clarity, we define our ideal concept of a pB and cB that we strive to classify our sub-samples into, recognising that we do not actually have the information necessary to attain this distinction unambiguously.

~

\noindent
$\bullet$ {\bf classical-bulge (cB)}: high-stellar density, dynamically hot system, with high velocity dispersion, and low rotational velocity (often inert, gas-free and dust-free, but not exclusively).

\noindent
$\bullet$ {\bf pseudo-bulge (pB)}: a low-stellar density, dynamically cold system, with low velocity dispersion, and high rotational velocity (often star-forming, containing gas and/or dust, but not exclusively).

~

In our definition, a pB is taken literally to mean ``bulge-like'', and may include the combination of a number of secondary perturbations including bars, rings and extended planar orbits that we are here aggregating into a central combination of structures predominantly formed via secular processes (i.e., orbital migration and resonances, with some star-formation and accretion). The motivation for this is to map this classification to our two-component fitting approach, as we do not believe fitting with additional components is viable or stable at this level of signal-to-noise and spatial resolution. Following this definition, we now explore the mass-size ($M_*-\rm{R}_e$) plane, which matches directly to the stellar surface density, and allows us to be guided by the visual classifications of pB and cB made for GAMA and SDSS (based on the Kormendy relation; \citealt{Gadotti09}) galaxies at $z = 0$, as well as the distribution of our structural measurements, to attempt to select pB and cB structures in our sample. This method directly takes the bulge's stellar mass and effective radius into account rather than calculating the mean effective surface brightness within the effective radius ($<\mu_e>$) as in the Kormendy relation. We expect this method to reduce further propagation of uncertainties in $\rm{R}_e$ and $M_*$ in calculating the stellar surface density.

Note that we elect not to use a simple S\'ersic cut to separate pBs and cBs, as others have advocated, for a number of reasons. First, the bulge component S\'ersic index is fairly unstable, particularly given the uncertainty around the HST ACS PSF due to HST's ``breathing''. Second, dust can lower S\'ersic indices and also make the bulge appear larger (see e.g., \citealt{Pastrav13}). Our sample spans a broad redshift range where galaxies are also likely to become more dusty at higher-redshift (due to bandpass shifting, and higher star-formation rates). The fraction of massive galaxies with a dust-lane in the COSMOS region out to $z \sim 0.8$ is reported to be 80\% (\citealt{Sheth08}; \citealt{Holwerda12}). Third, the S\'ersic index is known to be wavelength dependent \citep{Kelvin14}, and hence a simple cut in a ``direct observable'' could introduce a redshift bias (due to bandpass shifting).
Fourth, several studies (e.g., \citealt{Gadotti09}; \citealt{Fisher16}) have shown that S\'ersic indices of pBs and cBs could be overestimated due to not modelling bars (e.g., \citealt{Fisher08,Fisher16}), and overlap when selecting via mean surface brightness (e.g., \citealt{Gadotti09}; using the Kormendy relation). This effect is also observed in our sample.

Finally, more recent results from the kinematic decomposition of disk galaxies with IFU observations have found no significant correlation between photometric S\'ersic index and kinematic properties (see e.g., \citealt{Krajnovic13}; \citealt{Zhu18}; \citealt{Schulze18}).  

\begin{figure*}
	\centering
	\includegraphics[width = \textwidth, angle = 0]{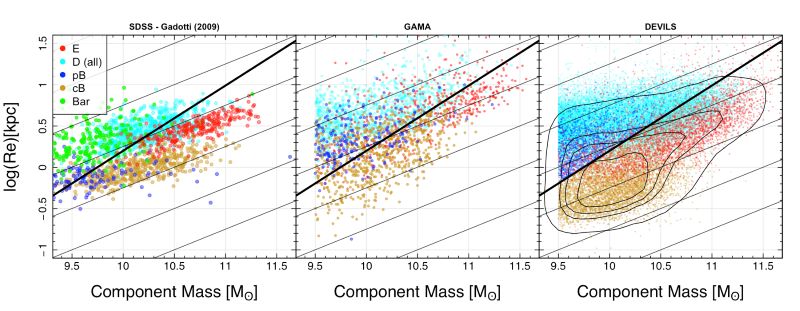}
	\caption{The size-mass plane for SDSS galaxies (left panel; \protect\cite{Gadotti09}), GAMA galaxies (middle panel; Hashemizadeh et al. in prep. and Casura et al. in prep.) and D10/ACS sample (right panel; this work). In the middle panel, the data is color coded based on our visual bulge classification of GAMA local galaxies (Driver et al. in prep). Black solid lines correspond to our pB-cB separation line; following the equation $\mathrm{log}(\rm{R}_e/\mathrm{kpc}) = 0.79 \mathrm{log}(M_*/M_\odot) - 7.7$. Faint gray lines indicate constant stellar mass densities equivalent to $\mathrm{log}(\Sigma) = 11, 10, 9, 8, 7, 6, 5$, from top to bottom. }
	\label{fig:Bulge_M_Re}
\end{figure*}

Figure \ref{fig:Bulge_M_Re} compares the $M_*-\rm{R}_e$ relation for elliptical (E: red), Disks (D: cyan; representing both pure disk systems and disk components) pBs (blue) and cBs (gold) for our D10/ACS galaxies (right panel) with those drawn from the local SDSS (left panel) and GAMA (middle panel) surveys. We also show the bar component (green) from \cite{Gadotti09} work of the SDSS galaxies. This Figure highlights bulge classification of SDSS galaxies based on the \cite{Kormendy77} relation (left panel; \citealt{Gadotti09}) and GAMA galaxies classified through our visual inspections (middle panel). In the right panel, we show our D10/ACS sample.

\begin{figure*}
\begin{subfigure} 
  \centering
  \includegraphics[width=.49\linewidth]{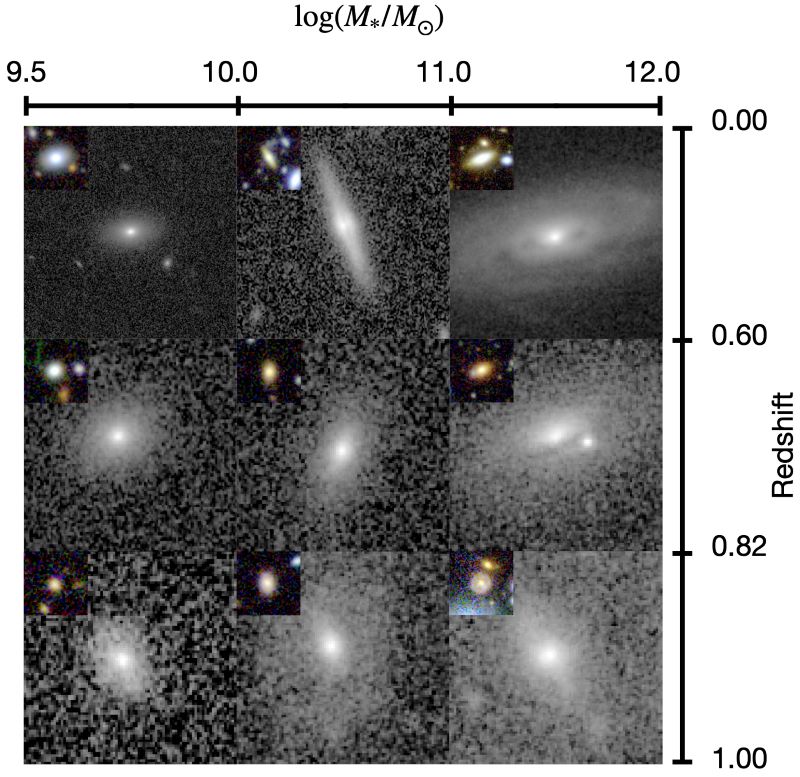}
\end{subfigure}
\begin{subfigure}
  \centering
  \includegraphics[width=.49\linewidth]{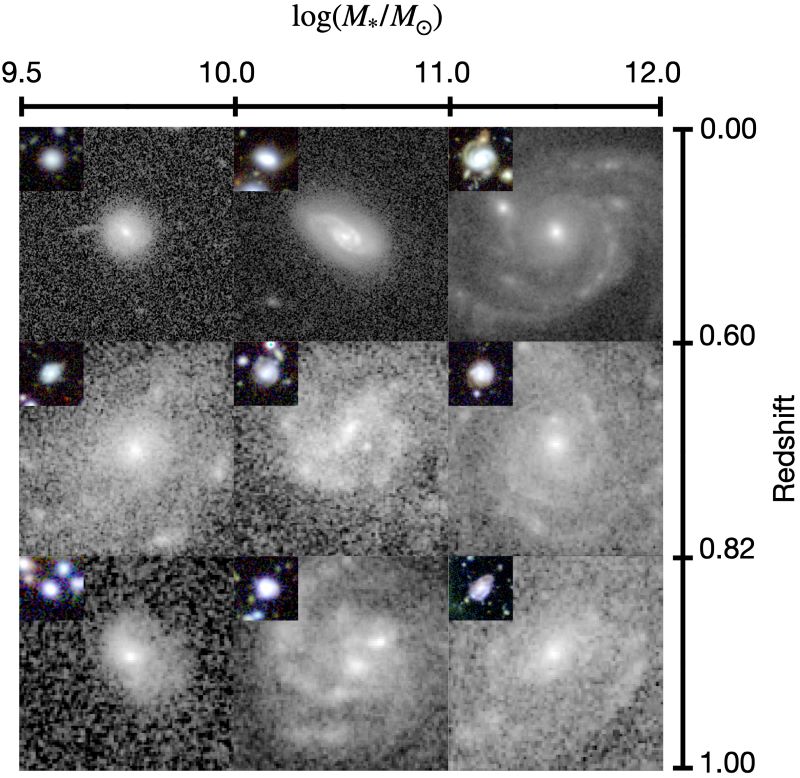}
\end{subfigure}
\caption{Random sample of galaxies harbouring a cB (left) and pB (right) as a function of redshift and stellar mass. The main image is cutout of galaxies in the ACS F814W filter while the inset color image is the SUBARU SuprimeCam gri combined image.}
\label{fig:pB_cB_stamps}
\end{figure*}



Note that lacking high-resolution colour information, we estimate the stellar mass of components by using our F814W bulge-to-total flux ratio (B/T), i.e., $M_*^\mathrm{Bulge} = \mathrm{B/T} \times M_*^\mathrm{Total}$ and $M_*^\mathrm{disk} = (1-\mathrm{B/T}) \times M_*^\mathrm{Total}$. The caveat here is that if the stellar population of the two components are different, then one can expect that the M/L are different, introducing errors into this method. However, this effect is unlikely to impact our results at the population scale.  

Figure \ref{fig:Bulge_M_Re} indicates that in the GAMA and SDSS data (left and middle panels) we see, despite obvious intermingling, a relatively clear demarcation between pB and cB. In the DEVILS data (right panel), we see a clumped population, which we identify as cBs (objects with higher stellar surface density), and a more dispersed population which we identify as pB (as one might expect from an amalgam of central perturbations, following our definition). We identify the line given by $\mathrm{log}(\rm{R}_e/\mathrm{kpc}) = 0.79 \mathrm{log}(M_*/M_\odot) - 7.7$, as providing a good demarcation across all three panels (surveys), and this is shown as the black lines on Figure \ref{fig:Bulge_M_Re} (essentially a cut slightly offset from a line of constant surface stellar mass density, shown as grey lines). 

Note that we show our visual pB/cB separation of GAMA galaxies in Figure \ref{fig:Bulge_M_Re} to highlight how our pB/cB separation line is guided by this data. However, going forward we will now consistently use the same pB/cB identification using the above cut for both the GAMA and the D10/ACS data. They possibly suffer from different systematic errors, e.g., how PSFs are made and how stable they are etc. 

Given this distinction between the two bulge structures we find the majority of bulges in the Universe, by number, to be pB. Overall, 58\% of our double component galaxies contain a pB, while 42\% of them have a cB. However, when we only consider components above our imposed stellar mass limit of $\mathrm{log}(M_*/M_\odot) > 9.5$, as we show in Figure \ref{fig:Bulge_M_Re}, we find that pBs and cBs constitute 31\% and 69\% of bulges, respectively. 

Finally, Figure \ref{fig:pB_cB_stamps} display a random set of our galaxies classified as cBD (classical-Bulge+Disk) and pBD (pseudo-Bulge+Disk), respectively, in regular bins of stellar mass and redshift. The Figures indicate that pBs typically lie in bluer, more star-forming systems than cBs, with their outer disks displaying more structure, i.e., spiral arms, star-formation regions etc.

\begin{figure*}
	\centering
	\includegraphics[width = \textwidth, angle = 0]{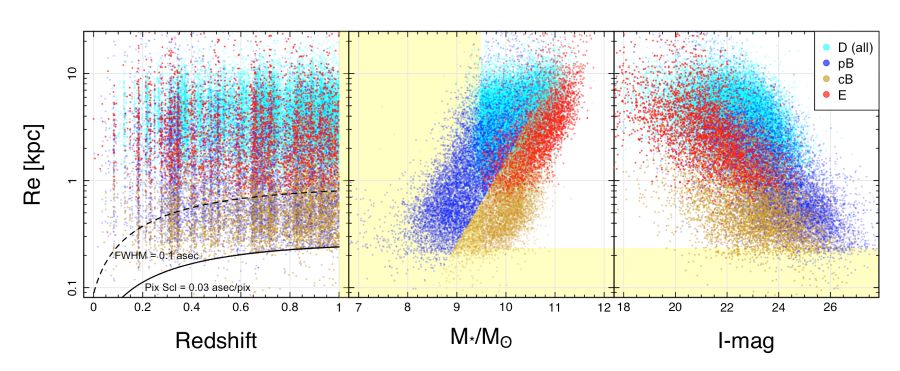}
    \caption{The relation between the effective radius, $\rm{R}_e$, of disks (cyan; including both pure disk systems and disk components), pB and cB (blue and gold) as well as ellipticals (red) with redshift, stellar mass and I-mag (left, middle and right panels, respectively). The curve in the left panel represents our imaging pixel scale ($0.03$ arcsec per pixel) converted into physical size. Vertical yellow boundaries in the left and middle panels represent our redshift and stellar mass limit, respectively. Horizontal yellow boundaries represent the completeness of our data in size, i.e., $\rm{R}_e$ = 0.235 kpc and stellar mass, i.e., $M_* > 10^{9.5} M_\odot$.}
	\label{fig:B_D_scatterPlot}
\end{figure*}

\begin{figure*}
\centering
\includegraphics[width=\textwidth]{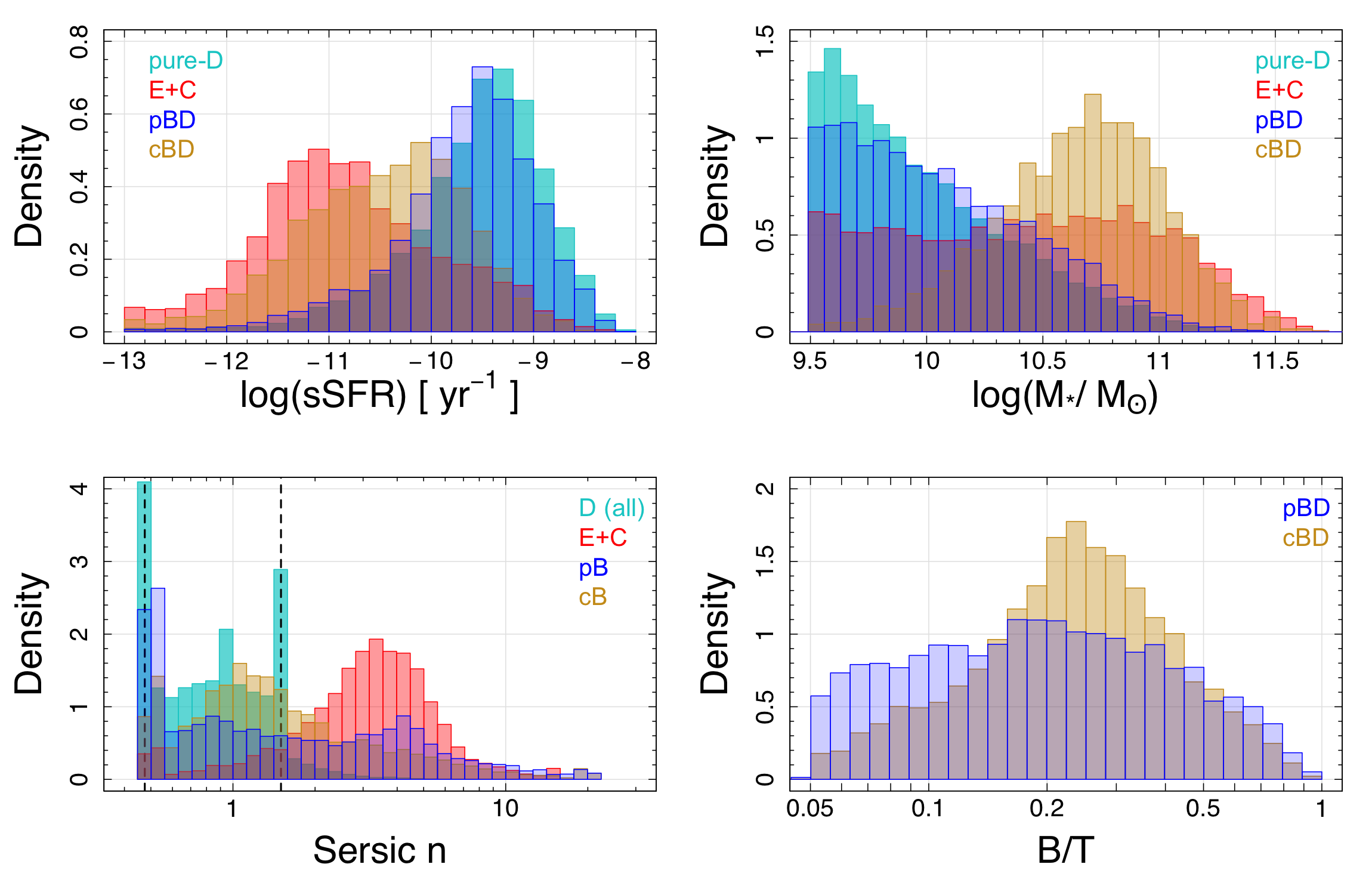}
\caption{PDF of the global sSFR, and stellar mass, as well as component S\'ersic index and B/T. Vertical dashed lines on the bottom left panel show our limits on the S\'ersic index of disk in double component fitting. A few bins seen beyond this buff=er represent our pure disk systems fitted by a single S\'ersic for which our buffer range is wider. See the text for more details. }
\label{fig:sSSFR_Mass}
\end{figure*}

\subsection{Discussion of our structural decompositions} 
\label{sec:str_dec_discussion}

Figure \ref{fig:B_D_scatterPlot} shows the relation between the physical half light radius, $\rm{R}_e$, of bulges and disks with redshift, stellar mass and I-mag. Note that we show the pixel size of HST/ACS, 0.03 arcsec/pixel (lower boundary in the left panel) to highlight that our measured structures, bulges in particular, are predominantly larger than the size of pixels, although we note a small number of unresolved bulges, at higher redshift, at very low stellar mass and at very faint apparent magnitudes. In this work, we will be limiting our studies to components with stellar masses $> 10^{9.5} M_{\odot}$ which removes most of the unresolved bulges. Given the 5$\sigma$ limiting depth of the COSMOS ACS F814W filter to be 27.2 (AB in a $0.''24$ diameter aperture, \citealt{Koekemoer07}) for point sources, the right panel of Figure \ref{fig:B_D_scatterPlot} indicates that all flux of our components are within the flux limit of the imaging data.

Figure \ref{fig:sSSFR_Mass} shows our final galaxy populations in various systemic observable or intrinsic parameter spaces. As expected, pure disk galaxies (D) are the least massive with the highest specific star-formation rates (sSFR). As one might also expect, elliptical galaxies (E) are the most massive with the lowest sSFR systems. This Figure also shows that disk galaxies containing a cB (cBD) do, in general, have lower sSFR, are more massive, and have higher B/T values than systems containing a pB (pBD). Figure \ref{fig:sSSFR_Mass} further indicates that ellipticals dominate the higher values of the systemic S\'ersic index (bottom left panel), $n_b \simeq 4$, indicating near de Vaucouleurs light profile (\citealt{deVaucouleurs48}), while disks occupy lower regions around $n_d \simeq 1$, indicating near exponential light profile. Interestingly, we do not find a significant discrimination between S\'ersic indices of pBs and cBs. In fact, we find cBs' S\'ersic index peaked around $n = 1$ and pBs' peaked at both $n = 1$ and 4.
Therefore, bulge S\'ersic indices, extend across the whole parameter space from 0.5 to 10, showing no clear correlation between the systemic S\'ersic index and the bulge morphology. We note that the systemic S\'ersic index does show some differentiation, but does not map well to bulge type, as noted earlier and reported in recent IFU studies (e.g., \citealt{Krajnovic13}; \citealt{Zhu18}; \citealt{Schulze18}; \citealt{Zhu20}; \citealt{Oh20}). We, however, do not rule out some uncertainties due to our pB-cB separation technique. 

Note that since we limit the range of S\'ersic indices of disks and bulges to $0.5 < n_d < 1.5$ and $0.5 < n_b < 20$, respectively, we find some fits trapped at lower or higher limits (see the bottom left panel of Figure \ref{fig:sSSFR_Mass}). One might decide to solve this by extending the buffer to give the mathematical modelling freedom to explore a wider space. Highlighting that not every mathematically-preferred optimised model is necessarily synonymous with the most physically valid ones, we decided to keep the parameters in a physically induced range following \cite{Cook19}. For example, one expects a stable disk to have a S\'ersic index close to unity. As a consequence of this buffer selection, we find $n_d$ histogram (cyan) also presenting two peaks on the boundaries ($n = 0.5$ and $1.5$, bottom left panel of Figure \ref{fig:sSSFR_Mass}). 

\begin{landscape}
\begin{figure*}
	\centering
	\includegraphics[width = 1.5\textwidth, height = 0.95\textwidth, angle = 0]{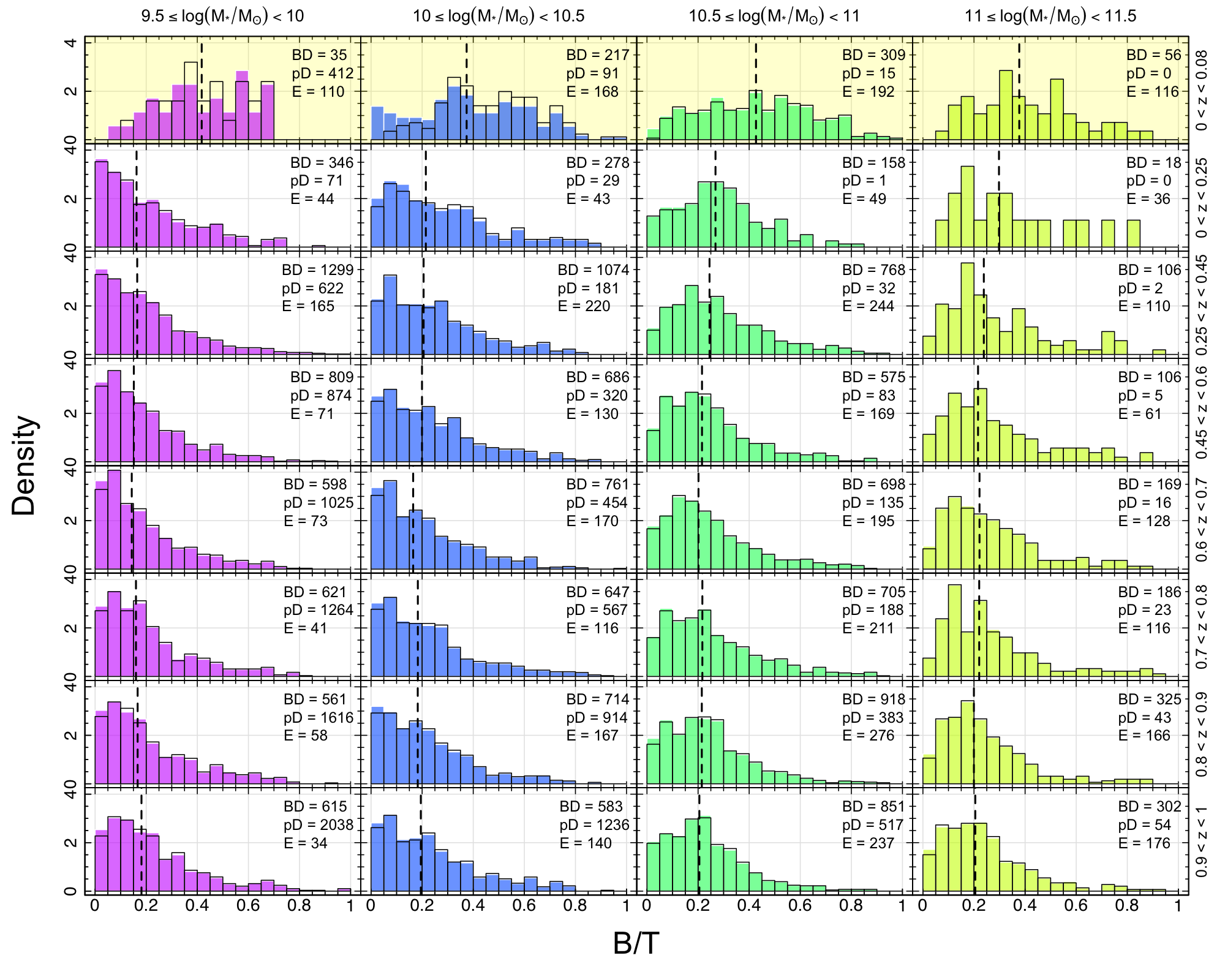}
	\caption{ The evolution of B/T as a function of stellar mass (columns) and redshift (rows). The inset numbers indicate the number of pure disk (pD), double-component (D) and elliptical (E) systems in each bin. The first row highlighted by yellow shows the histograms of B/T for $z=0$ GAMA galaxies. Dashed lines show the median values. Empty histograms with black borders represent systems with $\rm{R}_e > 0.25$ kpc while the background histograms show the total distribution in each bin.}
	\label{fig:BT}
\end{figure*}
\end{landscape}

\begin{figure*}
	\centering
	\includegraphics[width = 0.99\textwidth, angle = 0]{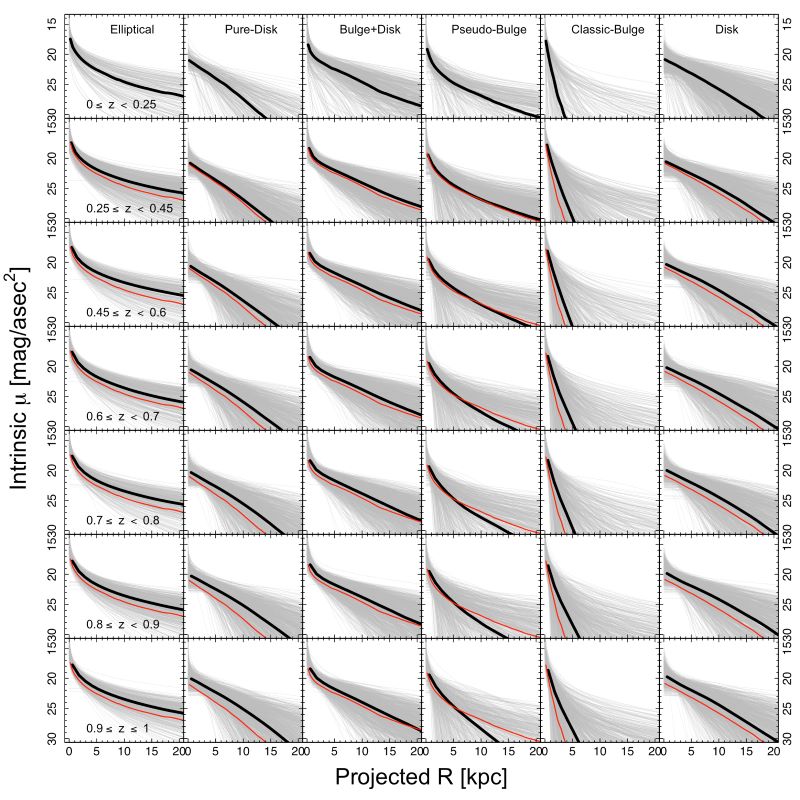}
	\caption{One-dimensional radial profile of different morphological types, as well as bulges and disks at different redshifts. Gray curves are 1000 profiles in each category (less in case of ellipticals). The overlaid thick profile is the median curve,
	while thin red curves represent the median profiles at our lowest redshift range, i.e., $0.0 < z < 0.25$. This plot only includes the component mass of $> 10^{9.5} M_\odot$. }
	\label{fig:R_profiles}
\end{figure*}

In Figure \ref{fig:BT}, we further inspect the correlation between B/T and stellar mass as well as redshift. We select our redshift binning extending from $z=0$ to $z=1$ similar to \cite{Hashemizadeh21}. The first row highlighted with yellow shows the B/T distribution of GAMA galaxies. The Figure shows that massive galaxies typically have more significant bulges, i.e., larger B/T. It also indicates that B/T is, however, stable throughout time. Most noticeable is the rise in lower B/T systems in the lowest mass bin, potentially this may be due to some classification bias with very small bulges at very low mass intervals at high-$z$ becoming harder to visually identify. However, we note the opposite trend in the most massive galaxies. The HST resolution is given by the black line on the left panel of Figure~\ref{fig:B_D_scatterPlot}, and in general very few bulges are at this limit, suggesting that the increase in low B/T systems at low-redshift may be genuine. However, we cannot fully rule out some other bias. Ultimately, according to our pixel size ($0.03$ arcsec), we are only able to resolve bulges with $R_e > 0.25$ kpc across all redshift intervals (see black line in Figure~\ref{fig:B_D_scatterPlot}). To explore whether this bias is significant, in Figure \ref{fig:BT} we also show the results if we impose a uniform $R_e > 0.25$ kpc limit as the black line histograms, and while we do see a modest change in the very low-B/T objects in the lowest mass and redshift bin, the change is modest, and hence we conclude that the growth in low mass bulges towards lower redshift is real.

Note that for the GAMA data in the lowest redshift bin we find a more extended B/T range with a larger median value of the B/T. We note that the GAMA decompositions are still under review and not yet published.

Finally, Figure \ref{fig:R_profiles} shows a random selection of 1D component profiles, with component masses above $10^{9.5}$M$_{\odot}$ and indicative of our science analysis sample. Note that pure-disk, here, refers to galaxies visually classified as a pure-disk morphology, while disk refers to the disk component of bulge+disk systems. We convert the apparent surface brightness to the intrinsic surface brightness (SB) by correcting for $(1+z)^4$ SB dimming. Thick black curves represent the median profile for each subset, and red curves show the redshift zero fit. Our initial impression, is that there appears to be a marginal contraction (fading) in almost all structures likely due to merging galaxies of all stellar masses here. To further explore this, a detailed analysis of the $M_*-\rm{R}_e$ by component will be presented in Hashemizadeh et al. (in prep.) and is outside the scope of this paper.

\section{The evolution of the SMF since \lowercase{$z$} $= 1$} 
\label{Sec:SMF_evol}

\begin{landscape}
\begin{figure}
	\centering
	\includegraphics[width = 1.7\textwidth, height = 0.95\textwidth, angle = 0]{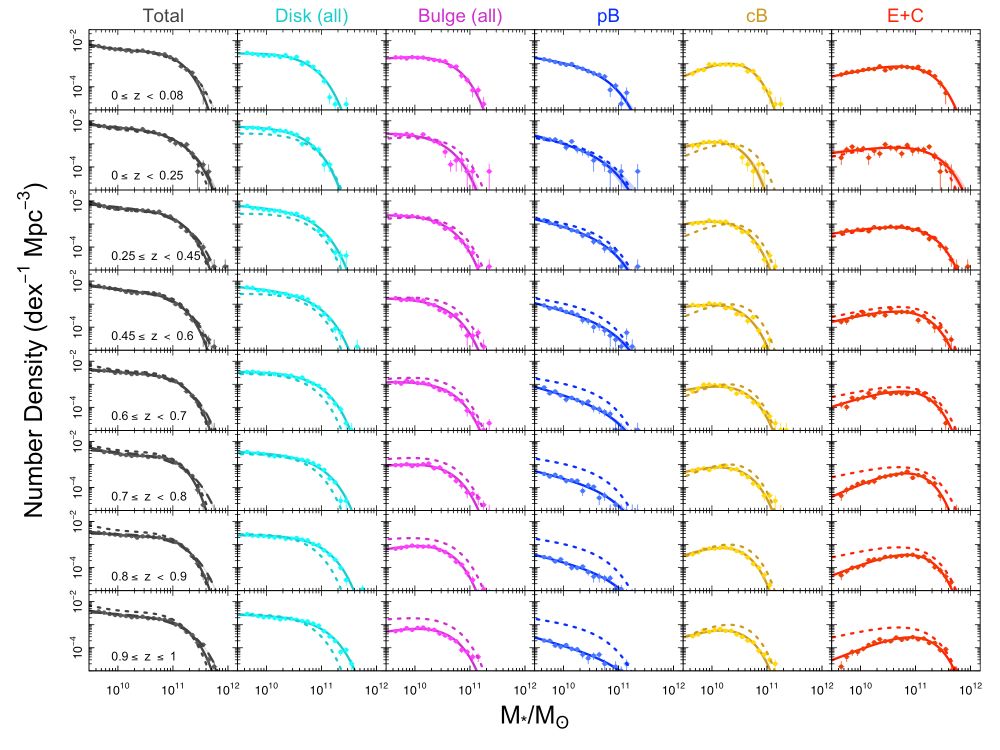}
	\caption{ The SMF of components in 8 redshift bins. The local GAMA SMFs are shown in the top row highlighted with yellow. Data points represent galaxy counts in each of equal-size stellar mass bins. Width of stellar mass bins are shown as horizontal bars on data points. Vertical bars show poisson errors. Shaded regions around the best fit curves are 68 per cent confidence regions. The over-plotted dotted curves represent the GAMA SMFs ($0 < z < 0.08$). Note that in the total SMFs (left column) solid lines represent our double Schechter fits to data that are very close to single Schechter fits (dashed curves).}
	\label{fig:Mfunc_Str}
\end{figure}
\end{landscape}

In \cite{Hashemizadeh21}, we showed that the volume-corrected distribution of morphologically subdivided stellar-mass for the D10/ACS sample is well described by single \cite{Schechter76} functions, as the mass range ($> 10^{9.5}M_{\odot}$) probed does not extend significantly beyond where a turn-up starts to be seen at around $10^{9.5}M_{\odot}$, while the global SMF is shown to fit well with double Schechter function (e.g., \citealt{Baldry08}; \citealt{Pozzetti10}; \citealt{Baldry12} and \citealt{Wright17}). In the present work, we therefore, use the same double Schechter function to fit our total SMF (solid black lines in Figure \ref{fig:Mfunc_Str}) although for completeness we also show our single Schechter fits as dashed black lines. Note that as can be seen in Figure \ref{fig:Mfunc_Str}, all components can be fitted with single Schechter functions at all redshifts.

To derive our stellar-mass functions (SMF), we make use of the \texttt{dftools} package implemented in R (see \citealt{Obreschkow18}). 
In all cases the fitted SMFs describe the data well, see Figure \ref{fig:Mfunc_Str}, which shows the evolution of both the total SMFs and the SMF broken into structural types of disks (all; including both pure disk systems and disk components), bulges (all), pB and cB and E+C (ellipticals+compacts). Note that we combine compact systems (Cs) with Es (i.e., E+C) because according to our visual inspections the C subcategory is dominated by unresolved and most likely spheroidal systems (see figure 21 in \citealt{Hashemizadeh21}). In Figure \ref{fig:Mfunc_Str}, each row represents a distinct redshift range extending from $z=0$ to 1, as indicated on the panel. As mentioned earlier, similar to \cite{Hashemizadeh21}, we select our redshift bins to be $z = 0.0,0.25,0.45,0.6,0.7,0.8,0.9,1.0$. For comparison, we also present the new local GAMA SMFs ($0.0 < z < 0.08$) in the top row of Figure \ref{fig:Mfunc_Str}. Note that we use our GAMA visual morphological classifications to inform our low-$z$ structural SMFs while the separation of pBs and cBs follows an identical procedure for both GAMA and DEVILS data as discussed in Section \ref{sec:pB_cB_dist}.

The total and elliptical SMFs are essentially identical to that shown in figure 12 in \cite{Hashemizadeh21}.
Our SMF values also include a correction for the large scale structure (LSS) along the COSMOS sight-line, i.e., under- and over-densities in the COSMOS field in different redshift bins, as described in section 4.2 in \cite{Hashemizadeh21}. In brief, we determine an LSS correction by forcing the total stellar-mass density (SMD) to match a smooth spline fit to the data of \cite{Driver18}. We then apply our LSS correction factors in each redshift interval to all SMDs by multiplying by the scale factor.     

\begin{figure}
	\centering
	\includegraphics[width = \textwidth, angle = 0]{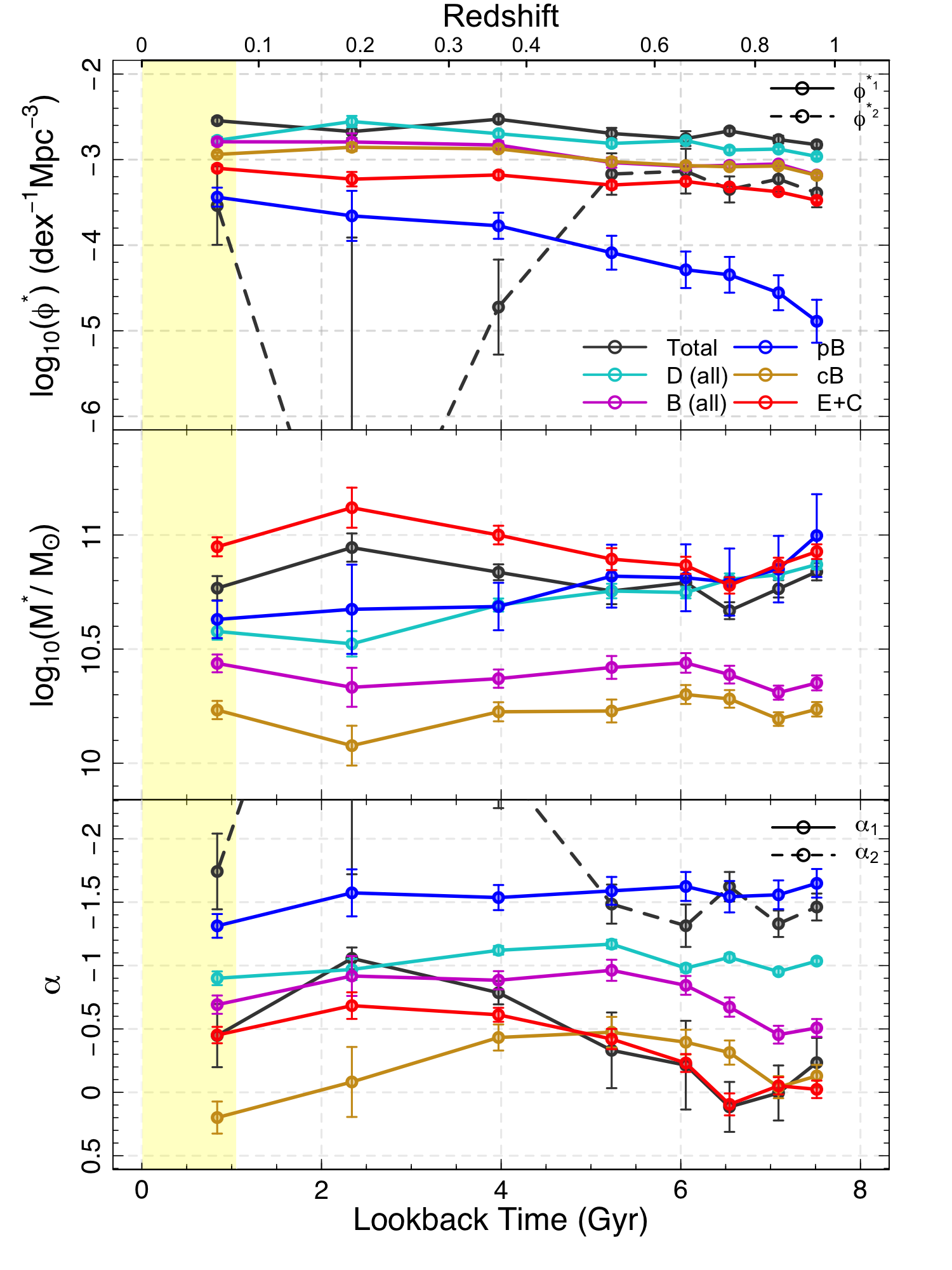}
	\caption{ The evolution of the best fit Schechter parameters from $z = 1$. Yellow band represents the time covered by our GAMA data. Dashed lines represent the parameters for our double Schechter fit to total and disk SMFs. For example, solid and dashed black lines in the top panel show the evolution of $\phi^{*}_{1}$ and $\phi^{*}_{2}$, respectively.  }
	\label{fig:Mfunc_par_evol}
\end{figure}

Figure \ref{fig:Mfunc_Str} highlights that the total SMF grows since $z = 1$ at both the low- and high-mass ends. We also see a similar increase with cosmic time in the low-mass end of the disk SMF, but a decrease in their intermediate- to high-mass end. Interestingly, the bulge component and ellipticals show stronger evolution with time with the pB's and cB's growing strongly and uniformly at all masses (internal secular processes and minor mergers?), and ellipticals predominantly at intermediate to lower-masses (major mergers?). Noticeable in the total data is the emergence of a bump and plateau in the mass function at lower redshifts. This has also been noted in \cite{Robotham14} and \cite{Wright18}. Physical interpretations will be discussed in Section \ref{sec:discussion}.
Finally, Figure \ref{fig:Mfunc_par_evol} shows the evolution of our best fit Schechter parameters as a function of lookback time for each component. The Schechter normalization parameter, $\phi^*$, of the total and disk population experiences a very slight increase since $z = 1$, while bulges' $\phi^*$ shows a small increase.
pBs occupy lower values and grow constantly over time while cBs and ellipticals experience a modest increase. Note that we also show the second parameters for our double Schechter functions (i.e., total and disk SMFs) as dashed lines. As expected, these are more fluctuated with larger errors.

The characteristic mass, $M^*$, also known as the knee or break mass of the total SMF is relatively stable while decreases for disks. The $M^*$ of all bulges and cBs evolves very little, while it decreases for pBs over the redshift range probed.

Lastly, $\alpha$, the faint end slope, is steepest for the pB population, but shows a modest decrease. The $\alpha$ of the total and disk population increases at all epochs except for the decline when transitioning to the GAMA data. It also increases for cB population by $z \sim 0.5$ and then declines to $z = 0$ while increasing for E+C types likely suggesting that low mass disks and bulges are growing/forming at all epochs. We report the tabulated version of our best Schechter function parameters in Table. \ref{tab:Str_MF_par}.

\begin{table*}
\centering
\caption{Best Schechter fit parameters of total and structural SMF in different redshift bins as well as the integrated SMD, i.e., $\mathrm{log}_{10}(\rho^*$) and the stellar baryon fraction, i.e., $f_s$. For completeness we report the $\rho^*$ values for integration over the stellar mass ranges of $0-\infty$ and $10^{9.5}-\infty$. $f_s$ is calculated using our main integration range ($0-\infty$).}
\begin{adjustbox}{scale = 0.43}
\begin{tabular}{lcccccccc}
\firsthline \firsthline \\
$z$-bins         &  $0.0 \leq z < 0.08$  &  $0.0 \leq z < 0.25$ &  $0.25 \leq z < 0.45$ &  $0.45 \leq z < 0.60$ &  $0.60 \leq z < 0.70$ &  $0.70 \leq z < 0.80$ &  $0.80 \leq z < 0.90$ &  $0.90 \leq z \leq 1.00$ \\ \\ \hline 
                   &   \multicolumn{8}{c}{\textbf{Total (Double Schechter)}}  \\
\cline{2-9} \\
$\mathrm{log}_{10}(\Phi^*_1)$      & $-2.55\pm0.04$  & $-2.67\pm0.08$   & $-2.53\pm0.03$ & $-2.69\pm0.07$ & $2.75\pm0.09$   & $-2.66\pm0.03$ & $-2.77\pm0.05$ & $-2.83\pm0.04$ \\ \\
$\mathrm{log}_{10}(M^*)$           & $10.77\pm0.05$  & $10.95\pm0.06$   & $10.84\pm0.03$ & $10.75\pm0.06$ & $10.79\pm0.06$  & $10.67\pm0.04$ & $10.76\pm0.04$ & $10.84\pm0.04$  \\ \\
$\alpha_1$                         & $-0.45\pm0.25$  & $-1.06\pm0.09$   & $-0.79\pm0.09$ & $-0.33\pm0.30$ & $-0.21\pm0.35$  & $0.12\pm0.20$  & $0.01\pm0.22$  & $-0.23\pm0.20$ \\ \\
$\mathrm{log}_{10}(\Phi^*_2)$      & $-3.54\pm0.46$  & $-8.40\pm4.48$   & $-4.72\pm0.55$ & $-3.17\pm0.24$ & $-3.13\pm0.26$  & $-3.35\pm0.15$ & $-3.23\pm0.15$ & $-3.39\pm0.17$   \\ \\
$\alpha_2$                         & $-1.74\pm0.30$  & $-4.77\pm3.05$   & $-2.61\pm0.37$ & $-1.48\pm0.15$ & $-1.31\pm0.17$  & $-1.62\pm0.12$ & $-1.33\pm0.10$ & $-1.46\pm0.11$  \\ \\ 
$\mathrm{log}_{10}(\rho^*) [10^{9.5}-\infty]$        & $8.245\pm0.08$   & $8.272\pm0.14$  & $8.235\pm0.09$ & $8.185\pm0.10$ & $8.183\pm0.13$ & $8.133\pm0.09$ & $8.143\pm0.09$ & $8.119\pm0.09$  \\ \\
$\mathrm{log}_{10}(\rho^*) [0-\infty]$               & $8.318\pm0.13$   & $8.288\pm0.24$  & $8.259\pm0.13$ & $8.231\pm0.23$ & $8.207\pm0.12$ & $8.192\pm0.11$ & $8.165\pm0.09$ & $8.151\pm0.09$  \\ \\
$f_s$                                                & $0.0349\pm0.0118$  & $0.0326\pm0.0239$  & $0.0305\pm0.0109$ & $0.0286\pm0.0201$ & $0.0271\pm0.0086$ & $0.0261\pm0.0077$ & $0.0246\pm0.0057$ & $0.0238\pm0.0055$ \\ \\ \hline
                    &   \multicolumn{8}{c}{\textbf{Total (Single Schechter)}}  \\
\cline{2-9} \\
$\mathrm{log}_{10}(\Phi^*)$         & $-2.74\pm0.03$   & $-2.74\pm0.06$   & $-2.74\pm0.03$ & $-2.81\pm0.03$ & $-2.79\pm0.03$ & $-2.91\pm0.03$ & $-2.86\pm0.02$ & $-2.98\pm0.2$ \\ \\
$\mathrm{log}_{10}(M^*)$            & $10.98\pm0.03$   & $10.99\pm0.05$   & $11.00\pm0.03$ & $10.97\pm0.03$ & $10.98\pm0.02$ & $11.03\pm0.02$ & $11.02\pm0.02$ & $11.09\pm0.02$ \\\\
$\alpha$                            & $-1.09\pm0.03$   & $-1.14\pm0.05$   & $-1.16\pm0.02$ & $-1.16\pm0.03$ & $-1.03\pm0.02$ & $-1.11\pm0.02$ & $-0.99\pm0.02$ & $-1.10\pm0.02$ \\ \\ 
$\mathrm{log}_{10}(\rho^*)[10^{9.5}-\infty]$          & $8.246\pm0.07$  & $8.272\pm0.11$   & $8.284\pm0.08$  & $8.185\pm0.08$  & $8.183\pm0.09$ & $8.133\pm0.09$  & $8.144\pm0.09$   & $8.120\pm0.09$ \\\\
$\mathrm{log}_{10}(\rho^*) [0-\infty]$                & $8.266\pm0.07$  & $8.296\pm0.11$  & $8.310\pm0.08$ & $8.212\pm0.08$ & $8.199\pm0.09$ & $8.153\pm0.09$ & $8.156\pm0.09$ & $8.137\pm0.09$  \\ \\
$f_s$                                                 & $0.0310\pm0.0061$  & $0.0332\pm0.0093$  & $0.0343\pm0.0064$ & $0.0274\pm0.0061$ & $0.0266\pm0.0059$ & $0.0239\pm0.0048$ & $0.0241\pm0.0052$ & $0.0230\pm0.0049$ \\ \\ \hline
                    &   \multicolumn{8}{c}{\textbf{Disk (all)}}  \\
 \cline{2-9} \\
$\mathrm{log}_{10}(\Phi^*)$         & $-2.96\pm0.02$  & $-2.55\pm0.07$   & $-2.70\pm0.04$ & $-2.81\pm0.04$ & $-2.78\pm0.03$ & $-2.89\pm0.03$ & $-2.88\pm0.02$ & $-2.96\pm0.02$ \\ \\
$\mathrm{log}_{10}(M^*)$            & $10.87\pm0.02$  & $10.52\pm0.06$   & $10.69\pm0.03$ & $10.75\pm0.03$ & $10.75\pm0.03$ & $10.81\pm0.03$ & $10.83\pm0.02$ & $10.87\pm0.02$ \\ \\
$\alpha$                            & $-1.04\pm0.02$  & $-0.97\pm0.08$   & $-1.12\pm0.03$ & $-1.17\pm0.04$ & $-0.98\pm0.03$ & $-1.06\pm0.03$ & $-0.95\pm0.02$ & $-1.04\pm0.02$ \\ \\
$\mathrm{log}_{10}(\rho^*)[10^{9.5}-\infty]$          & $7.754\pm0.07$   & $7.923\pm0.11$   & $7.990\pm0.08$ & $7.954\pm0.08$ & $7.943\pm0.09$ & $7.908\pm0.09$ & $7.920\pm0.09$ & $7.896\pm0.09$  \\ \\
$\mathrm{log}_{10}(\rho^*)[0-\infty]$                 & $7.781\pm0.07$   & $7.961\pm0.11$   & $8.032\pm0.08$ & $7.997\pm0.08$ & $7.966\pm0.09$ & $7.935\pm0.09$ & $7.937\pm0.09$ & $7.917\pm0.09$  \\ \\
$f_s$                                                 & $0.0101\pm0.0018$ & $0.0154\pm0.0043$  & $0.0181\pm0.0038$ & $0.0167\pm0.0036$ & $0.0155\pm0.0038$ & $0.0145\pm0.0034$ & $0.0145\pm0.0033$ & $0.0139\pm0.0031$ \\ \\ \hline
                    &   \multicolumn{8}{c}{\textbf{Bulge (all)}}  \\
 \cline{2-9} \\
$\mathrm{log}_{10}(\Phi^*)$         & $-3.18\pm0.03$  & $-2.79\pm0.10$   & $-2.83\pm0.05$ & $-3.03\pm0.06$ & $-3.08\pm0.05$ & $-3.07\pm0.04$ & $-3.05\pm0.03$ & $-3.18\pm0.03$ \\ \\
$\mathrm{log}_{10}(M^*)$            & $10.35\pm0.03$  & $10.33\pm0.09$   & $10.37\pm0.04$ & $10.42\pm0.05$ & $10.44\pm0.04$ & $10.39\pm0.04$ & $10.31\pm0.03$ & $10.35\pm0.03$ \\ \\
$\alpha$                            & $-0.51\pm0.07$  & $-0.92\pm0.16$   & $-0.88\pm0.07$ & $-0.96\pm0.08$ & $-0.84\pm0.07$ & $-0.67\pm0.08$ & $-0.45\pm0.07$ & $-0.51\pm0.07$ \\ \\
$\mathrm{log}_{10}(\rho^*)[10^{9.5}-\infty]$          & $7.577\pm0.07$   & $7.469\pm0.11$ & $7.473\pm0.08$ & $7.329\pm0.09$ & $7.297\pm0.10$ & $7.250\pm0.09$ & $7.190\pm0.09$ & $7.103\pm0.09$  \\ \\
$\mathrm{log}_{10}(\rho^*)[0-\infty]$                 & $7.599\pm0.07$   & $7.520\pm0.12$ & $7.516\pm0.08$ & $7.377\pm0.09$ & $7.329\pm0.10$ & $7.273\pm0.09$ & $7.207\pm0.09$ & $7.120\pm0.09$  \\ \\
$f_s$                                                 & $0.0067\pm0.0012$ & $0.0056\pm0.0017$  & $0.0055\pm0.0012$ & $0.0040\pm0.0009$ & $0.0036\pm0.0009$ & $0.0031\pm0.0008$ & $0.0027\pm0.0006$ & $0.0022\pm0.0005$ \\ \\ \hline
                    &   \multicolumn{8}{c}{\textbf{Pseudo-Bulge}}  \\
 \cline{2-9} \\
$\mathrm{log}_{10}(\Phi^*)$         & $-3.44\pm0.11$  & $-3.66\pm0.3$   & $-3.77\pm0.2$ & $-4.09\pm0.20$ & $-4.29\pm0.21$ & $-4.35\pm0.21$ & $-4.56\pm0.20$ & $-4.89\pm0.25$ \\ \\
$\mathrm{log}_{10}(M^*)$            & $10.63\pm0.08$  & $10.67\pm0.2$   & $10.69\pm0.1$ & $10.82\pm0.14$ & $10.81\pm0.15$ & $10.79\pm0.15$ & $10.85\pm0.15$ & $11.00\pm0.18$ \\ \\
$\alpha$                            & $-1.31\pm0.09$  & $-1.57\pm0.2$   & $-1.54\pm0.1$ & $-1.59\pm0.11$ & $-1.62\pm0.11$ & $-1.54\pm0.12$ & $-1.56\pm0.11$ & $-1.65\pm0.11$ \\ \\
$\mathrm{log}_{10}(\rho^*)[10^{9.5}-\infty]$          & $7.226\pm0.08$  & $7.148\pm0.13$   & $7.032\pm0.09$ & $6.899\pm0.10$ & $6.707\pm0.11$ & $6.590\pm0.11$ & $6.453\pm0.11$ & $6.338\pm0.11$     \\ \\
$\mathrm{log}_{10}(\rho^*)[0-\infty]$                 & $7.312\pm0.09$  & $7.335\pm0.31$   & $7.194\pm0.13$ & $7.066\pm0.15$ & $6.899\pm0.23$ & $6.735\pm0.18$ & $6.597\pm0.16$ & $6.513\pm0.22$     \\ \\
$f_s$                                                 & $0.0035\pm0.0008$ & $0.0036\pm0.0037$  & $0.0026\pm0.0009$ & $0.0020\pm0.0011$ & $0.0013\pm0.0007$ & $0.0009\pm0.0004$ & $0.0007\pm0.0005$ & $0.0006\pm0.0005$     \\ \\ \hline
                    &   \multicolumn{8}{c}{\textbf{Classical-Bulge}}  \\
 \cline{2-9} \\
$\mathrm{log}_{10}(\Phi^*)$         & $-2.94\pm0.02$  & $-2.86\pm0.05$   & $-2.87\pm0.03$ & $-3.02\pm0.04$ & $-3.07\pm0.03$  & $-3.08\pm0.03$ & $-3.07\pm0.02$ & $-3.18\pm0.02$ \\ \\
$\mathrm{log}_{10}(M^*)$            & $10.23\pm0.04$  & $10.08\pm0.09$   & $10.23\pm0.04$ & $10.23\pm0.05$ & $10.30\pm0.04$  & $10.28\pm0.04$ & $10.19\pm0.03$ & $10.24\pm0.03$ \\ \\
$\alpha$                            & $0.20\pm0.13$   & $-0.08\pm0.28$   & $-0.43\pm0.10$ & $-0.47\pm0.12$ & $-0.40\pm0.10$  & $-0.31\pm0.10$  & $-0.04\pm0.09$  & $-0.13\pm0.09$ \\ \\
$\mathrm{log}_{10}(\rho^*)[10^{9.5}-\infty]$          & $7.332\pm0.08$   & $7.192\pm0.12$ & $7.280\pm0.09$ & $7.131\pm0.09$ & $7.168\pm0.10$ & $7.143\pm0.10$ & $7.103\pm0.09$ & $7.021\pm0.09$     \\ \\
$\mathrm{log}_{10}(\rho^*)[0-\infty]$                 & $7.336\pm0.08$   & $7.208\pm0.13$ & $7.301\pm0.09$ & $7.154\pm0.09$ & $7.183\pm0.10$ & $7.156\pm0.10$ & $7.111\pm0.09$ & $7.030\pm0.09$     \\ \\
$f_s$                                                 & $0.0036\pm0.0007$ & $0.0027\pm0.0009$  & $0.0034\pm0.0008$ & $0.0024\pm0.0006$ & $0.0026\pm0.0007$ & $0.0024\pm0.0006$ & $0.0022\pm0.0005$ & $0.0018\pm0.0004$     \\ \\ \hline
                   &   \multicolumn{8}{c}{\textbf{Elliptical + Compact}}  \\
 \cline{2-9} \\
$\mathrm{log}_{10}(\Phi^*)$         & $-3.10\pm0.04$  & $-3.23\pm0.08$   & $-3.18\pm0.04$ & $-3.30\pm0.04$ & $-3.25\pm0.03$ & $-3.32\pm0.02$ & $-3.38\pm0.02$ & $-3.48\pm0.02$ \\ \\
$\mathrm{log}_{10}(M^*)$            & $10.95\pm0.04$  & $11.12\pm0.09$   & $11.00\pm0.04$ & $10.89\pm0.05$ & $10.87\pm0.04$ & $10.78\pm0.04$ & $10.87\pm0.03$ & $10.93\pm0.03$ \\ \\
$\alpha$                            & $-0.45\pm0.06$  & $-0.68\pm0.10$   & $-0.61\pm0.05$ & $-0.42\pm0.08$ & $-0.23\pm0.07$ & $0.10\pm0.09$ & $-0.05\pm0.07$ & $-0.02\pm0.07$ \\ \\
$\mathrm{log}_{10}(\rho^*)[10^{9.5}-\infty]$          & $7.795\pm0.08$  & $7.840\pm0.14$   & $7.766\pm0.09$ & $7.546\pm0.10$ & $7.577\pm0.10$ & $7.478\pm0.10$ & $7.484\pm0.10$ & $7.447\pm0.10$ \\ \\
$\mathrm{log}_{10}(\rho^*)[0-\infty]$                 & $7.796\pm0.08$  & $7.843\pm0.14$   & $7.769\pm0.09$ & $7.548\pm0.10$ & $7.578\pm0.10$ & $7.479\pm0.10$ & $7.484\pm0.10$ & $7.448\pm0.10$ \\ \\
$f_s$                                                 & $0.0105\pm0.0022$ & $0.0117\pm0.0042$  & $0.0099\pm0.0023$ & $0.0059\pm0.0015$ & $0.0064\pm0.0017$ & $0.0051\pm0.0013$ & $0.0051\pm0.0013$ & $0.0047\pm0.0012$     \\ \\ \hline

\lasthline
\end{tabular}
\end{adjustbox}
\label{tab:Str_MF_par}
\end{table*}

\section{The Evolution of the SMD since \lowercase{$z$} $= 1$} 
\label{sec:rho}

\begin{landscape}
\begin{figure*}
	\centering
	\includegraphics[width = 1.3\textwidth, angle = 0]{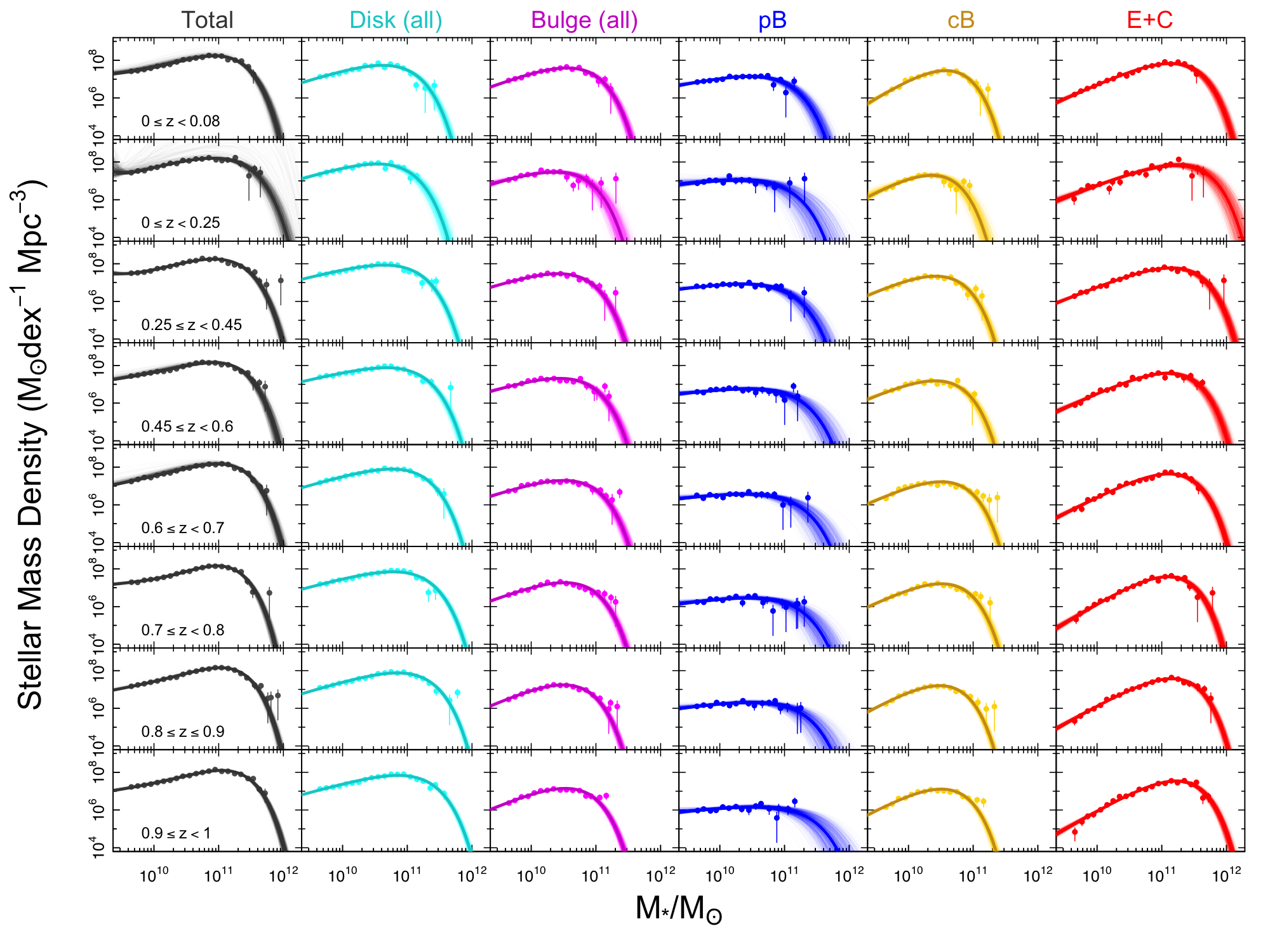}
	\caption{The total and structural SMD distributions in different redshifts. Redshift bins are the same as Figure \ref{fig:Mfunc_Str}. The local GAMA SMDs are shown in the top row highlighted with yellow. Points and lines indicate $M \phi(M)$, where $\phi(M)$ is our Schechter function fit. The shade transparent regions represent the error range calculated by 1000 times sampling of the full posterior probability distribution of the fit parameters.}
	\label{fig:Mdens}
\end{figure*}
\end{landscape}

Having derived the SMF of different galaxy components, we can now determine the stellar mass density (SMD) distribution for each population and finally calculate the total integrated stellar mass locked in each component. 
The distribution of the SMDs are shown on Figure \ref{fig:Mdens}, and highlight the evolution of the SMD distributions for our sample from $z=1$ to $z=0$. Shown in color shaded regions of Figure \ref{fig:Mdens} are the errors obtained from 1000 random samplings of the full posterior probability distribution from the best fit Schechter function parameters, and these are used to estimate the errors on our SMD measurements. 
The SMDs are, in all cases, well bounded within the stellar mass range, implying that the majority of stellar mass is captured at all redshifts. 
To construct the total SMD, $\rho^*$, for each type, we integrate under these distributions over all stellar masses from 0 to $\infty$. We note that our measurements of $\rho^*$ are generally robust to the integration range and in Table \ref{tab:Str_MF_par} we report both the $\rho^*$ derived from integrating from $10^{9.5}$ to infinity and when extrapolating to zero mass. In almost all case the extrapolated portion contains less than 5\% of the measured SMD, except in the case of pBs where it rises to 33\% at higher redshifts. This supports our assertion that the majority of mass is captured by the galaxies studied in our analysis.

\begin{figure*}
	\centering
	\includegraphics[width=\textwidth]{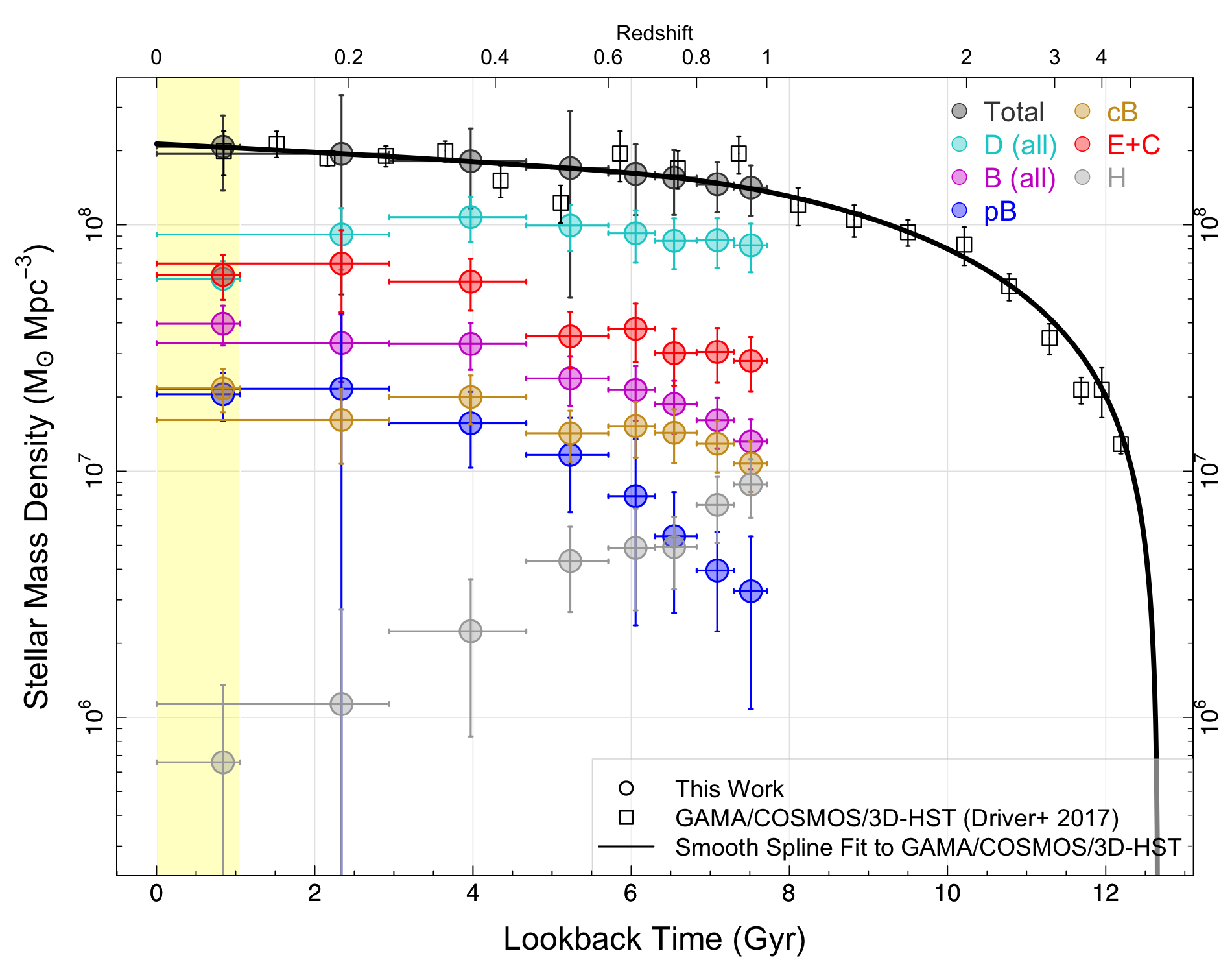}
    \caption{The evolution of the total and structural SMD, $\rho_{*}$, in the last 8 Gyr of the cosmic age. Colour codes are similar to previous plots. Vertical bars on the points show all errors including, fits and Poisson errors together with the classification and the cosmic variance error within the associated redshift bins taken from \protect\cite{Driver10}. Horizontal bars show the redshift ranges, while the data points are plotted at the mean redshift. The correction for the cosmic large scale structure is applied to the SMD in each redshift, as discussed in text and \protect\cite{Hashemizadeh21}. The local SMDs from GAMA are highlighted with yellow band. Note that we combine Cs with Es (E+C) here. See the text for more details. }
	\label{fig:MassBuildUp}
\end{figure*}

Figure \ref{fig:MassBuildUp} shows the evolution of the extrapolated $\rho_*$ values for each component. As a reminder, the SMDs include the correction for the large scale structure as described in \cite{Hashemizadeh21}. We also show the measurement from local galaxies as obtained from our GAMA measurements using identical methods, classifiers and techniques highlighted by a yellow transparent band (see Hashemizadeh et al. 2021 - PhD thesis). The error bars include all sources of uncertainties including classification, Poisson and fitting errors as well as the Cosmic Variance (CV) obtained as described in \cite{Hashemizadeh21}. 

\begin{figure}
	\centering
	\includegraphics[width = \textwidth, angle = 0]{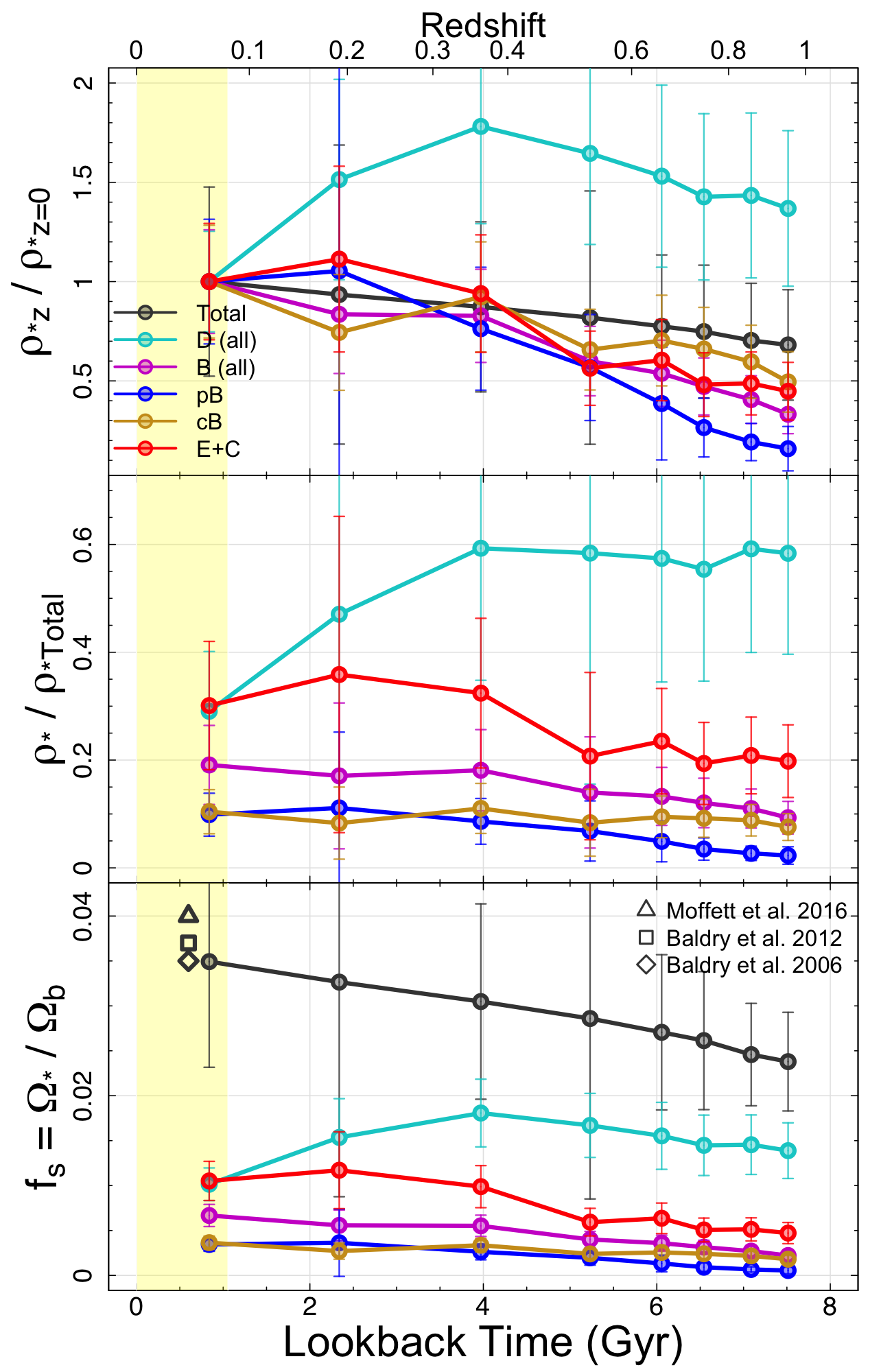}
	\caption{Top panel: variation of the stellar mass density, $\rho_{*}$, indicating the fraction of final stellar mass density (at $\overline{z} = 0$) assembled or lost by each redshift, i.e., $\rho_{*z}/\rho_{*z=0}$. Middle panel: shows the ratio of the structural SMDs to the total SMD at each redshift, i.e., $\rho_{*}/\rho_{*Total}$.
	Bottom panel: shows the evolution of the baryon fraction in form of stars in each galaxy component, i.e., $f_s = \Omega_*/\Omega_b$. Yellow band represents the time covered by our GAMA data.}
	\label{fig:fs}
\end{figure}

By splitting the total SMD into separate contributions of bulges and disks, we find that the disk component dominates the SMD of the Universe at all epochs except $z=0$ ($\sim 50\%$ of the total SMD, on average; see middle panel of Figure \ref{fig:fs}). The SMD of disks increases by a factor of $\sim 1.3$ over the interval $z = 1.0$ to $z \sim 0.35$, then declines by a factor of $\sim 0.57$ to $z = 0$. The SMD of (all) bulges experiences more significant growth at all epochs increasing by a factor of $\sim 3$ from $z = 1$ to 0. 

Subdividing our bulges into pB and cB, we find that cBs dominate over pBs in terms of the mass density, particularly at $z > 0.35$. 
However, we note that the pB SMD grows dramatically ($\sim \times 6.3$) from $z = 1$ to 0, while cBs grow consistently by a factor of $\sim 1.86$ to $z \sim 0.35$ and thereafter flatten.
If pBs can be considered redistributed disk material then there is some justification to fold the pB SMD measurements into the disk category. In this case the disk+pB contribution at $z=0$ constitutes $40\%$ of the total SMD.
Finally, we note that the elliptical population also grows by a factor of $\sim \times 2.23$ over this time interval. 

We note a slight disjoint of bulge mass build-up between GAMA and DEVILS. This is likely due to a significant difference between imaging quality of GAMA KiDS versus DEVILS HST and/or PSF modeling differences resulting in more significant bulges in KiDS data. This effect directly reflects into higher B/T values in KiDS data (see the distribution of the B/T in Figure \ref{fig:BT}). We report our SMDs, $\rho^*$, for all structures at all redshift ranges in Table \ref{tab:Str_MF_par}.   

To summarise these findings, Figure \ref{fig:fs} shows the evolution of the SMD as fractional changes (i.e., $z \sim 0$ ($\rho_{*z}/\rho_{*z=0}$) for the different structures (top panel), together with the ratio of the SMD of each component to the total (middle panel). The bottom panel shows the evolution of the stellar baryon fraction, i.e., $f_s = \Omega_*/\Omega_b$ where $\Omega_* = \rho_*/\rho_c$ and $\Omega_b = 0.0493$ \citep{Planck20} with the critical density of the Universe assumed to be $\rho_c = 1.21 \times 10^{11} \mathrm{M}_{\odot} \mathrm{Mpc}^{-3}$ at our GAMA median redshift ($\overline{z} \sim 0.06$) in a 737 cosmology. We report all the $f_s$ values at higher redshifts in Table \ref{tab:Str_MF_par}.

We find that, unsurprisingly, all galaxy components except disk grow in stellar mass density from $z = 1$ to $0$. pBs show the largest fractional mass growth, but overall pBs still contribute the least to the total SMD ($\sim 2-11$\% at z = 1-0; see Figure~\ref{fig:fs}). The middle and lower panels of Figure \ref{fig:fs} indicate that disks have the largest contribution to the total SMD and the stellar baryon fraction at all epochs, but interestingly, they have decreased their contribution to the total SMD over the last 4 billion years. This declining contribution is mirrored as an increased significance in the elliptical and classical bulge populations that show comparable growth. At face value it appears that $z = 1-0.36$ represents the phase where disk population grows more slower than bulges and a period of both secular evolution (growing pseudo-bulges) and the growth and/or emergence of both ellipticals and classical bulges (see their contributions to the total SMD in the middle panel of Figure \ref{fig:fs}).

\section{Discussion}
\label{sec:discussion}

Section \ref{sec:rho} reports the factual measurements of our structural stellar mass densities. Here we try to provide some interpretation of these measurements in the context of galaxy formation. However, before commencing, it is worth highlighting the various caveats in play. 
\begin{description}

\item[$\bullet$]
Firstly, we have identified that the HST PSF is less stable than one would like, which will introduce errors in bulge shape measurements and in particular the recovered S\'ersic indices. However fluxes and sizes should generally be more immune to imprecise PSF modelling. 
\item[$\bullet$]
Secondly, bulge-disk decomposition is also fraught with concerns over the minimisation algorithm becoming trapped in a local minimum or guided by the initial conditions. Our new Bayesian MCMC code (ProFit), is specifically designed to overcome this and Figures~\ref{fig:Conv_test_HST} \& \ref{fig:Conv_test_GAMA} suggest this is the case. 
\item[$\bullet$]
Thirdly, whether a galaxy requires decomposition into two-components (or one-component is appropriate to capture the true radial profile) is reliant on our eyeball classifications from \cite{Hashemizadeh21}. While these demonstrate greater than 90 per cent consistency across our three classifiers for all redshifts and masses, we cannot rule out systematic biases with redshift that are influencing all our classifiers in the same way. Certainly the smoothness of the data suggests random errors are not dominating and the consistency of the classifications, while not ruling out some bias, would suggest it is secondary and likely modifying but not driving the trends seen. 
\item[$\bullet$]
Fourthly, we accept that our bulge measurements are likely measuring ``bulge complexes'' and our classification process is most likely sifting the bulges into pseudo or classical based on the dominant component. Many of our bulges likely contain multiple components, e.g., bars, peanut/boxy bulges, nuclear disks, nuclei, and classical bulges. In due course it may be possible, with higher resolution higher signal-to-noise or IFU data to disentangle further, however here we believe that taking the simple approach of classifying the dominant component will introduce less uncertainty than trying to fit multiple components given the variable PSF and that we are working at the resolution limit. 
\item[$\bullet$]
Fifthly, no attempt has been made to correct for the influence of dust attenuation which we know is more severe at higher redshifts due to elevated star-formation and at shorter wavelengths because of stronger attenuation. In due course, with JWST mid-IR observations of selected galaxies, this issue could be explored. 
\item[$\bullet$]
Finally, we note that our stellar mass measurements for bulges and disks are necessarily based on applying a simple B/T multiplier due to having only a single HST band. In reality, some bulges and disks will have a range of mass-to-light ratios and we expect that this will introduce significant errors in individual galaxies and a modest, e.g., 10 per cent, bias in our aggregated masses. 

\end{description}

The above caveats could combine or cancel in ways that may have a significant impact on our measured values but we don't believe that they are likely to drive the general trends we see in Figures \ref{fig:MassBuildUp} \& \ref{fig:fs}. One aspect that gives us confidence that this might be the case is the relatively smooth transition from the redshift trends seen in the DEVILS data to the GAMA data and in general we see that the GAMA data (shown in the yellow band) is consistent with extrapolations of the redshift trends seen in the DEVILS data. While these studies have used identical methodologies and codes the data quality is dramatically different (i.e., $1''$ vs $0.03''$ resolution), and strong biases dependent on the data quality would likely lead to discontinuities. All of these caveats provide rich potential for further study in future years with facilities such as JWST, ESO MUSE and KMOS. Moving forward we acknowledge these caveats but for the remainder of this section, assume that they are not driving the trends that we see. 

As our backdrop we are aware that the cosmic star-formation rate and galaxy major merger rates are both in significant decline by $\sim \times 6$ and $\sim \times 3$ since $z=1$ respectively (see \citealt{Driver18}, and \citealt{Robotham14}). Various studies (e.g., \citealt{vanderWel14}; \citealt{Trujillo11}; etc) have also reported a significant growth in galaxy sizes with decreasing redshift, and a relatively smooth and modest evolution in the overall stellar mass function (e.g., \citealt{Wright18}).

To now add to this overall picture, we find from Figure \ref{fig:MassBuildUp} a rise and fall in the total disk mass (peaking at $z=0.35$) and a consistent rise in the SMDs of the Elliptical and bulge populations. We note the Hard and Compact populations, which together contains relatively minimal mass is likely reflecting an increasing merger rate with lookback time (for Hards), as well as difficulties in the classification of low-mass systems and limiting resolution. However, given the mass involved we can for the moment ignore these classes. It is also notable that the rise of the Ellipticals, compacts and classical bulges is fairly similar, with the rise in the pB class the strongest, especially at high-$z$ and flattening at low-$z$. 

\begin{figure}
	\centering
	\includegraphics[width = \columnwidth, angle = 0]{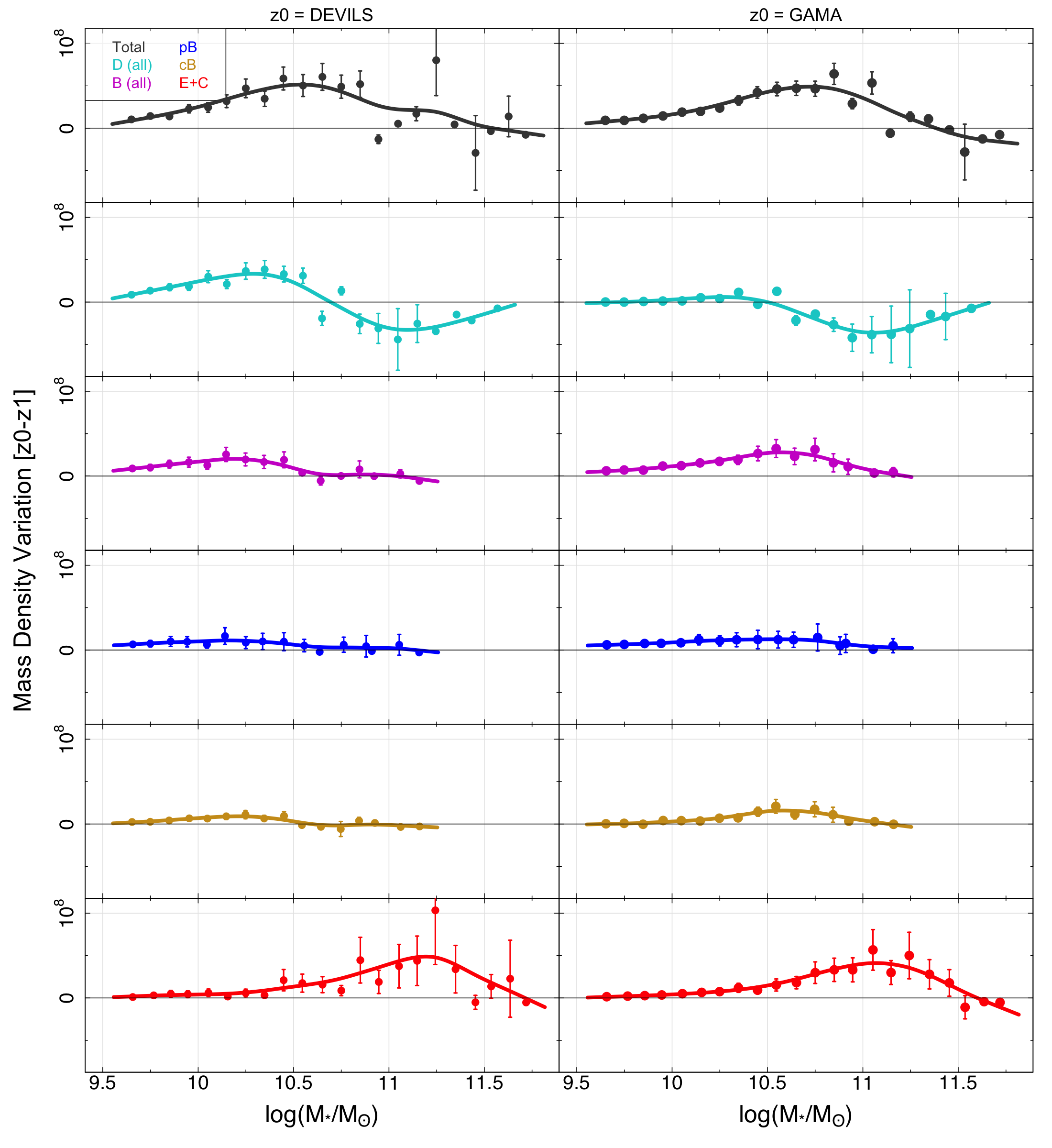}
	\caption{The variation of the SMD distribution (SMD$[z=0]-$SMD$[z=1]$) as a function of stellar mass. Left column shows the variation when we take DEVILS lowest redshift bin ($0.0 \le z < 0.25$), while right panel indicates the variation when we take GAMA data as the lowest redshift bin ($0.0 < z \le 0.08$). Solid lines are smooth splines with degree of freedom of 6 fitted to the data points. }
	\label{fig:SMD_var}
\end{figure}

To go one step further we can also look at how the SMD distributions have evolved with redshifts. We show this on Figure~\ref{fig:SMD_var} by plotting the ratio of the SMD distribution for the highest and lowest redshift bins. Left-side panels show the lowest-$z$ DEVILS data as the denominator (where statistics are poor but data and methodologies identical), and the right-side panels use GAMA on the denominator (where statistics are good and methodology similar, but data quite different). In general, the left and right-side panels show a very similar picture. Generally, from Figure~\ref{fig:SMD_var} (top panel), we see that the overall mass growth is around the knee of the mass-function and fairly broad (i.e., $10^{10}-10^{11}$M$_{\odot}$) and consistent with the findings of \cite{Wright18} and \cite{Robotham14}. Now looking by components we see that the most interesting behaviour in the disks, where mass is lost at the high-mass end and gained at the lower mass-end. This loss at the high mass-end appears to be mirrored by the gain in mass at the Elliptical bright-end. This would seem to provide quite compelling evidence for Elliptical growth through the major merger of high-mass disk systems, i.e. these disk mergers form Ellipticals moving mass from one class to the other. In general bulge growth while significant in Figure~\ref{fig:MassBuildUp} is less so in absolute terms (reflecting the log scaling in Figure~\ref{fig:MassBuildUp}). Nevertheless we can see the growth in both classical and pseudo-bulges is skewed slightly towards the lower mass end indicative of more subtle low-mass secular processes including accretion and redistribution.

\begin{figure}
	\centering
	\includegraphics[width= \columnwidth]{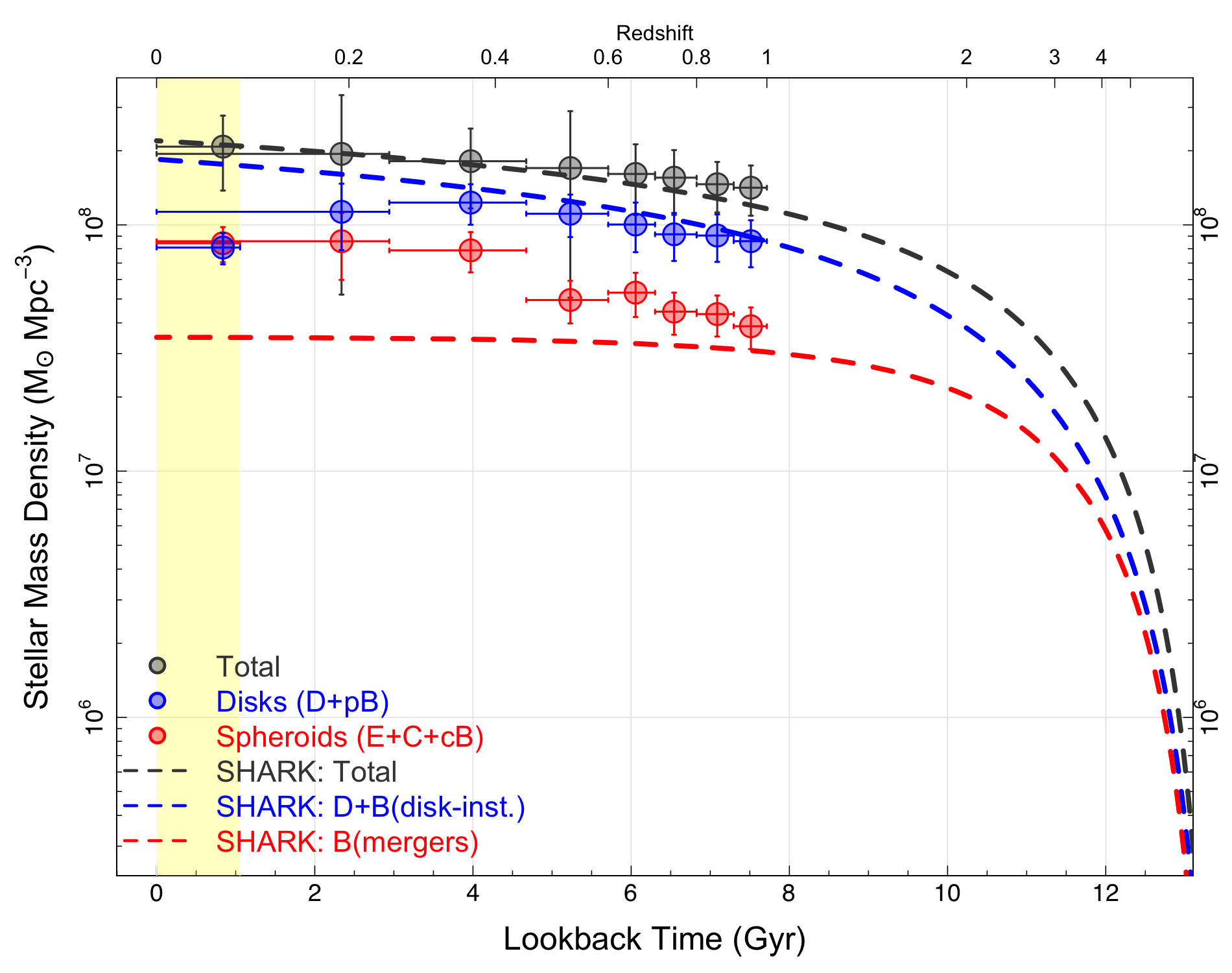}
    \caption{The evolution of the total SMD, $\rho_{*}$, together with the the SMD locked in all disks (D+pB; blue) and all spheroids (E+C+cB; red). For comparison, we show the SMD evolution from SHARK semi-analytic model in dashed lines. Dashed black line shows the total SMD which is in good agreement with our LSS-corrected SMDs. Dashed blue and red lines represent the evolution of all disks+bulges formed through disk instabilities, noted as D+B(disk-inst.) and bulges formed through mergers, noted as B(mergers), respectively.}
	\label{fig:MassBuildUp_disks_sph}
\end{figure}

Finally, we also look to compare our results to the predictions from the SHARK semi-analytic model (SAM; \citealt{Lagos18b}; \citealt{Lagos19}) and the forensic study of the spectral energy distributions of nearby bulges and disks by \cite{Driver13}. Figure \ref{fig:MassBuildUp_disks_sph} compresses our results into all disk structures (i.e., D+pB) and all spheroid structures (E+C+cB), generally trying to reflect structures built by dynamically cold (infall, secular and minor mergers), or dynamically hot processes (i.e., major mergers and collapse), respectively. 
Overlain are the predictions for bulge and disk evolution from the SHARK SAM with the dashed blue line representing the SMD of disks plus bulges formed through disk instabilities, and the dashed red line representing the SMD of all spheroids formed through mergers. We find that although our disks' SMD is in relatively good agreement with the SHARK SAM at $z > 0.4$, they diverge at later times, and the general trend of spheroids is inconsistent throughout.

Figure \ref{fig:MassBuildUp_disks_sph} also appears to be strongly inconsistent with the two-phase formation scenario presented by \cite{Driver13} (see their figure 5). As mentioned in Section \ref{sec:intro}, \cite{Driver13} argued for a transition at $z \approx 1.7$ with hot processes dominating at earlier times and cold processes dominating at later times. From this they predicted that at $z < 1.7$ spheroids should experience a flat and stable SMD with disk growth occurring around some subset of the spheroids. This is clearly the opposite of the trends that we see on Figure~\ref{fig:MassBuildUp_disks_sph}. However, we note a strong caveat in that we are comparing data derived from a forensic analysis (Driver et al.) with that derived from a core-sample analysis (as shown here and in most comparative studies). One speaks to the formation history of the stellar populations while the second to the appearances of the structures defined by a stellar population. Hence, it is potentially conceivable that an old population can be transformed from say a disk structure to a spheroid structure and a forensic study would show no evolution while a core-sample study would show a dramatic evolution. As nearby studies advocate that the stellar populations of Ellipticals are old, this would suggest that this may well be the case, i.e., while the stars that make up Elliptical systems formed very early, the structural form of an Elliptical galaxy has only emerged relatively late through presumably a dry-merger event.

The final conclusion is that to reconcile the current results with the SHARK and forensic-analysis we require a picture in which the late-time Universe is dominated by significant late-time dry-mergers at high-mass, combined with significant secular evolutionary processes at low mass. To further quantify the physical processes at work, we now need to fold in additional parameters and in particular explore the mass-size plane and its evolution. This will be explored further in Hashemizadeh et al. (in prep.)

\section{Summary and Conclusions} 
\label{sec:summary}

In this work, we have performed a 2D photometric structural decomposition of D10/ACS galaxies using high-resolution HST imaging. The goal is to provide a catalogue of credible structural measurements for galaxies at $z \leq 1$ and with $\mathrm{log}(M_*/M_\odot) \geq 9.5$ as defined in \cite{Hashemizadeh21}. In these two papers, we therefore provide the complex morphological and structural breakdown of the COSMOS galaxies. Catalogues are available within the DEVILS database, and will be released as part of the periodic DEVILS data releases. Here we summarize our results as follows:

\begin{description}

\item[$\bullet$] By performing several tests on the {\sc Tiny Tim} PSFs for HST/ACS we find that PSF uncertainties remain at the 10-20 per cent level, particularly within the central pixel, and attribute this to a combination of HST's periodic ``breathing'' (thermal expansion/contraction) and the difficulty in identifying the object centroid to very high precision. 

\item{$\bullet$} We use our decomposition pipeline, {\sc GRAFit} and the structural analysis code {\sc ProFit} (\citealt{Robotham17}) to fit three profile types to each of $\sim 35,000$ galaxy in the D10/ACS sample: a single S\'ersic, a S\'ersic+S\'ersic, and a S\'ersic+exponential disk. While we are unable to find a clear automated process for selecting the optimal profile, we find sensible outcomes when using our previous visual classifications to determine whether we should adopt a 1 or 2 component profiles. Where possible, we adopt double S\'ersic for the two component case, unless the bulge size exceeds the disk size, where we revert to a S\'erisc+exponential and confirm via visual inspection the veracity of the fit. 

\item{$\bullet$} The {\sc ProFit} Bayesian code together with the MCMC optimizations are demonstrated to be robust to initial conditions, and to provide good fits to more than 95 per cent of the sample, with poor fits generally arising when we have highly visually distorted or double-cored objects. This fraction of difficult to fit cases increases with look-back time.

\item{$\bullet$} We find that bulges generally fall into two categories, with a notable proportion forming a tight sequence in the stellar mass-half light radius ($M_*- \rm{R}_e$)-plane, and the remaining bulge systems scattered broadly across this plane. We attribute the compact distribution to be classical bulges, and the more dispersed distribution to be bulge-complexes which we refer to as pseudo-bulges (literally ``bulge-like'' or ``inner-disk'') and our selection is consistent with previous SDSS and GAMA results.

\item{$\bullet$} We report the B/T distributions in mass and redshift intervals, and find relatively strong trends in both directions, with what appears to be a strong emergence of low-B/T components in low mass systems at low redshifts. This appears to be robust to the evolution in our physical size limit at high-$z$ (i.e., $B_{\rm{R}_e} \sim 0.25$ kpc).

\item[$\bullet$] Moving from high- to low-redshift, the evolution of the stellar mass function for galaxy components reveals an enhancement in the low-mass end while a modest growth in the high-mass end of the SMF as reported in \cite{Hashemizadeh21}. Subdividing by structural component we find:
\begin{description}
\item{(i)} Disk components increase their mass at low-mass end, while showing a decrease at intermediate- and high-mass regions.
\item{(ii)} pBs and cBs experience significant growth at all mass intervals.
\item{(iii)} Ellipticals show strong growth in their intermediate- and low-mass end of their SMF, and minimal evolution in their high mass end.
\end{description}


~

\item[$\bullet$] We report the distribution of stellar mass density and its growth from $z=1$ to $z=0$ and find:
\begin{description}
\item{(i)} Disks dominate the stellar mass density at all redshifts. 
The population increases their mass in  $z = 1-0.35$ and decreases gradually since then, their contribution to the total SMD declines from $\sim 60$ per cent to $\sim 32$ per cent over the whole redshift range. This appears to suggest an end to the epoch of disk growth.
\item{(ii)} The pB population grows in stellar mass by a factor of $\sim 6.3$ from $z = 1$ to $z = 0.0$. 
\item{(iii)} The cB population grows in stellar mass by a factor of $\sim 1.86$ from $z = 1$ to $z = 0.35$ and flattens its growth rate since then.
\item{(iv)} The Es contribute the most in the total stellar mass after disk population, growing in stellar mass by a factor of $\sim 2.23$ from $z = 1$ to $z = 0$.
\end{description}
\end{description}

We conclude that by performing a robust structural decomposition of D10/ACS galaxies using high-resolution HST imaging data we appear to unveil a Universe in which disk formation and growth has completed ($z = 1-0.35$), and then stalled/stabilised ($ z < 0.35$). The latter Universe is dominated by secular processes building pseudo-bulges, and ongoing minor and probably major mergers, consolidating and building mass in spheroidal structures (E's and cB's). The exact role of minor and major mergers is still unclear and somewhat hard to constrain in this study but a key goal of the DEVILS program. In addition, a critical factor we are unable to address here is the identification of the precise mechanisms or the role that may be played by dust attenuation. A key question is whether classical bulges are growing, or emerging as dust is dissipated. Future studies will explore the evolving dust content. In Hashemizadeh et al. (in prep.) we will study the evolution of the $M_*-\rm{R}_e$ relations that will enable us to better distinguish between major mergers, minor mergers and secular and in-situ processes. Perhaps more critical is the need to directly measure the minor and major merger rates, which requires completion of the DEVILS redshift measurements, as well as understanding the neutral gas supply and accretion which will be unveiled through joint DEVILS MIGHTEE/LADUMA analysis. 
        
        
\section*{Acknowledgements}
DEVILS is an Australian project based around a spectroscopic campaign using the Anglo-Australian Telescope. The DEVILS input catalogue is generated from data taken as part of the ESO VISTA-VIDEO \citep{Jarvis13} and UltraVISTA \citep{McCracken12} surveys. DEVILS is part funded via Discovery Programs by the Australian Research Council and the participating institutions. The DEVILS website is \href{https://devilsurvey.org}{https://devilsurvey.org}. The DEVILS data is hosted and provided by AAO Data Central (\href{https://datacentral.org.au}{https://datacentral.org.au}). This work was supported by resources provided by The Pawsey Supercomputing Centre with funding from the Australian Government and the Government of Western Australia. We also gratefully acknowledge the contribution of the entire COSMOS collaboration consisting of more than 200 scientists. The HST COSMOS Treasury program was supported through NASA grant HST-GO-09822. SB and SPD acknowledge support by the Australian Research Council's funding scheme DP180103740. MS has been supported by the European Union's  Horizon 2020 research and innovation programme under the Maria Sk\l odowska-Curie (grant agreement No 754510), the National Science Centre of Poland (grant UMO-2016/23/N/ST9/02963) and by the Spanish Ministry of Science and Innovation through Juan de la Cierva-formacion program (reference FJC2018-038792-I). ASGR and LJMD acknowledge support from the Australian Research Council's Future Fellowship scheme (FT200100375 and FT200100055, respectively).

This work was made possible by the free and open R software (\citealt{R-Core-Team}).
A number of figures in this paper were generated using the R \texttt{magicaxis} package (\citealt{Robotham16b}). This work also makes use of the \texttt{celestial} package (\citealt{Robotham16a}) and \texttt{dftools} (\citealt{Obreschkow18}).

\subsection{Data Availability} 
\label{subsec:data_access}

The catalogues used in this paper are \texttt{D10VisualMoprhologyCat}, described in \cite{Hashemizadeh21}, and \texttt{DEVILS\_BD\_Decomp} and is held on the DEVILS database managed by AAO Data Central (\href{https://datacentral.org.au}{https://datacentral.org.au}). The catalogues are currently only available for the DEVILS team, but will be made publicly available in a future DEVILS data release.  All imaging data are in the public domain and were downloaded from the the NASA/IPAC Infrared Science Archive (IRSA) web-page: \href{https://irsa.ipac.caltech.edu/data/COSMOS/images/acs\_2.0/I/}{irsa.ipac.caltech.edu/data/COSMOS/images/acs\_2.0/I/}. The main tools used in this study are {\sc ProFit} \citep{Robotham17} version 1.3.3 (available at: \href{https://github.com/ICRAR/ProFit}{https://github.com/ICRAR/ProFit}) and ProFound \citep{Robotham18} version 1.3.4 (available at: \href{https://github.com/asgr/ProFound}{https://github.com/asgr/ProFound}). We used {\sc Tiny Tim} version 6.3 to generate the HST/ACS Point Spread Function (PSF). We further use {\sc LaplacesDemon} version 1.3.4 implemented in {\sc R} available at: \href{https://github.com/LaplacesDemonR/LaplacesDemon}{https://github.com/LaplacesDemonR/LaplacesDemon}. Our structural decomposition pipeline, {\sc GRAFit}, is available at: \href{https://github.com/HoseinHashemi/GRAFit}{https://github.com/HoseinHashemi/GRAFit}. 

\graphicspath{{images/ChapterFive/}}

\chapter[DEVILS: Mass-size Relation]{Deep Extragalactic VIsible Legacy Survey (DEVILS): The evolution of the mass-size relation of bulges and disks since $z = 1$.}
\label{ch:5}

This Chapter is the draft of a paper prepared to be submitted to the Monthly Notices of the Royal Astronomical Society with myself as the first author.

\section{Introduction}
\label{sec:ch5_intro}

The size and mass evolution of galaxies are arguably two of most important aspects of galaxy formation and evolution, making the scaling relation between these two parameters an informative fundamental plane. The distribution of galaxies on this plane gives insight into their angular momentum and hence their dark matter halos and their assembly history (see e.g., \citealt{Mo97,Mo98}; \citealt{Jiang19}). The mass and angular momentum of disks are argued to be linked to the mass and angular momentum of their halos which are in turn tightly correlated with the density of the Universe at time when the halo was formed (e.g., \citealt{Dalcanton97}). Consequently, this scenario concludes that halos that formed earlier in cosmic time are denser than halos formed at later times that in turn impacts the mass-size relation (e.g., \citealt{Trujillo06}). In addition, galaxies stellar mass-size relations ($M_*-R_e$) changes depending whether built by major or minor mergers (see \citealt{Trujillo07,Trujillo11}). 
Therefore, studying the fundamental plane of mass-size for different galaxy types and components can potentially shed light on how galaxies have reached their current size and mass.  

Disk and spheroid galaxies are shown to follow different $M_*-R_e$ relation (e.g., \citealt{Freeman70}; \citealt{Kormendy79}; \citealt{Kauffmann03}). 
\cite{Shen03} explored the $M_*-R_e$ relation of local galaxies in the Sloan Digital Sky Survey (SDSS, \citealt{York00}) and found that early- and late-type galaxies (ETG and LTG, respectively) follow different relations between their size and stellar mass with LTGs with $M_*/M_\odot \gtrsim 10^{10.6}$ following $R \propto M^{0.4}$ and less massive LTGs following $R \propto M^{0.15}$ while ETGs follow a steeper relation of $R \propto M^{0.56}$. 
Finding a steeper relation for ETGs, they examined two models of (I) single merger event versus (II) multiple mergers for the formation of ETGs and find that the latter scenario can reproduce the observed $M_*-R_e$ relation better than the former so they propose that ETGs are the likely remnants of a sequence of mergers with each merger increasing the size of the galaxy. 
\cite{Lange16} studied the $M_*-R_e$ relation of $z = 0$ galaxies from the Galaxy And Mass Assembly (GAMA, \citealt{Driver11}) survey. They further investigated the relation for galaxies of different morphological types as well as for disks and bulges using structural analysis based on GALFIT (\citealt{Peng02,Peng10}).

The evolution of the $M_*-R_e$ relation has been to interest of both observers and simulators as a key meeting ground between easy observables ($M_*, R_e$) and parameters traced in numerical simulations (e.g., $M_H$). This has become more feasible with the development of deeper photometric and spectroscopic surveys such as Cosmic Assembly Near-infrared Deep Extragalactic Legacy Survey (CANDELS, \citealt{Grogin11}; \citealt{Koekemoer11}) and the Cosmic Evolution Survey (COSMOS, \citealt{Scoville07}). The state-of-the-art study of the evolution of the $M_*-R_e$ with redshift is that performed by \cite{vanderWel14} where they utilized the Hubble Space Telescope (HST) WFC3 imaging of CANDELS and explored the $M_*-R_e$ relations of early- and late-type galaxies between $0 < z < 3$ (also see \citealt{vanderWel12} for their single S\'ersic fit measurements using {\sc GALFIT}, \citealt{Peng02,Peng10}). They highlighted that ETGs evolve faster ($R_{\mathrm{eff}} \propto (1+z)^{-1.48}$) than LTGs ($R_{\mathrm{eff}} \propto (1+z)^{-0.75}$) while the late-type population is larger than the early-type population at all redshifts (\citealt{vanderWel14}). More recently, \cite{Mowla19} combined a sample of high-mass galaxies in the COSMOS-Drift And SHift (COSMOS-DASH) with the 3D-HST/CANDELS sample of \cite{vanderWel14} and provided an updated version of the evolution of the $M_*-R_e$ relation for early- and late-type galaxies.      

The size evolution of disk and spheroid dominated galaxies with redshift is argued to have a direct link to their evolutionary history suggesting that disks that follow a flatter size evolution might have grown through gas infall (e.g., \citealt{Cayon96}; \citealt{Bouwens97}; \citealt{Lilly98}; \citealt{Ravindranath04}; \citealt{Barden05}) while spheroid dominated galaxies following steeper size evolution (e.g., \citealt{vanderWel14} and \citealt{Shen03}) are expected to grow through mostly major mergers (\citealt{White91}). 

As mentioned above, the evolution of the $M_*-R_e$ relation has so far been studied for galaxies in two broad categories (e.g., ETG/LTG). Despite the importance of this fundamental plane the evolution of the $M_*-R_e$ relation of galaxies of different morphological types, and more specifically for bulges and disks components, has not been thoroughly studied in the literature. 
In the present work, we make use of our morphological classifications (as described in Chapter \ref{ch:3}) together with our bulge-disk decomposition analysis (as described in Chapter \ref{ch:4}), to investigate the evolution of the $M_*-R_e$ relation of galaxies of different morphological types as well as disks and bulges since $z = 1$. 

\begin{landscape}
\begin{figure*} 
	\centering
	\includegraphics[width = 1.5\textwidth, height = 0.97\textwidth]{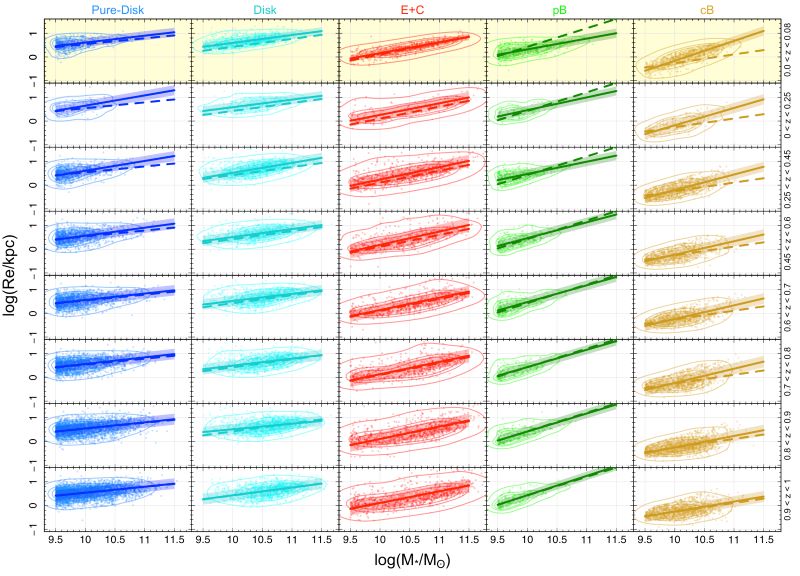}
	\caption{The evolution of the $M_*-R_e$ relations at $0 < z < 1$. Contours represent quantiles showing the levels containing $50\%$, $68\%$ and $95\%$ of the data. Solid lines are fits to the data and dashed lines represent the fit to the data at $z \sim 1$. Transparent regions around lines show the intrinsic scatter along y-axis ($1 \sigma$). Note that the stellar mass on the x-axis represent the component mass not the total mass of host galaxies. The first row highlighted with yellow transparent band shows the redshift range covered by GAMA data (i.e., $0 < z < 0.08$).}
	\label{fig:MRe}
\end{figure*}
\end{landscape}

\section{The evolution of the $M_*-R_e$ relation}
\label{sec:DEVILS_MRe}

In this section, we aim to measure the evolution of the $M_*-R_e$ relation of different morphological types together with that of disks and bulges, separately.
Similar to other chapters, we use our visual morphological classifications and separate our sample into pure-disk, elliptical+compact (E+C, following Chapter \ref{ch:4}) and double-component (BD) morphological types. We subdivide the BD systems into disk and bulge components using our bulge-disk de-compositions (as described in Chapter \ref{ch:4}) with bulges then divided into pseudo- and classical-bulges (pB and cB) according to their surface stellar mass densities as described in \ref{sec:pB_cB_dist}. Note that we use the same pB/cB separation method for both D10/ACS and the $z \sim 0$ GAMA samples. 

Figure \ref{fig:MRe} shows the $M_*-R_e$ relations of our D10/ACS sample together with $z \sim 0$ GAMA galaxies highlighted with yellow colour (see Chapter \ref{ch:GAMA}). Note that similar to Chapter \ref{ch:GAMA} we use the $R_e$ of GAMA galaxies contained within segments as recommended by \cite{Casura-inprep} and stellar masses as measured by \cite{Bellstedt20b}. Likewise, we use the same B/T in r-band (within segments) to calculate the stellar mass of bulges and disks of BD systems, i.e., $M_*^\mathrm{Bulge} = \mathrm{B/T} \times M_*^\mathrm{Total}$ and $M_*^\mathrm{disk} = (1-\mathrm{B/T}) \times M_*^\mathrm{Total}$. 

Following \cite{Lange16} and \cite{Shen03} we adopt a power law function to fit the $M_*-R_e$ relations:

\begin{equation}
    \mathrm{log}(R_e/kpc) = a\,\mathrm{log}(M_*/M_\odot)-b,
    \label{eq:MRe}
\end{equation}

\noindent where $R_e$ is the half-light radius or effective radius in unit of kpc and $M_*$ represents the stellar mass of each component. We then use the {\sc HyperFit} package \citep{Robotham15} to fit Equation \ref{eq:MRe} to the data, adopting a Markov Chain Monte Carlo (MCMC) minimisation method using a Componentwise Hit-And-Run Metropolis (CHARM) algorithm with $10,000$ iterations. See \cite{Robotham15} for more details about {\sc HyperFit}. 

Figure \ref{fig:MRe} shows both the distribution of the data in the $M_*-R_e$ plane and the fit to data. The transparent region around the best fit lines represent the $1 \sigma$ intrinsic scatter along the y-axis. The over-plotted dashed lines are $M_*-R_e$ relations at $z \sim 1$ to highlight the evolution across time.   

\begin{figure*} 
	\centering
	\includegraphics[width = \textwidth]{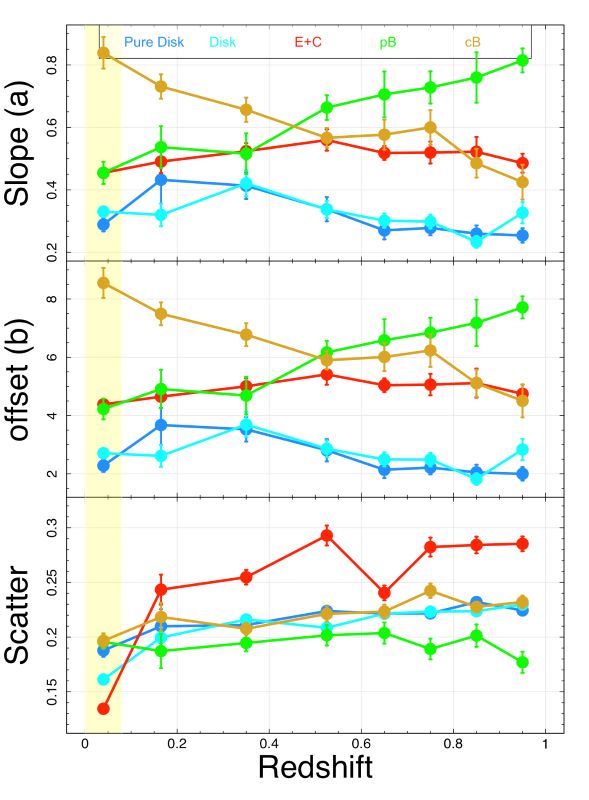}
	\caption{The evolution of the slope (a; top), offset (b; middle) and intrinsic scatter (bottom) of the $M_*-R_e$ relations. The yellow transparent band shows the redshift range covered by GAMA data. }
	\label{fig:MRe_scatter}
\end{figure*}

In general, the evolution of the $M_*-R_e$ relation shows that disks are larger than all other structures at all epochs while at later times ellipticals start to become larger at the high-mass end. Further, while pure-disks and disk structures follow similar relations, at a given $M_*$ disks located in bulge+disk systems are relatively more massive than pure-disk systems. Figure \ref{fig:MRe} also highlights that at a given mass cBs are smaller and more massive than pBs at all redshifts, implying that cBs are denser structures, which is expected by our pB/cB definitions. 

Investigating Figure \ref{fig:MRe}, we find that the $M_*-R_e$ relation of pure-disk systems (blue) and disk components (cyan) are consistent with modest discernible evolution from $z = 1$ to $z \sim 0$. This result is in agreement with other studies for example \cite{Ravindranath04}, \cite{Barden05}, \cite{Mosleh17}.   
We also show this trend in Figure \ref{fig:MRe_scatter} where we plot the evolution of the best fit slope (a), offset (b) and scatter of the $M_*-R_e$ relations and highlight how pure-disk systems follow the same best fit parameters as disk components across all the redshift range suggesting that, as expected, they have similar origins. 

The E+C population experience what is consistent with no evolution in their $M_*-R_e$ relation from $z = 1$ to 0. According to Hashemizadeh et al. (in review), the stellar mass growth in E+C population is dominated by the growth in their high-mass end, indicating that these systems move along their $M_*-R_e$ relation. This is consistent with other studies showing a significant growth of early-type galaxies in size (see e.g., \citealt{Navarro00}; \citealt{Trujillo06}; \citealt{vanderWel08}; \citealt{vanderWel14}).

Figure \ref{fig:MRe_scatter} further summarizes the above results and displays the evolution of the slope (a), offset (b) and intrinsic scatter, $\sigma$, of the $M_*-R_e$ relations (top and bottom panels, respectively).
As shown in the top panel of Figure \ref{fig:MRe_scatter}, pBs (green) have the steepest relation by $z \sim 0.45$ before cBs (gold) become the steepest relation. We note that pBs experience a flattening with cosmic time that is, interestingly, mapped to a significant steepening of cB relation. The top panel of this figure also shows that disks (both pure-disks; blue and disk components; cyan) have flattest relations with lowest offset (middle panel) while Es experience a somewhat unchanged slope ($a \sim 0.5$) and offset ($b \sim 5$).
We note that, despite the evolution and transition between morphological types due for example to mergers, accretions, bulge formation, etc; see \citealt{Hashemizadeh21} and Hashemizadeh et al., in review the $M_*-R_e$ relations are relatively stable and hence well established at $z = 1$. 



We calculate the scatter of the $M_*-R_e$ distributions along the y-axis, i.e., $\mathrm{log}(R_e)$ (bottom panel), and we find that the scatter of all structures are unchanged across redshift (with an average value of $\sim 0.2-0.25$; see Table \ref{tab:MRe_evol} for exact values). At this stage we cannot entirely rule out whether the scatters of our pB/cB $M_*-R_e$ relations could be a by-product of our structural decomposition and/or pB/cB separation. However, the stability of the other component classes would suggest it is intrinsic. Figure \ref{fig:MRe_scatter} also shows that Es have a slightly broader $M_*-R_e$ distribution than other structures and its scatter decreases with time while pBs have the lowest scatter.    

We report the best fit parameters of Equation \ref{eq:MRe} to each structure at different redshift bins in Table \ref{tab:MRe_evol}. 


\begin{landscape}
\begin{table*}
\centering
\caption{Best fit parameters of Equation \ref{eq:MRe} at different redshift bins. }
\begin{adjustbox}{scale = 0.7}
\begin{tabular}{lcccccccc}
\firsthline \firsthline \\
$z$-bins      &  $0.0 \leq z < 0.08$  &  $0.0 \leq z < 0.25$ &  $0.25 \leq z < 0.45$ &  $0.45 \leq z < 0.60$ &  $0.60 \leq z < 0.70$ &  $0.70 \leq z < 0.80$ &  $0.80 \leq z < 0.90$ &  $0.90 \leq z \leq 1.00$ \\ \\ \hline                     
                    &   \multicolumn{8}{c}{\textbf{Pure-Disk}}  \\
\cline{2-9} \\
$a$           & $0.364\pm0.048$   & $ 0.403\pm0.086$ & $0.432\pm0.055$ & $0.346\pm0.039$ & $0.318\pm0.041$ & $0.285\pm0.024$ & $0.261\pm0.022$ & $0.257\pm0.026$ \\ \\
$b$           & $3.022\pm0.475$   & $3.391\pm0.843$  & $3.715\pm0.541$ & $2.889\pm0.390$ & $2.619\pm0.405$ & $2.285\pm0.237$ & $2.066\pm0.224$ & $2.034\pm0.263$ \\ \\
scatter       & $0.190\pm0.006$   & $0.207\pm0.015$  & $0.212\pm0.006$ & $0.224\pm0.005$ & $0.223\pm0.004$ & $0.222\pm0.004$ & $0.232\pm0.003$ & $0.225\pm0.003$ \\ \\ \hline 
                    &   \multicolumn{8}{c}{\textbf{Disk Component}}  \\
\cline{2-9} \\
$a$         & $0.327\pm0.016$ & $0.354\pm0.034$  & $0.390\pm0.029$ & $0.333\pm0.023$ & $0.327\pm0.034$ & $0.332\pm0.040$ & $0.253\pm0.038$ & $0.330\pm0.043$  \\ \\
$b$         & $2.674\pm0.165$ & $2.968\pm0.353$  & $3.366\pm0.300$ & $2.811\pm0.239$ & $2.768\pm0.353$ & $2.849\pm0.418$ & $2.024\pm0.406$ & $2.870\pm0.459$  \\ \\
scatter     & $0.161\pm0.003$ & $0.201\pm0.009$  & $0.215\pm0.005$ & $0.208\pm0.005$ & $0.223\pm0.005$ & $0.225\pm0.005$ & $0.225\pm0.004$ & $0.230\pm0.005$  \\ \\ \hline 
                    &   \multicolumn{8}{c}{\textbf{Pseudo-Bulge}}  \\
\cline{2-9} \\ 
$a$          & $0.447\pm0.038$  & $0.521\pm0.078$ & $0.481\pm0.033$ & $0.726\pm0.095$ & $0.710\pm0.048$ & $0.722\pm0.054$ & $0.744\pm0.060$ & $0.849\pm0.061$ \\ \\
$b$          & $4.147\pm0.377$  & $4.756\pm0.770$ & $4.347\pm0.324$ & $6.786\pm0.933$ & $6.628\pm0.467$ & $6.785\pm0.536$ & $7.027\pm0.600$ & $8.044\pm0.599$  \\\\
scatter      & $0.195\pm0.006$  & $0.186\pm0.015$ & $0.193\pm0.008$ & $0.205\pm0.011$ & $0.204\pm0.009$ & $0.189\pm0.010$ & $0.200\pm0.010$ & $0.179\pm0.010$ \\ \\\hline 
                    &   \multicolumn{8}{c}{\textbf{Classical-Bulge}}  \\
\cline{2-9} \\ 
$a$           & $0.820\pm0.036$  & $0.779\pm0.076$ & $0.677\pm0.046$ & $0.599\pm0.044$ & $0.558\pm0.047$ & $0.647\pm0.066$ & $0.498\pm0.055$ & $0.405\pm0.042$ \\ \\
$b$           & $8.355\pm0.369$  & $7.972\pm0.765$ & $6.981\pm0.465$ & $6.224\pm0.445$ & $5.822\pm0.470$ & $6.708\pm0.666$ & $5.238\pm0.555$ & $4.306\pm0.426$  \\ \\ 
scatter       & $0.195\pm0.006$  & $0.221\pm0.013$ & $0.209\pm0.005$ & $0.223\pm0.007$ & $0.223\pm0.006$ & $0.245\pm0.007$ & $0.228\pm0.005$ & $0.231\pm0.005$ \\ \\ \hline 
                    &   \multicolumn{8}{c}{\textbf{Elliptical+Compact}}  \\
\cline{2-9} \\
$a$           & $0.440\pm0.006$   & $0.471\pm0.025$ & $ 0.480\pm0.018$ & $0.574\pm0.043$ & $0.533\pm0.034$ & $0.492\pm0.030$ & $0.505\pm0.035$ & $0.496\pm0.034$ \\ \\
$b$           & $4.227\pm0.068$   & $4.446\pm0.266$ & $4.547\pm0.193$ & $5.560\pm0.451$  & $5.204\pm0.354$ & $4.776\pm0.313$ & $4.939\pm0.373$ & $4.856\pm0.355$  \\ \\ 
scatter       & $0.134\pm0.004$   & $0.240\pm0.013$ & $0.252\pm0.006$ & $0.295\pm0.010$  & $0.241\pm0.007$ & $0.280\pm0.008$ & $0.282\pm0.007$ & $0.286\pm0.007$ \\ \\ \hline 

\lasthline
\end{tabular}
\end{adjustbox}
\label{tab:MRe_evol}
\end{table*}
\end{landscape}

\section{Comparison with the literature}
\label{sec:literature_comp}

\begin{figure*} 
	\centering
	\includegraphics[width = \textwidth]{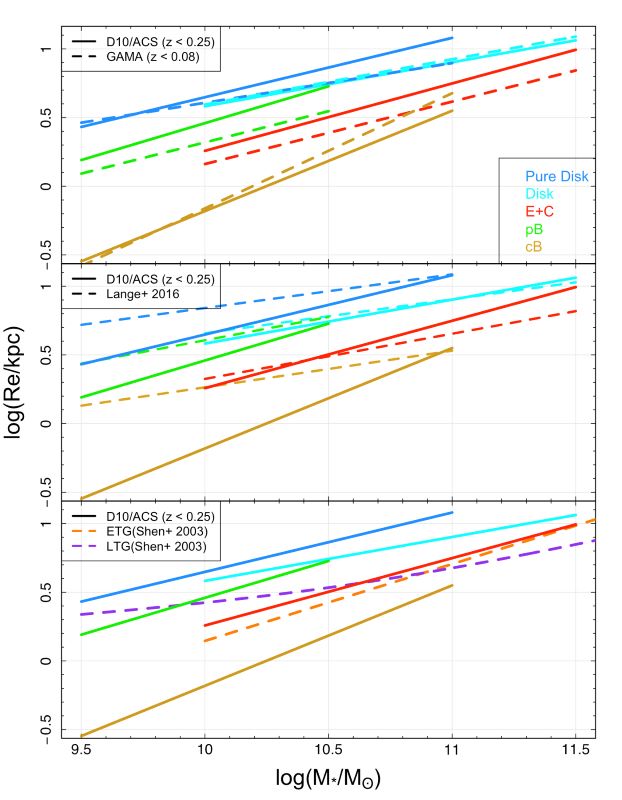}
	\caption{Comparison of our GAMA and DEVILS low-$z$ $M_*-R_e$ relations (left and right panels, respectively) compared with \cite{Lange16} and \cite{Shen03} measurements (dashed and dotted lines, respectively). Note that dashed green and gold lines represent late- and early-type bulges in \cite{Lange16} that might not be comparable to our pB and cB classification. }
	\label{fig:MRe_Lange_Shen}
\end{figure*}

In this section, we compare our $M_*-R_e$ relations with key literature results at both low and high redshifts. 
Figure \ref{fig:MRe_Lange_Shen} shows the comparison of our D10/ACS low-$z$ ($z < 0.25$) $M_*-R_e$ relations with our GAMA sample (as described in Section \ref{sec:GAMA_MRe}, \citealt{Casura-inprep}) as well as with previous works by \cite{Lange16} for GAMA galaxies and \cite{Shen03} for SDSS galaxies. 

As shown in the top panel of Figure \ref{fig:MRe_Lange_Shen}, we find that despite the difference in redshift ranges ($z < 0.25$ for the D10/ACS versus $z < 0.08$ for GAMA) the $M_*-R_e$ relation of both pure-disk population and disk components of the D10/ACS are in good agreement with the GAMA relations. We find that at a given stellar mass the D10/ACS elliptical galaxies are larger in size than those of the GAMA sample, although they both follow relatively similar slope. The relation of our cB population is also consistent with GAMA while pB population is slightly more contracted in GAMA than in D10/ACS. We note again that we are comparing two very distinct types of data with significant difference in resolution, PSF etc. Moreover, the structural analysis of GAMA and D10/ACS have been done by different groups with different settings and pipelines. Therefore, one might naturally expect some mismatch between these two data-sets. These effects remain under investigation, and such work is outside the scope of this thesis.     

We can also compare our D10/ACS low-$z$ $M_*-R_e$ relations with the earlier measurements of GAMA galaxies presented by \cite{Lange16} (shown in the middle panel of Figure \ref{fig:MRe_Lange_Shen}). The structural divisions of \cite{Lange16} are not directly comparable with our classification, however, for completeness, we compare our pure-disk, disk component, pB, cB and E+C systems with their late-type disk, early-type disk, late-type bulge, early-type bulge and elliptical, respectively. We find that our $M_*-R_e$ relations of disk components and ellipticals are in good agreement with \cite{Lange16}. However, the relation for other structures are inconsistent likely due to our different classification techniques than that in \cite{Lange16}. This difference is more obvious in bulge structures where, as expected, morphological classification and structural analysis are important. This large difference in bulge relations (both cBs and pBs) are expected as with the HST high-resolution imaging we are able to resolve much smaller bulges at this low redshift range leading to steeper $M_*-R_e$ relations.

In the lower panel of Figure \ref{fig:MRe_Lange_Shen}) we compare our low-$z$ D10/ACS $M_*-R_e$ relations with the \cite{Shen03} relation for early- and late-type galaxies (ETG and LTG, respectively) and based on SDSS data. Their ETG/LTG classification is based on S\'ersic index (n) with $n < 2.5$ and $n > 2.5$ representing LTGs and ETGs, respectively. Again, their broad distinction is not comparable with our morphological classifications, but we find a relatively good agreement between their ETGs and our Es.    

\begin{figure*} 
	\centering
	\includegraphics[width = 1.1\textwidth]{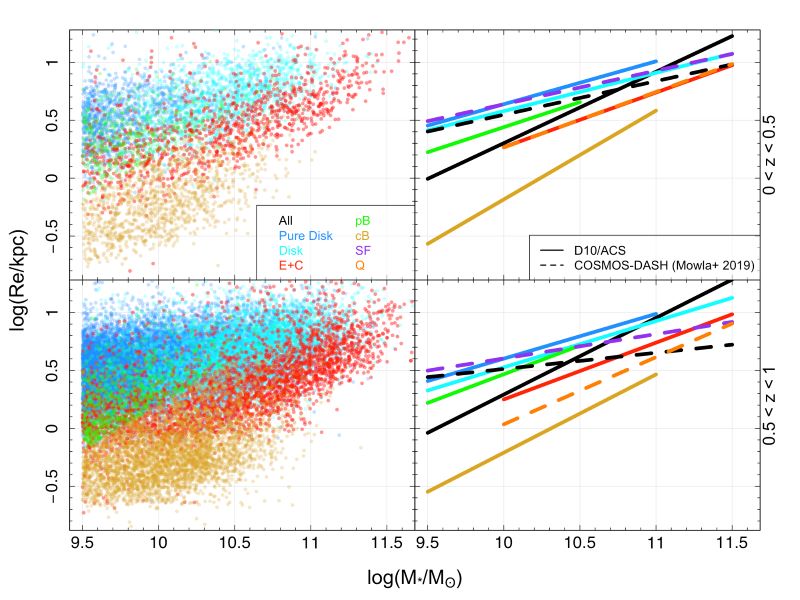}
	\caption{Comparison of our $M_*-R_e$ relations (solid lines) in two redshift ranges of $0.0 < z < 0.5$ and $0.5 < z < 1$ with that of COSMOS-DASH (dashed lines) from \cite{Mowla19}. COSMOS-DASH includes the $M_*-R_e$ relations for star-forming (purple) and quiescent (orange) galaxies.}
	\label{fig:MRe_COMOS_DASH}
\end{figure*}

Finally, in Figure \ref{fig:MRe_COMOS_DASH}, we compare our D10/ACS $M_*-R_e$ relations with higher-$z$ relations from \cite{Mowla19} obtained from a combination of COSMOS-DASH data and previous \cite{vanderWel14} data in CANDELS. \cite{Mowla19} separate their sample into two classes of star-forming (SF) and quiescent (Q) galaxies which is not the same as our morphological categories but we show this comparison for completeness. Here, similar to \cite{Mowla19}, we split our sample into two broad redshift bins of $0.0 < z < 0.5$ and $0.5 < z < 1$ and highlight that their total $M_*-R_e$ relations in both redshift bins are flatter than our D10/ACS. The relation of our E+C population is in excellent agreement with their quiescent galaxies (orange dashed lines) in $0 < z < 0.5$ while offsets in $0.5 < z < 1$. We find that their relation for star-forming systems (purple dashed lines) are consistent with our disk populations.

\section{Size evolution of structures}
\label{sec:size_evol}

\begin{figure*} 
	\centering
	\includegraphics[width = \textwidth]{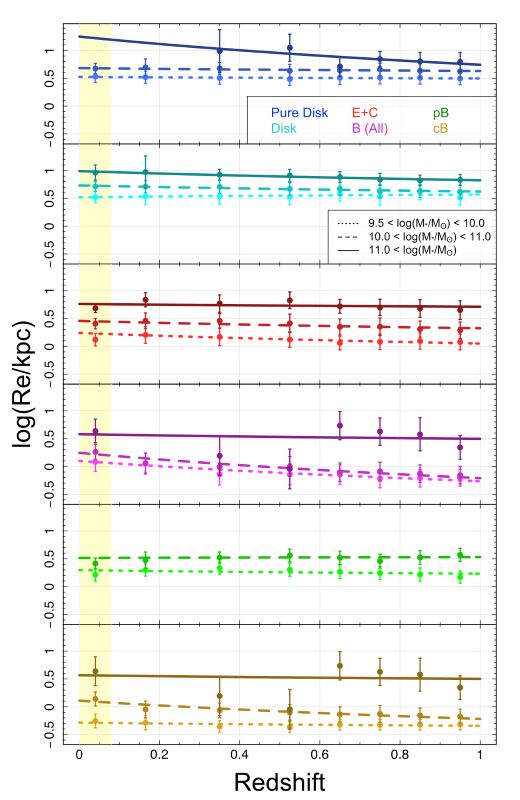}
	\caption{The size evolution of structures since $z = 1$ in three stellar mass bins shown as dotted ($9.5 \leq \mathrm{log}(M_*/M_\odot) < 10.0$,), dashed ($10.0 \leq \mathrm{log}(M_*/M_\odot) < 11.0$) and solid ($11.0 \leq \mathrm{log}(M_*/M_\odot)$) lines. Note that the stellar mass here is the component stellar mass not the total. Data points represent median sizes in each $z$-bin while lines show the results of our parameterized fit of Equation \ref{eq:Re_z} to data points. See Table \ref{tab:Re_z} for best fit parameters.  }
	\label{fig:Re_z}
\end{figure*}

With the mass-size measurements of our D10/ACS sample in place, we now explore the variation of the size of each of the above structures over time. Figure \ref{fig:Re_z} shows this evolution with data points representing the median size of each structure per redshift bin per stellar mass bin. We bin our structures into three bins of component stellar mass, as: $9.5 \leq \mathrm{log}(M_*/M_\odot) < 10.0$, $10.0 \leq \mathrm{log}(M_*/M_\odot) < 11.0$ and  $11.0 \leq \mathrm{log}(M_*/M_\odot)$. 

Following the literature (see e.g., \citealt{vanderWel14}; \citealt{Mosleh17}; \citealt{PaulinoAfonso17}; \citealt{Marshall19}) we fit the size evolution as a function of $z$ with the following function:  

\begin{equation}
    R_e/kpc = \alpha\,(1+z)^\beta,
    \label{eq:Re_z}
\end{equation}

\noindent where $R_e$ is the effective radius in units of kpc and $z$ is redshift. Table \ref{tab:Re_z} summarizes the best fit parameters of Equation \ref{eq:Re_z} to our data. We use a Levenberg-Marquardt nonlinear least-squares
algorithm implemented in the \texttt{minpack.lm} package in {\sc R} for fitting the above equation to our data.

Figure \ref{fig:Re_z} shows that regardless of the type of structure, as expected, more massive components are larger in size than less massive ones throughout the redshift range. For example, the high-mass Es (red) are on average $\sim 3$ times larger than the low-mass ones while high-mass disk components (cyan) are on average $\sim 2$ times larger than low-mass ones. This dependence of mass and size of structures is obviously the direct reflection of the slope of the $M_*-R_e$ relations that we discussed in Section \ref{sec:DEVILS_MRe}, implying that the steeper the relation, the more mass dependent the size is.  

Interestingly, as shown in Figure \ref{fig:Re_z}, we find a minimal size evolution since $z = 1$ for all structures. This unchanged trend is still valid for structures of different stellar mass. 
The best fit parameters reported in Table \ref{tab:Re_z} show that although the growth in size is small, the slope of our parameterized fits, $\beta$, seems to have some correlation with the stellar mass with more massive structures growing slightly faster than lower mass ones (\citealt{Robotham14}). For example, the disk component (cyan) has $\beta \sim -0.16 \pm 0.05$ in their low stellar mass bin versus $\beta \sim -0.55\pm0.07$ in their high-mass range. 
We note that as expected we do not find any pB in the largest stellar mass bin ($\mathrm{log}(M_*/M_\odot) > 11$) while we find only 14 cB systems in this mass regime. Due to this small number of objects we have large errors on data and hence not clear whether the trend is real. In spite of the large uncertainties, this trend is consistent with minimal size evolution.  
However, we do not rule out some effects of our pB/cB distinction procedure on this trend. We also show the size evolution of all bulges (magenta) showing that lower mass bulges, in general, grow by a factor of $\sim 7$ in size since $z = 1$.

\begin{table*}
\centering
\caption{Best fit parameters of Equation \ref{eq:Re_z} for different bins of component stellar mass. Note the we don't find any pB in the last mass bin. }
\begin{adjustbox}{scale = 0.9}
\begin{tabular}{lccc}
\firsthline \firsthline \\
mass bins      &  $9.0 \leq \mathrm{log}(M_*/M_\odot) < 10.0$  &  $10.0 \leq \mathrm{log}(M_*/M_\odot) < 11.0$ &  $11.0 \leq \mathrm{log}(M_*/M_\odot)$ \\ \\ \hline                
                    &   \multicolumn{3}{c}{\textbf{Pure-Disk}}  \\
\cline{2-4} \\
$\alpha$           & $3.342\pm0.086$    & $4.811\pm0.126$ & $17.753\pm7.366$  \\ \\
$\beta$            & $-0.093\pm0.054$   & $-0.171\pm0.056$ & $-1.684\pm0.866$  \\ \\ \hline
                   &   \multicolumn{3}{c}{\textbf{Disk Component}}  \\
\cline{2-4} \\
$\alpha$           & $3.318\pm0.233$    & $5.411\pm0.123$  & $9.783\pm0.295$  \\ \\
$\beta$            & $0.157\pm0.154$    & $-0.365\pm0.054$ & $-0.547\pm0.067$  \\ \\ \hline
                   &   \multicolumn{3}{c}{\textbf{Bulge (All)}}  \\
\cline{2-4} \\
$\alpha$           & $1.259\pm0.070$    & $1.759\pm0.128$  & $3.802\pm1.536$  \\ \\
$\beta$            & $-1.211\pm0.149$   & $-1.513\pm0.219$ & $-0.276\pm0.854$  \\ \\ \hline
                   &   \multicolumn{3}{c}{\textbf{Pseudo-Bulge}}  \\
\cline{2-4} \\
$\alpha$           & $1.974\pm0.178$    & $3.272\pm0.528$ & $-$  \\ \\
$\beta$            & $-0.209\pm0.204$   & $0.056\pm0.310$ & $-$  \\ \\ \hline
                   &   \multicolumn{3}{c}{\textbf{Classical-Bulge}}  \\
\cline{2-4} \\
$\alpha$           & $0.517\pm0.030$    & $1.286\pm0.095$   & $3.685\pm1.648$  \\ \\
$\beta$            & $-0.188\pm0.125$   & $-1.091\pm0.198$  & $-0.219\pm0.919$  \\ \\ \hline
                   &   \multicolumn{3}{c}{\textbf{Elliptical+Compact}}  \\
\cline{2-4} \\
$\alpha$           & $1.740\pm0.097$    & $2.864\pm0.181$  & $5.778\pm0.638$  \\ \\
$\beta$            & $-0.633\pm0.121$   & $-0.440\pm0.156$ & $-0.166\pm0.266$  \\ \\ 
\lasthline
\end{tabular}
\end{adjustbox}
\label{tab:Re_z}
\end{table*}

\section{Summary and conclusion}

In this chapter, using our structural analysis, we have investigated the evolution of the stellar mass-size relation since $z = 1$. We find close-to-no-variation in the $M_*-R_e$ relation for almost all structures indicating that in spite of ongoing although declining star-formation, mass evolution, morphological transitions and mergers the $M_*-R_e$ relations only modestly varies. This implies that in general evolution moves galaxies along their $M_*-R_e$ relations. Although the results of Chapter \ref{ch:4} fairly moderate mass movement. 

We find that in agreement with other studies (e.g., \citealt{Trujillo06}; \citealt{Lange16}) our E+C class and bulge components follow steeper $M_*-R_e$ relations than disk structures indicating that at a given stellar mass Es and bulges are smaller than disk structures but also that they likely build up via distinct evolutionary mechanisms. Note that this is valid until $\mathrm{M_*}/\mathrm{M}_\odot < 10^{11}$, while more massive ellipticals are on average larger than disks and potentially the largest structures. This are likely the cDs in clusters cores.

Our results show that the slope of the $M_*-R_e$ relations of pure disk systems at all redshift ranges is very consistent with that of disk components of bulge+disk systems (see Figures \ref{fig:MRe} and \ref{fig:MRe_scatter}) suggesting the same origin and evolutionary pathway for all disks regardless of the presence or absence of a bulge component.
In addition, in agreement with other studies (e.g., \citealt{Ravindranath04} and \citealt{Barden05}), we find a redshift independent $M_*-R_e$ relation for disks with a slope of $\sim 0.3$. This is also in agreement with several studies that in a broader classification have shown that the $M_*-R_e$ of late-type galaxies evolves only very little in $z < 1$ (see e.g., \citealt{Lilly98}; \citealt{Dutton11}; \citealt{vanderWel14}; \citealt{Mosleh17}).  

Exploring the variation of the $M_*-R_e$ relation of the E+C galaxies we find that this relation experience a what is consistent with no change while we have shown in Chapter \ref{ch:4} (Hashemizadeh et al., in review) that the majority of the evolution in stellar mass happens at their high mass end (see Figures \ref{fig:MRe} and \ref{fig:MRe_scatter}). Note, however, that Figure \ref{fig:Re_z} shows that low-mass E+C systems seem to grow slightly faster in size than high-mass ones. 
This further confirms our results in Chapters \ref{ch:3} and \ref{ch:4} (see Figures \ref{fig:Mfunc_6z} and \ref{fig:SMD_var}) suggesting that, in agreement with \cite{Robotham14}, at $z < 1$ minor mergers are the dominant driver of the growth/formation of E systems. This interpretation is in agreement with other observational (e.g., \citealt{Trujillo11}) and theoretical (e.g., \citealt{Naab09}; \citealt{Xie15}) studies identifying minor mergers to play the most important role in the growth of E galaxies. 

In addition, classical bulges also present a steepening with time, whilst pseudo-bulges experience only a modest flattening (contraction). We find different $M_*-R_e$ relation for cBs than that of Es likely suggesting that they have different origins. This further highlights our argument that in-situ and secular star formation are the dominant processes in bulge formation in the $z < 1$ Universe given that we concluded in Chapter \ref{ch:4} that pBs have the largest growth rate in their stellar mass (by a factor of $6.3$ see Figure \ref{fig:MassBuildUp}).

We also investigated the size evolution of structures (Figure \ref{fig:Re_z}) and concluded that as expected more massive structures are larger in size, too, with only a modest variation since $z = 1$. We showed that although disk structures grow a little in size, bulges (of $\mathrm{log}(M_*/M_\odot) < 11$, in particular) grow slightly and eventually by a factor of $\sim 3$ at $z \sim 0$.  

In summary, we find a surprisingly unchanged size and $M_*-R_e$ relation over the last $\sim 8$ Gry. This lack of evolution particularly in disk structures is consistent with our previous results, suggesting that evolution is predominantly along the respective $M_*-R_e$ relations. In effect, the scaling relations lay down the trail along which galaxies have evolved. Each trail hence requiring a distinct path, i.e. an inside-out growth and evolution in which the size of galaxies grows as they grow in stellar mass. In addition minor mergers seem to be responsible for the growth of elliptical systems at least since $z < 1$. However, further investigations, numerical simulations in particular, is required to confirm these results and put them into a comprehensive physical picture. 


\chapter[Summary and Future Works]{Summary and Future Works}

\section{Summary}

In this thesis, we have used high-resolution imaging data from the Hubble Space Telescope in the D10 (COSMOS) region and performed a visual morphological classification and a 2D photometric structural decomposition of $\sim 44,000$ galaxies up to $z \sim 1$ and stellar mass of $M_*/M_\odot > 10^9$ (D10/ACS sample). These analyses have led to the realisation of two catalogues of \textbf{DEVILS\_D10MorphCat} and \textbf{DEVILS\_D10StructureCat}, both currently available to team members on the DEVILS website\footnote{\href{https://devilsurvey.org}{https://devilsurvey.org}}. We also do similar analysis for GAMA galaxies at $z < 0.08$ to build our $z \sim 0$ benchmark sample. Using these analyses we explore the evolution of the stellar mass function and the stellar mass density together with the stellar mass-size relations of each morphological type and structure (bulges and disks). We summarize our results below.

\subsection{The stellar mass evolution of the morphological types}

In our morphological classification we visually sub-classify our D10/ACS sample into the following 5 categories: bulge+disk (BD), pure-disk (D), elliptical (E), compact (C) and hard (H) systems. Below we summarize our investigation of the evolution of the SMF and integrated SMD of the above morphological types since $z = 1$.

\begin{description}

  \item[$\bullet$] Morphological classification of galaxies beyond $z > 1$ becomes increasingly hard as we find that this epoch is dominated by interacting and disrupted systems at least at the wavelengths we probe. For at these wavelengths we expect to sample more star-forming regions (rest-frame UV) at redshifts beyond $z > 1$.
  
  \item[$\bullet$] Our total SMF can be well described by a single Schechter function given that our stellar mass range only extends to $M_*/M_\odot = 10^9$. The evolution of the total SMF shows an increase at both low- and high-mass ends suggesting that there are two distinct physical processes in place at $z < 1$.
  
  \item[$\bullet$] Dividing the global SMF into the contribution from different morphological types we find that BD systems experience a significant growth at their low-mass end while their high-mass end is unchanged throughout the redshift range suggesting that the evolution of BD systems is dominantly driven by secular and/or in-situ evolution. We find D systems to increase at their low-mass end and decrease at their high-mass end again indicating the in-situ star-formation in disk systems leading to the formation of central components (bulges), thus transitioning high-$z$ pure-disks into BD systems at later times (witnessed by the decrease of the high-mass end of their SMF). Our E galaxies experience an increase at all stellar masses, although modestly at their high-mass end. We interpret this as the effect of mergers, both minor and major, in the growth of E systems.
  
  \item[$\bullet$] The evolution of the integrated SMD reveals that approximately two-thirds of the total stellar mass of the Universe today was in place at $z \sim 1$ with the remaining $\sim 33\%$ formed during the last 8 Gyr. We find that the major contributor in the total stellar mass of the Universe at all redshifts (at least by $z=1$) is BD systems, increasing by a factor of $\sim 2$ in stellar mass density since $z = 1$, while E systems grow their SMD by a factor of $\sim 1.6$. On the contrary, D population gradually lose their stellar mass eventually by $\sim 40\%$ at $z \sim 0.25$.
  
  \item[$\bullet$] Our toy model based on the mass transfer between morphological types and some assumptions suggests that a large fraction of the stellar mass in the $z < 1$ Universe is forming through in-situ star-formation while a somewhat constant merger rate also transfers the mass into elliptical systems. 

\end{description}

\subsection{The evolution of bulges and disks}

Our results based on the evolution of the morphological types suggest that bulge+disk systems, hence bulges, are dominantly built through in-situ and secular processes. To further explore this, we now make use of our structural decompositions to decode the evolution of bulges and disks separately. In addition, we distinguish between pseudo- and classical-bulges with a cut in the stellar mass-size plane where we find two distinct populations of bulges in the parameter space.  
Below we summarize our main results.  

\begin{description}

\item[$\bullet$] The evolution of the SMF of bulges and disks show no evolution at the high-mass end of the SMF of disk structures while a significant growth at their low-mass end. Bulge components, both pBs and cBs, experience an enhancement in their low- and high-mass ends. We note that we find a minimal disjoint between D10/ACS and GAMA SMFs, which is likely driven by 0.2 dex smaller stellar mass measurements in GAMA.

\item[$\bullet$] We find that although on average $\sim 50\%$ of the stellar mass of the Universe at $z < 1$ is bounded in disk structures, the mass growth of disk structures is declining, suggesting an end to the era of disk growth.

\item[$\bullet$] While Es contribute more than bulges to the overall SMD of the Universe, pBs evolve faster than other structures with a growth of a factor of 3 in their stellar mass density between $0.2 < z < 1$. We find cB and E populations to grow in stellar mass by a factor of $\sim 1.5$ and $\sim 1.7$, respectively.

\end{description}

\subsection{The evolution of the $M_*-R_e$ relations}

Our explorations of the build-up of the stellar mass in different galaxy components proposes a Universe (at $z < 1$) in which the growth rate of disks has slowed down while secular and in-situ evolution together with minor mergers are in play to build bulges and bulge-like structures. To further confirm this scenario we studied the important scaling relation of $M_*-R_e$ for different structures and their evolution since $z = 1$. 
Below we summarize our findings from this analysis:

\begin{description}

\item[$\bullet$] Our results are consistent with very little/no evolution in the $M_*-R_e$ relations of the structures. However, we find that the relation for E galaxies becoming slightly flatter with time, suggesting that minor mergers likely play the most important role in the growth of E systems at least for $z < 1$. 

\item[$\bullet$] We also show that E's relations follow a steeper slope than other structures including disks except for pBs at all redshifts. Therefore, at a given stellar mass Es and bulges are smaller than disks suggesting different evolutionary pathways for these structures.       

\item[$\bullet$] We confirm that the slope of the $M_*-R_e$ relation of all disks in the Universe are similar and unchanged throughout the redshift range regardless of them being bulge-less or hosting a bulge, suggesting the same formation and evolutionary history for all disks.

\item[$\bullet$] cBs' relation also shows a minimal variation while we confirm that pBs have the steepest $M_*-R_e$ relations at all epochs and become steeper with time.  

\item[$\bullet$] The size variation of structures of different stellar masses with time shows that more massive structures are larger while we find a minimal growth in size since $z = 1$.

\item[$\bullet$] Therefore, we quantify a somewhat stable $M_*-R_e$ scaling relation for all structures. When coupled with our previous conclusions this suggests an inside-out growth scenario. 

\end{description}

\newpage
\section{Future Works}
This study has focused on the morphological and structural evolution of galaxies since $z = 1$. However, other questions are yet to be explored.

\subsection{Kinematic decomposition of intermediate-$z$ galaxies.}

We also wish to propose an IFS survey of a sub-sample of our galaxies at intermediate redshifts, $0.3 < z < 0.5$, aiming at further exploring the origins of bulge formation. For this study we will propose high-resolution IFU observations (our target instrument is high-resolution MUSE in Narrow Field Mode) of a sample of our disk galaxies with pseudo- and classical-bulges. This observations will allow us to explore the origin of the dichotomy in bulges at intermediate redshifts ($0.3 \le z \le 0.5$). This observations will extend the studies in the local Universe to higher redshifts testing whether the local scaling relations are still available at higher-$z$ or we find any signs of evolution indicating that the relations are redshift dependant.  
\\
We will then investigate some important scaling relations, for example between bulges' S\'ersic index and $v_{max}/\sigma$. Investigating these relations will enable us to further quantify whether the S\'ersic index is a proper photometric representative for different bulges. Confirmation of the S\'ersic index as a proxy for geometry, level of rotation/dispersion velocity and stellar population of different bulge structures would be a key outcome of this survey. Having an estimate of the velocity dispersion of bulges, we will be able to also explore other important scaling relations such as the Faber-Jackson relation and the canonical plane of $v_{max}/\sigma$ vs ellipticity ($\epsilon$).

This study will potentially allow us to measure the gas phase metallicities via spectral lines ([OII], [OIII] and H$\beta$). If we observe strong enough lines, we will ultimately be able to investigate the position of the pseudo-bulges and classical bulges on the mass-metallicity plane and quantify whether they follow the known M-Z relation or are outliers. Along with the $\epsilon-v_{max}/\sigma$ relation, if bulges (classical-bulges in particular) are confirmed to be outliers it might be an indication of different formation pathway other than what typical elliptical galaxies are thought to take.

\subsection{The morphological evolution of galaxies in environments}

The intermediate redshift range that we studied in this thesis represents a key phase in the evolution of the Universe
where almost half the stellar mass is formed and large structures such as groups, clusters and filaments
emerge. Having our structural catalogue of tens of thousands of galaxies in this epoch, we plan to study the relationship between environments and morphology and light profiles of galaxies. For such study we need a highly complete spectroscopic sample and a robust group catalogue. As the DEVILS survey progresses we will achieve higher completeness in the COSMOS/DEVILS (D10) region, eventually reaching $> 95\%$ completeness.  
This high completeness will facilitate a robust environmental study of galaxies in groups and clusters. 

\subsection{Does astronomy still need photometric bulge-disk decomposition?}

In the past 15 years, the Integral Field Spectroscopy (IFS) methods have broadly come into astronomy allowing us to explore the spectral features of objects over a 2D field of view. With the recent advent and continuous extension of Integral Field Unit (IFU) instruments on large telescopes such as SAURON on the William Herschel Telescope (WHT), MUSE, SINFONI, VMOS and KMOS on the ESO Very Large Telescope (VLT), MaNGA on the SDSS and SAMI on the Anglo-Australian Telescope (AAT) one might imagine that kinematic decomposition is adequate and more robust compared with the photometric decomposition. Although the 3D perspective of galaxies from IFS gives insight into the dynamical and kinematical status of galaxies and their internal structures, the current methods come at high cost and therefore have so far been applicable to small samples of galaxies. In addition, their relatively low resolution makes the current IFS techniques inefficient for high-$z$ galaxies while a thorough study of galaxy evolution requires investigations at higher redshifts too.  

Photometric structural decomposition however enables us to fit large sample of galaxies from high redshifts down to the local Universe at relatively low cost. Current and future large ground- and space-based telescopes such as ELT\footnote{The Extremly Large Telescope}, HST\footnote{The Hubble Space Telescope}, JWST\footnote{James Webb Space Telescope}, Euclid and LSST\footnote{The Large Synoptic Survey Telescope; The Vera C. Rubin Observatory} will provide us with high resolution imaging data of millions of galaxies in different wavebands, making photometric structural decomposition to continue to be the ideal method for exploring galaxy formation and evolution.       

\subsection{Automatic classification and decomposition}

In the upcoming era of large astronomical data acquired using large telescopes such as LSST, JWST and Euclid conventional methods of galaxy morphological classifications and bulge-disk decomposition will be inefficient. Visual morphological classification of millions of galaxies with new surveys would be unrealistic and perhaps next to impossible. Even citizen science projects such as Galaxy Zoo (\citealt{Lintott08}; \citealt{Willett17}) and AstroQuest have their own issues and might not necessarily fit into the timeline of all future scientific projects. 

Recent development of machine learning and deep learning algorithms and particularly convolutional neural networks (CNN) in the last decade is likely the most suitable candidate for processing of such big astronomical data. Currently, some studies have started to develop CNN algorithms for automatic galaxy morphological classification as well as fast structural decomposition (see e.g., \citealt{Aniyan17}; \citealt{Ghosh19}; \citealt{Walmsley20}). More investment in this field is required in near future. Our results in this thesis have formed a state-of-the-art sample of morphological classification and bulge-disk decomposition that we aim to use in the future as a gold training set of CNN networks.  

\subsection{Fitting high-$z$ galaxies.}

In agreement with other studies, from our visual inspection of $z > 1$ galaxies we find that this epoch is dominated by gas rich clumpy galaxies. We wish to use programs and pipelines developed by us to fit these galaxies with some innovative models such as a disk with multiple bulges. Investigating the evolution of these models will give insights into the structural evolution of galaxies. This would enable us to test other notions of bulge formation, for example the scenario in which bulges formed through the coalescence of gas clumps at high redshifts.

\cleardoublepage
\addcontentsline{toc}{chapter}{References}
\bibliographystyle{mnras}
\bibliography{library}
\appendix
\graphicspath{{images/ChapterTwo/}}
\chapter{GAMA Morphological Classification Error}
\label{App:GAMA_morph_err}

To highlight how our morphological classifications in GAMA by different classifiers affects the stellar mass function, in this appendix, we show the SMFs outcomes from each of our three independent classifications.   

\begin{sidewaysfigure}
   \centering
\begin{subfigure}
  \centering
  \includegraphics[width=0.3\textwidth, height=0.5\textwidth]{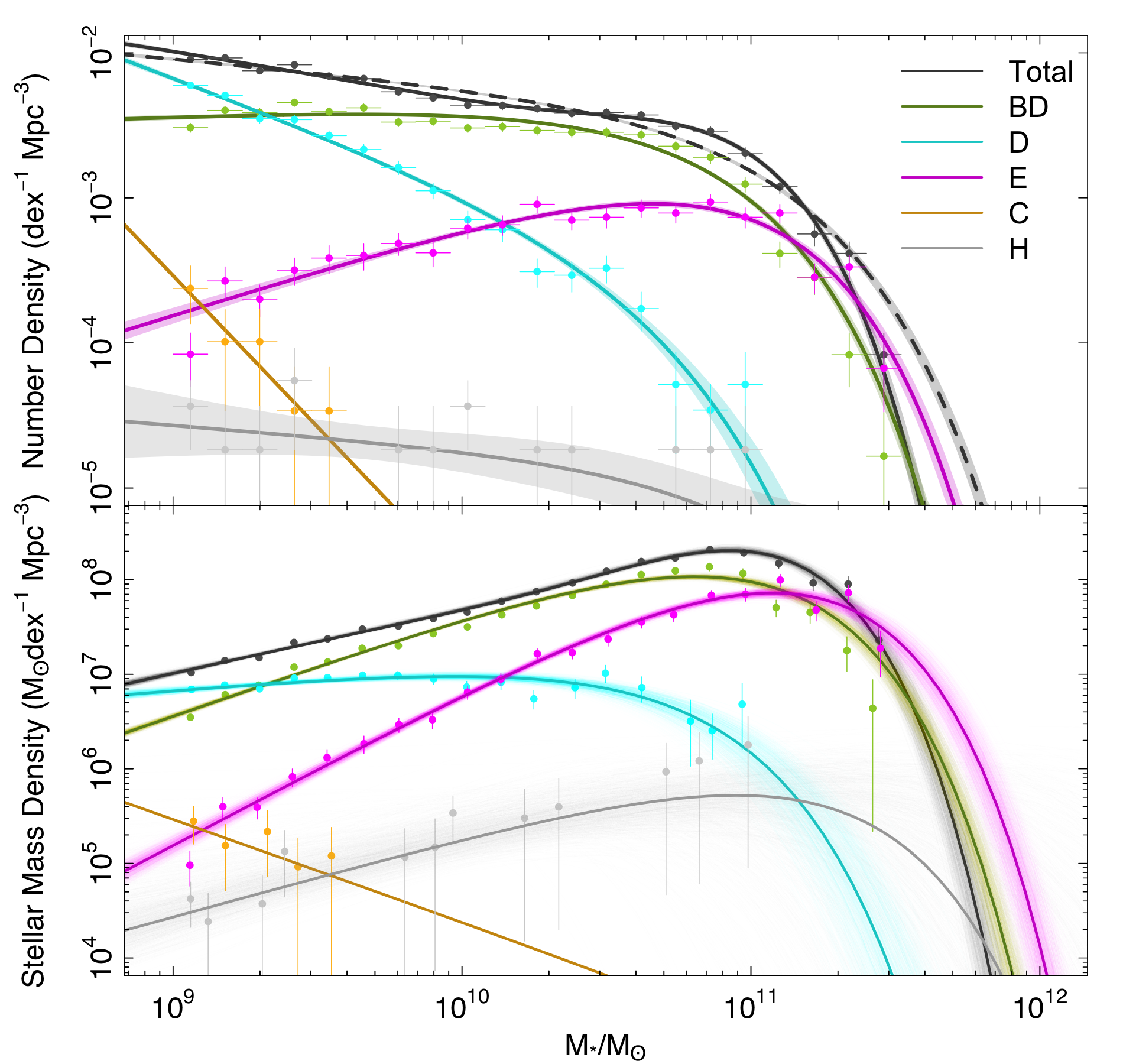}
  \label{subfig:GAMA_SimonClass}
\end{subfigure}%
\begin{subfigure}
  \centering
  \includegraphics[width=0.3\textwidth, height=0.5\textwidth]{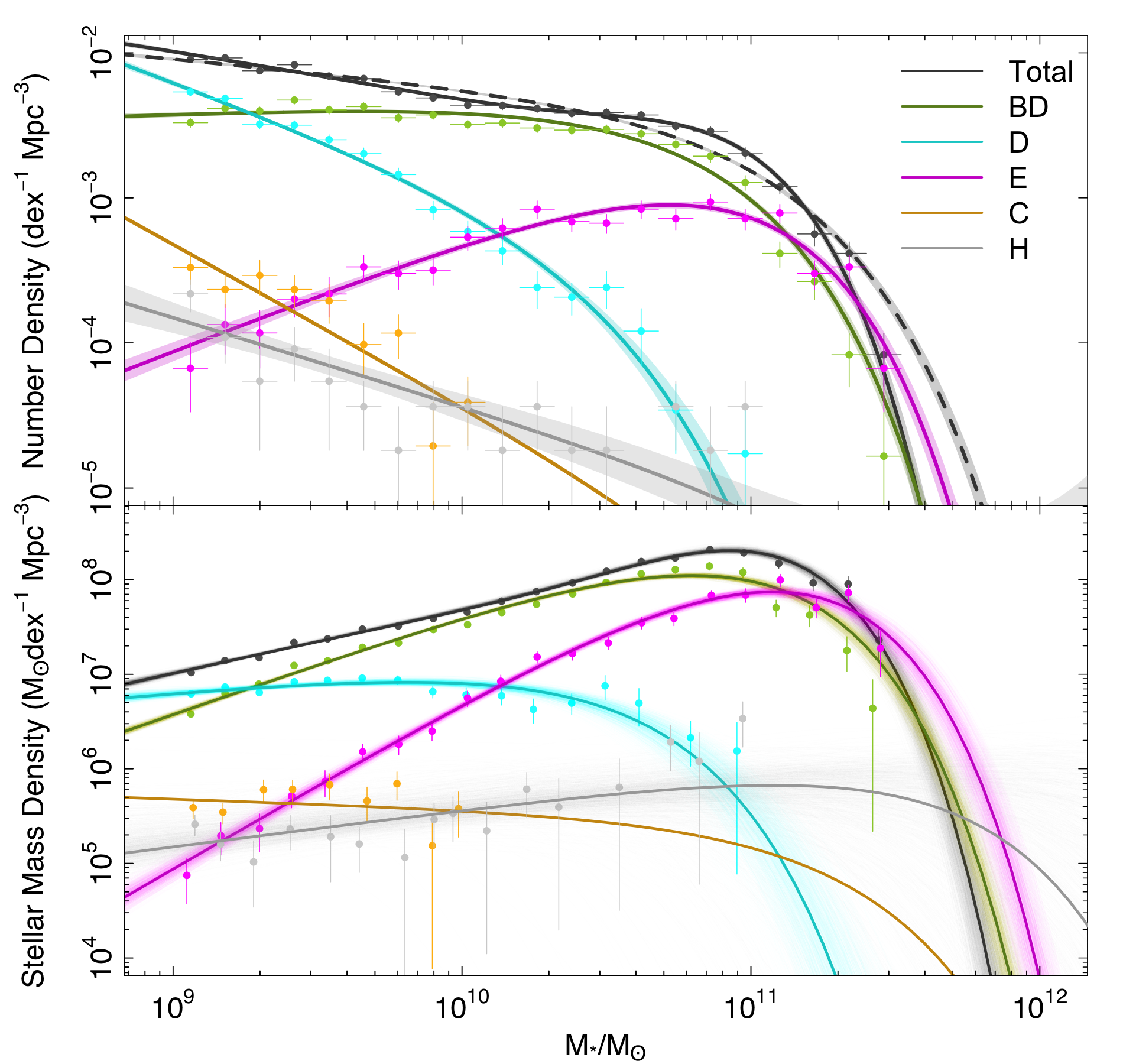}
  \label{subfig:GAMA_LukeClass}
\end{subfigure}
\begin{subfigure}
  \centering
  \includegraphics[width=0.3\textwidth, height=0.5\textwidth]{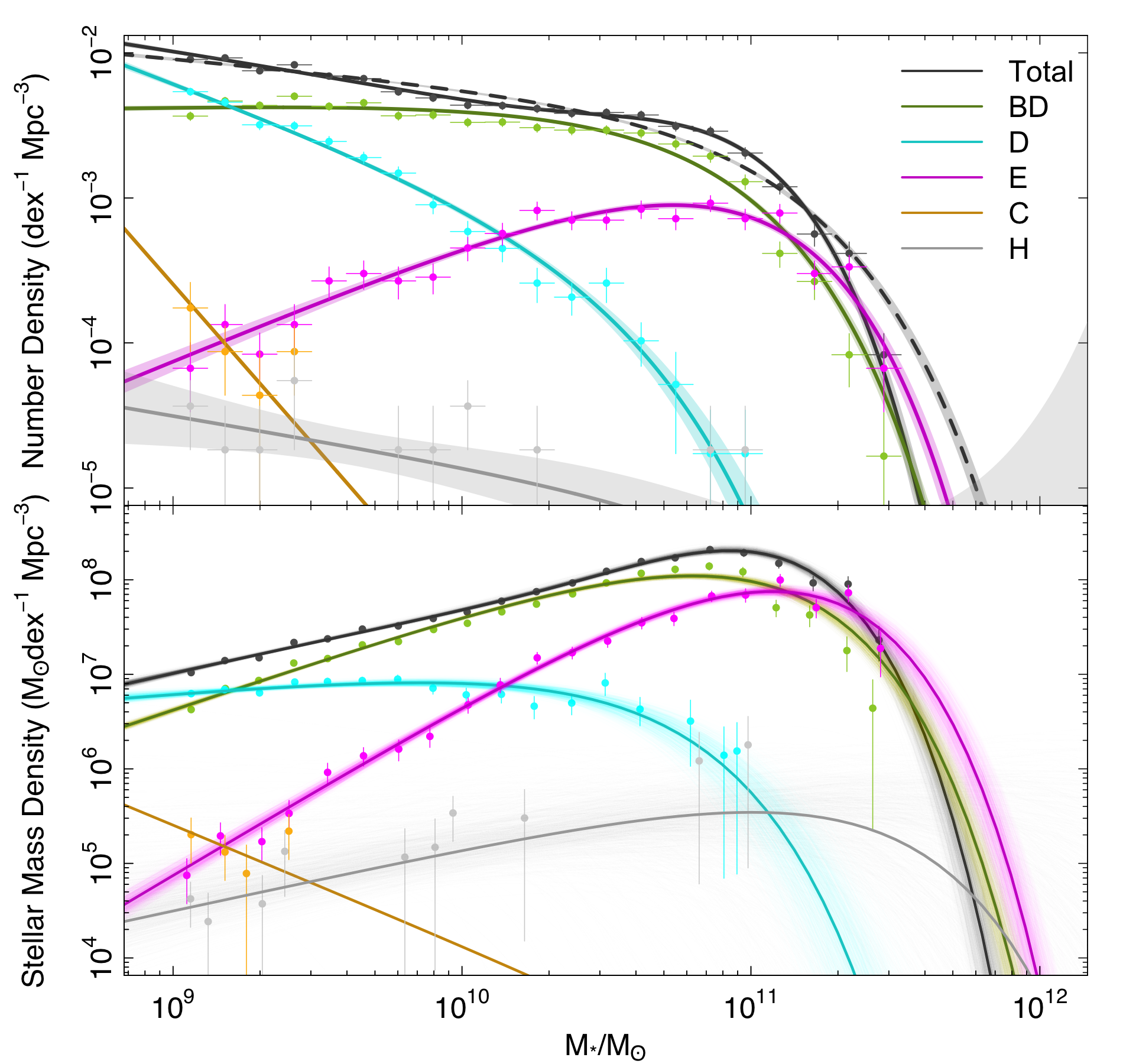}
  \label{subfig:GAMA_SabineClass}
\end{subfigure}
\caption{The morphological stellar mass function (SMF) and distribution of the stellar mass density (SMD) as obtained from three of our independent morphological classifications. Left: classification by Simon Driver, middle: classification by Luke Davies, right: classification by Sabine Bellstedt.}
\label{fig:GAMA_morph_ClassErr}
\end{sidewaysfigure}

\begin{sidewaysfigure}
   \centering
\begin{subfigure}
  \centering
  \includegraphics[width=0.3\textwidth, height=0.5\textwidth]{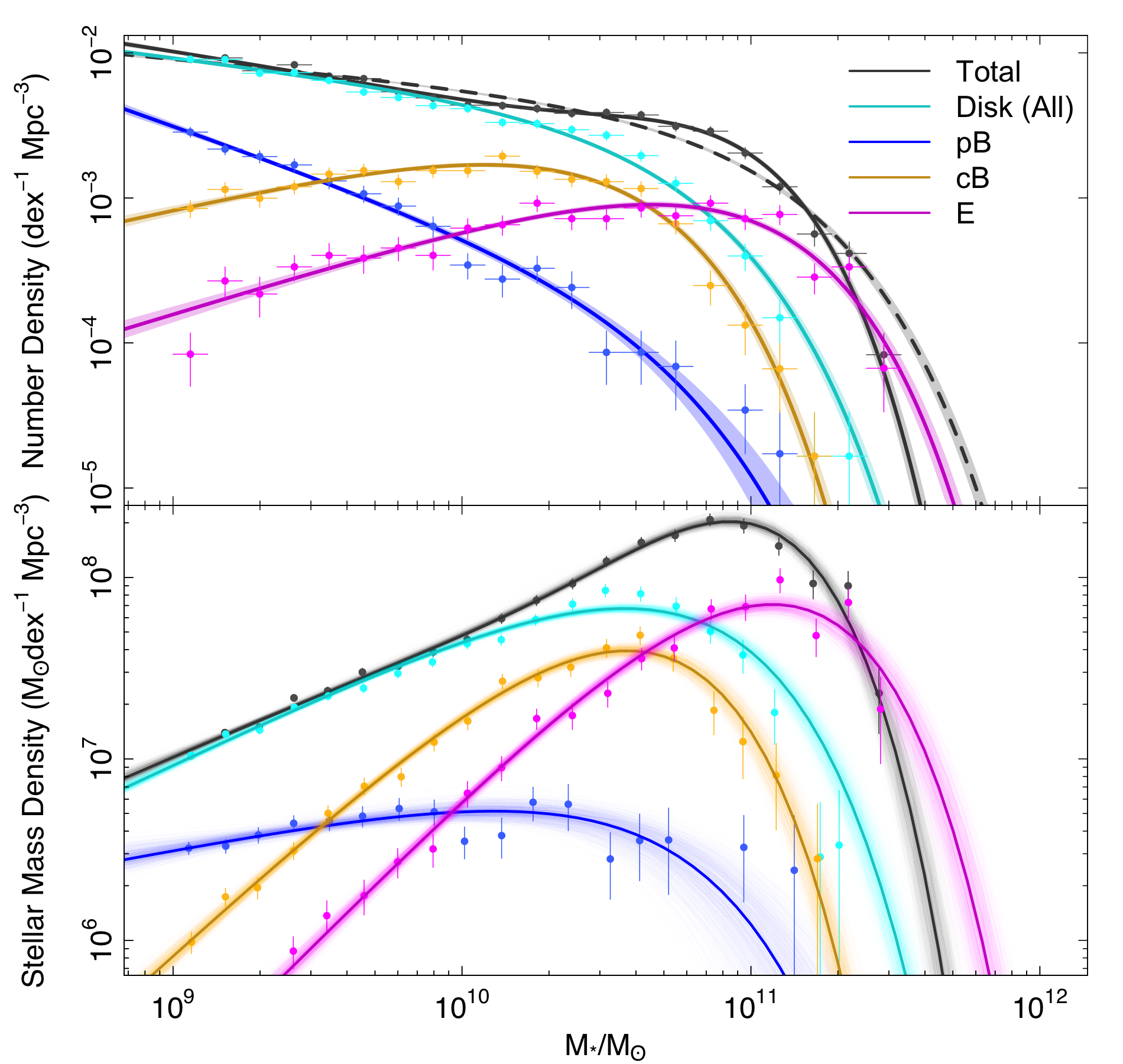}
  \label{subfig:GAMA_SimonClass}
\end{subfigure}%
\begin{subfigure}
  \centering
  \includegraphics[width=0.3\textwidth, height=0.5\textwidth]{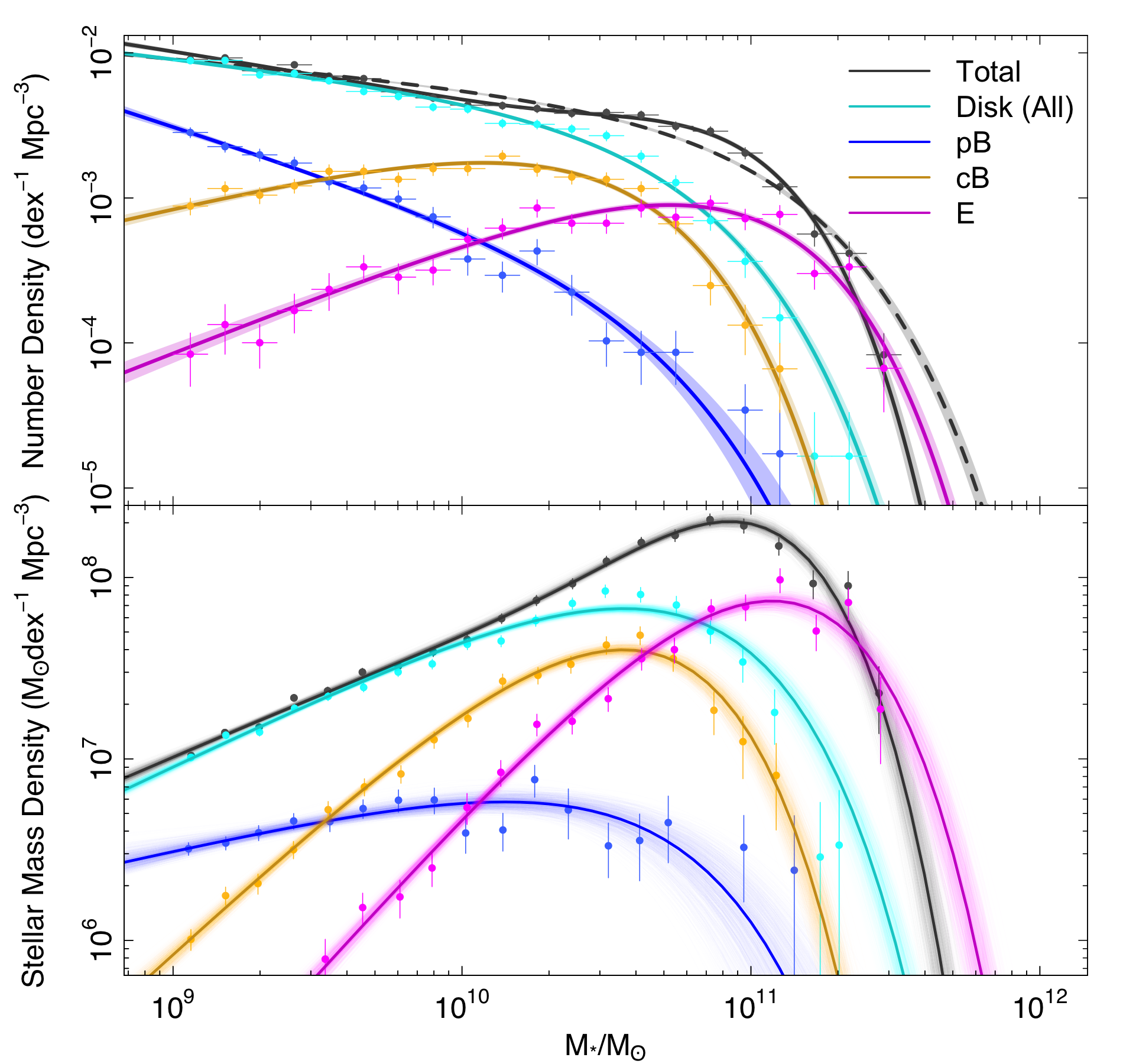}
  \label{subfig:GAMA_LukeClass}
\end{subfigure}
\begin{subfigure}
  \centering
  \includegraphics[width=0.3\textwidth, height=0.5\textwidth]{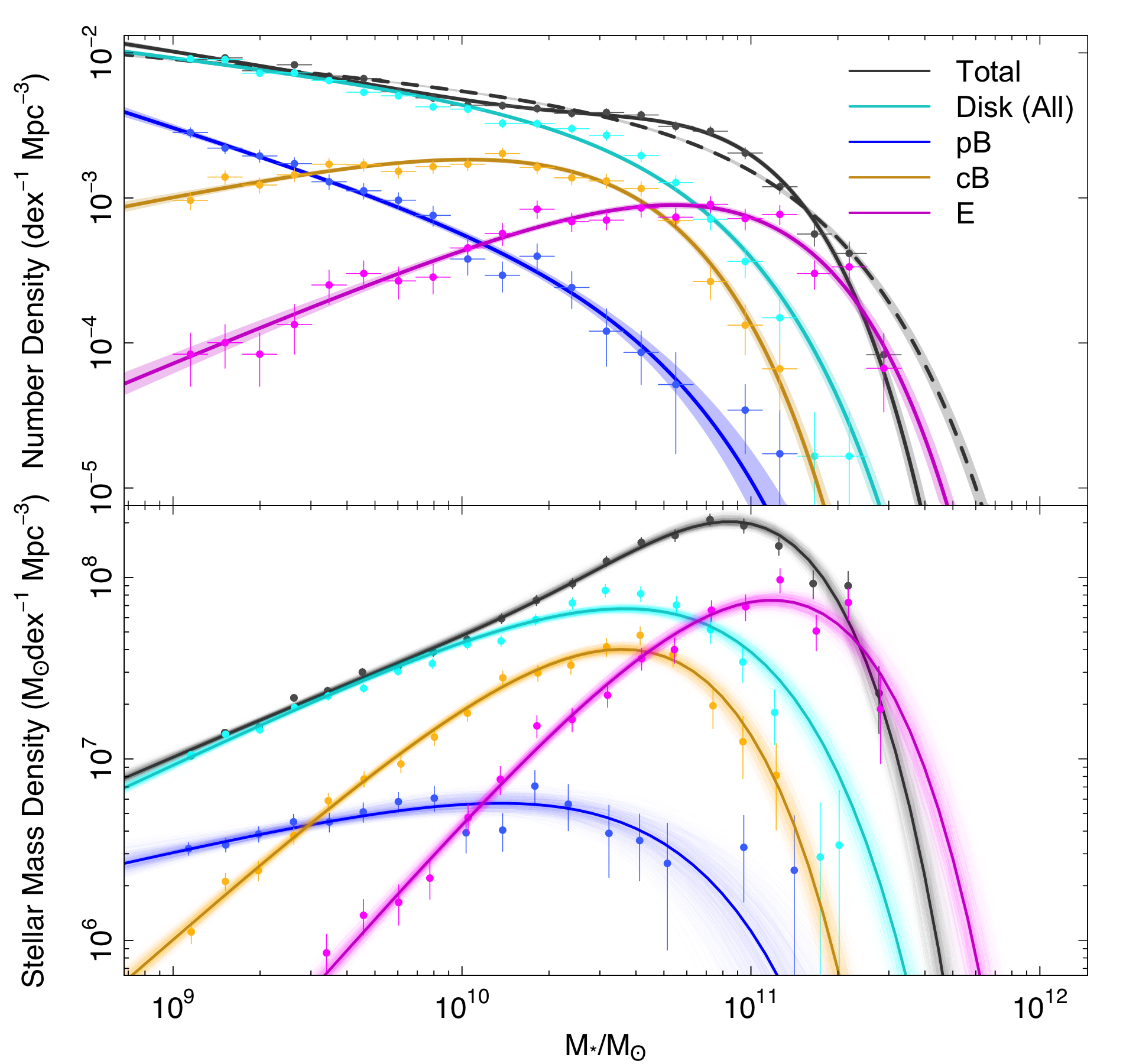}
  \label{subfig:GAMA_SabineClass}
\end{subfigure}
\caption{The structural stellar mass function (SMF) and distribution of the stellar mass density (SMD) as obtained from three of our independent morphological classifications. Left: classification by Simon Driver, middle: classification by Luke Davies, right: classification by Sabine Bellstedt.}
\label{fig:GAMA_str_ClassErr}
\end{sidewaysfigure}

\graphicspath{{images/ChapterTwo/}}
\chapter{GAMA Cosmic Variance Error}
\label{App:GAMA_CV}

As mentioned in Section \ref{sec:GAMA_err_budget}, we calculate the cosmic variance in the GAMA field by calculating the SMFs and SDMs in each of the three fields (G09, G12, G15), separately. Below we show the SMFs in each of the above fields.

\begin{sidewaysfigure}
   \centering
\begin{subfigure}
  \centering
  \includegraphics[width=0.3\textwidth, height=0.5\textwidth]{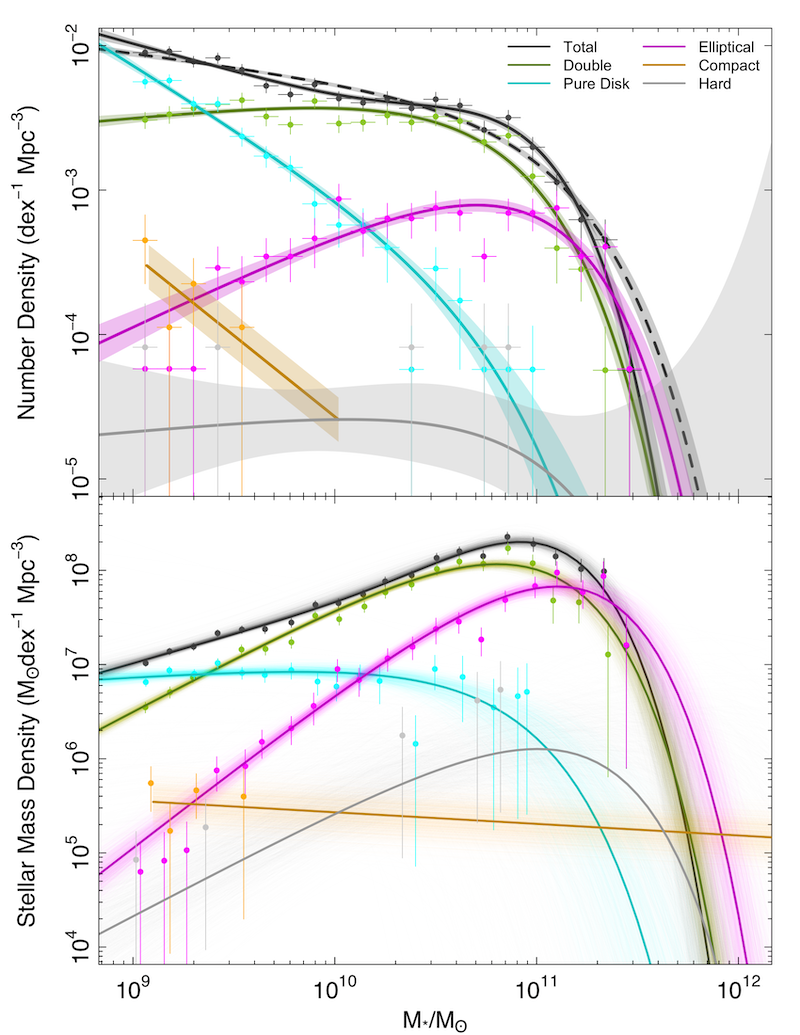}
  \label{subfig:GAMA_morph_G09}
\end{subfigure}%
\begin{subfigure}
  \centering
  \includegraphics[width=0.3\textwidth, height=0.5\textwidth]{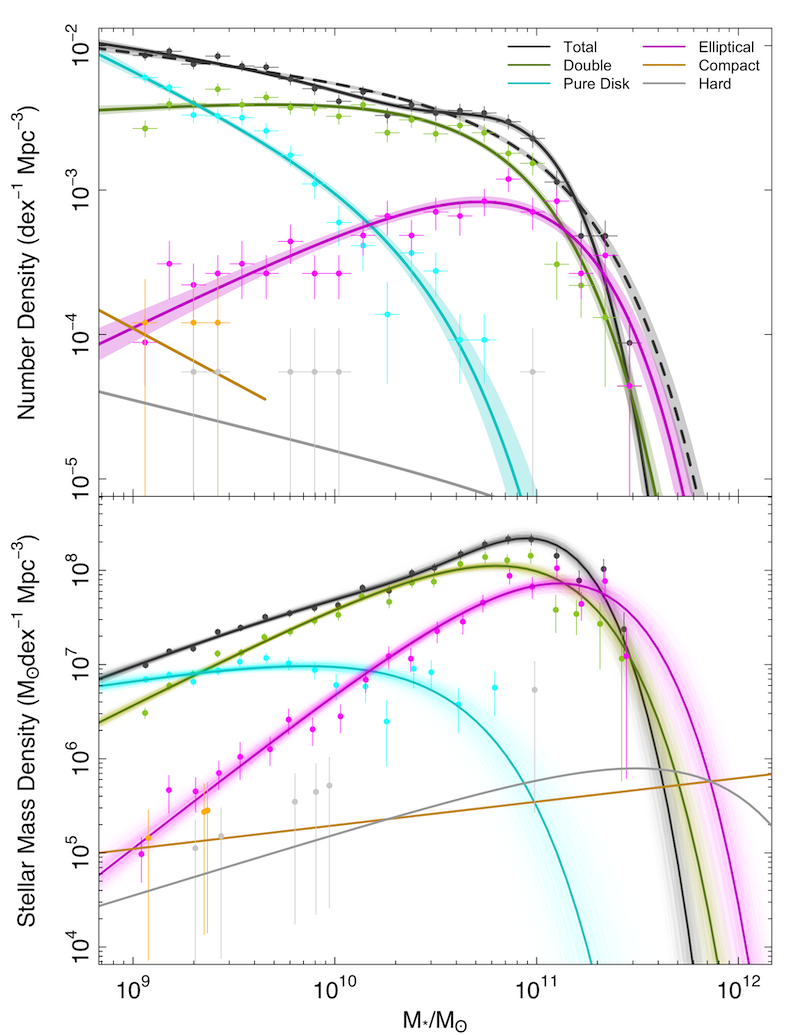}
  \label{subfig:GAMA_morph_G12}
\end{subfigure}
\begin{subfigure}
  \centering
  \includegraphics[width=0.3\textwidth, height=0.5\textwidth]{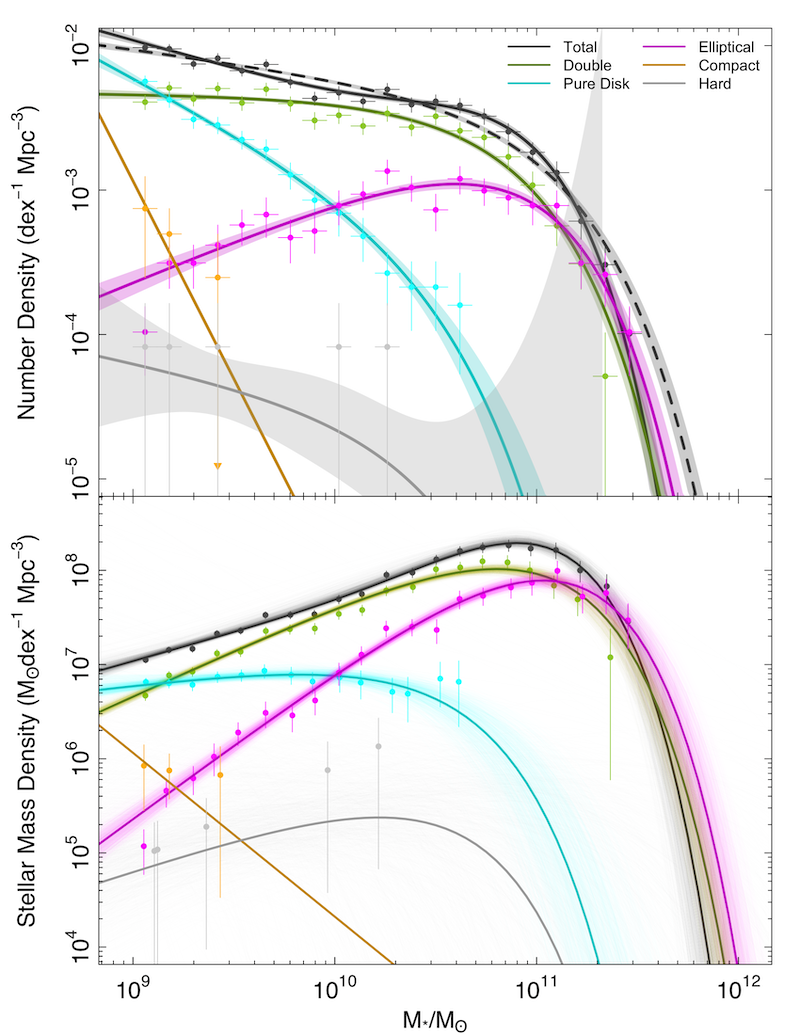}
  \label{subfig:GAMA_morph_G15}
\end{subfigure}
\caption{The morphological SMFs and SMDs measured in three GAMA regions that we study in this work; G09 (left), G12 (middle) and G15 (right).}
\label{subfig:GAMA_morph_3reg}
\end{sidewaysfigure}

\begin{sidewaysfigure}
   \centering
\begin{subfigure}
  \centering
  \includegraphics[width=0.3\textwidth, height=0.5\textwidth]{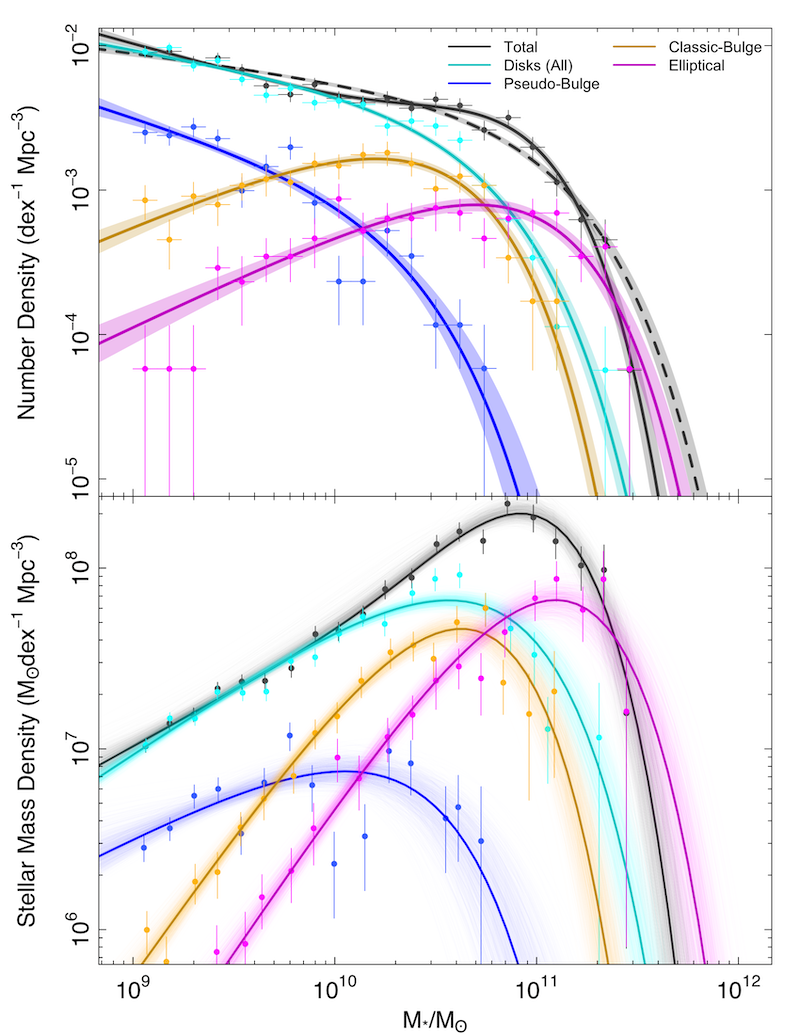}
  \label{subfig:GAMA_str_G09}
\end{subfigure}%
\begin{subfigure}
  \centering
  \includegraphics[width=0.3\textwidth, height=0.5\textwidth]{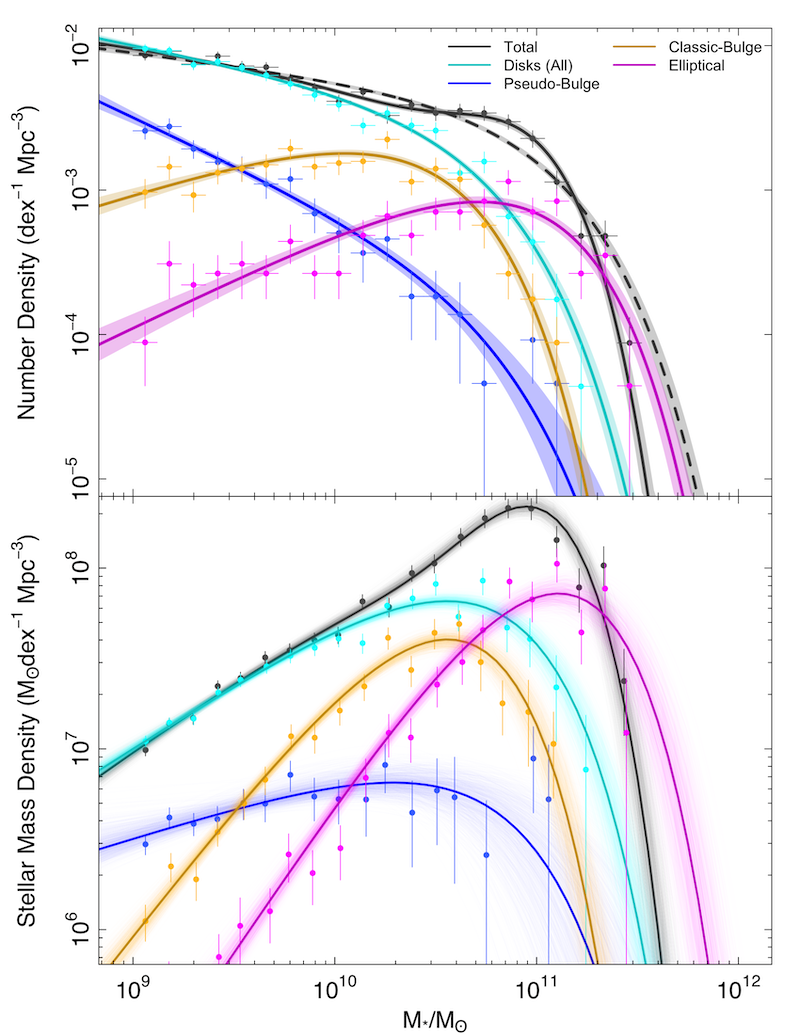}
  \label{subfig:GAMA_str_G12}
\end{subfigure}
\begin{subfigure}
  \centering
  \includegraphics[width=0.3\textwidth, height=0.5\textwidth]{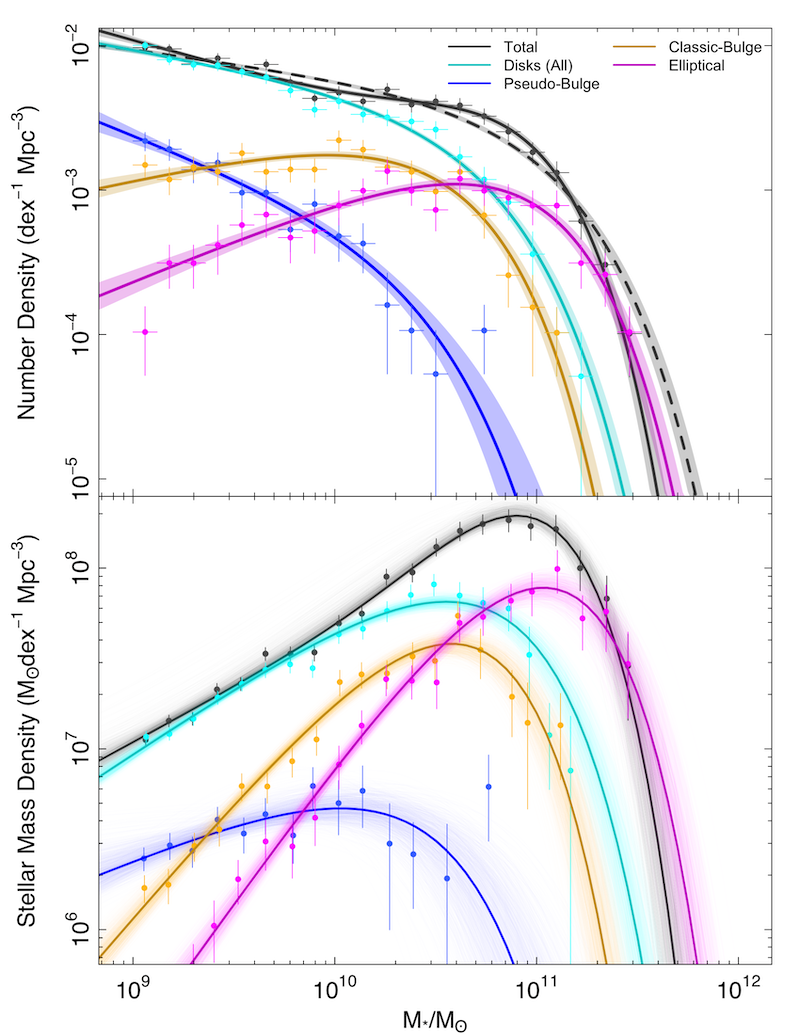}
  \label{subfig:GAMA_str_G15}
\end{subfigure}
\caption{The structural SMFs and SMDs measured in three GAMA regions that we study in this work; G09 (left), G12 (middle) and G15 (right).}
\label{subfig:GAMA_str_3reg}
\end{sidewaysfigure}
\graphicspath{{images/ChapterThree/}}
\chapter{Random samples of the morphological types}
\label{sec:App3_1}

Here we show 49 random galaxies in each of our morphological categories (BD,D,E,H,C). We show our stamps in both HST/F814W filter and Subaru $gri$.  

\begin{figure*}
\centering
\begin{subfigure}
  \centering
  \includegraphics[width=0.8\textwidth]{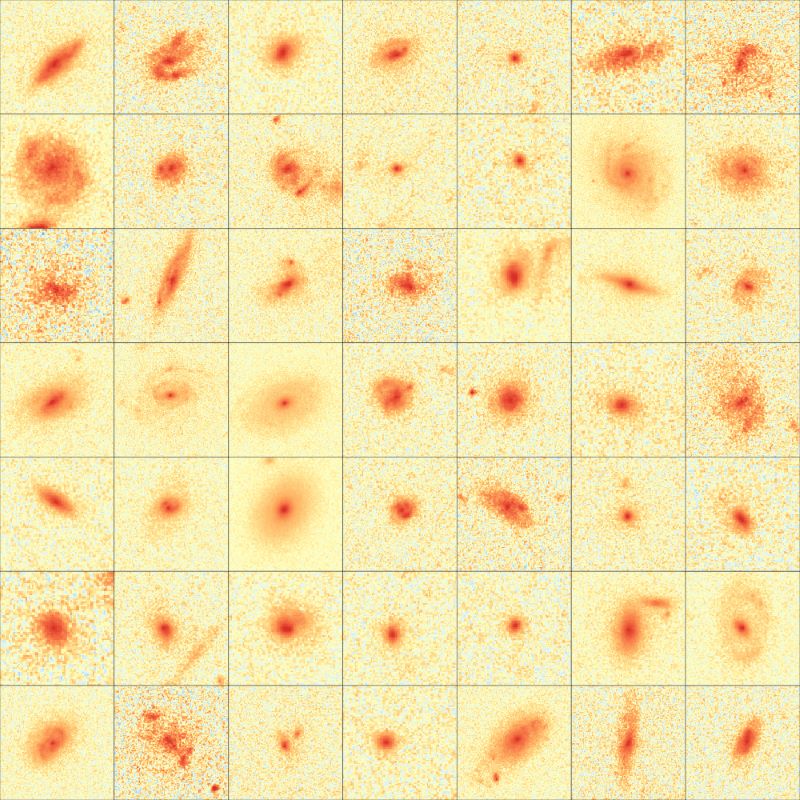}
  \label{subfig:DoubleStamps_F814W}
\end{subfigure}

\begin{subfigure}
  \centering
  \includegraphics[width=0.8\textwidth]{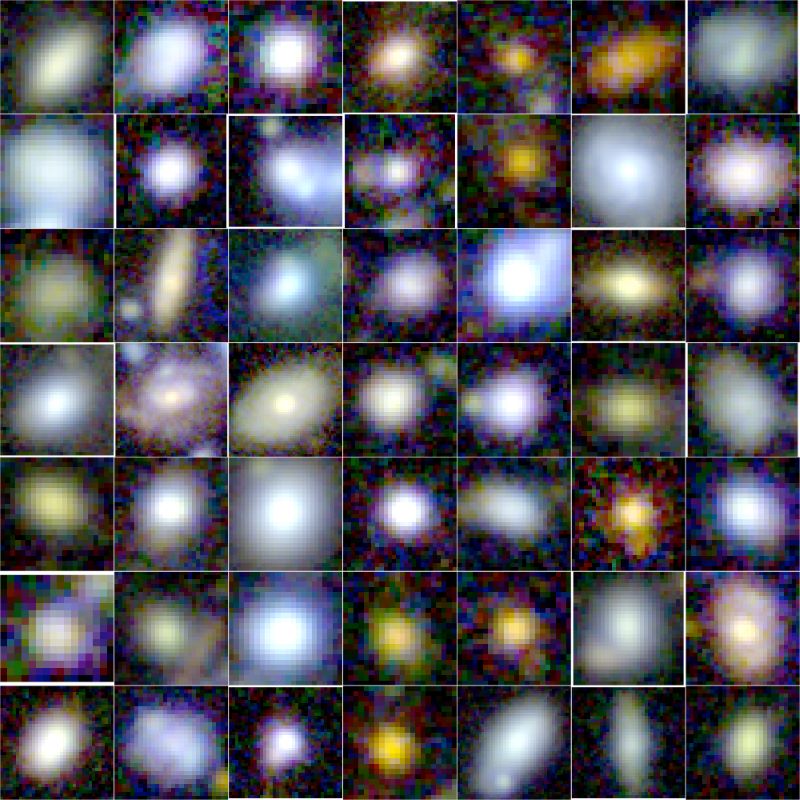}
  \label{subfig:DoubleStamps_RGB}
\end{subfigure}
\caption{Random sample of double component systems (BD). Top panels: ACS/F814W image. Bottom panels: SUBARU $gri$ combined image.}
\label{fig:contact_sheets_double}
\end{figure*}

\begin{figure*}
\centering
\begin{subfigure}
  \centering
  \includegraphics[width=0.8\textwidth]{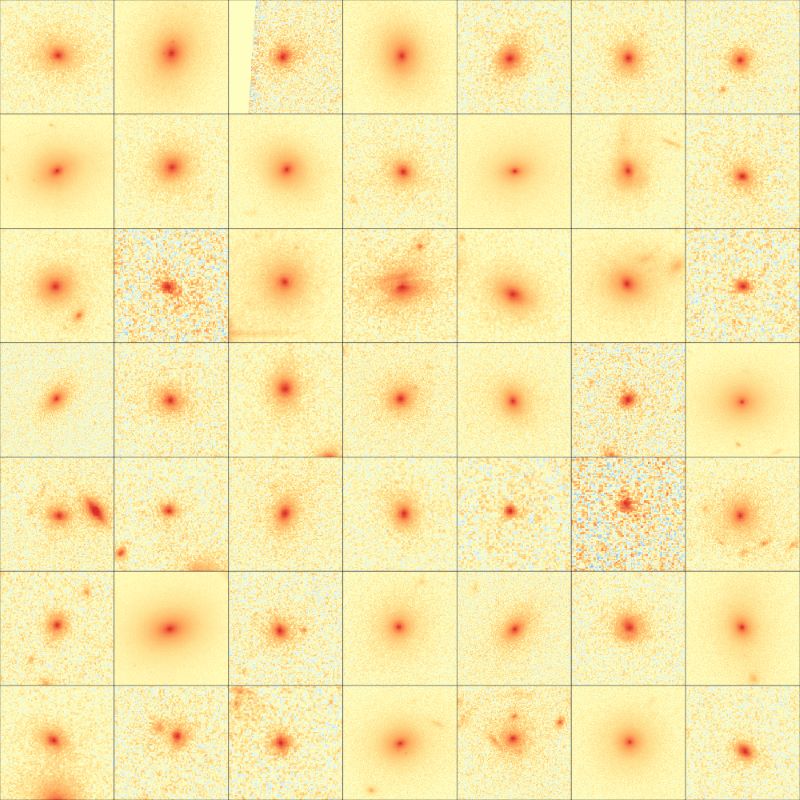}
  \label{subfig:EllipStamps_F814W}
\end{subfigure}

\begin{subfigure}
  \centering
  \includegraphics[width=0.8\textwidth]{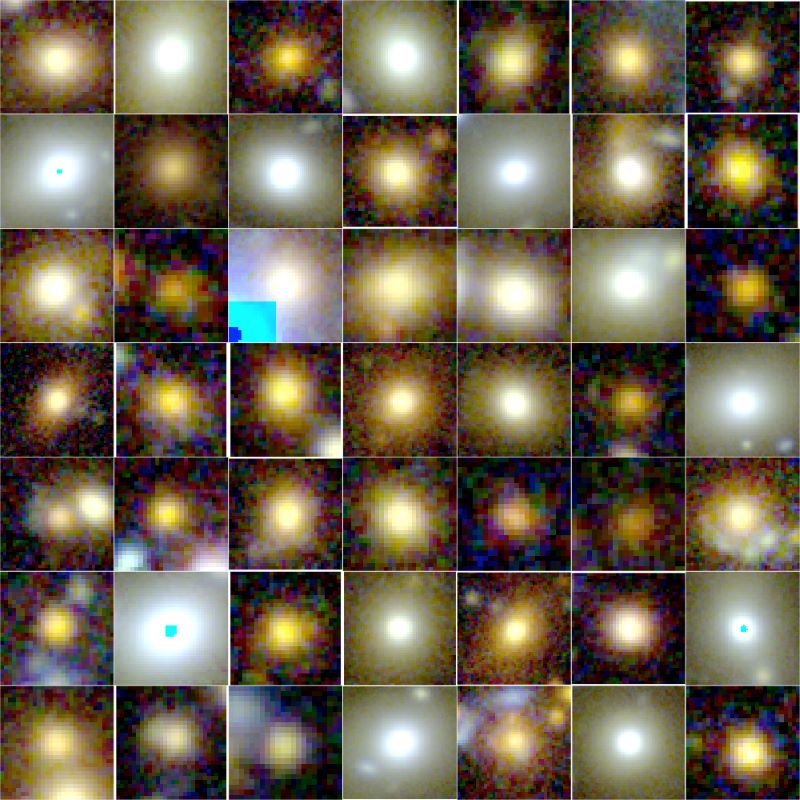}
  \label{subfig:EllipStamps_RGB}
\end{subfigure}
\caption{Random sample of elliptical systems (E). Top panels: ACS/F814W image. Bottom panels: SUBARU $gri$ combined image.}
\label{fig:contact_sheets_ellip}
\end{figure*}

\begin{figure*}
\centering
\begin{subfigure}
  \centering
  \includegraphics[width=0.8\textwidth]{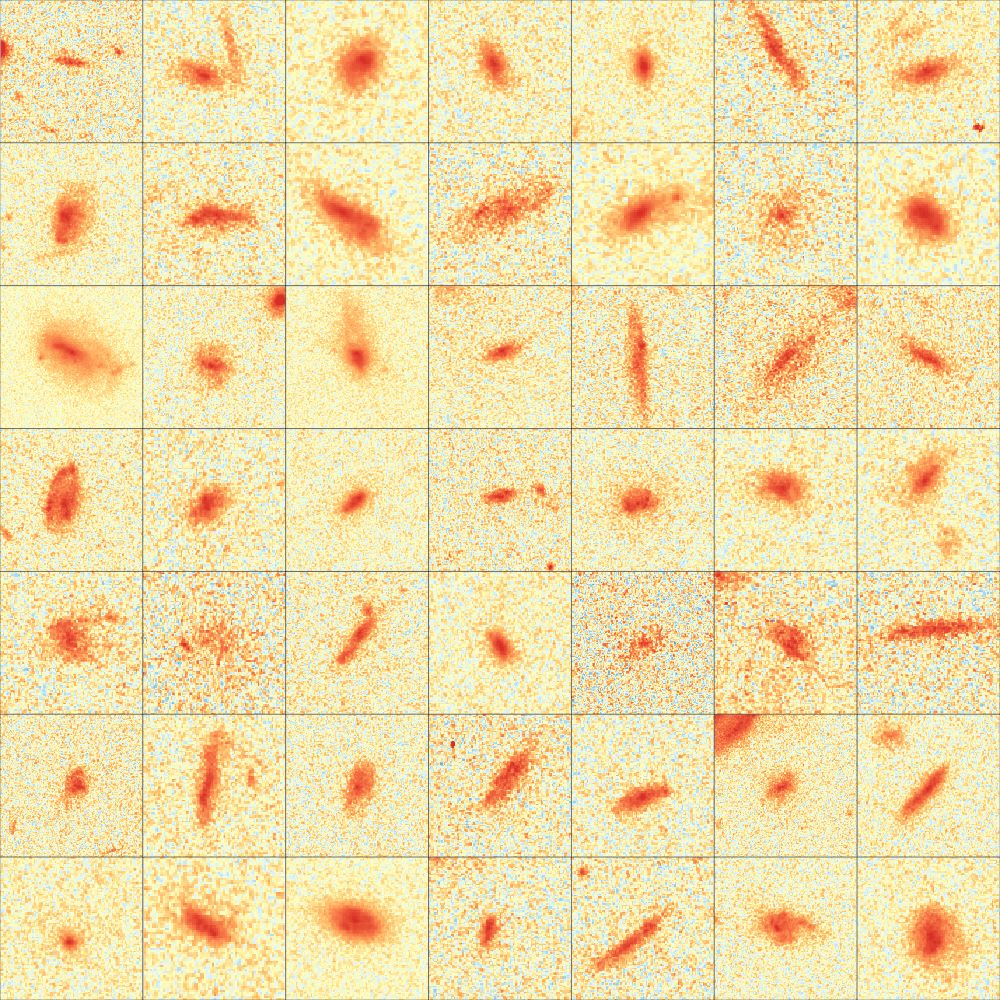}
  \label{subfig:DiskStamps_F814W}
\end{subfigure}

\begin{subfigure}
  \centering
  \includegraphics[width=0.8\textwidth]{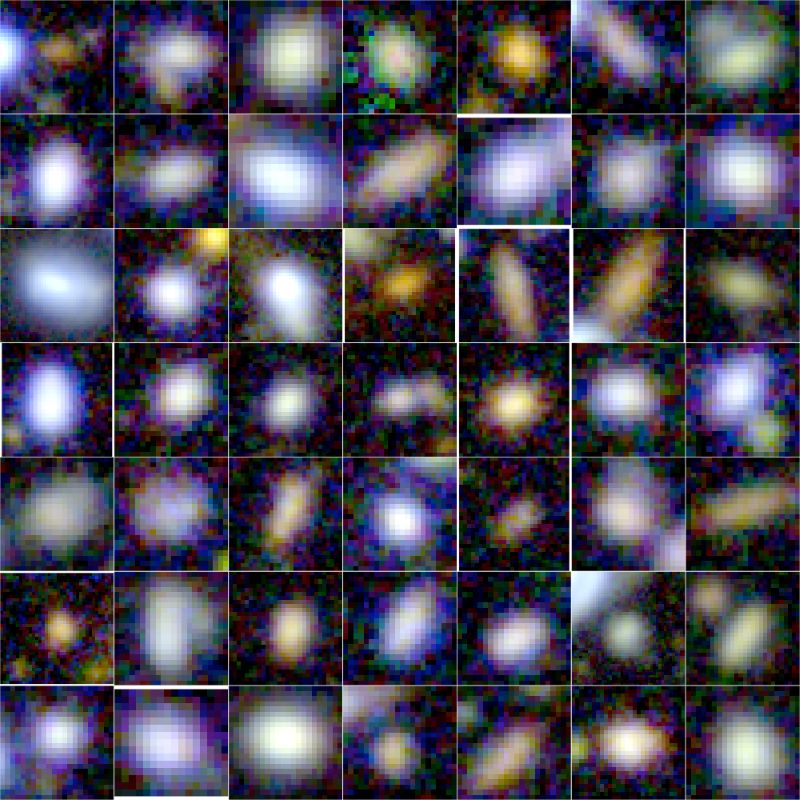}
  \label{subfig:DiskStamps_RGB}
\end{subfigure}
\caption{Random sample of pure disk systems (D). Top panels: ACS/F814W image. Bottom panels: SUBARU $gri$ combined image.}
\label{fig:contact_sheets_disk}
\end{figure*}

\begin{figure*}
\centering
\begin{subfigure}
  \centering
  \includegraphics[width=0.8\textwidth]{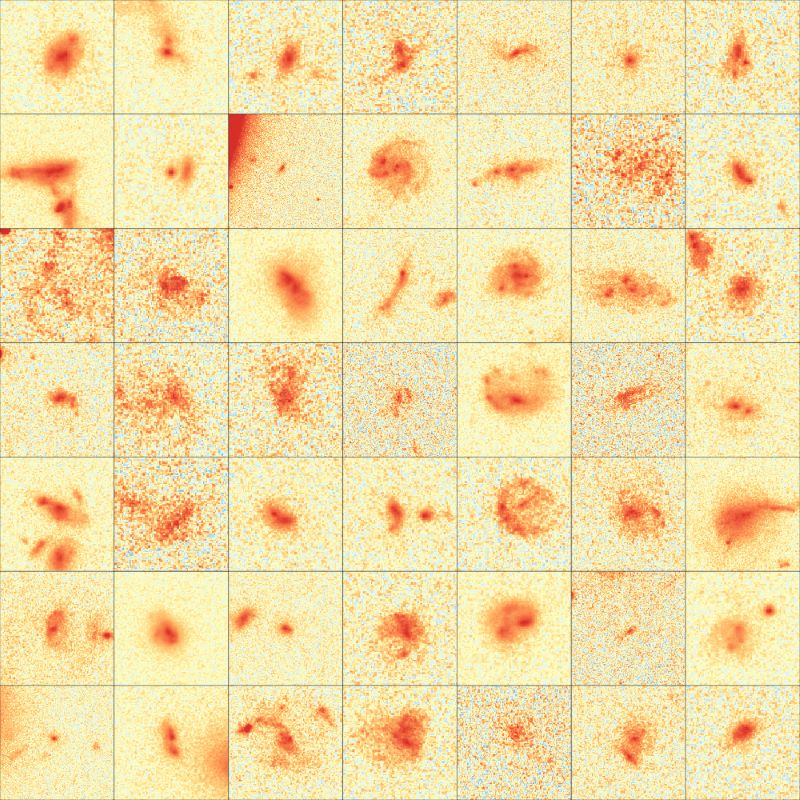}
  \label{subfig:HardStamps_F814W}
\end{subfigure}

\begin{subfigure}
  \centering
  \includegraphics[width=0.8\textwidth]{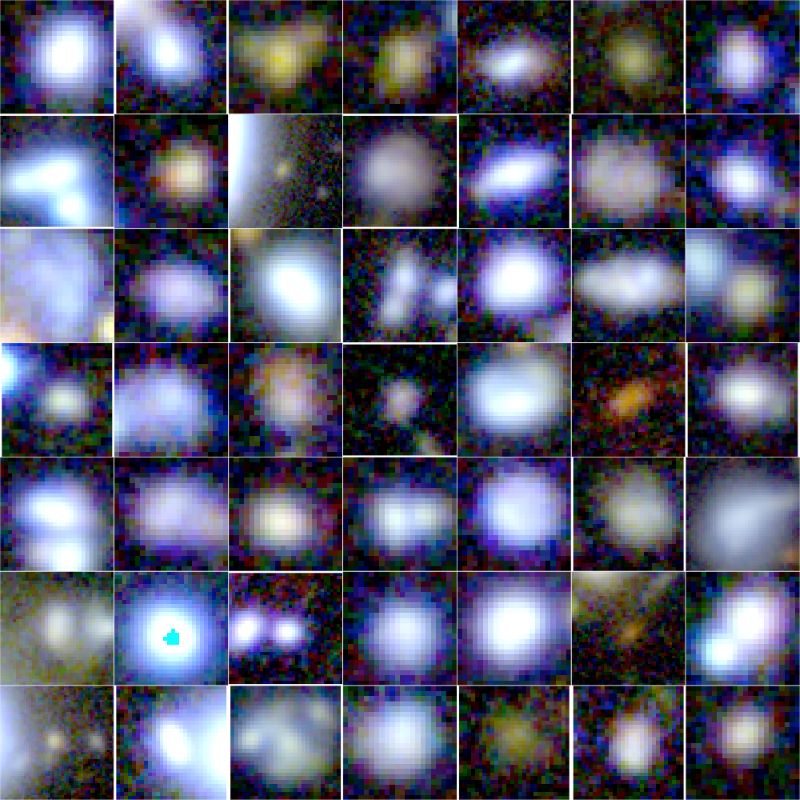}
  \label{subfig:HardStamps_RGB}
\end{subfigure}
\caption{Random sample of complex systems (\textit{hard}, H). Top panels: ACS/F814W image. Bottom panels: SUBARU $gri$ combined image.}
\label{fig:contact_sheets_hard}
\end{figure*}

\begin{figure*}
\centering
\begin{subfigure}
  \centering
  \includegraphics[width=0.8\textwidth]{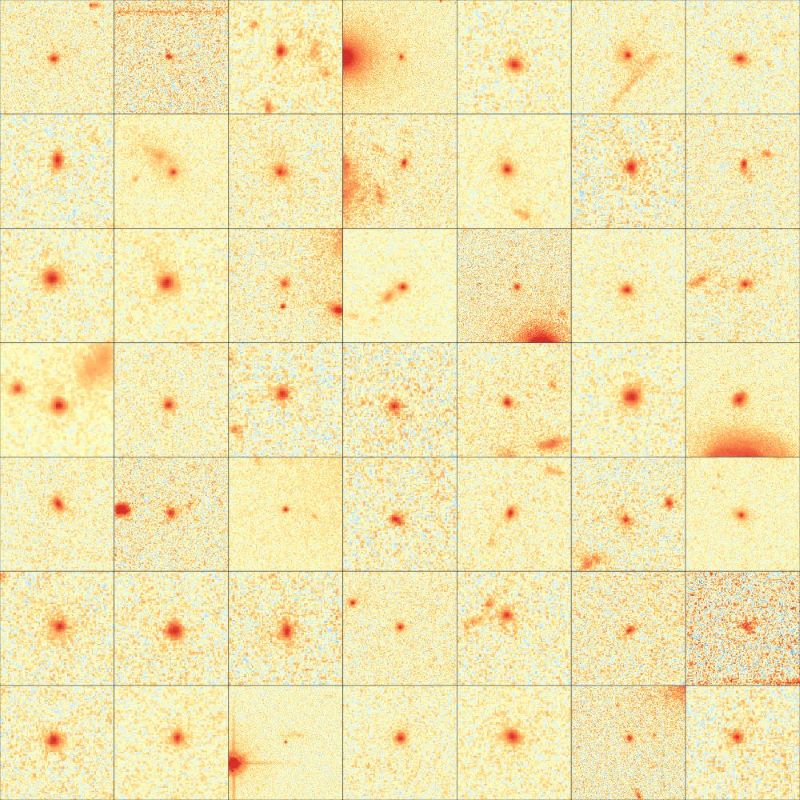}
  \label{subfig:CompStamps_F814W}
\end{subfigure}

\begin{subfigure}
  \centering
  \includegraphics[width=0.8\textwidth]{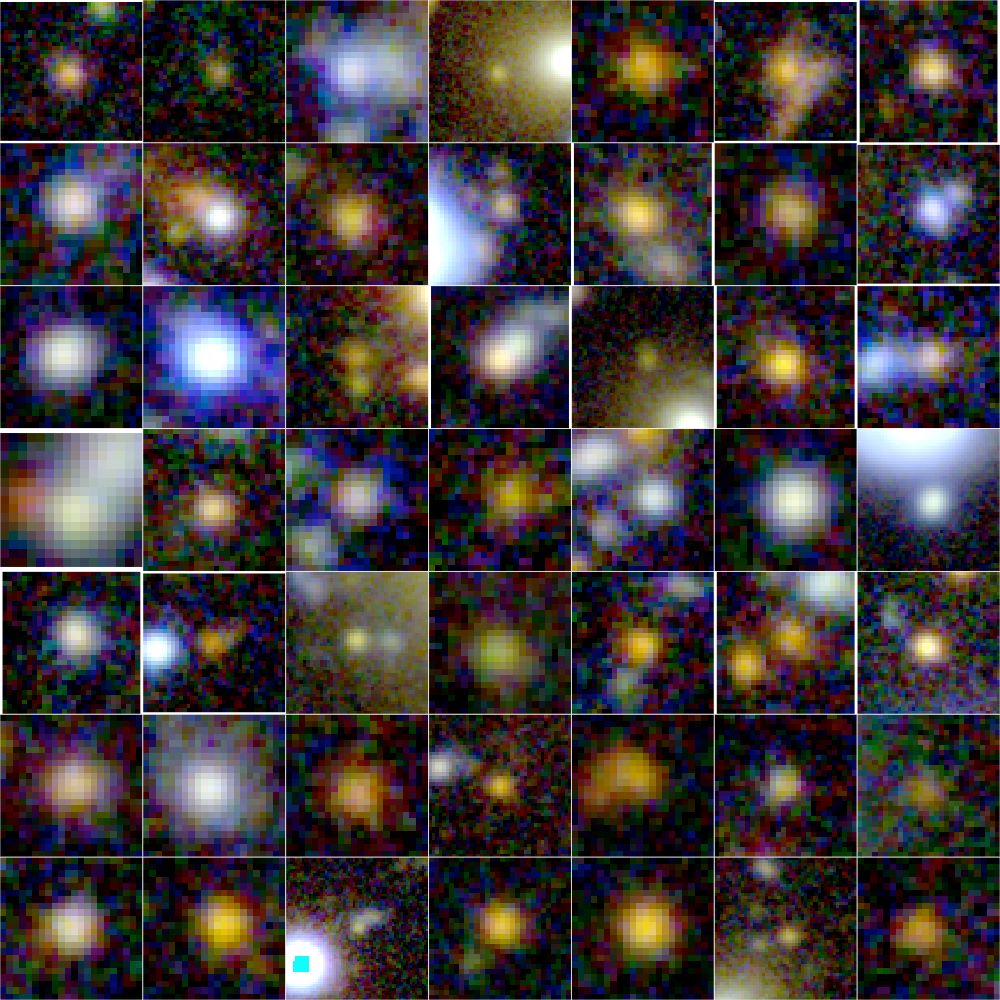}
  \label{subfig:CompStamps_RGB}
\end{subfigure}
\caption{Random sample of low angular sized systems (\textit{compact}, C). Top panels: ACS/F814W image. Bottom panels: SUBARU $gri$ combined image.}
\label{fig:contact_sheets_comp}
\end{figure*}

\graphicspath{{images/ChapterThree/}}
\chapter{The non-LSS-corrected evolution of the SMD}

\label{sec:SMD_evol_noLSS}

\begin{figure*}
\centering
  \centering
  \includegraphics[width=\textwidth]{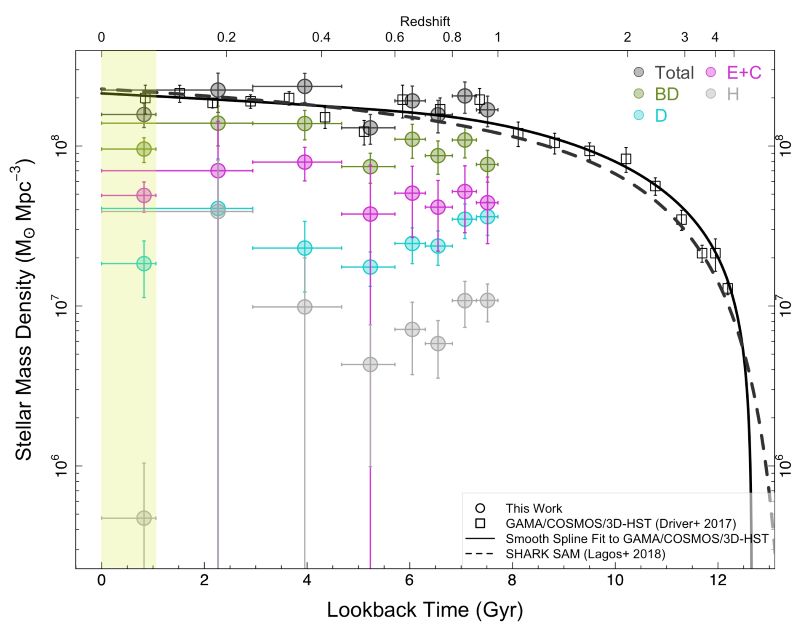}
  \caption{The evolution of the stellar mass density (SMD) of total and morphological types before applying the LSS corrections.}
  \label{fig:SMD_evol_noLSS}
\end{figure*}

Figure \ref{fig:SMD_evol_noLSS} shows the evolution of the integrated stellar mass density, $\rho_*$ before we apply our large scale structure corrections (reported in Table \ref{tab:rho} and shown in Figure \ref{fig:LSS_cor}). This is to further confirm that the corrections do not derive the overall trends that we find in Figure \ref{fig:MassBuildUp_LSS} as explained in Section \ref{sec:rho}.

\graphicspath{{images/ChapterFour/}}
\chapter{Selection of stars for PSF subtraction}
\label{sec:star_sel}

\begin{figure}
	\centering     
	\includegraphics[width=\linewidth]{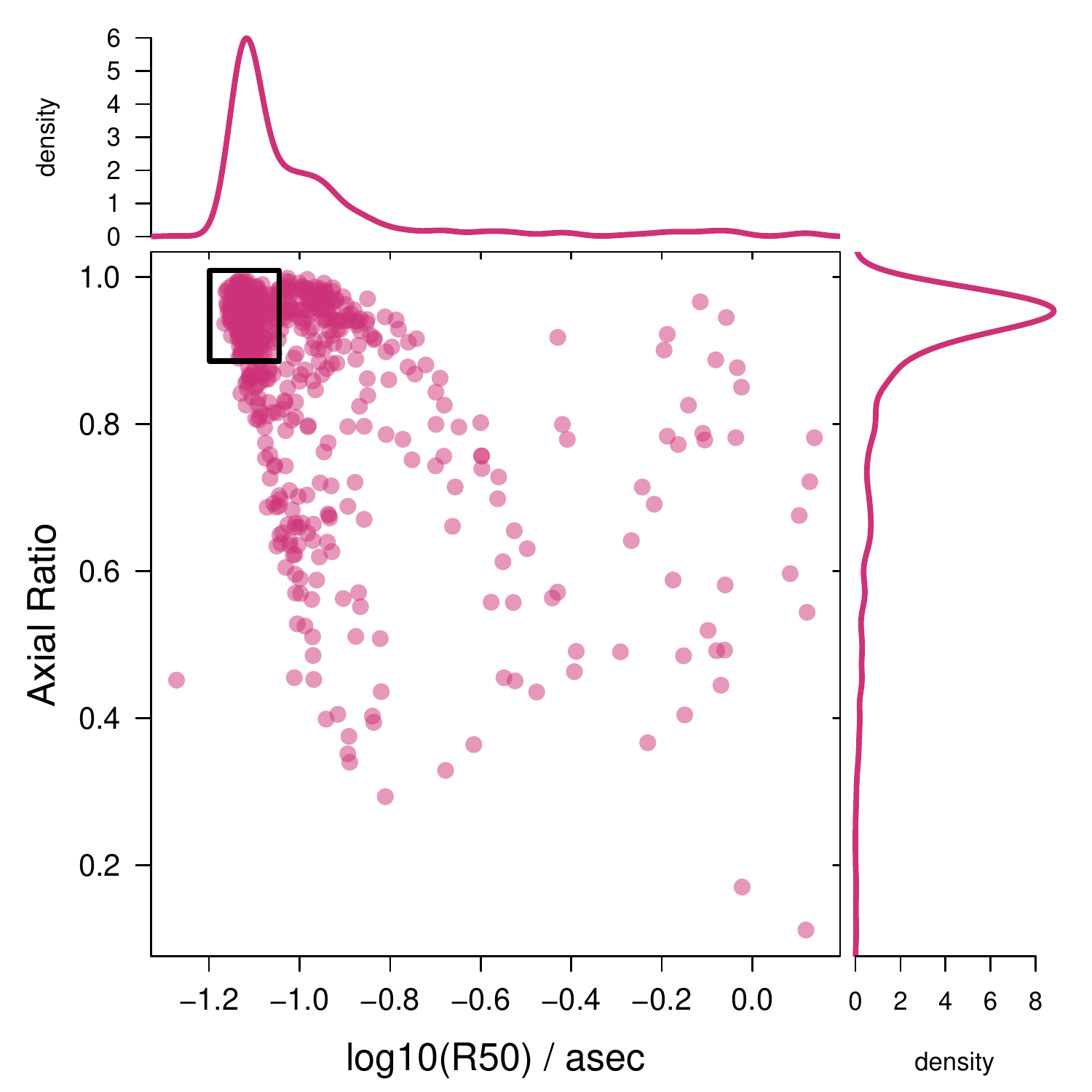}
	\caption{ The main plot shows ellipticity as a function of half light radius (R50) of $\sim700$ random selected unsaturated bright stars. Upper and right marginal histograms show the probability density of R50 and axial ratio, respectively. For star subtraction from PSF, we select 5 random stars in the black rectangle ensuring the selected stars are within the most frequent size (R50 $\sim0.07''$, peak of distribution), and most likely single stars (axial ratio $> 0.9$). 
}
	\label{fig:5star_sel}
\end{figure}

To test the accuracy of our modelled PSF, we select 5 random stars based on their R50 and axial ratio.
The main panel in Figure \ref{fig:5star_sel} shows axial ratio versus R50 for $~700$ stars identified in the mosaic frame. The small black rectangular area is where we randomly select our five stars for which we subtract the corresponding PSFs as presented in Figure \ref{fig:starsub_5star}.

\graphicspath{{images/ChapterFour/}}
\chapter{Star-PSF subtraction from a star in raw ACS frames.}
\label{sec:star_psf_raw}

\begin{figure*} 
	\centering
	\includegraphics[width=\textwidth]{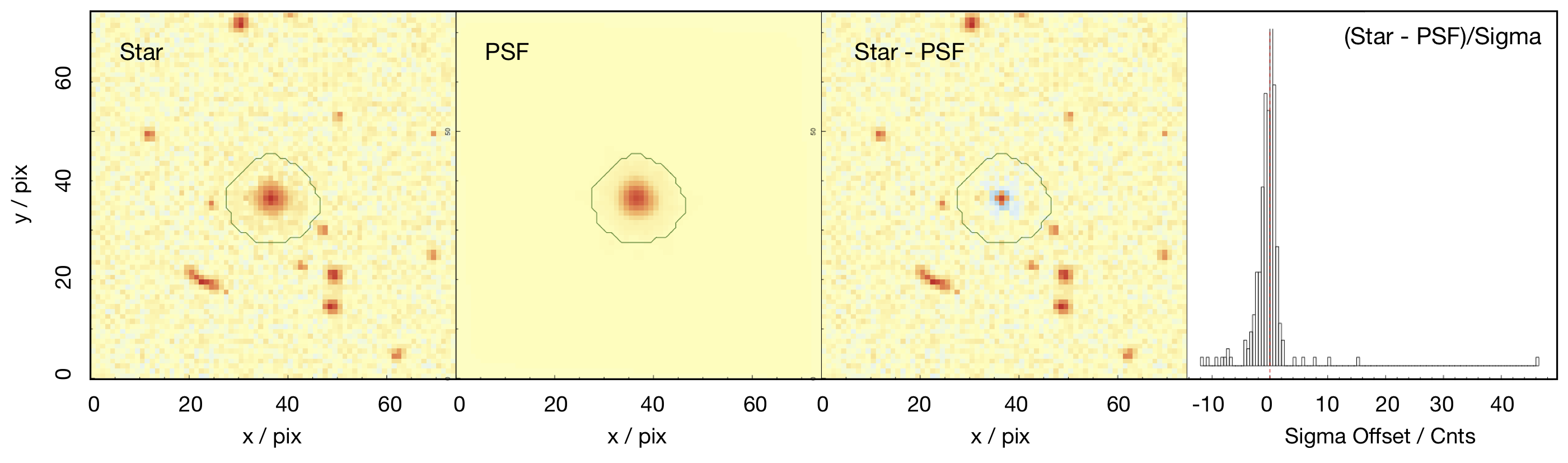}
	\caption{PSF subtraction from a star in a raw HST/ACS frame before cosmic ray rejection, stacking and sub-sampling. Columns are the same as Figure \ref{fig:starsub_5star}. Most of the sources around the star are cosmic rays.}
	\label{fig:starsub_raw}
\end{figure*}

In Section \ref{sec:PSF_test}, we described our method for stacking four PSFs generated by {\sc Tiny Tim}. Here we perform a star subtraction from stars in the HST/ACS raw frames to confirm that the accuracy of our final PSF is not influenced by our stacking procedure. Figure \ref{fig:starsub_raw} shows this process where we subtract a star in raw frame (first panel) from our PSF directly out of {\sc Tiny Tim} (second panel) implying that the residual is still present at the centre (third and fourth panels). 

\graphicspath{{images/ChapterFour/}}
\chapter{Stationary MCMC chain}
\label{sec:MCMC_chain}

The corner plot of the stationary MCMC chain of our double S\'ersic model for D103274337, showing the MCMC chain for each parameter as a scatter plot (top-left corner) alongside their contour version diametrically opposite (i.e., lower-left corner). We also present the diagonal one-dimensional marginalized distribution of the sampling chain.  

\begin{figure*} 
	\centering
	\includegraphics[width=\textwidth]{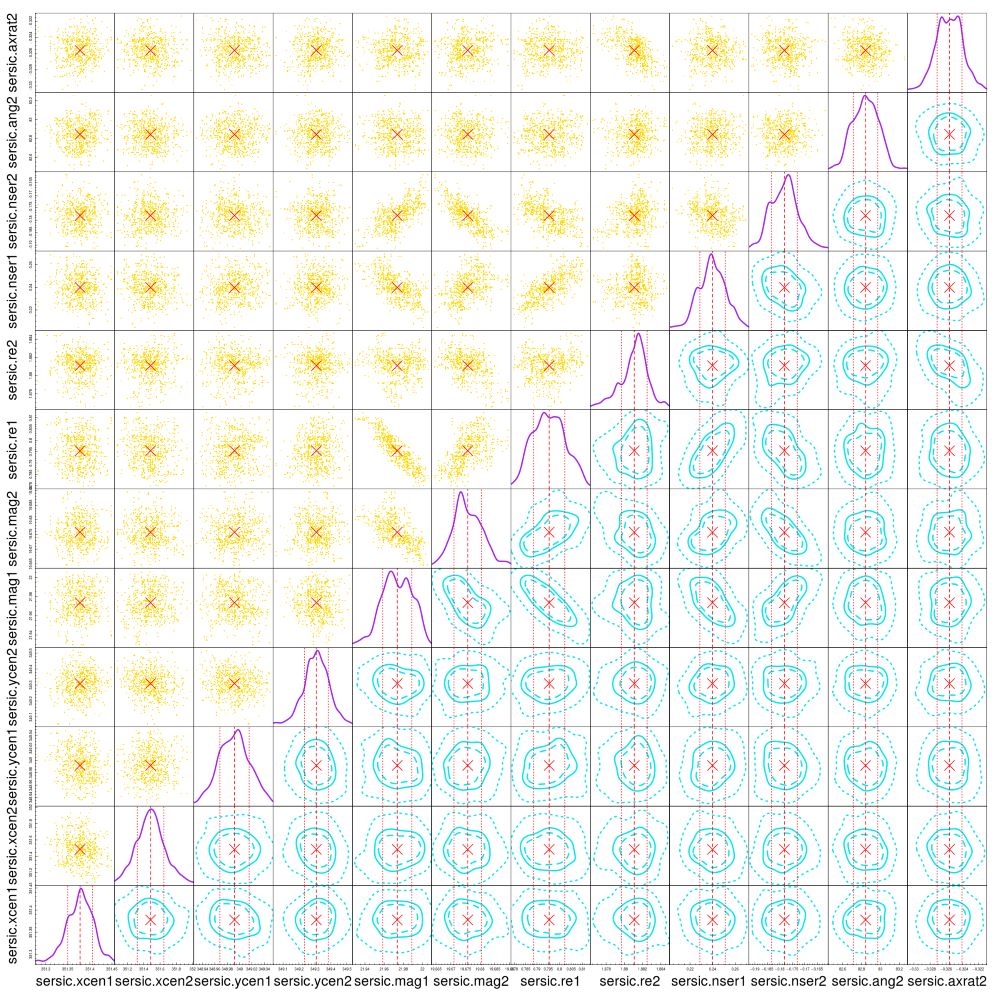}
	\caption{ The corner plot of the stationary MCMC chain of our fitting for D103274337. In this case we fit 12 parameters. The top left corner of the plot shows the scatter sample while the lower right corner shows the contour version of the sample. Dashed, solid and dotted contours contain 50, 68 and 95 per cent of the data, respectively. The diagonal density plots show the one-dimensional marginalized distribution of the sample for each parameter. }
	\label{fig:triPlot}
\end{figure*}

\graphicspath{{images/ChapterFour/}}
\chapter{Parameter convergence test on a GAMA galaxy}
\label{sec:conv_GAMA}

To further quantify the accuracy of our MCMC sampling we perform our analysis on the worst case of \cite{Lange16} for which they find the most diverged results, i.e., G32362. Similar to Section \ref{sec:MCMC} we start from a set of initial conditions and find that all fits well converge to a global solution.

\begin{figure*}
	\centering
	\includegraphics[width = \textwidth, angle = 0]{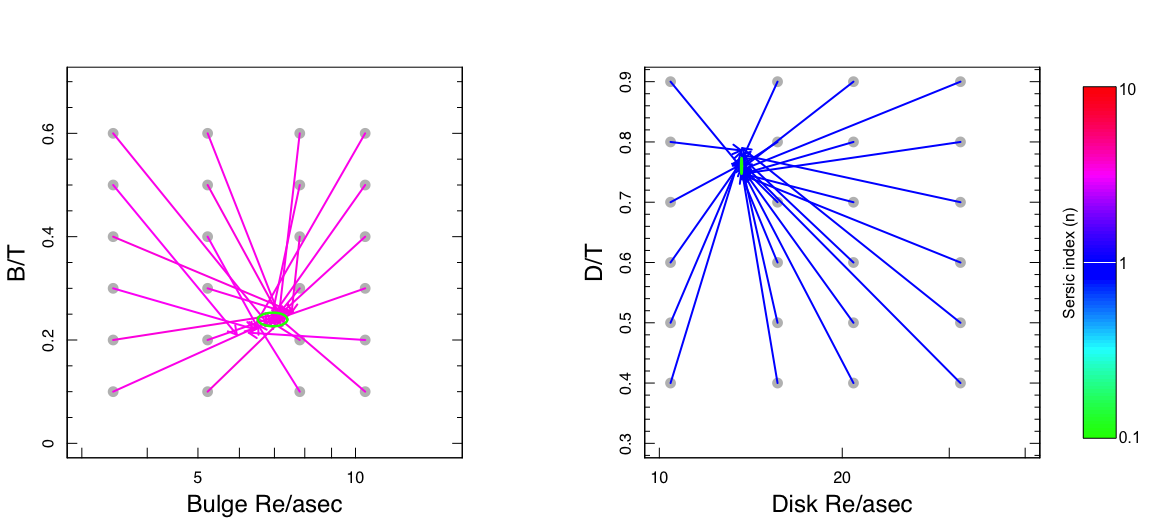}
	\caption{ An MCMC convergence test on GAMA galaxy G32362.}
	\label{fig:Conv_test_GAMA}
\end{figure*}


\graphicspath{{images/ChapterFour/}}

\chapter{The non-LSS-corrected evolution of the SMD} 
\label{sec:MassBuildUp_noLSS}

\begin{figure*}
\centering
  \centering
  \includegraphics[width=\textwidth]{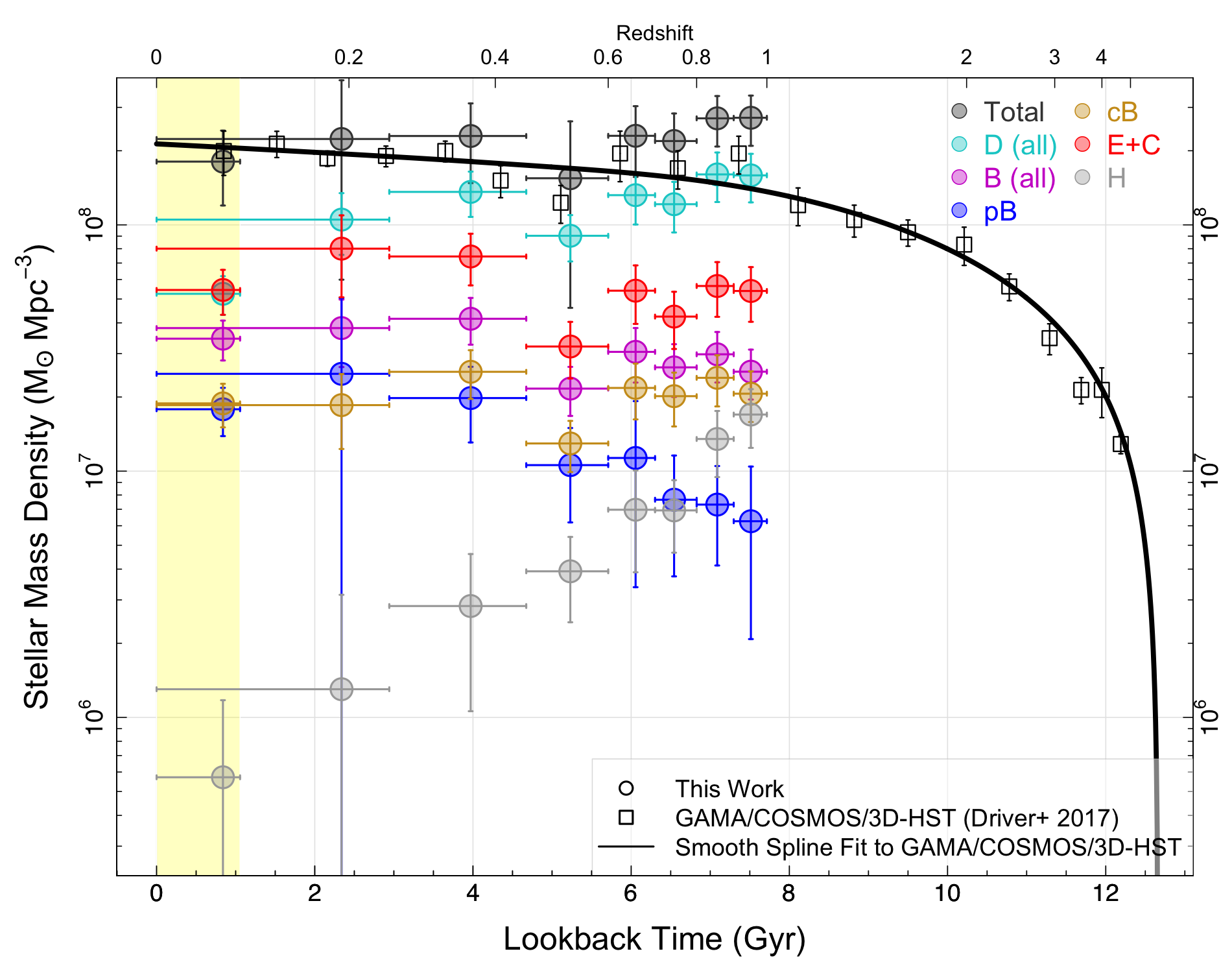}
  \caption{The evolution of the stellar mass density (SMD) of total and morphological types before applying the LSS corrections. Highlighted region shows the epoch covered by the GAMA data.}
  \label{fig:MassBuildUp_noLSS}
\end{figure*}

Figure \ref{fig:MassBuildUp_noLSS} shows the evolution of the integrated stellar mass density, $\rho_*$, before we apply our large scale structure corrections. This is to further confirm that the corrections do not derive the overall trends that we find in Figure \ref{fig:MassBuildUp} as explained in Section \ref{sec:rho}.

\chapter{{\sc GRAFit} Documentation}
\label{App:GRAFit_doc}



\section*{Supplementary material}

# GRAFIT DOCUMENTATION

# Contents

# GRAFit (R package)

**GRAFit** is an automatic modular pipeline for galaxy structural decomposition (also known as bulge-disk decomposition) originally designed for the Hubble Space Telescope (HST) imaging data, the Advanced Camera for Surveys (ACS) in particular. However, it can be used for any standard astronomical imaging data. **GRAFit** is able to fit galaxies with different combination of models including double Sersic profile and single Sersic profile.

## Installation

### Install R

Get **R** for OSX (Mac), Linux and Windows from: https://cloud.r-project.org

### Install ProFit

For fitting process **GRAFit** uses **ProFit**, a Bayesian galaxy profile fitting tool.

Get **ProFit** with:

```
install.packages('devtools')
library(devtools)
install_github("ICRAR/ProFit")
library(ProFit)
```

See ProFit if you faced any issues with installation and/or dependencies.

### Install ProFound

Before we model a galaxy we need an approximate estimation of some photometrical parameters such as magnitude, central positions, position angle, etc. **GRAFit** will parse

these values to **ProFit** as initial inputs of our model. **GRAFit** uses **ProFound** for photometry, source extraction and generating segmentation and mask maps.

Get **ProFound** with:

```
devtools::install_github("asgr/ProFound")
library(ProFound)
```

This should work properly, and the dependencies should be installed automatically.

See ProFound if you faced any issues with installation.

### Install Tiny Tim

To simulate the optical effects on our model we need to convolve the with a Point Spread Function (PSF). **GRAFit** uses **Tiny Tim** for generating HST PSF.

Get **Tiny Tim** from here

The user can also provide the PSF in their working directory or any other directories and provide the direction to them. This way **GRAFit** will prioretize the provided PSFs and will not generate new PSFs anymore.

### Install GRAFit

You are now ready to install **GRAFit** from GitHub with:

```
devtools::install_github("hoseinhashemi/GRAFit")
library(GRAFit)
```

### Dependencies

Most of the required dependencies should be installed by this stage. In case any of the dependencies are failed to install, you can install them with.

```
install.packages(c('fftw', 'R2Cuba', 'RColorBrewer', 'EBImage',
                    'LaplacesDemon', 'imager', 'magicaxis',
                    'FITSio', 'data.table'))
```

The segmentation map required for **GRAFit** is generated by **ProFound** using **EBImage** package. If this package was not installed by this stage install it with:

```
install.packages("BiocManager")
BiocManager::install("EBImage")
```

### Modules

**GRAFit** comes with a number of modules. For more details about each module and get the documentation for functions run the help operator in R environment with `?functionname` or `help(functionname)`. For example:

```
?GRAFitDynamo_v2
help(GRAFitDynamo_v2)
```

## Example

**GRAFit** is originally designed for a large number of galaxies distributed between one or more imaging frames, although there is no restriction for fitting one galaxy in a single cutout. In principle, you need to interact with only one **GRAFit** function; `GRAFitMaster`. This function calls all other modules and dependencies when/where it needs during the process.

You should provide:

- a working directory where the outputs will be saved.

- a directory where your imaging data is stored.

- If any, a directory where you have stored your pre-generated Point Spread Function (PSF) for each object. Otherwise, you may leave a PSF in your `wrk_dir` or alternatively GRAFit will generate a HST PSF using **Tiny Tim**. In the later you will beed to provide the PSF specifications required by **Tiny Tim**.

For a simple run, I provide a folder including a COSMOS HST/ACS frame together with the associated PSF and an input catalogue consisting of a random galaxy. Download the whole folder from here and put it where you wish in you local directory, suppose in your Desktop.

Then run the code below for a basic example of **GRAFit** fitting.

```
inCat = read.csv('~/Desktop/GRAFit_test_Run/Input.csv')

GRAFitMaster(wrk_dir = "~/Desktop/GRAFit_test_Run/",
             data_dir = "~/Desktop/GRAFit_test_Run/",
             threadMode = 0,
             ncores = 1,
             nComp= 2,
             optimMode = 'optim',
             object_list = inCat)
```

- Note that `object_list` should include at least the RA and DEC of your target galaxy.

This run uses a relatively simple `optim` optimizer to save you some time. More robust optimization that **GRAFit** provides is MCMC for which you need to specify optimMode = 'MCMC'. See `GRAFitMaster` documentation for more details and options.

This run will genrate a folder named with the galaxy ID (or RA/DEC if ID is not provided in inCat) containing all outputs associated with this galaxy such as figures of initial and optimized models, 1D profiles etc. In addition it will generate a master catalogue (Master-Cat.csv) in you `wrk_dir` directory including all fitting parameters as well as a ProCat.csv including the results of photometric measurements by **ProFound**. Another directory could be also generated containing work spaces for which you need to specify `keep_wrk_space = TRUE`.

**Contributor(s)**

Hosein Hashemizadeh

**License**

L-GPL3



\printindex{}
\end{document}